# History-Deterministic Parity Automata: Games, Complexity, and the 2-Token Theorem

by

# Aditya Prakash

**Thesis**

Submitted to the University of Warwick

for the degree of

**Doctor of Philosophy in Computer Science**

# Department of Computer Science

January 2025

# Contents

# List of Figures

# Acknowledgments

I am grateful to the Chancellor's International Scholarship at the University of Warwick for funding my PhD, and to the Centre for Discrete Mathematics and Its Applications (DIMAP) at the University of Warwick for funding several of my conference participations.

The progression of my PhD so far would not have been possible without several outstanding individuals who helped me in various way.

I am extremely grateful to have Marcin Jurdziński as my PhD supervisor. He has been kind and encouraging throughout, and I left all meetings with him feeling more confident and reassured about my research direction; this made a huge difference. His patient feedback on this thesis and my writings have been invaluable. His approach to problem-solving and writing, which focuses on clarity of concepts and boiling down proofs to its essentials is inspiring and something I wish to continue work towards.

I am grateful to Dmitry Chistikov and Patrick Totzke for agreeing to examine this thesis.

Thejaswini has been my collaborator ever since I started at Warwick. She has been infinitely patient and kind in listening to so many of my wrong and impulsive ideas, even on problems she is not working on, and has been the first ear for most of my right ones. Working with her so extensively has been a constant joy and I have learnt a lot from it, especially for my first conference submission when I was just finding my feet. In addition, she has been an excellent resource. Whenever I faced a non-scientific problem at work, Thejaswini had been through it much before me, so I never had to look for a solution alone. Thank you so much, Thejaswini.

Karoliina has been awesome to work with. My collaboration with her made the last phase of my PhD exhilarating. I thank her for sharing her time and expertise with me during several online meetings and various conferences, for inviting me to Marseille twice, and for all the wise advice that she has generously shared with me.

The researchers that I have collaborated with have all been patient and kind towards me, and working with them has been a pleasure. Thanks to Antonio, Corto, Denis, Karoliina, Marcin, Rohan, Thejaswini, Tom, and Udi.

Thanks to Tom for hosting me at IST Austria several times, and to Thejaswini and the researchers at ISTA for sharing their time and knowledge with me. Thanks to the participants of Autobóz 2024 for being lovely, and thanks to the organisers—Aliaume, Chana, and Marie Fortin—for their excellent organisation and for inviting me. I thank Udi for inviting me to the Reichman University, even though I was unable to visit him.

It has been a pleasure updating Denis, Michał, and Udi with various research discoveries throughout my PhD, and I thank them for listening to me, for sharing my excitement, and for their feedback.

Several individuals at Warwick helped me in various ways. Thanks to my doctoral advisors Ramanujan and Ranko for all their valuable advice that they shared with me during my annual reviews and over the various interactions I had with them. Thanks to Alex,

Dmitry, Finnbar, Henry, Neha, Sarah, Thejaswini, and Tuva for the many discussions we had on automata.

I give the warmest of thanks to the friends I made during my PhD, for all the support and fun. Ahad, Hugo, Marcel, Finnbar, Greg, Namrata, Ninad, Patrick, Thejaswini, and Zhihao made my start at Warwick less overwhelming. Iman and Stas were always there with their constant warmth. Ahad and Thejaswini hosted me in Vienna so many times. Matteo and Somak were lovely, and their cats Lucrezia and Ludovico hosted me for the most wonderful stay at their place. The support from Finnbar, Neha, Samarth, Sarah, Thejaswini, and Tuva was critical in keeping me sane while I was writing my thesis.

Thanks to Arjun, Arya, Ashwini, Dia, Paul, Khushi, and Urshita, who stayed connected with me despite distance and timezone constraints. Thanks to my family for their love and constant support throughout.

# Declarations

I declare that this thesis was composed by myself, that the work contained herein is my own except where explicitly stated otherwise in the text. This work has not been submitted for any other degree or processional qualification. Parts of the work in this thesis is from joint collaborations and is published or under submission in related conferences. The following articles form parts of this thesis.

[Pra24] Checking History-Determinism is NP-hard for Parity Automata, FoSSaCS 2024.
Chapter 8 contain results from this work.

[AJP24] Lookahead Games and Efficient Determinisation of History-Deterministic Büchi automata, with Rohan Acharya and Marcin Jurdziński, ICALP 2024.
Chapter 5 contain results from this work.

[LP24] The 2-Token Theorem: Recognising History-Deterministic Parity Automata Efficiently, with Karoliina Lehtinen, submitted to STOC 2025.
Chapters 4, 6 and 7 contain results from this work.

The following works were carried out during the development of this thesis but does not form a part of the thesis. Some techniques that appear in this thesis had originally appeared first in these works. They are mentioned below.

[PT23] On History-Deterministic One-Counter Nets, with K. S. Thejaswini, FoSSaCS 2023.
Definition 5.21 is inspired from this work.

[BHLP24] History-Determinism vs Fair Simulation, with Udi Boker, Thomas A. Henzinger, and Karoliina Lehtinen, CONCUR 2024.
Lemma 5.19 first appeared in this work.

[CIK$^+$24] On the Minimisation of Deterministic and History-Deterministic Generalised Co(Büchi) Automata, with Antonio Casares, Olivier Idir, Denis Kuperberg, and Corto Mascle, to appear in CSL 2025.

The proof of Theorem 7.21 is based on the proof of Lemma 36 in [CIK$^+$24].

# Abstract

History-deterministic automata are a restricted class of nondeterministic automata where the nondeterminism while reading an input can be resolved successfully based on the prefix read so far. History-deterministic automata are exponentially more succinct than deterministic automata, while still retaining some of the algorithmic properties of deterministic automata, especially in the context of reactive synthesis.

This thesis focuses on the problem of checking history-determinism for parity automata. Our main result is the 2-token theorem, due to which we obtain that checking history-determinism for parity automata with a fixed parity index can be checked in PTIME. This improves the naive EXPTIME upper bound of Henzinger and Piterman that has stood since 2006. More precisely, we show that the so-called 2-token game, which can be solved in PTIME for parity automata with a fixed parity index, characterises history-determinism for parity automata. This game was introduced by Bagnol and Kuperberg in 2018, who showed that to decide if a Büchi automaton is history-deterministic, it suffices to find the winner of the 2-token game on it. They conjectured that this 2-token game based characterisation of history-determinism extends to parity automata. Boker, Kuperberg, Lehtinen, and Skrzypczak showed in 2020 that this conjecture holds for coBüchi automata as well. We prove Bagnol and Kuperberg's conjecture that the winner of the 2-token game characterises history-determinism on parity automata.

We also give a polynomial time determinisation procedure for history-deterministic Büchi automata, thus solving an open problem of Kuperberg and Skrzypczak from 2015. This result is a consequence of our proof of the 2-token theorem.

Finally, we show NP-hardness for the problem of checking history-determinism for parity automata when the parity index is not fixed. This is an improvement from the lower bound of solving parity games shown by Kuperberg and Skrzypczak in 2015.

# Abbreviations

| | |
|---:|---|
| **MSO** | Monadic Second Order |
| **LTL** | Linear Lemporal Logic |
| **HD** | History-Deterministic |
| **COCOA** | Chain Of CoBüchi Automata |
| **SD** | Semantically-Deterministic Automata |
| **CR** | Coreachable |
| **WCR** | Weakly Coreachable |
| **SCC** | Strongly Connected Component |
| **DAG** | Directed Acyclic Graph |

# Part I

# Stretch

# Chapter 1

# Introduction

## 1.1 Summary of the contributions

We start by briefly describing the main contributions of this thesis, for the convenience of the readers who are familiar with the theory of automata over infinite words. We will then motivate, contextualise, and detail automata over infinite words in Section 1.2, history-determinism in Section 1.3, and our contributions in Section 1.4.

This thesis focuses on the problem of checking history-determinism for parity automata. Parity automata are automata that take as input infinite words over a finite alphabet. They have a finite state space, and each transition is labelled by an alphabet and a natural number that we call priority of that transition. An example of a parity automaton is shown in Fig. 1.1.

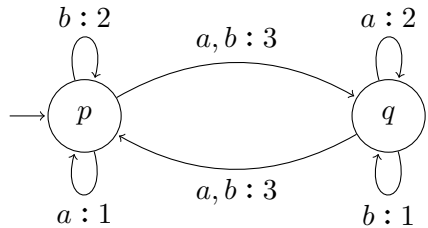

Figure 1.1: A parity automaton

An infinite run in a parity automaton is accepting if the *lowest priority occurring infinitely often* amongst the transitions of that run is even. An infinite word is accepted by a parity automaton if there is an accepting run of the automaton over the word, and the language of a parity automaton is the set of words that it accepts.

History-deterministic (HD) parity automata are nondeterministic parity automata, in which a run can be constructed 'on-the-fly' while reading a word, where the nondeterminism between available transitions are resolved based on the prefix read so far, without the knowledge of the rest of the word. This resolution of nondeterminism must be done such that on every word accepted by the automaton, the run produced is an accepting run.

Equivalently, history-determinism of an automaton can be characterised by Eve winning the history-determinism game (HD game, for short), which is a two-player turn-based game between Eve and Adam. The HD game of an automaton starts with an Eve's token at the initial state and proceeds in infinitely many rounds. In each round, Adam selects a letter and then Eve moves her token along an outgoing transition on that letter from the token's current state. Thus, in the limit of a play of the HD game, Adam builds a word and Eve builds a run on her token on that word. We say that Eve wins that play if either her token's run is accepting or Adam's word is rejected. An automaton then is history-deterministic if and only if Eve wins the HD game on that automaton.

The above definition of history-deterministic automata was introduced by Henzinger and Piterman in 2006, but the concept of history-determinism has independent origin in the equivalent definition of what we now call 'good-for-trees' automata [BKKS13], studied by Kupferman, Safra, and Vardi in 1996 [KSV06]. The concept of history-determinism was also studied independently by Colcombet in the setting of cost automata [Col09].

History-deterministic parity automata are exponentially more succinct than deterministic parity automata [KS15] and, at the same time, they enjoy good algorithmic properties similar to deterministic parity automata: the problem of language containment reduces to deciding simulation, while Church's reactive synthesis problem for when the specification is given by an HD automaton can be reduced (in LOGSPACE) to solving a parity game [HP06]. History-deterministic parity automata are also used in constructing canonical representations of $\omega$-regular languages [AK22, ES22]. Furthermore, the problem of deciding whether an automaton is history-deterministic, which we tackle in this thesis, is LOGSPACE interreducible with the *good-enough realisability* problem ([AK20] and [BL23b, Page 23]), a variant of the reactive synthesis problem.

These theoretical results on history-determinism in parity automata have also inspired recent efforts to use HD automata in practice [IK19, EK24b, EK24a]. The notion of history-determinism has been extended to models other than parity automata as well, such as infinite state systems and quantitative automata [BL23b].

Despite these growing efforts on history-determinism over the last decade, the computational complexity of the problem of deciding history-determinism in parity automata has been open until now. A naive algorithm that directly solves the HD games takes EXPTIME since it involves determinising the automaton [HP06]. In contrast, the best lower bound for the problem of deciding history-determinism so far has been of the complexity of solving parity games [KS15], a problem that is in NP ∩ coNP and that can be solved in quasipolynomial time [CJK[+]22].

Polynomial time algorithms have been known for deciding history-determinism for certain subclasses of parity automata, however. Kuperberg and Skrzypczak in 2015 used the so-called Joker game to give a PTIME algorithm to decide history-determinism of coBüchi automata—these are parity automata whose priorities are all 1 or 2.

This was followed in 2018 by Bagnol and Kuperberg giving a PTIME algorithm to decide history-determinism for Büchi automata [BK18]—these are parity automata whose priorities are all 0 or 1. Their algorithm relied on solving the so-called 2-token game. This game is similar to the HD game, where we have Adam constructing a word letter-by-letter and Eve constructing a run on her token transition-by-transition but, in addition, Adam has two distinguishable tokens on which he is also constructing runs transition-by-transition. In each round of the 2-token game, Adam selects a letter, then Eve selects a transition on her token, and then Adam selects a transition on each of his 2-tokens. In the limit of a play of the 2-token game, Adam builds a word, Eve builds a run on her token on that word and Adam builds a run on each of his two tokens, also on that word. The winning condition for Eve is the following: if either of the runs of Adam's tokens is accepting, then the run of Eve's token is accepting.

Bagnol and Kuperberg showed that for every Büchi automaton, Eve wins the 2-token game on it if and only if it is HD. Thus, to decide if a Büchi automaton is HD, it suffices to solve the 2-token game on it, and doing so takes polynomial time. Bagnol and Kuperberg conjectured that this 2-token game based characterisation of history-determinism is true for all of parity automata: we call this the 2-*token conjecture*. For parity automata with a fixed number of priorities, the 2-token game can be solved in PTIME, and in PSPACE if the number of priorities is not fixed.

The 2-token conjecture was also proved for coBüchi automata by Boker, Kuperberg, Lehtinen, and Skrzypczak in 2020 [BKLS20]. Their argument, while similar to that of Bagnol and Kuperberg's for Büchi automata, was much more involved. Crucially, neither their argument nor Bagnol and Kuperberg's argument generalised to show the 2-token conjecture for parity automata.

The main contribution of this thesis is the proof that the 2-token conjecture is true for parity automata. Thus, as a consequence, we obtain that history-determinism can be checked in PTIME for parity automata with a fixed number of priorities, and in PSPACE otherwise. As further corollaries, we also obtain similar upper bounds for the problems of deciding history-determinism of alternating parity automata [BKLS20], as well as the good-enough-realisability problem for when the specification is given by a deterministic parity automaton [BL23b, Page 23].

**The 2-Token Theorem.** *For every nondeterministic parity automaton $\mathcal{A}$, Eve wins the 2-token game on $\mathcal{A}$ if and only if $\mathcal{A}$ is history-deterministic. Thus, the problem of deciding history-determinism is in* PTIME *for parity automata with a fixed number of priorities, and in* PSPACE *if the number of priorities is part of the input.*

On the way to showing the 2-token theorem, we also give a proof of the 2-token theorem for coBüchi automata that is much simpler than that of [BKLS20].

Additionally, our proof of the 2-token theorem shows that history-deterministic Büchi automata can be converted to a language-equivalent deterministic Büchi automata in polynomial time: this solves an open problem of Kuperberg and Skrzypczak from 2015 [KS15].

Finally, we show that the problem of deciding history-determinism for parity automata when the parity index is not fixed is NP-hard, thus improving upon Kuperberg and Skrzypczak's lower bound of solving parity games from 2015 [KS15].

## 1.2 Automata and games

### 1.2.1 Automata over infinite words

Automata over infinite words are a natural extension of finite state automata whose goal is to reason about systems that run indefinitely. For instance, these can describe properties such as the safety condition, where termination of the system is undesirable, or liveness, where we require all processes to eventually make progress.

Automata over infinite words have two independent origins. Muller, in 1963, introduced what we now call *Muller automata*. A Muller automaton has the same structure as a finite state automaton, with finitely many states, and transitions that are each a directed edge between two states, labelled by a letter from a finite alphabet. In addition, each transition is also labelled by some colour from a finite subset of colours $C$. The acceptance condition for infinite runs in this automaton is described by a set $\mathcal{F}$ consisting of nonempty subsets of $C$, which we call *accepting subsets*. An infinite run in this automaton is accepting if the set of colours that occur infinitely often in that run is a set in $\mathcal{F}$. An infinite word is accepted by this automaton if there is an accepting run of the automaton over this word. The language recognised by this automaton is the set of accepted words.

Independently, Büchi in 1962 introduced automata over infinite words in the context of logic [Bü62]. He extended the equivalence between finite automata and MSO over finite strings to showing the equivalence between MSO logic over natural numbers and a model of finite automata over infinite words that we now call Büchi automata. Büchi's result of the equivalence of MSO formulas over naturals and Büchi automata was the crucial ingredient in his proof of showing decidability for MSO over the naturals.

Büchi automata are a restricted variant of Muller automata, where each transition is labelled either accepting or rejecting, i.e., the colours in a Büchi automaton are given by $C = \{\text{accepting}, \text{rejecting}\}$. An infinite run in a Büchi automaton is accepting if there is some accepting transition that is visited infinitely often. Or equivalently, the set $\mathcal{F} \subseteq \mathcal{P}(C)$ of accepting subsets are the nonempty subsets containing accepting, i.e., $\mathcal{F} = \{\{\text{accepting}, \text{rejecting}\}, \{\text{accepting}\}\}$. Thus, a language that is recognised by a Büchi automaton is also recognised by a Muller automaton. The converse also holds when the automata are nondeterministic, and thus, nondeterminstic Büchi automata and nondeterministic Muller automata are equally expressive. We will elaborate on this point in Section 1.2.3.

The conversion of logical formulas in MSO over naturals to automata also extends to the practical logical formulations, such as linear temporal logic (LTL) and its variants. LTL was introduced by Pnueli in 1977 who suggested that Prior's tense operators [Pri55] can be used for checking the correctness of non-terminating programs. The satisfiability problem and the model-checking problem (does a given program satisfy a given LTL formula?) can be solved by converting the LTL formula to a Büchi automaton [PR89, VW94], albeit with an unavoidable double-exponential blowup. Satisfiability then involves checking nonemptiness of the resulting automaton, while the model-checking problem corresponds to a language inclusion problem. For a modern treatment of these results, we refer the reader to [Kup18].

### 1.2.2 Games and the synthesis problem

The problem of Church synthesis [Chu57], also known as reactive synthesis, has the goal of constructing a program that is correct for a given specification or determining that no such program exists.

More concretely, it represents the interaction of a program with its environment as an infinite game, where in each round $i$, for every natural number $i$, the environment produces an input letter $a_i$ from an input alphabet $I$, to which the program must respond with a letter $b_i$ from an output alphabet $O$. Then, the goal of the program is to guarantee that in the limit, the word $a_0 b_0 a_1 b_1 \ldots$ that describes the result of this interaction is in the specification—a language of infinite words over the alphabet $(I \times O)$. For example, in the synthesis of a scheduler whose task is to allocate resources to clients, the inputs might be requests for resources, the outputs might be resource allocations, and the specification might define what satisfactory schedule are. These could be conditions such as every request is being eventually answered and the priorities of these requests are being respected and so on. A winning strategy in such a game corresponds to a program that guarantees the specification, whatever the behaviour of its environment. This program or the winning strategy is represented by a *transducer*: an automaton that takes an infinite word over $I$ and outputs an infinite word over $O$ on-the-fly. The *realisability problem* asks, given a specification $S$, whether such a transducer representing a winning strategy exists. The synthesis problem asks to produce a winning strategy if it exists, or determine that no such strategy exists.

**Games on graphs.** The problems of reactive synthesis and realisability are intimately connected to the rich theory of 2-player games on graphs. These games are played over directed graphs, where each vertex is owned by either Adam or Eve. An infinite play of the game starts with a token at a designated initial vertex, and in each round, the owner of the vertex the token is at chooses an outgoing edge from that vertex and the token is moved along that edge to the target of that edge. Winning conditions for either players in such a game can be given by, for instance, a Muller condition. That is, the edges of the graph are each labelled by a colour from a finite set of colours $C$, and we have a fixed set of accepting subsets $\mathcal{F}$ specifying a Muller condition. We say Eve wins a play in this game if the set of colours occurring infinitely often amongst the edges in that play is a set in $\mathcal{F}$, and Adam wins otherwise. We say Eve (resp. Adam) wins the game if Eve (resp. Adam) if she (resp. he) has a strategy that ensures she wins all plays in the game. Muller games were shown to be finite-memory determined by Büchi and Landweber [BL69], i.e., either Eve or Adam has a finite-memory winning strategy, and this strategy can be represented by a finite state transducer, whose input is the edges of the play of the game chosen by the opponent, and the output is the edges of the winning player.

**Solving the synthesis problem.** Let us return to the problems of reactive synthesis and realisability. These problems were solved by Büchi and Landweber in 1969 [BL69] for when the specification is given by an MSO formula, or equivalently, a nondeterministic Büchi automaton. The realisability problem for this specification thus corresponds to a game where the winning condition is given by a nondeterministic Büchi automaton—such games are called $\omega$-regular games. The synthesis problem requires us to construct a winning strategy for Eve if one exists.

The crucial ingredient behind Büchi and Landweber's solution was the result of McNaughton that nondeterministic Büchi automata can be determinised into a deterministic Muller automata [McN66]. Thus, to find the winner and their winning strategy in an $\omega$-regular game, we can convert that game to a Muller game via McNaughton's determinisation procedure, and then find the winner and their winning strategy in the resulting Muller game.

These works of McNaughton [McN66] and of Büchi and Landweber [BL69] on synthesis, logic, and its connections to automata [Bü62] were the igniter for the rich and ever expanding field of using automata and games for reactive synthesis as well as for reasoning about logic. Rabin, in 1969, extended the connection between MSO logic over the naturals and automata over infinite words to show the equivalence of MSO logic over trees and automata over infinite trees [Rab69]. These conversions from logic to automata were extended to practical logical formulations as well, notably the linear temporal logic introduced by Pnueli [Pnu77, PR89], its variants of computation tree logics CTL [CE81, QS82] and CTL* [EH86], and the modal $\mu$-calculus introduced by Kozen [Koz83]. Over the years, algorithms for model-checking and synthesis in these logics have established themselves as successful tools in both theory and industry [CHVB18].

This development has been closely accompanied by the development of the theory of automata and games. Simpler proofs of Büchi and Landweber's result of finite-memory determinacy [GH82, EJ91, DJW97], and faster algorithms to solve Muller games have emerged. These developments have been largely driven by the study of various acceptance conditions related to the Muller acceptance conditions, with the parity acceptance condition we describe next taking a central position.

### 1.2.3 Automata and games with parity acceptance conditions

We mentioned briefly earlier that nondeterministic Muller automata are as expressive as nondeterministic Büchi automata, and McNaughton's result that every nondeterministic Büchi automaton can be converted to a language-equivalent deterministic Muller automaton. Thus, in terms of expressivity, the classes of nondeterministic Büchi automata, nondeterministic Muller automata, and deterministic Muller automata are all equivalent.

Deterministic Büchi automata express a strictly smaller subclass of languages than nondeterministic Büchi automata, however. Consider the classical example of the nondeterministic Büchi automaton shown in the Fig. 1.2, where the `accepting` transitions are double-arrowed.

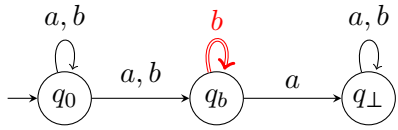

Figure 1.2: A nondeterministic Büchi automaton for which there is no language equivalent deterministic Büchi automaton. Doubled arrows indicate `accepting` transitions.

The language $L$ that this automaton recognises is the set of infinite words over $\{a,b\}^*$ that contain finitely many $a$'s, and there is no deterministic Büchi automaton that recognises $L$. The language $L$ can be recognised by a deterministic coBüchi automaton, however.

In a coBüchi automaton, each transition is labelled either `accepting` or `rejecting` similarly to a Büchi automaton, but a run is accepting if and only if it contains finitely many `rejecting` transitions. The language recognised by the automaton in Fig. 1.2 can be recognised by the deterministic coBüchi automaton shown in the figure below, where the `rejecting` transitions are represented by dashed arrows.

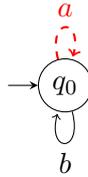

Figure 1.3: A deterministic coBüchi automaton that is language-equivalent to the nondeterministic Büchi automaton in Fig. 1.2.

In terms of expressivity, nondeterministic coBüchi and deterministic coBüchi automata are equivalent, due to Miyano and Hayashi's breakpoint construction [MH84]. Nondeterministic (or equivalently, deterministic) coBüchi automata express a strictly smaller class of languages than nondeterministic Büchi automata.

Parity acceptance condition is a generalisation of the Büchi and coBüchi acceptance conditions. In a parity automaton, each transition is labelled by a natural number, which we call the *priority* of the transition. A run in a parity automaton is accepting if *the least priority occurring infinitely often in the run is even.*

Büchi and coBüchi automata are subclasses of parity automata. Every Büchi automaton is a parity automaton where `accepting` transitions have priority 0 and `rejecting` transitions have priority 1, and similarly, every coBüchi automaton is a parity automaton where `accepting` transitions have priority 2 and `rejecting` transitions have priority 1.

Every nondeterministic parity automaton is a nondeterministic Muller automaton, where the set of colours $C$ are the priorities occurring in that automaton, and the set of accepting subsets $\mathcal{F}$ is the set of nonempty subsets of $C$ in which the lowest priority is even. Furthermore, every deterministic Muller automaton can be converted to an equivalent deterministic parity automaton, as was shown by Gurevich and Harrington using the data structure called the latest appearance record [GH82]. Another way to transform Muller automata to parity automata is to use Zielonka trees [DJW97]. McNaughton's original proof for the determinisation of a nondeterministic Büchi automaton into deterministic Muller automaton can, with some additional care, be modified to yield a deterministic parity automaton as well [McN66].

Thus, nondeterministic Büchi automata, deterministic and nondeterministic parity automata, and deterministic and nondeterministic Muller automata are all equivalent in terms of expressivity. The expressivity of automata with the various acceptance conditions we have discussed so far is shown in the Fig. 1.4.

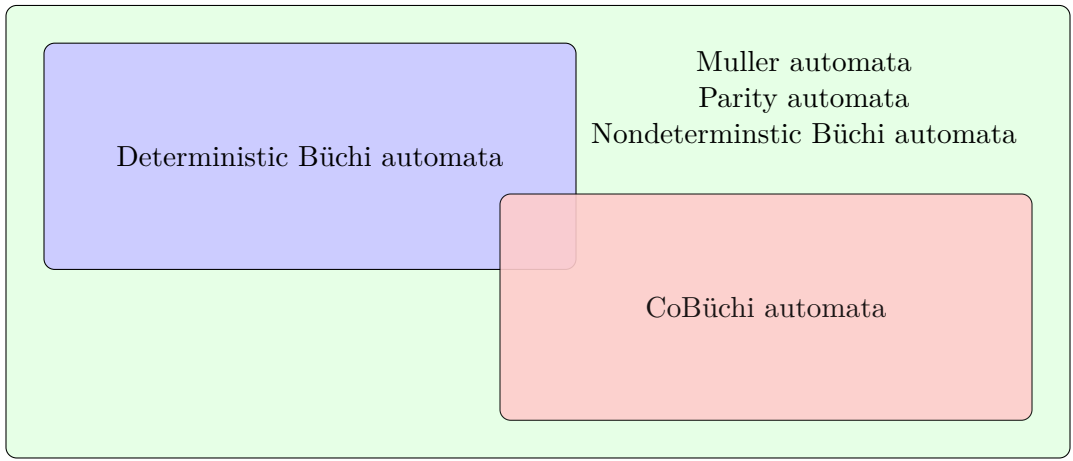

Figure 1.4: The venn diagram for the expressivity of the mentioned automata classes. Where unspecified, deterministic X automata are as expressive as nondeterministic X automata.

There are several other acceptance conditions in the literature such as Rabin, Streett, and Emerson-Lei, with varying trade-offs for complexities and succinctness. Let us also mention $\omega$-regular automata, where the acceptance condition is given by the language of a nondeterministic Büchi automata. Deterministic automata with the acceptance conditions just mentioned are as expressive as nondeterministic Büchi automata.

Amongst the acceptance conditions with this property, i.e., where deterministic automata with those acceptance conditions capture the class of $\omega$-regular languages, the parity acceptance condition is the simplest: every parity automaton can be efficiently transformed to an equivalent Muller, Rabin, Streett, Emerson-Lei, or an $\omega$-regular automaton without any additional blow up in its state space.

**Parity games.** This simplicity of the parity acceptance condition in automata extends to games. Parity games are positionally determined, i.e., exactly one of Eve or Adam has a winning strategy, and the winner has a winning strategy that chooses the outgoing edge only based on the current vertex the token is at [EJ91]. The problem of solving parity games is in NP ∩ coNP (and even in UP ∩ coUP [Jur98]), and in PTIME for when the number of priorities is fixed. Following the breakthrough result of Calude, Jain, Khoussainov, Li, and Stephan from 2017, parity games can be solved in quasipolynomial time, i.e., time $n^{\mathcal{O}(\log d)}$ for a parity game with $n$ vertices and in which all priorities are at most $d$ [CJK$^+$22].

We will focus mostly on parity automata in this thesis due to their simplicity, but our main result extends to all of $\omega$-regular automata.

We call a parity automaton whose priorities are in the interval $[i, j] = \{i, i+1, i+2, \ldots, j\}$ for two natual numbers $i < j$ as an $[i, j]$ automaton. Note that decreasing each of the priorities of transitions by 2 does not change the acceptance of each run, so we will assume $i = 0$ or $1$. Then, $[i, j]$ is called the *parity index* of the automaton. A Büchi automaton is a $[0, 1]$ automaton, while a coBüchi automaton is a $[1, 2]$ automaton.

The expressivity of parity automata based on their parity indices forms an alternating hierarchy, as shown in the figure below. This hierarchy collapses to the level $[0, 1]$ or Büchi for nondeterministic automata, but it is strict for deterministic automata [Wag79].

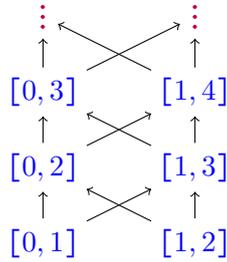

Figure 1.5: The parity index hierarchy. Büchi and coBüchi automata correspond to $[0, 1]$ and $[1, 2]$-automata respectively, and they are at the bottom of the index hierarchy.

We end this section by briefly mentioning some references for where the readers can find recent presentations of the results mentioned in this section. For the results of Mc-Naughton on determinisation of Büchi automata, Büchi and Landweber's result on the finite-determinacy of Muller games, and the positional determinacy of parity games, we refer the reader to Chapters 1 and 2 of Bojanczyk's book [Boj18]. For connections between automata and logic, we refer the reader to the article of Thomas [Tho97], and another article of his for connections between automata and games to the synthesis problem [Tho08]. We also point the reader to Boker's survey for details and references on automata translations [Bok18], and in particular, his site where he has organised the references and state blow-up for automata translations between different acceptance conditions [Bok12].

## 1.3 History-deterministic automata

### 1.3.1 History-determinism and related notions

Nondeterministic automata are exponentially more succinct than deterministic automata, yet this comes at an algorithmic cost. In order to solve the reactive synthesis problem for when the specification is given by a nondeterministic automaton, most solutions involve determinising the nondeterministic automaton, which takes exponential time. This was the key motivation of Henzinger and Piterman behind introducing history-deterministic automata [HP06].

History-deterministic automata are an intermediate model between deterministic and nondeterministic automata, where the nondeterminism that occurs while reading a word can be successfully resolved on-the-fly, just based on the prefix read so far. More concretely, the history-determinism of an automaton is characterised by Eve winning the following two-player game, which we call the history-determinism game or HD game. The HD game on an automaton starts with an Eve's token at the initial state, and then each round proceeds as follows: Adam plays a letter from the input alphabet, and Eve responds by moving her token along an outgoing transition from the token's current position over that letter. The game proceeds in infinitely many rounds, and in the limit, Adam builds an infinite word and Eve builds a run over that word. Eve wins a play if either her run is accepting, or Adam's word is not in the language of the automaton. That is, for Eve to win a play of the HD game, Eve needs to construct an accepting run whenever Adam's word is accepting. We say that an automaton is history-deterministic (HD) if Eve has a winning strategy in the HD game on that automaton.

As an example, consider the HD coBüchi automaton shown in Fig. 1.6 below, which is over the alphabet $\Sigma = \{a, b, c\}$.

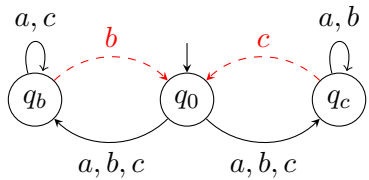

Figure 1.6: Example of a history-deterministic coBüchi automaton [CCFL24].

This automaton accepts a run if the rejecting transitions that are represented by dashed arrows are seen finitely often. This is the case for a run $\rho$ satisfies either of the following two conditions.

1. The run $\rho$ eventually stays in state $q_b$, and hence eventually $b$ does not appear in the word being read.

2. The run $\rho$ eventually stays in the state $q_c$, and hence eventually $c$ does not appear in the word being read.

It follows that the automaton accepts the set of words in which the letter $b$ or the letter $c$ occur finitely often. Note that the automaton is not deterministic because there is nondeterminism on all the letters in the initial state. It is history-deterministic, however, because Eve has the following strategy to resolve the non-determinism: whenever her token is at the initial state, Eve alternates between moving her token transitions to the state $q_b$ and $q_c$. This strategy doesn't need to know what the word looks like, but is guaranteed to produce an accepting run on every word accepted by the automaton.

**Good-for-gameness.** Henzinger and Piterman introduced history-deterministic automata in 2006, and they called them as 'good-for-games' automata for the following reason. Consider a game $\mathcal{G}$ where the winning objective for Eve is given by a parity automaton $\mathcal{A}$, i.e., the edges of $\mathcal{G}$ are each coloured by a colour from a finite subset $C$, and $\mathcal{A}$ is a parity automaton over the finite alphabet $C$. Eve wins a play of this game if the word in $C^\omega$ consisting of colours occurring along the edges in that play is accepted by $\mathcal{A}$.

If $\mathcal{A}$ is deterministic, then in order to solve $\mathcal{G}$, it suffices to solve the game $\mathcal{G} \times \mathcal{A}$ obtained by taking the product of $\mathcal{G}$ and $\mathcal{A}$. The priorities of $\mathcal{G} \times \mathcal{A}$ are inherited from the priorities of $\mathcal{A}$, while the ownership of vertices is inherited from $\mathcal{G}$. Thus, $\mathcal{G} \times \mathcal{A}$ is a parity game. If $\mathcal{A}$ is a nondeterministic parity automaton, however, then we need to determinise $\mathcal{A}$ in order to solve $\mathcal{G}$.

But if $\mathcal{A}$ is history-deterministic, this determinisation procedure can be avoided: we can take the 'product' of $\mathcal{G}$ and $\mathcal{A}$ to obtain an equivalent game $\mathcal{G} \times \mathcal{A}$. Here, we require Eve to resolve the nondeterminism arising from $\mathcal{A}$ in $\mathcal{G} \times \mathcal{A}$, which is possible due to $\mathcal{A}$ being HD. The priorities of transitions and ownership of the vertices in $\mathcal{G} \times \mathcal{A}$ are then defined similarly to when $\mathcal{A}$ is deterministic. Note that we don't need the knowledge of Eve's winning strategy in the HD game on $\mathcal{A}$, just knowing that $\mathcal{A}$ is history-deterministic allows us to solve $\mathcal{G}$ efficiently, by instead solving the parity game $\mathcal{G} \times \mathcal{A}$.

We call the above property of history-deterministic automata we have just described good-for-gameness. This property of HD automata makes them especially useful for the reactive synthesis problem: indeed, if the specification is given by a history-deterministic parity automata, then the reactive synthesis problem can be solved as efficiently as parity games: PTIME for parity automata with a fixed number of priorities, and in quasipolynomial time when the number of priorities is a part of the input [CJK$^+$22].

**Good-for-trees**  While Henzinger and Piterman introduced the definition of history-deterministic automata that we have presented, history-determinism has several independent origins. HD automata are the same as what we now call good-for-trees automata [BKKS13], which where originally studied in the works of Kupferman, Safra, and Vardi from 1996 [KSV06].

**Guidability.**  Additionally, Colcombet and Löding introduced the notion of guidable parity automata over infinite trees [CL08], that is, unlike deterministic parity tree automata, as expressive as nondeterministic tree automata. This notion, when restricted to automata on words, coincides with history-deterministic automata.

**History-determinism vs good-for-games.**  The term history-determinism that we use here was coined by Colcombet in the setting of cost automata [Col09]. Since Henzinger and Piterman have originally called these automata good-for-games automata, research on HD automata that followed in the 2010s and early 2020s referred to HD automata as good-for-games automata. Boker and Lehtinen in 2021 suggested that good-for-games automata should be defined as automata that have the property of good-for-gameness [BL21]. Good-for-gameness and history-determinism are equivalent notions over parity automata, but this is not true in all settings, like in pushdown automata [LZ22]. Thus, recent papers over the last three years make this distinction, and furthermore, use the term HD parity automata instead of good-for-games parity automata: we do the same.

### 1.3.2 History-deterministic parity automata

Despite the introduction of history-deterministic parity automata by Henzinger and Piterman in 2006, it was not known for a while whether they are more succinct than deterministic parity automata. In fact, it was conjectured that every history-deterministic parity automaton is determinisable-by-pruning, that is, it contains a language-equivalent deterministic subautomaton [Col12, Conjecture 8]. This was quickly disproven by Boker, Kuperberg, Kupferman, and Skrzypczak, however, as they gave a counterexample [BKKS13]. The example automaton in Fig. 1.6 is one such automaton, i.e., it is history-deterministic but not determinisable-by-pruning.

This was followed by Kuperberg and Skrzypczak showing in 2015 that HD coBüchi automata are exponentially more succinct than deterministic coBüchi automata [KS15]. In addition, they showed that checking whether a nondeterministic coBüchi automaton is HD can be decided in polynomial time, and every HD Büchi automaton can be converted into a language-equivalent deterministic Büchi automaton with a quadratic state space blow-up.

Expressivity wise, history-deterministic parity automata follow a similar trend to deterministic parity automata. History-deterministic $[i,j]$ automata are exactly as expressive as deterministic $[i,j]$ automata [BKS17]. To see this, note that every deterministic automaton is also a history-deterministic automaton. For the other direction, if $\mathcal{H}$ is a history-deterministic automaton, then Eve has a finite-memory strategy in the HD game on $\mathcal{H}$ that constructs an accepting run on every word accepted by $\mathcal{H}$. The deterministic automaton $\mathcal{D}$ obtained by 'taking a product' of $\mathcal{H}$ with Eve's strategy is then language equivalent to $\mathcal{H}$.

The results of Kuperberg and Skrzypczak from 2015 were the starting point of numerous lines of research, with the complexity of recognising history-deterministic automata being a central one. We will elaborate on the efforts in this direction in Section 1.3.3.

**Canonicity and Minimisation** In 2019, Abu Radi and Kupferman showed that HD coBüchi automata can be minimised in polynomial time, when the acceptance is based on transitions. They extended their result in 2020 to further suggest a canonical HD coBüchi automata for languages recognised by nondeterministic (or equivalently, HD) coBüchi automata [AK22].

This was a striking result. While canonical deterministic automata exist and can be efficiently found for regular languages over finite words [Hop71], such a canonical form taking the shape of an automaton has not been known for fragments of $\omega$-regular languages more complicated than safety or reachability languages. Furthermore, it is unlikely that a canonical form for coBüchi languages can be given by deterministic coBüchi automata [AK22, Theorem 4.15].

Abu Radi and Kupferman's result also allowed for Ehlers and Schewe to give a canonical representation for $\omega$-regular languages [ES22]. This representation is not by an automaton, but rather a structure that is now called chain of coBüchi automata or COCOA [EK24b].

Recall that we consider automata where acceptance is based on transitions. A reader familiar with automata over infinite words would know that state-based acceptance is more common. The questions of minimising history-deterministic [Sch20] or deterministic automata [Sch10], with coBüchi, Büchi, or parity acceptance conditions is NP-hard, however, if the automata we consider have state-based acceptance. Thus, Abu Radi and Kupferman's results also highlight that transition-based acceptance conditions are better suited for reasoning about $\omega$-regular automata. We refer the reader to [Cas23, Chapter VI] for more arguments in favour of automata with transition-based acceptance.

It is open whether every $\omega$-regular automaton has a canonical representation given by a history-deterministic automaton. The questions of minimisation for automata with transition-based acceptance conditions, beyond coBüchi, on deterministic or history-deterministic automata are also open.

**Reactive synthesis.** The key bottleneck to the reactive synthesis problem when the specification is given by an LTL formula is the construction of a deterministic parity automaton, which incurs a doubly exponential blow-up. While HD parity automata can be exponentially more succinct than deterministic parity automata, there are currently no constructions that exploits this succinctness, that is, all known constructions of HD parity automata are determinisable-by-pruning.

Nevertheless, we have seen a couple of positive results that successfully use HD automata for the synthesis problem. Firstly, Iosti and Kuperberg have described a construction of HD coBüchi automata from EvLTL, a fragment of LTL describing what they call 'eventually safe' properties. Their construction has the double-exponential blowup in the worst case complexity too, but constructs HD coBüchi automata that are strictly more succinct than language-equivalent deterministic coBüchi automata in several cases [IK19].

Secondly, Ehlers and Khalimov have used chain of coBüchi automata, or COCOA, introduced by [ES22] towards the synthesis problem for LTL as well. They have described a construction that converts an LTL specification into COCOA without going through a deterministic parity automaton [EK24a], and they have also described how to leverage COCOA for the synthesis problem in [EK24b].

**History-determinism beyond nondeterministic parity automata.** Motivated by the existing rich theory of history-determinism on parity automata, researchers have studied history-determinism on quantitative automata [BL23a] and various infinite state systems such as pushdown automata [LZ22, GJLZ24], timed automata [BHL$^+$24], and counter systems [PT23, BPT23]. Additionally, a notion of history-determinism has also been defined for alternating automata [BL19].

Typical questions studied there include expressivity, succinctness, comparison with related notions such as good-for-gameness [BL21] and guidability [BHLP24], and the decision problem of checking if a given automaton is HD. The solutions for the last problem borrow from techniques introduced to recognise HD parity automata, which we will present next. For readers who would like to know more about history-determinism on models beyond parity automata, we refer to the survey of Boker and Lehtinen [BL23b].

### 1.3.3 Recognising history-deterministic parity automata

Despite the attention the topic of history-determinism has received over the last decade, the complexity of recognising HD parity automata, i.e., deciding history-determinism for parity automata, has, up till now, remained stubbornly open.

> **HD Problem.** Given a parity automaton $\mathcal{A}$, decide if $\mathcal{A}$ is history-deterministic.

To see why this is a hard problem, consider a direct approach where to decide if an automaton is history-deterministic, we attempt to directly solve the HD game on that automaton. The winning condition for Eve is that either the run of her token is accepting, or Adam's word is rejected. This is an $\omega$-regular objective, and a direct approach to solve the HD game involves determinising the automaton. This takes exponential time, and was the original approach given by Henzinger and Piterman in 2006 [HP06].

But, in contrast, the only lower bound the problem of checking history-determinism had up till now was of solving parity games [KS15], which, recall, can be solved in quasipolynomial time [CJK$^+$22] and is in NP ∩ coNP.

An early alternative approach of Kuperberg and Skrzypczak gave a PTIME algorithm for checking history-determinism for coBüchi automata [KS15], which involves the following steps on an input coBüchi automaton $\mathcal{C}$ based on the 'so-called' Joker game.

1. First, decide if Eve wins the Joker game on $\mathcal{C}$. If Eve loses, then $\mathcal{C}$ is not HD.

2. If Eve wins the Joker game, then construct a language-equivalent HD automaton $\mathcal{H}$.

3. Check if $\mathcal{C}$ simulates $\mathcal{H}$. If this is the case, then $\mathcal{C}$ is HD, and otherwise $\mathcal{C}$ is not HD.

**2-token game.** Then in 2018, Bagnol and Kuperberg suggested a novel approach to give a PTIME algorithm for deciding history-determinism of Büchi automata [BK18]. Their algorithm involved solving the so-called 2-token game on the Büchi automaton to decide if it is HD. The 2-token games are similar to HD games, with Adam building a word letter-by-letter and Eve building a run on her token transition-by-transition, but in addition, Adam has 2 tokens on which he also builds a run transition-by-transition.

Concretely, the 2-token game starts with Eve's token and each of Adam's two tokens on the initial state of an automaton, and in each round:

1. Adam selects a letter,

2. Eve moves her token along a transition on that letter,

3. Adam moves each of his two tokens along transitions on that letter.

Thus, in a play of the 2-token game, Adam constructs a word, Eve constructs a run on her token on that word, and Adam also constructs a run on each of his two tokens on that word. The winning condition for Eve is the following: if at least one of the runs of Adam's token is accepting, then the run on Eve's token is accepting.

Let us look at the 2-token game and the HD game from Adam's perspective: the 2-token game requires Adam to produce a witness accepting run for the accepted word, so the 2-token game is harder for Adam. That is, if Eve wins the HD game on an automaton, then she wins the 2-token game on that automaton as well: her strategy in the 2-token game is simply to pick transitions on her token according to a winning strategy in the HD game, ignoring the moves of Adam on his tokens.

Bagnol and Kuperberg showed that the converse also holds for Büchi automata, that is, if Eve wins the 2-token game on a Büchi automaton, then that automaton is HD. Thus, to decide if a Büchi automaton $\mathcal{B}$ is HD, it suffices to check if Eve wins the 2-token game on $\mathcal{B}$, which can be done in polynomial time.

**Theorem 1.1** ([BK18])**.** *For every Büchi automaton $\mathcal{A}$, Eve wins the $2$-token game on $\mathcal{A}$ if and only if $\mathcal{A}$ is HD.*

We can define, for every positive natural number $k$, the $k$-token game similar to how we defined the 2-token game. Bagnol and Kuperberg's key insight for their result was that 2-token games and $k$-token games have the same winner for $k \geq 2$ [BK18, Theorem 14].

**Theorem 1.2** ([BK18])**.** *For every parity automaton $\mathcal{A}$, Eve wins the $2$-token game on $\mathcal{A}$ if and only if for each $k \geq 2$, Eve wins the $k$-token game on $\mathcal{A}$.*

Before we discuss why Theorem 1.2 is true, we note that 1-token games have a different winner from the HD games [BK18, Lemma 8]: the example automaton presented in Fig. 1.2 demonstrates this, which we re-illustrate below for convenience.

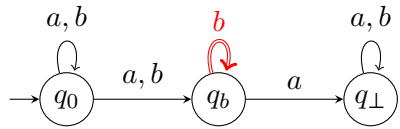

This automaton is not HD, since Adam has the following winning strategy: he chooses the letter $b$ in each round of the HD game for as long as Eve's token is in $q_0$. If Eve moves to $q_b$, then Adam chooses the letter $a$, and then always chooses letter $b$. This way, Adam's word contains at most one $a$ and hence is in the language of the automaton, while Eve's run is always rejecting.

Eve wins the 1-token game on it, however. Her strategy is to have her token stay in the state $q_0$ till Adam's token moves to $q_b$, and then have her token move to $q_b$ in the next round.

Let us continue towards showing that for every automaton $\mathcal{A}$ and $k \geq 2$, the 2-token game on $\mathcal{A}$ and $k$-token game on $\mathcal{A}$ have the same winner (Theorem 1.2). We prove this for $k = 3$, using which the reader should be able to prove it for all $k \geq 2$. It is clear that if Eve wins the 3-token game on $\mathcal{A}$, then she also wins the 2-token game on $\mathcal{A}$. For the other direction, we describe a winning strategy for Eve in the 3-token game that uses a winning strategy of Eve in the 2-token game. In the 3-token game, Eve stores an additional token in her memory, which we call the memory token. She picks transitions in the memory token by using a winning strategy in the 2-token game against Adam's second and third token. Eve chooses transitions on her token by playing the 2-token game against Adam's first token and her memory token. If either of Adam's tokens produces an accepting run, then either Adam's first token or Eve's memory token produces an accepting run, and hence Eve's token produces an accepting run.

The fact that 2-token games are equivalent to $k$-token games for every $k \geq 2$ (Theorem 1.2) was the crucial insight in Bagnol and Kuperberg's proof of the result that Eve winning the 2-token game characterises history-determinism for Büchi automata (Theorem 1.1). Their proof used contradiction, and it went as follows. They argued that if Adam wins the HD game on a Büchi automaton and Eve wins the 2-token game on it, then Adam can win the $k$-token game for some $k$ that is doubly-exponential in the size of the automaton, and hence also the 2-token game.

Bagnol and Kuperberg conjectured that their result extends to parity automata, i.e., for every parity automaton $\mathcal{A}$, Eve wins the 2-token game on $\mathcal{A}$ if and only if $\mathcal{A}$ is HD. We call this the 2-token conjecture.

> **2-Token Conjecture.** ([BK18]) For every parity automaton $\mathcal{A}$, Eve wins the 2-token game on $\mathcal{A}$ if and only if $\mathcal{A}$ is HD.

In 2020, Boker, Kuperberg, Lehtinen, and Skrzypczak showed that the 2-token conjecture holds for coBüchi automata as well [BKLS20]. Their proof used a similar template to that of Bagnol and Kuperberg's proof, but was much more involved. They also argued that 2-token games can be solved in PTIME for parity automata when the parity index is fixed. Thus, proving the 2-token conjecture would also imply that history-determinism can be checked in PTIME for parity automata when the parity index is fixed.

They also showed that if the 2-token conjecture is true for nondeterministic parity automata, then this implies the correctness of a similar 2-token based algorithm to decide history-determinism for alternating HD automata, which runs in PTIME if the parity index is fixed [BKLS20, Page 11].

A similar statement is true for $\omega$-regular automata too [CIK$^+$24, Lemma 36]. That is, if the 2-token conjecture is true for nondeterministic parity automata, then the 2-token conjecture holds for $\omega$-regular automata as well.

## 1.4 Contributions

The main contribution of this thesis is a proof of Bagnol and Kuperberg's conjecture that 2-token games characterise history-determinism in parity automata.

**Contribution I: the 2-Token theorem**

> **The 2-Token Theorem.** *For every nondeterministic parity automaton $\mathcal{A}$, Eve wins the 2-token game on $\mathcal{A}$ if and only if $\mathcal{A}$ is history-deterministic. Thus, the problem of deciding history-determinism is in* PTIME *for parity automata with a fixed number of priorities, and in* PSPACE *if the number of priorities is part of the input.*

The upper bounds inplied by 2-token theorem for checking history-determinism of nondeterministic parity automata also apply to alternating parity automata [BKLS20, Page 11]. Similarly, another easy consequence the 2-token theorem for parity automata is that the 2-token game based characterisation of history-determinism for parity automata extends to $\omega$-regular automata [CIK$^+$24, Lemma 36].

We prove the 2-token theorem by induction on the parity index of the automata. This involves two induction steps, one for adding a most significant odd priority, which we call the even-to-odd induction step (Theorem A), and the other for adding a most significant even priority, which we call the odd-to-even induction step (Theorem B).

**Theorem A.** *Let $K > 0$ be a natural number, such that for every $[0, K]$ (or equivalently, $[2, K+2]$) automaton $\mathcal{A}$, Eve wins the 2-token game on $\mathcal{A}$ if and only if $\mathcal{A}$ is HD. Then, for every $[1, K+2]$ automaton $\mathcal{A}$, Eve wins the 2-token game on $\mathcal{A}$ if and only if $\mathcal{A}$ is HD.*

**Theorem B.** *Let $K > 1$ be a natural number, such that for every $[1, K]$ automaton $\mathcal{A}$, Eve wins the 2-token game on $\mathcal{A}$ if and only if $\mathcal{A}$ is HD. Then, for every $[0, K]$ automaton $\mathcal{A}$, Eve wins the 2-token game on $\mathcal{A}$ if and only if $\mathcal{A}$ is HD.*

Our proofs of Theorems A and B, when restricted to the cases of Büchi and coBüchi automata, yield alternate proofs for the 2-token game based characterisations of history-determinism, and stronger game based characterisations of history-determinism than the 2-token theorem.

**Contribution II: Joker games for Büchi and coBüchi automata**

Kuperberg and Skrzypczak used Joker games in 2015 to give the first algorithm to decide history-determinism for coBüchi automata [KS15] that runs in polynomial time. The Joker game on an automaton is similar to the 1-token game on it, but in addition, Adam has the power to (finitely many times) 'play Joker' and move his token along an outgoing transition from the penultimate position of Eve's token. More concretely, in each round of the Joker game, where Eve's and Adam's tokens are at some states $q$ and $p$ respectively,

1. Adam selects a letter $a$;

2. Eve selects an $a$-transition from $q$ to $q'$ on her token;

3. Adam either selects an $a$-transition from $p$ to $p'$ on his token, or he selects an $a$-transition from $q$ to $p''$ and moves his token to $p''$.

The winning condition for Eve is the following: either the run of Eve's token is accepting, or the sequence of transitions on Adam's token does not satisfy the parity-condition, or Adam has played Joker infinitely many times. That is, if Adam has played finitely many Jokers and the priorities occurring in his sequence of transitions (note that the moves of his token do not necessarily constitute a run) satisfies the parity condition, then Eve's winning condition requires the run of her token to be accepting.

While Kuperberg and Skrzypczak used Joker games to give a algorithm for recognising history-deterministic coBüchi automata in PTIME, they left the question of whether Eve winning the Joker game characterises history-determinism or not open.

We answer this question in positive, showing that Eve winning the Joker game on a Büchi or coBüchi automaton characterises history-determinism for that automaton.

**Theorem C.** *For every Büchi or coBüchi automaton $\mathcal{A}$, Eve wins the Joker game on $\mathcal{A}$ if and only if $\mathcal{A}$ is history-deterministic.*

As a consequence of Theorem C, we get alternate proofs for the 2-token game based characterisations of history-determinism for Büchi and coBüchi automata. Our proof for coBüchi automata in particular is much simpler than that of [BKLS20]. Indeed, our chapter on coBüchi automata (Chapter 4), which proves Theorem C for coBüchi automata, is the shortest in this thesis, just over four pages long.

For Büchi automata, our proof of Theorem C instead solves an open problem of Kuperberg and Skrzypczak from 2015 concerning determinisation of history-deterministic Büchi automata.

**Contribution III: polynomial-time determinisation of HD Büchi automata**

A challenge towards generalising Bagnol and Kuperberg's proof for the 2-token theorem for Büchi automata, or Boker, Kuperberg, Lehtinen, and Skrzypczak's for coBüchi automata, is that their proofs are not constructive. That is, starting from a winning strategy for Eve in the 2-token game on an automaton, they do not construct a winning strategy for Eve in the HD game on that automaton. This is noted in the survey of Boker and Lehtinen [BL23b, Page 20], who conjectured that a constructive proof might also help towards showing the 2-token conjecture for parity automata.

Our proof of Theorem C is constructive. In particular for Büchi automata on which Eve wins the Joker game, the strategy we construct (in PTIME) has linear size, and therefore we get a PTIME determinisation procedure for HD Büchi automata that has a quadratic state-space blowup. Kuperberg and Skrzypczak had given a determinisation procedure for HD Büchi automata with a quadratic state space blow up too [KS15], but they left the question of whether this determinisation can be done in PTIME open.

**Theorem D.** *There is a polynomial-time procedure that converts every HD Büchi automaton with n states into a language-equivalent deterministic Büchi automaton with $n^2$ states.*

We note that we do not yet know if the quadratic state-space blowup to convert a HD Büchi automaton into a deterministic Büchi automaton is necessary, or if there is an HD Büchi automaton that has strictly fewer states than every language-equivalent deterministic Büchi automaton.

**Contribution IV: lookahead games**

Analogous to Bagnol and Kuperberg's result that the 2-token game and the *k*-token game on the same automaton have the same winner for every $k \geq 2$ [BK18], we show that 1-token games and *k*-lookahead games have the same winner on the same automaton for every $k \geq 0$. For each natural number $k \geq 0$, the *k*-lookahead game is the 1-token game, in which Adam's transition on his token is delayed by *k* steps, thus giving him a lookahead of *k*. We prove that the 1-token game and the *k*-lookahead game on the same automaton have the same winner.

**Theorem E.** *For every parity automaton $\mathcal{A}$, Eve wins the 1-token game on $\mathcal{A}$ if and only if she wins the k-lookahead game on $\mathcal{A}$ for all $k \geq 0$.*

The 1-token game is syntactically equivalent to the 0-lookahead game. With Theorem E as a key tool, we can show that the 1-token game characterises history-determinism on semantically-deterministic Büchi automata. These are automata in which, from every state and every letter, all outgoing transitions from that state labelled on that letter lead to language-equivalent states [RK23].

**Theorem F.** *For every semantically-deterministic Büchi automaton $\mathcal{A}$, $\mathcal{A}$ is history-deterministic if and only if Eve wins the 1-token game on $\mathcal{A}$.*

This is a stronger statement that Theorem C for Büchi automata, and thus it yields yet another proof for the 2-token game based characterisation of history-determinism on Büchi automata. We note that the assumption of semantic determinism is necessary: recall that we argued in Section 1.3.3 that the automaton in Fig. 1.2 is such that Eve wins the 1-token game on it, but it is not HD [BK18, Lemma 8].

**Contribution V: when the parity index is not fixed**

We show that the problem of deciding history-determinism for parity automata, when the parity index is not fixed, is NP-hard. Our reduction for NP-hardness of deciding history-determinism also shows that the problem of deciding simulation between two parity automata is NP-hard and hence NP-complete [CHP07]. Our proof also shows NP-hardness for deciding if Eve wins the 1-token game on a given parity automaton, and for deciding if Eve wins the 2-token game on a given parity automaton.

**Theorem G.** *The following problems are* NP*-hard.*

1. *Given a parity automaton $\mathcal{A}$, decide if $\mathcal{A}$ is HD.*

2. *Given a parity automaton $\mathcal{A}$, decide if Eve wins the 2-token game on $\mathcal{A}$.*

*The following problems are* NP*-complete.*

1. *Given a parity automaton $\mathcal{A}$, decide if Eve wins the 1-token game on $\mathcal{A}$.*

2. *Given parity automata $\mathcal{A}$ and $\mathcal{B}$, decide if $\mathcal{B}$ simulates $\mathcal{A}$.*

Simulation is a relation between parity automata that is coarser than language inclusion, i.e., if $\mathcal{B}$ simulates $\mathcal{A}$, then $L(\mathcal{A}) \subseteq L(\mathcal{B})$ [HKR02]. Simulation between two parity automata can be characterised by the simulation game. These games are like 1-token games, but with Eve and Adam moving their tokens in different automata, and Adam picking a transition on his token before Eve picks a transition on her token.

More concretely, the simulation game of $\mathcal{A}$ by $\mathcal{B}$ starts with an Eve's token at the initial state of $\mathcal{B}$ and Adam's token at the initial state of $\mathcal{A}$. In each round,

1. Adam selects a letter and moves his token along a transition on that letter in $\mathcal{A}$,

2. Eve moves her token along a transition on that letter in $\mathcal{B}$.

This way, Adam constructs a run on his token in $\mathcal{A}$, and Eve a run on her token in $\mathcal{B}$, both on the same word chosen by Adam. We say that Eve wins a play of this game if either her token's run is accepting or Adam's token's run is rejecting. If Eve has a winning strategy in the simulation game of $\mathcal{A}$ by $\mathcal{B}$, then we say that $\mathcal{B}$ simulates $\mathcal{A}$.

For history-deterministic automata, the relation of language inclusion is equivalent to simulation [BHL+24, Theorem 3.4]. That is, if $\mathcal{A}$ is a nondeterministic parity automaton and $\mathcal{H}$ is an HD parity automaton, then $L(\mathcal{A}) \subseteq L(\mathcal{H})$ if and only if $\mathcal{H}$ simulates $\mathcal{A}$. This makes HD automata useful for model checking, since deciding simulation (NP-complete) is easier than the problem of deciding language inclusion (PSPACE-complete) for nondeterministic parity automata. We show that for HD automata, we can decide language inclusion even faster, in quasipolynomial time, by reducing the problem to solving parity games.

**Theorem H.** *There is an algorithm that on given input parity automaton $\mathcal{A}$ with $n_1$ states and $d_1$ priorities and an HD parity automaton $\mathcal{H}$ with $n_2$ states and $d_2$ priorities, both over the same alphabet $\Sigma$, decides whether $L(\mathcal{A}) \subseteq L(\mathcal{H})$ in time*

$$(n_1 \cdot d_1 \cdot n_2 \cdot d_2 \cdot |\Sigma|)^{\mathcal{O}(\log d_2)}.$$

### 1.4.1 An overview of techniques

We briefly discuss some key ideas that occur in our proof of the 2-token theorem.

We will write that two automata $\mathcal{A}$ and $\mathcal{B}$ are *simulation-equivalent* if $\mathcal{A}$ simulates $\mathcal{B}$ and $\mathcal{B}$ simulates $\mathcal{A}$. If $\mathcal{A}$ and $\mathcal{B}$ are simulation-equivalent, then $\mathcal{A}$ and $\mathcal{B}$ are language-equivalent, and $\mathcal{A}$ is HD if and only if $\mathcal{B}$ is HD.

A recurring pattern of arguments used in our proofs is the following: starting from an automaton on which Eve wins the 2-token game, we will make certain modifications to get, as an intermediate step, a simulation-equivalent automaton on which Eve still wins the 2-token game, and which has some additional desirable properties. Due to these properties, it will be easier to show that the new automaton is HD, and it will follow from simulation-equivalence that the automaton we started from is HD as well.

Towards elaborating on these modifications, we will need token games that start at different states in the automaton. For states $q, p_1, p_2, \ldots, p_k$ in an automaton $\mathcal{A}$ for some $k \geq 1$, we will write that Eve wins the $k$-token game from $(q; p_1, p_2, \ldots, p_k)$ in $\mathcal{A}$, if Eve wins the $k$-token game on $\mathcal{A}$ in which her token starts at the state $q$, and Adam's $k$ tokens start at the states $p_1, p_2, \ldots, p_k$. We call states $p_1, p_2, \ldots, p_k$ *coreachable* in $\mathcal{A}$ if there is a finite word $u$ such that there are runs from the initial state $q_0$ to each of the $p_i$s on $u$. We call the transitive closure of the coreachable relation as *weakly coreachable*. Note that weak coreachability is an equivalence relation between the states of the automaton.

One of the crucial modifications that we use throughout our thesis allows us to modify an automaton on which Eve wins the 2-token game into another automaton on which *Eve wins the $k$-token game from everywhere* for every $k \geq 1$, i.e., Eve wins the $k$-token game from every configuration of states that can be reached in the $k$-token game from the initial configuration where all tokens start at the initial state.

**Theorem I.** *Let $\mathcal{A}$ be a parity automaton on which Eve wins the $2$-token game. Then, there is a simulation-equivalent subautomaton $\mathcal{B}$ of $\mathcal{A}$ such that the following three conditions hold.*

1. *For each $k \geq 1$, Eve wins the $k$-token game from everywhere in $\mathcal{B}$.*

2. *For states $q, p_1, p_2, \ldots, p_k$ that are weakly coreachable in $\mathcal{B}$, Eve wins the $k$-token game from $(q; p_1, p_2, \ldots, p_k)$ in $\mathcal{B}$.*

3. *If $\mathcal{B}$ is history-deterministic, then so is $\mathcal{A}$.*

For proving he even-to-odd induction step (Theorem A) and the odd-to-even induction step (Theorem B), starting with an automaton on which Eve wins the 2-token game, we will make modifications to obtain a simulation-equivalent automaton satisfying certain properties as an intermediate step, so that Eve still wins the 2-token game on it.

For the even-to-odd induction step (Theorem A), we call this intermediate property *1-safe double-coverage*. Let $\mathcal{A}$ be a $[1, K + 2]$ automaton for some $K \geq 1$. The property of 1-safe double-coverage is based on the 2-token game between the states of the 2-approximation of $\mathcal{A}$, which we denote by $\mathcal{A}_{>1}$. The automaton $\mathcal{A}_{>1}$ is a $[2, K + 2]$ automaton that is obtained by preserving all transitions of $\mathcal{A}$ with priorities in $[2, K+2]$, and redirecting transitions of priority 1 to a rejecting sink state.

We then say that the automaton $\mathcal{A}$ has 1-safe double-coverage if for each state $p$ in $\mathcal{A}$, there is another state $q$ weakly coreachable to $p$ in $\mathcal{A}$, such that Eve wins the 2-token game from $(q; p, p)$ in $\mathcal{A}_{>1}$.

> **Automata with 1-safe double-coverage.** We say a $[1, K+2]$ automaton $\mathcal{A}$ with $K \geq 0$ has the property of 1-safe double-coverage if for each state $p$ in $\mathcal{A}$, there is another state $q$ weakly coreachable to $p$ in $\mathcal{A}$, such that Eve wins the 2-token game from $(q; p, p)$ in $\mathcal{A}_{>1}$.

Then we prove Theorem A in the following steps.

1. We first show that if Eve wins the 2-token game on a $[1, K+2]$ automaton $\mathcal{A}$, then we can modify the automaton $\mathcal{A}$ into another simulation-equivalent automaton $\mathcal{A}'$ that has 1-safe double-coverage and on which Eve wins the 2-token game from everywhere.

2. Then we use the induction hypothesis, of Eve winning the 2-token game characterising history-determinism on $[0, K]$ or $[2, K+2]$ automaton, to deduce that $\mathcal{A}'$ is history-deterministic, and hence so is $\mathcal{A}$.

For the odd-to-even induction step (Theorem B), we have an analogous property called *0-reach double-covering*. For a $[0, K]$ automaton $\mathcal{A}$ for some $K \geq 1$, we define the 1-approximation of $\mathcal{A}$, denoted $\mathcal{A}_{>0}$, as the automaton obtained by preserving all transitions of $\mathcal{A}$ with priority at least 1, and redirecting all transitions of priority 0 to an accepting sink state.

We then say that $\mathcal{A}$ has 0-reach double-covering if for each state $p$ in $\mathcal{A}$, there is another state $q$ weakly coreachable to $p$ in $\mathcal{A}$, such that Eve wins the 2-token game from $(p; q, q)$ in $\mathcal{A}_{>0}$.

> **Automata with 0-reach double-covering.** We say a $[0, K]$ automaton $\mathcal{A}$ with $K \geq 1$ has the property of 0-reach double-covering if for each state $p$ in $\mathcal{A}$, there is another state $q$ weakly coreachable to $p$ in $\mathcal{A}$, such that Eve wins the 2-token game from $(p; q, q)$ in $\mathcal{A}_{>0}$.

We draw the readers' attention to the fact that the roles of $q$ and $p$ in the 2-token games in the definitions of 0-reach double-covering and 1-safe double-coverage are reversed. Nevertheless, Theorem B follows a similar template to for Theorem A: starting with a $[0, K]$ automaton $\mathcal{A}$ on which Eve wins the 2-token game, we convert $\mathcal{A}$ to a simulation-equivalent automaton $\mathcal{A}'$, such that Eve wins the 2-token game on $\mathcal{A}'$ and $\mathcal{A}'$ has 0-reach double-covering. We then argue, using the induction hypothesis, that $\mathcal{A}'$ is HD, and thus so is $\mathcal{A}$.

For the even-to-odd induction step (Theorem A), the structural modification to obtain 1-safe double-coverage involves relabelling of priorities while preserving the acceptance of each run, and uses a refinement of classical techniques that have been used previously [KS15, BKLS20]. On the other hand, the odd-to-even induction step (Theorem B) is more technically involved, and uses the concepts of Zielonka trees [DJW97] and progress measures [Jur00, Wal02].

## 1.5 Organisation of the thesis

The rest of this thesis is organised as follows. We start by introducing some necessary background on automata and games in Chapter 2. Then, in Chapter 3, we introduce history-determinism, token games, and results on them, which we will use throughout this thesis. We also show Theorem I in this chapter, which allows us to assume, without loss of generality, when trying to show that a parity automaton on which Eve wins the 2-token game is HD, that Eve wins the 2-token game from everywhere.

We then continue on to giving alternate proofs for the 2-token theorem for coBüchi automata in Chapter 4, and for Büchi automata in Chapter 5. These chapters are also where we prove the result of Joker game characterising history-determinism on coBüchi and Büchi automata (Theorem C). Chapter 5 on Büchi automata also contains the polynomial-time determinisation procedure of HD Büchi automata (Theorem D), introduces lookahead games (Theorem E), and proves the 1-token game characterisation of history-determinism on semantically-deterministic Büchi automata (Theorem F).

Chapters 6 and 7 contain the main contribution of this thesis, which is proving the 2-token theorem. In Chapter 6, we prove the even-to-odd induction step (Theorem A), and in Chapter 7, we prove the odd-to-even induction step (Theorem B). Chapters 4 and 5 can be seen as a warm up towards Chapters 6 and 7 respectively. We end Chapter 7 with some immediate consequences of the 2-token theorem.

Chapter 8 then focuses on history-deterministic parity automata when the parity index is not fixed. In this chapter, we show NP-hardness for deciding history-determinism in parity automata, as well as NP-hardness for checking simulation between two parity automata (Theorem G). We also give a quasipolynomial algorithm to decide the language-inclusion problem for history-deterministic parity automata (Theorem H).

We end this thesis with Chapter 9, where we discuss problems that remain open, and directions for future work that have emerged due to this thesis.

## 1.6 Reading tips

This thesis is written to allow for both non-linear and linear reading, given the reader is familiar with the contents of Chapter 2 and Chapter 3. Since Chapters 4 and 5 also serve the purpose of warm ups to Chapters 6 and 7 respectively, we will inevitably draw parallels to the content of Chapters 4 and 5 in Chapters 6 and 7, but the latter chapters are self-contained both in mathematical rigour and intuition.

# Chapter 2

# Background on Games and Automata

In this chapter, we define the concept of games and automata with $\omega$-regular acceptance conditions, which we will use throughout this thesis. We also detail on two concepts from games on graphs that will be useful to us.

1. Zielonka trees [DJW97], which we use in Section 3.5.2 to give upper bounds on the complexity of solving token games, in Chapter 7 for our odd-to-even induction step, and in giving a quasipolynomial algorithm for language-containment problem over HD parity automata (Theorem H in Chapter 8).

2. A restricted variant of the concept of ranks in parity games [Büc83, Wal02], which we use in Chapters 5 and 7.

## 2.1 Games on graphs

**Basic notations.** Throughout this thesis, we will use $\mathbb{N}$ to denote the set of natural numbers $\{0, 1, 2, \dots\}$. For two natural numbers $i < j$, we use the interval $[i, j]$ to denote the set $[i, j] = \{i, i+1, \dots, j\}$. An *alphabet* $\Sigma$ is a finite set of *letters*. We use $\Sigma^*$ and $\Sigma^\omega$ to denote the set of words of finite and countably infinite length over $\Sigma$, respectively. A language $L$ is a subset of $\Sigma^\omega$. For a finite word $u$ and a language $L$, we denote $u^{-1}L$ to be the language
$$u^{-1}L = \{w \mid uw \in L\}.$$

**Game arenas.** An *arena* is a directed graph $G = (V, E)$ with vertices partitioned into $V_\forall$ and $V_\exists$ between two players Adam and Eve, respectively. Additionally, a vertex $v_0 \in V$ is designated as the initial vertex. We say that vertices in $V_\exists$ are *owned* by Eve and those in $V_\forall$ are owned by Adam. Unless stated otherwise, the graphs of the arenas are finite. We assume that each vertex in $V$ has at least one outgoing edge.

A *play* on this arena is an infinite path starting with a token at $v_0$, which is constructed in infinitely many rounds. In each round, the player who owns the vertex on which the token is currently placed chooses an outgoing edge, and the token is moved along this edge to the target of that edge for another round of play. The movement of this token creates an infinite path in the arena, which we call a play. A *finite play* is a finite prefix of a play.

A *game* $\mathcal{G}$ consists of an arena $G = (V, E)$ together with a *winning condition* given by a language $L \subseteq E^\omega$. We say that Eve *wins a play* $\rho$ of $\mathcal{G}$ if $\rho$ is in $L$, and Adam wins otherwise. A *strategy* for Eve in such a game $\mathcal{G}$ is a function from the set of finite plays that end at an Eve's vertex to an outgoing edge from that vertex. An Eve's strategy is *winning* if every play produced in which she chooses edges according to this strategy is winning for her. We say that Eve *wins the game* if she has a winning strategy. Winning strategies are defined for Adam analogously, and we say that Adam wins the game if he has a winning strategy.

We say that Eve (resp. Adam) wins from $v$ in $\mathcal{G}$ if she (resp. he) wins the game $\mathcal{G}$ with its initial vertex at $v$. The set of vertices which Eve (resp. Adam) wins from is called the *winning region* of Eve (resp. Adam).

**Winning conditions.**

We now describe the various winning conditions for games on graphs, given as languages $L \subseteq E^\omega$, which we will deal with in this thesis.

**Parity condition.** Let $G = (V, E)$ be a finite directed graph equipped with a *priority function* $\pi : E \to \mathbb{N}$ that assigns each edge with a natural number, which we call the *priority* of that edge. We say that an infinite path in $G$ satisfies the $\pi$-parity condition, or just parity condition when $\pi$ is clear from the context, if the least priority occurring infinitely often along the run is even.

A parity condition is easily *complemented*. Given a priority function $\pi$ as above, consider the priority function $\pi' := \pi + 1$ that is obtained by increasing all the labels by 1. Then, an infinite path satisfies $\pi'$ if and only if it does not satisfy $\pi$.

**Parity games.** A parity game $\mathcal{G}$ consists of a game arena $G = (V, E)$ with a priority function $\pi : E \to [0, d]$ for some natural number $d$. The winning condition $L \subseteq E^\omega$ for $\mathcal{G}$ consists of paths that satisfy the $\pi$-parity condition. When an edge $e = v \to v'$ in $\mathcal{G}$ has the priority $p$, we will sometimes write $v \xrightarrow{p} v'$ to indicate this.

Parity games are positionally determined [EJ91], i.e., in every parity game $\mathcal{G}$, either Eve or Adam has a winning strategy, and this winning strategy only depends on the current position the token is at. We say that a positional winning strategy for Eve (resp. Adam) is *uniform* if it is a strategy using which Eve (resp. Adam) wins the game $\mathcal{G}$ with its initial vertex as any vertex in Eve's (resp. Adam's) winning region. In a parity game $\mathcal{G}$, both players Eve and Adam have uniform positional strategies that ensure that they win from $\mathcal{G}$ whenever the initial vertex is in their winning region [EJ91].

A parity game $\mathcal{G}$ with $n$ vertices and priorities in $[0, d]$ can be *solved*, i.e., the winning regions of Eve and Adam and their positional winning strategies from their winning regions can be found, in time $n^{\mathcal{O}(\log d)}$ [CJK$^+$22].

**Theorem 2.1** ([CJK$^+$22]). *There is an algorithm that, on a parity game with $n$ vertices and priorities in $[0, d]$, finds the winning regions and positional winning strategies of the players from their winning regions in time $n^{\mathcal{O}(\log d)}$.*

**Multi-dimensional parity games.** Multi-dimensional parity games were introduced by Chatterjee, Henzinger, and Piterman [CHP07].

A *$k$-dimensional parity condition* for $k \geq 1$ on a graph $G = (V, E)$ is similar to a parity condition, but we now have $k$ priority functions $\pi_i : E \to [0, d]$ for each $i \in [1, k]$ defining $k$ parity conditions. An infinite path $\rho$ in the graph $G$ is said to satisfy the $(\pi_1 \vee \pi_2 \vee \cdots \vee \pi_k)$ condition if the following holds: *$\rho$ satisfies the $\pi_i$-parity condition for some $i \in [1, k]$*.

A *$k$-dimensional parity game* $\mathcal{G}$, consists of a game arena $G = (V, E)$ together with $k$ priority functions $\pi_i$ for each $i \in [1, k]$. The winning condition $L \subseteq E^\omega$ for $\mathcal{G}$ consists of paths that satisfy the $(\pi_1 \vee \pi_2 \vee \cdots \vee \pi_k)$ condition.

For all fixed positive natural numbers $k \geq 2$, checking if Eve wins a $k$-dimensional parity game $\mathcal{G}$ is an NP-complete problem [CHP07]. The lower bound of NP-hardness was shown by Chatterjee, Henzinger, and Piterman [CHP07, Lemma 1]. And the upper bound of NP follows from the fact that $k$-dimensional parity games are a subclass of Rabin games [CHP07, Page 158], and deciding if Eve wins Rabin games is in NP [Eme85]. In fact, if Eve wins a $k$-dimensional parity game or a Rabin game, then Eve has a uniform positional winning strategy [Eme85] from all vertices in her winning region, which can be verified in PTIME. Rabin games, like parity games, are also determined, i.e., in each Rabin game, either Eve or Adam has a winning strategy.

**Theorem 2.2.** *Let $k$ be a fixed positive natural number. Given a $k$-dimensional parity game $\mathcal{G}$, every vertex in $\mathcal{G}$ is in the winning region of Eve or Adam, and Eve has a uniform positional winning strategy from her (possibly empty) winning region in $\mathcal{G}$. The problem of deciding if the initial vertex of $\mathcal{G}$ is in her winning region is NP-complete.*

**Implication games.** Implication games are a reformulation of 2-dimensional parity games that will be convenient for us. An *implication condition* on a graph $G = (V, E)$ is specified by two priority functions $\pi_1 : E \to [0, d_1]$ and $\pi_2 : E \to [0, d_2]$. An infinite path $\rho$ in the graph $G$ is said to satisfy the $(\pi_1 \Rightarrow \pi_2)$ condition if the following holds:

*if $\rho$ satisfies the $\pi_1$-parity condition, then $\rho$ satisfies the $\pi_2$-parity condition.*

An implication game $\mathcal{G}$ consists of an arena $G = (V, E)$ together with two priority functions $\pi_1 : E \to [0, d_1]$ and $\pi_2 : E \to [0, d_2]$. The winning condition $L \subseteq E^\omega$ for $\mathcal{G}$ consists of paths that satisfy the $(\pi_1 \Rightarrow \pi_2)$ condition.

Implication games can easily be converted to 2-dimensional parity games and vice versa. This is due to the fact that the logical statements of $(A \vee B)$ and $(\neg A \Rightarrow B)$ are equivalent, and that parity conditions can be easily complemented. This conversion shows that the results on multi-dimensional parity games in Theorem 2.2 are also true for implication games.

**Lemma 2.3.** *Given an implication game $\mathcal{G}$, every vertex in $\mathcal{G}$ is in the winning region of Eve or Adam, and Eve has a uniform positional winning strategy from her (possibly empty) winning region in $\mathcal{G}$. The problem of deciding if the initial vertex of $\mathcal{G}$ is in her winning region is* NP*-complete.*

**Muller games.** A *Muller condition* on a graph $G = (V, E)$ is given by a function $\mathsf{col} : E \to C$ that labels each edge by a colour from a finite set of colours $C$, and a set $\mathcal{F} \subseteq \mathcal{P}(C)$ of nonempty *accepting subsets* of $C$. We say that an infinite path $\rho$ satisfies the $(C, \mathcal{F})$ Muller condition if the set of colours occurring infinitely often in $\rho$ is in $\mathcal{F}$. When an edge $e = v \to v'$ in $\mathcal{G}$ has the priority $c$, we will sometimes write $v \xrightarrow{c} v'$ to indicate this.

A *Muller game* $\mathcal{G}$ consists of an arena $G = (V, E)$ together with a colouring function $\mathsf{col} : E \to C$ and a Muller condition $(C, \mathcal{F})$. The winning condition $L \subseteq E^\omega$ for $\mathcal{G}$ consists of paths that satisfy the $(C, \mathcal{F})$ Muller condition.

We note that the games we have discussed above can be reformulated as Muller games. For $k$-dimensional parity games, for instance, we have $C = [0, d]^k$, and the set $\mathcal{F}$ consists of nonempty subsets $S$ of $C$, such that for some $i$ in $[1, k]$ the least number occurring in the $i^{th}$ component of elements in $S$ is even.

Every Muller game $\mathcal{G}$ can be converted into a parity game $\mathcal{G}'$ such that $\mathcal{G}$ and $\mathcal{G}'$ have the same winner. We will discuss this conversion in detail in Section 2.4.

## 2.2 Automata over infinite words

The automata that we deal with are over infinite words. We will predominantly deal with parity automata and its subclasses of Büchi, coBüchi, safety, and reachability automata, which we define next.

**Parity automata.** An $[i, j]$ nondeterministic parity automaton, or $[i, j]$ automaton for short, is given by a 4-tuple $\mathcal{A} = (Q, \Sigma, \Delta, q_0)$. The automaton $\mathcal{A}$ is a directed graph with its vertices as a finite set of *states* $Q$, and edges that are each labelled by a *letter* from a finite alphabet $\Sigma$ and a *priority* in $[i, j]$. These edges are called *transitions*, and a transition from the state $q$ to $q'$ on the letter $a$ with a priority $c$ is denoted by $q \xrightarrow{a:c} q'$. A *run* of the automaton $\mathcal{A}$ on an infinite word $w$ is an infinite path that starts at the *initial state* $q_0$, and follows transitions consisting of the letters of $w$ in sequence. We say that this run is *accepting* if the *least priority occurring infinitely often* in the labels of its transitions is even. A word $w$ is accepted by $\mathcal{A}$ if $w$ has an accepting run, and the language of $\mathcal{A}$, denoted by $L(\mathcal{A})$, is the set of words accepted by $\mathcal{A}$. We equivalently write that $L(\mathcal{A})$ is the *language recognised* by $\mathcal{A}$.

We will assume, unless otherwise stated, that every parity automaton $\mathcal{A}$ we deal with in this thesis is *complete*, i.e., for every state and letter, there is an outgoing transition in $\mathcal{A}$ from that state on that letter.

A parity automaton is said to be *deterministic* if for each state and letter, there is exactly one transition outgoing from that state on that letter.

For a state $q$ in $\mathcal{A}$, we use $(\mathcal{A}, q)$ to denote the automaton $\mathcal{A}$ with $q$ as the initial state.

**Büchi automata.** A Büchi automaton is a $[0, 1]$ automaton. We call transitions of priority 0 as accepting transitions, and transitions of priority 1 as rejecting transitions. A run $\rho$ of a Büchi automaton is accepting if $\rho$ contains infinitely many accepting transitions.

**CoBüchi automata.** A coBüchi automaton is a $[1, 2]$ automaton. We call transitions of priority 2 as accepting transitions, and transitions of priority 1 as rejecting transitions. A run $\rho$ of a coBüchi automaton is rejecting if $\rho$ contains finitely many rejecting transitions.

**Safety automata.** A safety automaton is a $[1, 2]$ automaton which has a unique rejecting sink state, on which we only have outgoing 1-priority transitions to itself on each letter, and all other transitions in the automaton have priority 2.

**Reachability automata.** A reachability automaton is a $[0, 1]$ automaton which has a unique accepting sink state, on which we only have outgoing 0-priority transitions to itself on each letter, and all other transitions in the automaton have priority 1.

We note that every safety automaton can be equivalently represented as $[0, 1]$ automaton where transitions of priority 2 are relabelled to have priority 0, and analogously reachability automata can be represented as a $[1, 2]$ automaton where transitions of priority 0 are relabelled to have priority 2. This modification does not change the acceptance of any run, since there are no cycles in safety or reachability automata that contain transitions of both odd and even priority.

**Muller automata.** A Muller automaton $\mathcal{M} = (Q, \Sigma, \Delta, q_0, C, \mathcal{F})$ is similar to a parity automaton, but the transitions $\Delta$ are each labelled by a *colour* in $C$ instead of a priority. The acceptance of a run is given by the $(C, \mathcal{F})$ Muller condition, where $\mathcal{F}$ is a set of nonempty subsets of $C$. A run in $\mathcal{M}$ is accepting if it satisfies the $(C, \mathcal{F})$ Muller condition, i.e., the set of colours seen infinitely often is a set in $\mathcal{F}$.

Parity automata can be easily reformulated as a Muller automata: for an $[i, j]$ automaton, the equivalent Muller automaton has its colours as $C = [i, j]$ and the set of accepting subsets $\mathcal{F}$ consists of nonempty subsets of $C$ whose least element is even.

**Parity-index hierarchy.** We had discussed the parity-index hierarchy in the introduction, but let us recall it again here. For an $[i, j]$ automaton, note that if $i$ is at least 2, then we can decrease all priorities by 2 in that automaton without changing the acceptance of any run in that automaton. Thus, we assume without loss of generality that $i = 0$ or $i = 1$. Then, we say that $[i, j]$ is the parity index of that automaton. In terms of expressivity, parity automata have an alternating hierarchy in terms of its parity indices as shown in Fig. 2.1, which is a re-illustration of Fig. 1.5.

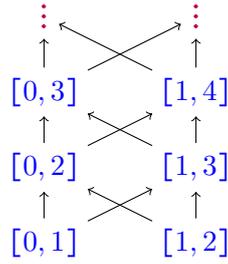

Figure 2.1: The parity index hierarchy.

Here, Büchi automata correspond to the $[0,1]$ level, while coBüchi automata correspond to the $[1,2]$ level. As mentioned in Chapter 1, the parity index hierarchy collapses to the Büchi level for nondeterministic automata, i.e., nondeterministic Büchi automata are as expressive as nondeterministic and deterministic parity automata, which are in turn as expressive as nondeterministic and deterministic Muller automata [Boj18, Theorem 1.2 and Lemma 2.6].

This is not the case for deterministic parity automata with a fixed parity index, however. Indeed, for $i = 0, 1$ and $j > i$ a natural number, define the $[i,j]$ language $L_{[i,j]}$ over the alphabet $[i,j]$ as the set of words in which the lowest natural number occurring infinitely often is even. Then, for every $l \geq 1$, there is no deterministic $[1, l+1]$ automaton recognising $L_{[0,l]}$, and similarly, there is no deterministic $[0,l]$ automaton recognising $L_{[1,l+1]}$ [Wag79].

For coBüchi automata, deterministic coBüchi automata are as expressive as nondeterministic coBüchi automata [MH84]. Deterministic Büchi automata, on the other hand, are strictly less expressive than nondeterministic Büchi automata, as we discussed in Section 1.2.3. How the different classes of automata compare in terms of expressivity were shown in Fig. 1.4, which we re-illustrate in Fig. 2.2 for readers' convenience.

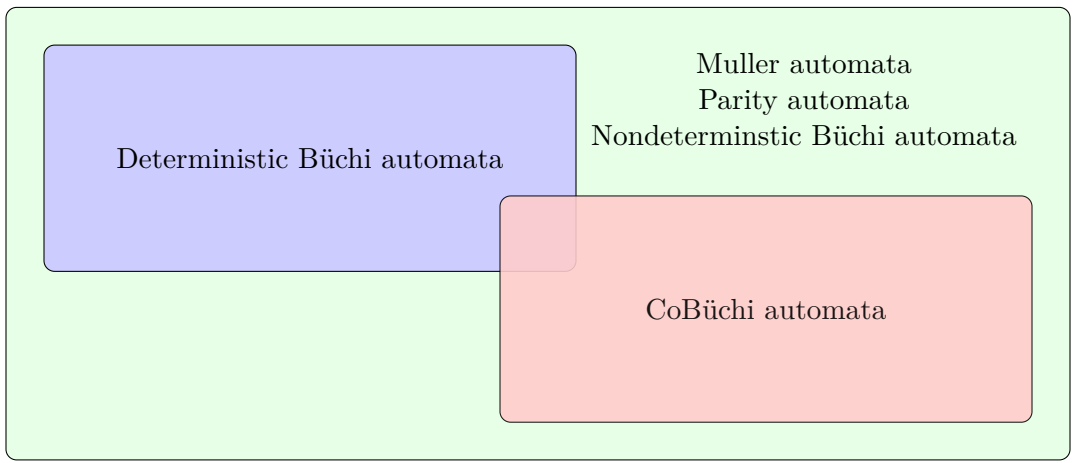

Figure 2.2: The Venn diagram for the expressivity of the mentioned automata classes. Where unspecified, deterministic X automata are as expressive as nondeterministic X automata.

**Theorem 2.4.** *The following classes of automata are equally expressive: nondeterministic Büchi automata, deterministic parity automata, nondeterministic parity automata, nondeterministic Muller automata, deterministic Muller automata.*

## 2.3 Finite-memory determinacy of $\omega$-regular games

In this section, we will use the result that nondeterministic Muller automata are as expressive deterministic parity automata to show finite-memory determinacy of Muller games. That is, we will show that in every Muller game $\mathcal{M}$, either Eve or Adam wins $\mathcal{M}$, and the winner has a finite-memory winning strategy. We will also define $\omega$-regular games and automata, and show finite-memory determinacy for $\omega$-regular games. This will enable us to show finite-memory determinacy of all the games that we discuss in this thesis.

**Theorem 2.5.** *Every Muller game or $\omega$-regular game has a winner, who has a finite-memory winning strategy [McN66, GH82].*

**Games and automata with $\omega$-regular acceptance conditions.** Given a finite directed graph $G = (V, E)$, an $\omega$-regular acceptance condition on the graph consists of a colouring $\mathsf{col} : C \to E$ of edges, and a nondeterministic Muller automaton $\mathcal{M}$ with its alphabet as $C$. We say that an infinite path $\rho$ in $\mathcal{G}$ satisfies the $(\mathsf{col}, \mathcal{M})$ $\omega$-regular condition if the word $w_C \in C^\omega$ consisting of colours of edges occurring in $\rho$ in sequence is accepted by $\mathcal{M}$.

An $\omega$-regular game $\mathcal{G}$ consists of a game arena $G = (V, E)$ together with a function $\mathsf{col} : E \to C$ that labels each edge with a colour from a finite set of colours $C$, and a Muller automaton $\mathcal{M}$ that is over the alphabet $C$. The winning condition $L \subseteq E^\omega$ for $\mathcal{G}$ consists of paths that satisfy the $(\mathsf{col}, \mathcal{M})$ acceptance condition. When an edge $e = v \to v'$ in $\mathcal{G}$ has the colour $c$, we will sometimes write $v \xrightarrow{c} v'$ to indicate this.

An $\omega$-regular automaton $\mathcal{A} = (Q, \Sigma, \Delta, q_0, C, \mathcal{M})$ is similar to a Muller automaton, with each transition labelled by a colour in $C$, but the acceptance condition is instead given by a nondeterministic Muller automaton $\mathcal{M}$ that is over the alphabet $C$. A run $\rho$ in $\mathcal{A}$ is accepting if it satisfies the $(\mathsf{col}, \mathcal{M})$ accepting condition, i.e., the sequence of colours occurring in the edges of $\rho$ is a word accepted by $\mathcal{M}$.

**Remark 1.** *We can equivalently define the acceptance conditions of $\omega$-regular automata to be given by deterministic Muller automata, nondeterministic or deterministic parity automata, or nondeterministic Büchi automata. Since these automata classes are all equivalent in expressivity, this does not change the expressivity of $\omega$-regular automata.*

Muller automata and $\omega$-regular automata are equally expressive. To prove this, we start by noting that every nondeterministic or deterministic Muller automaton can be easily represented as a language-equivalent nondeterministic or deterministic $\omega$-regular automaton.

Let $\mathcal{M}$ be a Muller automaton with its acceptance condition given by the $(C, \mathcal{F})$ Muller condition. Consider the $(C, \mathcal{F})$-*Muller language* $L_{C,\mathcal{F}}$ consisting of words $w$ of $C^\omega$ such that the set of colours occurring infinitely often in $w$ is a set in $\mathcal{F}$. The language $L_{C,\mathcal{F}}$ can then be recognised by a single state deterministic Muller automaton $\mathcal{M}_{C,\mathcal{F}}$ that has a transition $q \xrightarrow{c:c} q$ for each colour $c$ from its unique state $q$. The acceptance condition of $\mathcal{M}$ is the $(C, \mathcal{F})$ Muller condition.

Thus, the Muller automaton $\mathcal{M}$ is easily reformulated as a language-equivalent $\omega$-regular automaton, with identical structure and colours on transitions, and the accepting condition given by $\mathcal{M}_{C,\mathcal{F}}$.

For the other direction, let $\mathcal{A} = (Q, \Sigma, \Delta, q_0, C, \mathcal{M})$ be a nondeterministic $\omega$-regular automaton, where $\mathcal{M}$ is a nondeterministic Muller automaton. We will construct a nondeterministic parity automaton $\mathcal{P}$ that is language-equivalent to $\mathcal{A}$. Towards this goal, let $\mathcal{D}$ be a deterministic parity automaton that is language-equivalent to $\mathcal{M}$.

Then, we construct the automaton $\mathcal{P}$ by composing $\mathcal{D}$ with $\mathcal{A}$. More concretely, the states of $\mathcal{P}$ are pairs of states $(q, d)$, where $q$ is a state in $\mathcal{A}$ and $d$ is a state in $\mathcal{D}$. We have the transition $(q, d) \xrightarrow{a:p} (q', d')$ in $\mathcal{P}$, where $p$ is a natural number and $a$ is a letter in $\Sigma$, if there are transitions $q \xrightarrow{a:c} q'$ in $\mathcal{A}$ and a transition $d \xrightarrow{c:p} d'$ in $\mathcal{D}$. The initial state of $\mathcal{P}$ is the pair of initial states of $\mathcal{A}$ and $\mathcal{D}$.

The automata $\mathcal{P}$ and $\mathcal{A}$ are language-equivalent. If there is an accepting run $\rho_A$ over a word $w$ in $\mathcal{A}$ then there is an unique accepting run $\rho_D$ in $\mathcal{D}$ over the sequence of colours occurring in $\rho_D$. Consider the run $\rho_P$ in $\mathcal{P}$ over $w$ then, where the first component follows the run $\rho$ while the second component follows $\rho_D$, and priorities are inherited from $\rho_D$. The run $\rho_P$ is accepting.

Conversely, if there is an accepting run $\rho_P$ in $\mathcal{P}$ over $w$ then there is a projection of run $\rho_P$ in the $\mathcal{A}$ component over $w$ that is accepting as well.

Thus, nondeterministic Muller automata and nondeterministic $\omega$-regular automata are equally expressive. Furthermore, note that the parity automaton $\mathcal{P}$ we construct in our conversion above is deterministic whenever the $\omega$-regular automaton $\mathcal{A}$ that we start with is deterministic. Thus, deterministic or nondeterministic $\omega$-regular automata are equally expressive to the class of nondeterministic Büchi automata and all classes of automata that we mentioned in Theorem 2.4.

**Theorem 2.6.** *Deterministic and nondeterministic $\omega$-regular automata are as expressive as nondeterministic Büchi automata, deterministic or nondeterministic parity automata, and deterministic or nondeterministic Muller automata.*

We call a language $\omega$-regular if it can be recognised by an automaton belonging to one of the classes of automata that is mentioned in Theorem 2.6 above.

The language-preserving conversion between $\omega$-regular automata and Muller automata can be easily adapted to winner-preserving conversions between $\omega$-regular games and Muller games. We leave this as an exercise to the reader.

**Finite-memory strategies in games.** Let $\mathcal{G} = (V, E)$ be a game. A finite-memory strategy for Eve in $\mathcal{G}$ is given by a transition system $\Psi = (M, m_0, \mu, \mathsf{nextmove})$, where $M$ is a finite set of memory states, $m_0$ is the initial memory state, and $\mu : M \times E \to M$ is the transition function. The function nextmove : $V_\exists \times M \to E$ is a function, such that for each vertex $v \in V_\exists$ and memory $m$, the edge nextmove$(v, m)$ is an outgoing edge from $v$.

Given the game $\mathcal{G}$ and a finite-memory strategy $\Psi$ for Eve as above, Eve plays according to $\Psi$ as follows. Initially her memory is at $m_0$. When her memory is $m$ and a play is at $v$, then

1. if $v$ is in $V_\exists$ then she picks the outgoing edge given $e = $ nextmove$(v, m)$,

2. and if $v$ is in $V_\forall$ then Adam picks an edge $e$.

Eve updates her memory according to $\mu(m, e)$.

Finite-memory strategies for Adam are defined similarly. We say a finite-memory strategy is winning for Eve (resp. Adam) if every play in which she (resp. he) plays according to her (resp. his) strategy as above is winning for Eve (resp. Adam).

**Remark 2.** *We will rarely describe strategies for the players in our games by giving a complete description of their memory, for the sake of conciseness and readability. Our description of strategies will instead be informal, but precise enough that our reader will be able to give a complete description of the memory if they wish.*

**Finite-memory determinacy.** Let us now prove the result of Theorem 2.5 that $\omega$-regular games (and hence Muller games) are finite-memory determined, i.e., in every $\omega$-regular game $\mathcal{G}$, either Eve or Adam has a finite-memory strategy that is winning. Towards this, let $\mathcal{G}$ be an $\omega$-regular game over the arena $G = (V, E)$, with a colouring function col : $C \to E$ and its winning condition given by the nondeterministic Muller automaton $\mathcal{M}$ over the alphabet $C$. Let $\mathcal{D}$ be a deterministic parity automaton that is language equivalent to $\mathcal{M}$.

Consider the parity game $\mathcal{G}'$ obtained by composing $\mathcal{D}$ with $\mathcal{G}$, similar to how we converted $\omega$-regular automaton into deterministic parity automata. More concretely, the vertices of $\mathcal{G}'$ are pairs of vertices $(v, d)$ where $v$ is a vertex in $\mathcal{G}$ and $d$ is a state in $\mathcal{D}$. We have an edge $(v, d) \xrightarrow{p} (v', d')$ in $\mathcal{G}'$ if $v \xrightarrow{c} v'$ is an edge in $\mathcal{G}$ and $d \xrightarrow{c:p} d'$ is a transition in $\mathcal{D}$. The ownership of the vertices in $\mathcal{G}'$ is inherited from $\mathcal{G}$. The initial vertex of $\mathcal{G}'$ is $(v_0, d_0)$, where $d_0$ is the initial state of $\mathcal{D}$ and $v_0$ is the initial vertex in $\mathcal{G}$.

We will show that the winner of $\mathcal{G}$ and $\mathcal{G}'$ are the same, and furthermore, the winner has a finite-memory strategy in $\mathcal{G}$ that uses their positional winning strategy in $\mathcal{G}'$ (recall from Theorem 2.1 that parity games are positionally determined).

Let us first show that Eve wins $\mathcal{G}$ if and only if she wins $\mathcal{G}'$. Note that for every play $\rho_G$ in $\mathcal{G}$, there is a unique play $\rho'_G$ in $\mathcal{G}'$, such that the first component of $\rho'_G$ is based on $\rho_G$, since the second component is determined due to $\mathcal{D}$ being a deterministic automaton. Conversely, every play $\rho'_G$ in $\mathcal{G}'$ can be converted to the play $\rho_G$ in $\mathcal{G}$ by projecting $\rho'_G$ in the $\mathcal{G}$ component. Thus, what we have just described is a bijection between plays. Further, note that $\rho_G$ is a winning play for Eve in $\mathcal{G}$ if and only if $\rho'$ is a winning play for Eve in $\mathcal{G}'$. Since ownership in $\mathcal{G}'$ is inherited from $\mathcal{G}$, it follows that Eve wins $\mathcal{G}$ if and only if Eve wins $\mathcal{G}'$. We can similarly argue that Adam wins $\mathcal{G}$ if and only if he wins $\mathcal{G}'$.

Since $\mathcal{G}'$ is determined, so is $\mathcal{G}$. If Eve (resp. Adam) wins $\mathcal{G}$ and hence $\mathcal{G}'$, then she (resp. he) has a positional winning strategy in $\mathcal{G}'$ since $\mathcal{G}'$ is a parity game. This positional strategy gives us a finite-memory winning strategy in $\mathcal{G}$, where the size of Eve's memory is the number of states in $\mathcal{D}$. Thus, if Eve (resp. Adam) wins $\mathcal{G}$, then she (resp. he) has a finite-memory winning strategy in $\mathcal{G}$. This proves Theorem 2.5.

**Theorem 2.5.** *Every Muller game or $\omega$-regular game has a winner, who has a finite-memory winning strategy [McN66, GH82].*

## 2.4 Zielonka trees

Gurevich and Harrington showed that every Muller game can be converted to a winner-equivalent parity game using a data structure that they called the latest appearance record (LAR) [GH82]. Dziembowski, Jurdziński, and Walukiewicz proposed an alternate conversion that involved Zielonka trees [DJW97], which we describe now. We will use Zielonka trees extensively in our odd-to-even induction step for the 2-token theorem (Theorem B in Chapter 7), to give a quasipolynomial time algorithm for the language inclusion problem of HD parity automata (Theorem H in Chapter 8), and to give upper bounds on the complexity of solving 2-token games in Section 3.5.2.

Let $\mathcal{M}$ be a Muller game with a $(C, \mathcal{F})$-Muller condition. We will construct a deterministic parity automaton $\mathcal{D}_{C,\mathcal{F}}$ that recognises the Muller language $L_{C,\mathcal{F}}$: that is, the set of infinite words over $C$ in which the set of colours that occur infinitely often is in $\mathcal{F}$. We will then convert $\mathcal{M}$ to a parity game $\mathcal{G}$ by taking the product of $\mathcal{M}$ with $\mathcal{D}_{C,\mathcal{F}}$: we will describe this later in this section.

The automaton $\mathcal{D}_{C,\mathcal{F}}$ is based upon the Zielonka tree of the $(C, \mathcal{F})$-Muller language, which we denote $\mathcal{Z}_{C,\mathcal{F}}$. The tree $\mathcal{Z}_{C,\mathcal{F}}$ is an ordered rooted tree, i.e., it has a designated root, and the children of every vertex in this tree are ordered from left to right. We will call the vertices of this tree as *nodes*. Each node $\nu$ of the Zielonka tree is labelled by a nonempty subset of $C$.

The root of $\mathcal{Z}_{C,\mathcal{F}}$ is labelled by $C$. For a node $\nu_X$ labelled by $X \subseteq C$, its children are nodes $\nu_Y$ labelled by distinct maximal nonempty subsets $Y$ of $X$ such that $Y \in \mathcal{F}$ if and only if $X \notin \mathcal{F}$. If there are no such subsets $Y$, then $\nu_X$ has no children in $\mathcal{Z}_{C,\mathcal{F}}$ and hence is a *leaf*.

For a $(C, \mathcal{F})$ Muller condition where $C = \{1, 2, 3, 4\}$ and

$$\mathcal{F} = \{\{1, 2, 3, 4\}, \{2, 3, 4\}, \{1, 2\}, \{2, 3\}, \{3, 4\}, \{1\}, \{2\}\},$$

the Zielonka tree $\mathcal{Z}_{C,\mathcal{F}}$ is as shown in Fig. 2.3.

In a Zielonka tree, we use depth of a node $\nu$, denoted by $\mathsf{depth}(\nu)$, to refer to the distance of $\nu$ from the tree. The depth of the root is 0, the depth of the root's children are all 1, and so on.

For the Zielonka tree $\mathcal{Z}_{C,\mathcal{F}}$ of a $(C, \mathcal{F})$-Muller condition, we let $\iota$ be 0 if $C$ is in $\mathcal{F}$, and otherwise, $\iota$ to be 1 if $C$ is not in $\mathcal{F}$. We then define the priority-depth of each node $\nu$ as

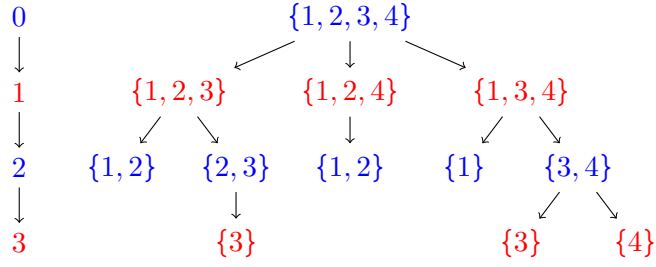

Figure 2.3: An example of a Zielonka tree

depth($\nu$) + $\iota$ and denote it by prioritydepth($\nu$). In the Zielonka tree shown in Fig. 2.3, the prioritydepth of each node is shown on the left, and it coincides with depth($\nu$) since $C$ is in $\mathcal{F}$.

Let us fix a $(C, \mathcal{F})$-Muller condition. We now describe a deterministic parity automaton $\mathcal{D}_{C,\mathcal{F}}$ that recognises $L_{C,\mathcal{F}}$, which we build using the Zielonka tree $\mathcal{Z}_{C,\mathcal{F}}$. The states of $\mathcal{D}_{C,\mathcal{F}}$ are the *branches* of $\mathcal{Z}_{C,\mathcal{F}}$; a branch is a path in $\mathcal{Z}_{C,\mathcal{F}}$ from the root to a leaf. The initial state can be any branch of $\mathcal{Z}_{C,\mathcal{F}}$.

We now describe the transitions in $\mathcal{D}_{C,\mathcal{F}}$. On the branch $\beta$ of $\mathcal{Z}_{C,\mathcal{F}}$ and the letter $c$, the outgoing transition from $\beta$ on $c$ is given via the following steps.

1. We find the node $\nu = support(\beta, c)$ in $\beta$ that has the largest prioritydepth amongst nodes whose label contains $c$. We call $\nu$ as the support of $c$ in $\beta$.

2. If $\nu$ is a leaf, then we have the transition $\beta \xrightarrow{c:d} \beta$ in $\mathcal{D}_{C,\mathcal{F}}$ where $d$ = prioritydepth($\nu$).

3. Otherwise, let $\chi$ be the child of $\nu$ in $\beta$. Recall that the children of $\nu$ are ordered from left to right. Define $rightsibling(\chi)$ to the be the leftmost child of $\nu$ that is strictly to the right of $\chi$. If $\chi$ is the rightmost child of $\nu$, then $rightsibling(\chi)$ is the leftmost child of $\nu$.

4. Let $\beta'$ be the leftmost branch in $\mathcal{Z}_{C,\mathcal{F}}$ that contains the node $rightsibling(\chi)$. Then, we have the transition $\beta \xrightarrow{c:d} \beta'$ in $\mathcal{D}_{C,\mathcal{F}}$ where $d$ = prioritydepth($\nu$) is the priority-depth of $\nu$.

This concludes our description of $\mathcal{D}_{C,\mathcal{F}}$.

**Lemma 2.7** ([CCF21, Proposition 3.9]). *For every Muller condition $(C, \mathcal{F})$ and a word $w$ in $C^\omega$, $w$ satisfies the $(C, \mathcal{F})$ Muller condition if and only if $w$ is accepted by $\mathcal{D}_{C,\mathcal{F}}$.*

The correctness of Lemma 2.7 does not depend on what we set the initial state of $\mathcal{D}_{C,\mathcal{F}}$ to be. The automaton $\mathcal{D}_{C,\mathcal{F}}$ allows us to convert a Muller game $\mathcal{M}$ with the winning condition as $(C, \mathcal{F})$-Muller condition to an equivalent parity game $\mathcal{G}$ by a product construction as follows. The vertices of $\mathcal{G}$ are given by $(m, \beta)$, where $m$ is a vertex in $\mathcal{M}$ and $\beta$ is a state in $\mathcal{D}_{C,\mathcal{F}}$. The game $\mathcal{G}$ contains the edge $(m, \beta) \xrightarrow{p} (m', \beta')$ if and only if $m \xrightarrow{c} m'$ is an edge in $\mathcal{M}$ and $\beta \xrightarrow{c:p} \beta'$ is the transition of $\mathcal{D}_{C,\mathcal{F}}$ from $\beta$ on $c$. The ownership of vertices in $\mathcal{G}$ is inherited from $\mathcal{M}$.

**Lemma 2.8.** *For games $\mathcal{M}$ and $\mathcal{G}$ as above, the following conditions are equivalent.*

1. *Eve wins from $m$ in $\mathcal{M}$.*

2. *Eve wins from $(m, \beta)$ in $\mathcal{G}$ for some branch $\beta$.*

3. *Eve wins from $(m, \beta)$ in $\mathcal{G}$ for all branches $\beta$.*

## 2.5 Ranks in parity games

Ranks of parity games are witnesses for winning strategies in parity games. We will use the concept of ranks in Chapter 5 and Chapter 7.

Büchi introduced ranks in 1983 for a more general class of game [Büc83], and they have been dubbed as signatures [Wal02] and progress measures [Jur00, JL17] over the years. Kuperberg and Skrzypczak called them ranks to give a determinisation procedure for history-deterministic Büchi automata [KS15], and we follow their precedent. We only require a restricted version of ranks for our purposes.

For a $[0, d]$ parity game $\mathcal{G}$, where $d \geq 1$ is a natural number, we define $\mathsf{rank}(v)$ of a vertex $v$ as the largest number $k \in \mathbb{N} \cup \{\infty\}$, such that Adam has a strategy $\sigma$ in $\mathcal{G}$ from $v$ that satisfies the following condition: in every play $\rho$ starting from $v$ where Adam is playing according to $\sigma$, at least $k$ many edges of priority 1 occur in $\rho$ before an edge of priority 0.

It is clear that if Eve wins the game $\mathcal{G}$ from everywhere (i.e., from all vertices), then $\mathsf{rank}(v)$ is bounded for all vertices in $\mathcal{G}$; otherwise, Adam has a strategy to ensure that the lowest priority occurring infinitely often is 1. Furthermore, the next result shows that there is a uniform positional winning strategy $\tau$ in $\mathcal{G}$ for Eve that ensures that every play from $v$ does not contain more than $\mathsf{rank}(v)$ edges of priority 1 before an edge of priority 0 occurs.

**Lemma 2.9** ([Wal02, Lemma 8])**.** *Let $\mathcal{G}$ be a $[0, d]$ parity game such that Eve wins from all vertices in $\mathcal{G}$. Then there is a uniform positional winning strategy $\tau$ for Eve in $\mathcal{G}$ that ensures the following: in every play from a vertex $v$ where Eve is playing according to her strategy $\tau$, at most $\mathsf{rank}(v)$ many edges of priority $1$ occur before an edge of priority $0$ occurs.*

We will call the strategy $\tau$ appearing in the above lemma as an *optimal strategy* of $\mathcal{G}$. Let us observe that ranks, in a sense, are monotonic.

**Proposition 2.10** (Monotonicity of ranks)**.** *Let $\mathcal{G}$ be a $[0, d]$ parity game where Eve wins from everywhere, and fix $\tau$ to be an optimal strategy in $\tau$. Then, for each vertex $u$ in $\mathcal{G}$, the following holds.*

1. *If $u \in V_\exists$ and $\tau(u) = u \xrightarrow{c} v$, then either $c = 0$, or $\mathsf{rank}(u) \geq \mathsf{rank}(v)$. This inequality is strict if $c = 1$.*

2. *If $u \in V_\forall$, then for all edges $u \xrightarrow{c} v$, either $c = 0$ or $\mathsf{rank}(u) \geq \mathsf{rank}(v)$. This inequality is strict if $c = 1$.*

Let $\mathcal{G}$ be a $[0, d]$ parity game on which Eve wins from everywhere. Then, we observe that $\mathsf{rank}(v)$ for each vertex $v$ is at most $|V|$, where $|V|$ is the number of vertices in $\mathcal{G}$.

**Proposition 2.11.** *In every parity game $\mathcal{G}$ where Eve wins from everywhere, the ranks of vertices in $\mathcal{G}$ are at most the number of vertices in $\mathcal{G}$.*

To prove Proposition 2.11, we let $\tau$ be an optimal positional strategy for Eve in $\mathcal{G}$. Then, any finite play $\rho$ where Eve is playing according to $\tau$ which starts and ends at the same vertex cannot have an edge of priority 1 but no edge of priority 0; otherwise, Adam can repeat his moves to build the play $\rho^\omega$ where Eve is playing according to $\tau$ but the lowest priority occurring infinitely often is 1.

The claim then follows from pigeonhole principle: if Adam has a strategy to witness more than $|V|$ many priority 1 edges from some vertex $v$ before a priority 0 edge then there is a finite play $\rho$, such that Eve is playing according to $\tau$ in $\rho$ and $\rho$ has a cycle consisting of a priority 1 edge but no priority 0 edge, which is a contradiction.

# Chapter 3

# History-Deterministic Automata and Token Games

In this chapter, we introduce history-deterministic automata, token games, and prove several results that we will use throughout this thesis. This includes the proof of Theorem I.

**Theorem I.** *Let $\mathcal{A}$ be a parity automaton on which Eve wins the $2$-token game. Then, there is a simulation-equivalent subautomaton $\mathcal{B}$ of $\mathcal{A}$ such that the following three conditions hold.*

1. *For each $k \geq 1$, Eve wins the $k$-token game from everywhere in $\mathcal{B}$.*

2. *For states $q, p_1, p_2, \ldots, p_k$ that are weakly coreachable in $\mathcal{B}$, Eve wins the $k$-token game from $(q; p_1, p_2, \ldots, p_k)$ in $\mathcal{B}$.*

3. *If $\mathcal{B}$ is history-deterministic, then so is $\mathcal{A}$.*

We start by introducing history-deterministic automata. We then introduce guidable automata in Section 3.1, and show that guidability and history-determinism are equivalent notions over parity automata (Theorem 3.9). We then introduce token games and Joker games in Section 3.2, which are games that have been used to give efficient algorithms to check history-determinism for Büchi and coBüchi automata [KS15, BK18, BKLS20]. We then prove Theorem I in Section 3.3, and prove an analogous result for Joker games in Section 3.4. We end this chapter with an analysis of the computational complexity of solving token games in Section 3.5.2.

History-deterministic (HD) automata are nondeterministic automata in which the nondeterminism that occurs while reading a word can be successfully resolved 'on-the-fly', just based on the word read so far. More concretely, history-determinism of an automaton is characterised by Eve winning the following history-determinism game.

**History-determinism game.** The history-determinism game (or HD game) of an automaton is a two-player turn-based game between Adam and Eve, who take alternating turns to select a letter and a transition in the automaton (on that letter), respectively. In the limit of a play of the HD game, the sequence of Adam's choices of letters constitutes an infinite word, and the sequence of Eve's choices of transitions constitutes a run on that word. Eve wins the game if her run is accepting or Adam's word is rejected.

**Definition 3.1** (History-determinism game). *Given a nondeterministic parity automaton $\mathcal{A} = (Q, \Sigma, \Delta, q_0)$, the* history-determinism game *on $\mathcal{A}$ is a two-player game between Eve and Adam that starts with an Eve's token at $q_0$ and proceeds for infinitely many rounds. For each $i \in \mathbb{N}$, round $i$ starts with Eve's token at a state $q_i$ in $Q$, and proceeds as follows.*

1. *Adam selects a letter $a_i \in \Sigma$;*
2. *Eve selects a transition $q_i \xrightarrow{a_i : p_i} q_{i+1} \in \Delta$ along which she moves her token. Eve's token then is at $q_{i+1}$ from where the round $(i+1)$ is played.*

*Thus, in the limit of a play of the HD game, Adam constructs a word letter-by-letter, and Eve constructs a run on her token transition-by-transition on that word. Eve wins such a play if the following condition holds: if Adam's word is in $L(\mathcal{A})$, then the run on Eve's token is accepting.*

We say that an automaton is history-deterministic (HD) if Eve has a winning strategy in the HD game on $\mathcal{A}$. We will call a winning strategy for Eve in the HD game on $\mathcal{A}$ as a *resolver*.

**Remark 3.** *We will focus primarily on parity automata and its subclasses in the rest of this thesis, so we focus our definitions of history-determinism games, and the definitions of various other games on automata we discuss in this chapter on parity automata. But these definitions can be lifted to define similar games on $\omega$-regular automata or Muller automata without much difficulty.*

## 3.1 Guidable automata

On parity automata, the notion of history-determinism is equivalent to the notions of good-for-trees [BKKS13, Proposition 1] and good-for-gameness [HP06, Theorem 3]. Another equivalent notion that will be useful for us is that of guidability, which was originally introduced by Colcombet and Löding on parity tree automata [CL08].

Guidable automata are automata for which language containment can be reduced to a simpler problem of deciding *simulation*. While language inclusion for nondeterministic parity automata is a computationally hard problem (PSPACE-complete) [KV98], simulation [HKR02] is a coarser relation that can be checked in PTIME if the parity index of automata is fixed, and is in NP otherwise.

For two parity automata $\mathcal{A}$ and $\mathcal{B}$, simulation is characterised by the following 2-player game that we call the *simulation game*.

**Definition 3.2** (Simulation game). *Given two parity automata $\mathcal{A} = (P, \Sigma, p_0, \Delta_A)$ and $\mathcal{B} = (Q, \Sigma, q_0, \Delta_B)$, the* simulation game *of $\mathcal{A}$ by $\mathcal{B}$ is a two-player game played between Eve and Adam as follows, with an Eve's token initially at $q_0$ and Adam's token initially at $p_0$. The game proceeds for infinitely many rounds, and in round $i$ when Eve's token is at $q_i$ and Adam's token is at $p_i$:*

1. *Adam selects a letter $a_i \in \Sigma$;*
2. *Adam moves his token along a transition $p_i \xrightarrow{a_i} p_{i+1}$ in $\mathcal{A}$;*
3. *Eve moves her token along a transition $q_i \xrightarrow{a_i} q_{i+1}$ in $\mathcal{B}$.*

*In the limit of a play of the simulation game, the letters selected by Adam in sequence form a word, while the sequence of Adam's selected transitions and the sequence of Eve's selected transitions form runs on that word in $\mathcal{A}$ and $\mathcal{B}$, respectively. We say Eve wins such a play if the following holds: if the run of Adam's token is accepting, then the run of Eve's token is accepting as well.*

If Eve has a strategy to win the above game, then we say that Eve wins the simulation game of $\mathcal{A}$ by $\mathcal{B}$, and that $\mathcal{B}$ *simulates* $\mathcal{A}$. For convenience, we will sometimes write $\mathsf{Sim}(\mathcal{B}, \mathcal{A})$ to denote the simulation game of $\mathcal{A}$ by $\mathcal{B}$.

We note that simulations game can be explicitly represented as a $(\pi_1 \Rightarrow \pi_2)$ implication game (see Section 2.1), where priorities of $\pi_1$ corresponds to the priorities of transitions that Adam's token takes, while priorities of $\pi_2$ corresponds to the priorities of transitions that Eve's token takes. Thus, simulation games, like implication games, are finite-memory determined, and if Eve wins the simulation game of $\mathcal{A}$ by $\mathcal{B}$, then she has a positional winning strategy in that game (Lemma 2.3).

**Proposition 3.3.** *For every two parity automata $\mathcal{A}$ and $\mathcal{B}$, either Eve or Adam wins $\mathsf{Sim}(\mathcal{B}, \mathcal{A})$, and the winner has a finite-memory winning strategy. Additionally, if Eve wins $\mathsf{Sim}(\mathcal{B}, \mathcal{A})$ then she has a positional winning strategy.*

Since deciding if Eve wins a given implication game is in NP (Lemma 2.3), we get that the same also holds for checking simulation.

**Proposition 3.4.** *Given two parity automata $\mathcal{A}$ and $\mathcal{B}$, deciding if $\mathcal{B}$ simulates $\mathcal{A}$, or equivalently, deciding if Eve wins the simulation game of $\mathcal{A}$ by $\mathcal{B}$, is in* NP.

Let us note that simulation is a coarser relation than language containment.

**Lemma 3.5.** *For every two parity automata $\mathcal{A}$ and $\mathcal{B}$, if $\mathcal{B}$ simulates $\mathcal{A}$ then $L(\mathcal{A}) \subseteq L(\mathcal{B})$.*

*Proof.* Let $\sigma$ be a winning strategy for Eve in the simulation game of $\mathcal{A}$ by $\mathcal{B}$. Let $w$ be a word in $L(\mathcal{A})$, and let $\rho_A$ be an accepting run over $w$ in $\mathcal{A}$. Consider the play of the simulation game where Adam builds the word $w$ and run $\rho_A$ on his token, while Eve plays according to $\sigma$. Then, Eve builds an accepting run $\rho_B$ on her token, implying $w$ is a word in $L(\mathcal{B})$. □

Note that the relation of simulation is reflexive, that is, every parity automaton simulates itself; for a parity automaton $\mathcal{A}$, Eve wins $\mathsf{Sim}(\mathcal{A}, \mathcal{A})$ by simply copying the transitions of Adam's token on her token. We next show that the relation of simulation is also transitive.

**Lemma 3.6.** *For every three parity automata $\mathcal{A}$, $\mathcal{B}$, and $\mathcal{C}$, if $\mathcal{B}$ simulates $\mathcal{A}$ and $\mathcal{C}$ simulates $\mathcal{B}$, then $\mathcal{C}$ simulates $\mathcal{A}$.*

*Proof.* Let $\sigma_{BA}$ be a winning strategy for Eve in $\mathsf{Sim}(\mathcal{B}, \mathcal{A})$, and $\sigma_{CB}$ be a winning strategy for Eve in $\mathsf{Sim}(\mathcal{C}, \mathcal{B})$. We describe a winning strategy $\sigma_{CA}$ for Eve in $\mathsf{Sim}(\mathcal{C}, \mathcal{A})$ as follows. Eve stores in her memory, a token that takes transitions in $\mathcal{B}$ and builds run on Adam's word in $\mathsf{Sim}(\mathcal{C}, \mathcal{A})$ simultaneously. In $\mathsf{Sim}(\mathcal{C}, \mathcal{A})$, she will choose transitions on her memory token in $\mathcal{B}$ according to her strategy $\sigma_{BA}$ against Adam's token in $\mathcal{A}$, and choose transitions on her token in $\mathcal{C}$ against her memory token via $\sigma_{CB}$. Then, if the run on Adam's token in $\mathcal{A}$ is accepting, the run produced by Eve's memory token in $\mathcal{B}$ is accepting since $\sigma_{BA}$ is a winning strategy for Eve in $\mathsf{Sim}(\mathcal{B}, \mathcal{A})$. Since $\sigma_{CB}$ is a winning strategy for Eve in $\mathsf{Sim}(\mathcal{C}, \mathcal{B})$, it follows that the Eve's run on her token in $\mathcal{C}$ is accepting as well. □

We now define guidable automata.

**Definition 3.7.** *A nondeterministic parity automaton $\mathcal{A}$ is said to be guidable if for all automata $\mathcal{B}$ with $L(\mathcal{B}) \subseteq L(\mathcal{A})$, $\mathcal{A}$ simulates $\mathcal{B}$.*

Note that every deterministic parity automaton $\mathcal{D}$ is guidable. Indeed, if $L(\mathcal{A}) \subseteq L(\mathcal{D})$ for some nondeterministic parity automaton $\mathcal{A}$, then Eve wins $\mathsf{Sim}(\mathcal{D}, \mathcal{A})$ by picking the unique transition on her token that is available to her. In the next two results, we will show that for every parity automaton $\mathcal{A}$, $\mathcal{A}$ is history-deterministic if and only if $\mathcal{A}$ is guidable.

**Lemma 3.8.** *For every history-deterministic parity automaton $\mathcal{H}$ and nondeterministic parity automaton $\mathcal{A}$, Eve wins $\mathsf{Sim}(\mathcal{H}, \mathcal{A})$ if and only if $L(\mathcal{A}) \subseteq L(\mathcal{H})$.*

*Proof.* If $\mathcal{H}$ simulates $\mathcal{A}$ then it is clear from Lemma 3.5 that $L(\mathcal{A}) \subseteq L(\mathcal{H})$. For the other direction, suppose that $L(\mathcal{A}) \subseteq L(\mathcal{H})$ and fix a winning strategy $\sigma_{HD}$ for Eve in the HD game on $\mathcal{H}$. Then Eve has the following winning strategy in $\mathsf{Sim}(\mathcal{H}, \mathcal{A})$, where she picks transitions on her token according to $\sigma_{HD}$ and Adam's letters, ignoring what Adam does on his token that is building a run in $\mathcal{A}$. If Adam's run on his token in $\mathsf{Sim}(\mathcal{H}, \mathcal{A})$ is accepting and is on the word $w$, then $w \in L(\mathcal{A}) \subseteq L(\mathcal{H})$, and thus Eve's run on her token is accepting since $\sigma_{HD}$ constructs an accepting run on every word in $L(\mathcal{A})$. □

Lemma 3.8 shows that history-deterministic automata are guidable. The next result show that the converse is also true.

**Theorem 3.9.** *For every parity automaton $\mathcal{A}$, $\mathcal{A}$ is history-deterministic if and only if $\mathcal{A}$ is guidable.*

*Proof.* We only need to show that guidable automata are HD, since the other direction follows from Lemma 3.8. Let $\mathcal{A}$ be a guidable automaton and $\mathcal{D}$ be a deterministic parity automaton that is language-equivalent to $\mathcal{A}$. Since $\mathcal{A}$ is guidable, $\mathcal{A}$ simulates $\mathcal{D}$.

Let $\sigma$ be a winning strategy for Eve in $\mathsf{Sim}(\mathcal{A}, \mathcal{D})$. We will use $\sigma$ to construct a winning strategy for in the HD game on $\mathcal{A}$ as follows. During the HD game on $\mathcal{A}$, Eve keeps a corresponding play of the simulation game of $\mathcal{D}$ by $\mathcal{A}$ in her memory, where Adam is playing the same letters as the HD game on $\mathcal{A}$ and choosing the unique transitions on his token in $\mathcal{D}$. Then Eve chooses transitions on her token in the HD game by playing in $\mathsf{Sim}(\mathcal{A}, \mathcal{D})$ against Adam's token in $\mathcal{D}$ using $\sigma$. This way, whenever Adam's word $w$ in the HD game is in $L(\mathcal{A})$, then the unique run in $\mathcal{D}$ on $w$ is accepting, and therefore Eve's run in the HD game on $\mathcal{A}$ must be accepting as well. □

Let us make explicit a few corollaries of Theorem 3.9.

**Corollary 3.10.** *If $\mathcal{A}$ is a nondeterministic parity automaton that simulates a language-equivalent history-deterministic automaton $\mathcal{H}$, then $\mathcal{A}$ is history-deterministic.*

*Proof.* Since simulation is transitive (Lemma 3.6) and all HD automata are guidable (Lemma 3.8), we note that if $\mathcal{A}$ simulates $\mathcal{H}$ and $\mathcal{A}$ is language-equivalent to $\mathcal{H}$ then $\mathcal{A}$ is guidable as well. It then follows from the equivalence of HD automata and guidable automata (Theorem 3.9) that $\mathcal{A}$ is HD. □

We say that two parity automata $\mathcal{A}$ and $\mathcal{B}$ are simulation-equivalent if $\mathcal{A}$ simulates $\mathcal{B}$ and $\mathcal{B}$ simulates $\mathcal{A}$.

**Corollary 3.11.** *If two parity automata $\mathcal{A}$ and $\mathcal{B}$ are simulation-equivalent, then $\mathcal{A}$ is HD if and only if $\mathcal{B}$ is HD.*

*Proof.* If $\mathcal{A}$ and $\mathcal{B}$ are simulation-equivalent, then they are language-equivalent (see Lemma 3.5). The proof then follows from Corollary 3.10. □

We end this section by showing finite-memory determinacy of HD games.

**Theorem 3.12.** *For every nondeterministic parity automaton $\mathcal{A}$, the history-determinism game on $\mathcal{A}$ has a winner and the winner has a finite-memory winning strategy.*

*Proof.* Let $\mathcal{D}$ be a deterministic parity automaton that is language-equivalent to $\mathcal{A}$. Consider the game $\mathsf{Sim}(\mathcal{A}, \mathcal{D})$, which we know is finite-memory determined (Proposition 3.3). We will argue that $\mathsf{Sim}(\mathcal{A}, \mathcal{D})$ and the HD game on $\mathcal{A}$ have the same winner, and in the process, show finite-memory determinacy of the HD game on $\mathcal{A}$.

Suppose that Eve wins $\mathsf{Sim}(\mathcal{A}, \mathcal{D})$, and let $\sigma_\exists$ be a positional winning strategy for Eve in $\mathsf{Sim}(\mathcal{A}, \mathcal{D})$; recall that such a strategy exists due to Proposition 3.3. Note that Adam's move on his token in $\mathsf{Sim}(\mathcal{A}, \mathcal{D})$ is determined by the choice of his letters since $\mathcal{D}$ is deterministic. Eve then has the following strategy $\sigma_{HD}$ in the HD game on $\mathcal{A}$ (similar to in proof of Theorem 3.9), where she keeps a corresponding play of the simulation game of $\mathcal{D}$ by $\mathcal{A}$ in her memory, where Adam is picking the same letters as in the HD game on $\mathcal{A}$ and choosing unique transitions on his token in $\mathcal{D}$. Then, Eve chooses a transition on her token in the HD game on $\mathcal{A}$ using $\sigma_\exists$. The strategy $\sigma_{HD}$ is a finite-memory strategy for Eve in the HD game on $\mathcal{A}$, where the size of memory is the size of $\mathcal{D}$.

This strategy is also winning for Eve, since whenever Adam's word $w$ is in $L(\mathcal{A}) = L(\mathcal{D})$, then the unique run of Adam's token in $\mathcal{D}$ on $w$ is accepting. Therefore Eve's run on her token in the $Sim(\mathcal{A}, \mathcal{D})$ in her memory, and hence her token's run in the HD game on $\mathcal{A}$ is accepting, as desired.

For the other direction, suppose that Adam has a finite-memory winning strategy $\sigma_\forall$ in $\mathsf{Sim}(\mathcal{A}, \mathcal{D})$. Adam can then win the HD game using the following strategy: he keeps in his memory a play of $\mathsf{Sim}(\mathcal{A}, \mathcal{D})$, and selects letters in the HD game and transition on his token in $\mathcal{D}$ (which is determined based on his letter) using his strategy in $\mathsf{Sim}(\mathcal{A}, \mathcal{D})$ against the run of Eve's token in the HD game on $\mathcal{A}$. This strategy is clearly a finite-memory strategy for Adam. It is also a winning strategy, since in every play resulting due to this strategy, the run of Eve's token in $\mathcal{A}$ will be rejecting, while Adam's word $w$ is in $L(\mathcal{A}) = L(\mathcal{D})$. Thus, either Eve or Adam wins the HD game on $\mathcal{A}$, and the winner has a finite-memory winning strategy in that game. □

The proof of Theorem 3.12 shows that to decide if a given parity automaton $\mathcal{A}$ is history-deterministic, it suffices to find the winner of a simulation game of a language-equivalent deterministic automaton by $\mathcal{A}$. Since determinisation takes exponential time, however, this procedure also takes exponential time. The games we present next have the objective of avoiding the bottleneck of determinisation.

## 3.2 Games used to check for history-determinism

### 3.2.1 Token games.

In 2018, Bagnol and Kuperberg introduced 2-token games and the conjecture that Eve wins the 2-token game on an automaton if and only if that automaton is HD. The 2-token games are similar to HD games, with Adam constructing a word letter-by-letter and Eve constructing a run on her token transition-by-transition, but in addition, Adam has two tokens, on which he is also constructing runs transition-by-transition. Eve wins a play of the 2-token game if the run on her token is accepting or both of the runs on Adam's tokens are rejecting.

**Definition 3.13** (2-token game). *The* 2-token game *on a nondeterministic parity automaton $\mathcal{A} = (Q, \Sigma, \Delta, q_0)$ is a two-player game that is played between the players Adam and Eve as follows, with an Eve's token and Adam's two tokens all initially at $q_0$. The game proceeds for infinitely many rounds, and in round $i$ where Eve's token is at $q_i$ and Adam's tokens are at $p_i^1$ and $p_i^2$:*

1. *Adam selects a letter $a_i \in \Sigma$;*
2. *Eve moves her token along a transition $q_i \xrightarrow{a_i : c_i} q_{i+1}$;*

3. Adam moves his first and second token along transitions $p_i^1 \xrightarrow{a_i:c_i^1} p_{i+1}^1$ and $p_i^2 \xrightarrow{a_i:c_i^2} p_{i+1}^2$ respectively.

*Thus, in a play of the 2-token game on $\mathcal{A}$, Eve constructs a run on her token and Adam constructs a run on each of his two tokens, all on the same word. Eve wins such a play if the following condition holds: if at least one of the runs of Adam's tokens is accepting, then the run on Eve's token is accepting. Otherwise, Adam wins that play.*

We say that Eve (resp. Adam) wins the 2-token game on $\mathcal{A}$ is she (resp. he) has a winning strategy in the above game. Let us note that 2-token games can be explicitly represented as Muller games. If the priorities of an automaton $\mathcal{A}$ are in the interval $[i, j]$, then the colours of the 2-token game represented as a Muller game are given by $C = [i, j] \times [i, j] \times [i, j]$, where the first component is for priorities occurring in the run of Eve's token, while the second and third components are for priorities occurring in the run of Adam's first and second tokens, respectively. The winning condition of the 2-token game on $\mathcal{A}$ can then be given by $(C, \mathcal{F})$, where $\mathcal{F}$ consists of nonempty subsets $S$ of $C$ in which the following holds: if the least priority occurring amongst the second component in elements of $S$ is even, or if the least priority occurring amongst the third component in elements of $S$ is even, then the least priority occurring amongst the first component in elements of $S$ is even as well.

Thus, since Muller games are finite-memory determined (Theorem 2.5), 2-token games are finite-memory determined as well. For every $k \geq 1$, we can define the $k$-token game similar to how we have defined the 2-token game, and argue that they are finite-memory determined as well.

**Proposition 3.14.** *For every parity automaton $\mathcal{A}$ and for all $k \geq 1$, the $k$-token game on $\mathcal{A}$ is finite-memory determined.*

Bagnol and Kuperberg showed the elegant result that if Eve wins the 2-token game on a parity automaton $\mathcal{A}$ then she wins the $k$-token game on $\mathcal{A}$ for all $k \geq 1$. That is, having more than 2 tokens in the token game does not give any additional power to Adam.

**Lemma 3.15** ([BK18]). *For every $k \geq 2$ and parity automaton $\mathcal{A}$, Eve wins the 2-token game on $\mathcal{A}$ if and only if Eve wins the $k$-token game on $\mathcal{A}$.*

*Proof.* Note that if Eve wins the $k$-token game on $\mathcal{A}$ for some $k \geq 2$, then Eve also wins the 2-token game on $\mathcal{A}$: Eve can use her winning strategy in the $k$-token game to play in the 2-token game against 2 tokens of Adam and $(k-2)$ tokens that are copying Eve's token. For the converse, we show that if Eve wins the 2-token game then she wins the 3-token game, using which our reader should be able to prove the lemma for every $k > 2$. Fix $\sigma_2$ to be a winning strategy for Eve in the 2-token game on $\mathcal{A}$. Eve then in the 3-token game on $\mathcal{A}$, stores in her memory an additional token, in which she plays the 2-token game using $\sigma_2$ against Adam's second and third token. Eve chooses transitions on her token by playing the 2-token game using $\sigma_2$ against Adam's first token and her memory token. If either of Adam's tokens produces an accepting run, then either Adam's first token or Eve's memory token produces an accepting run, and hence Eve's token produces an accepting run. □

In the above proof, we argued that if Eve wins the $k$-token game on an automaton, for some $k \geq 2$, then Eve wins the 2-token game on that automaton. We can similarly argue that if Eve wins the 2-token game on an automaton, then she also wins the 1-token game on that automaton. The converse is not true, however. The automaton in Fig. 1.2 that we re-illustrate in Fig. 3.1 below is an example that shows this.

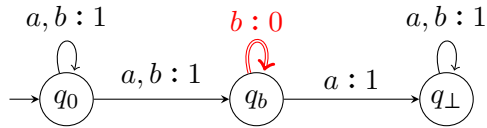

Figure 3.1: An automaton on which Eve wins the 1-token game but not the 2-token game

**Lemma 3.16.** *There is a parity automaton $\mathcal{A}$ on which Eve wins the 1-token game but not the 2-token game.*

*Proof.* Consider the Büchi automaton $\mathcal{A}$ shown in Fig. 3.1. We will show that Eve wins the 1-token game on $\mathcal{A}$ and Adam wins the 2-token game on $\mathcal{A}$.

To prove that Eve wins the 1-token game on $\mathcal{A}$, we consider the following strategy for Eve.

1. Eve takes the self-loops on her token until Adam's token is in the state $q_b$. If Adam's token never moves to $q_b$, then the run on Adam's token is rejecting and hence Eve wins the 1-token game.

2. When Adam moves his token to $q_b$, Eve in the next round moves her token to $q_b$.

It is easy to see then that if Adam's run on his token is accepting, then Eve's run on her token is accepting as well. Thus, we have described a winning strategy for Eve.

To see why Adam wins the 2-token game on $\mathcal{A}$, consider the following winning strategy for Adam in the 2-token game on $\mathcal{A}$.

1. Until Eve moves her token to the state $q_b$, Adam selects the letter $b$ in each round. Meanwhile, Adam moves his first token to $q_b$ in the first round, while his second token takes self loops on $b$ in $q_0$. If Eve's token never takes a transition to $q_b$, then the run of Eve's token is rejecting while the run of Adam's first token is accepting.

2. Otherwise, when Eve's token moves to the state $q_b$, Adam selects the letter $a$ in the next round, which forces Eve's token to the rejecting sink state $q_\perp$. Adam also moves his second token to the state $q_b$ in this round. For the rest of the play, Adam picks the letters $b$. This way, the run on Adam's second token is accepting while Eve's run on her token is rejecting.

□

Note that the 1-token games on parity automata can be explicitly represented as $(\pi_1 \Rightarrow \pi_2)$ implication game, where priorities of $\pi_1$ correspond to priorities of transition that Adam's token takes, while priorities of $\pi_2$ corresponds to priorities of transitions that Eve's token takes. Thus, the results of implication games (Lemma 2.3) also extend to 1-token games on parity automata.

**Proposition 3.17.** *For every parity automaton $\mathcal{A}$, if Eve wins the 1-token game on $\mathcal{A}$ then she wins using a positional winning strategy. Furthermore, the problem of deciding if Eve wins the 1-token game on $\mathcal{A}$ is in* NP.

### 3.2.2 Joker games

Kuperberg and Skrzypczak introduced Joker games in 2015 to give a PTIME algorithm for checking history-determinism of coBüchi automata. These are similar to 1-token games, but additionally in each round, Adam can choose to play *Joker* and choose, on his token, a transition from Eve's token's penultimate position on. The winning condition for Eve is the following: if Adam's sequence of transitions on his token satisfies the parity conditions and Adam has played finitely many Jokers, then Eve's run on her token is accepting as well.

**Definition 3.18** (Joker games). *The* Joker game *on a parity automaton $\mathcal{A} = (Q, \Sigma, q_0, \Delta)$ starts with an Eve's token and Adam's token both initially at $q_0$. The game proceeds in infinitely many rounds, and in round $i$, where Eve's token is at $q_i$ and Adam's token is at $p_i$:*

1. *Adam selects a letter $a_i \in \Sigma$;*

2. *Eve moves her token along a transition $q_i \xrightarrow{a_i:c_i} q_{i+1}$;*

3. *Adam either moves his token along a transition $p_i \xrightarrow{a_i:c'_i} p_{i+1}$, or plays Joker and selects a transition $q_i \xrightarrow{a_i:c''_i} p_{i+1}$, and places his token at $p_{i+1}$.*

*The new position is $(q_{i+1}, p_{i+1})$, from where round $(i+1)$ begins.*

*Eve wins a play of the Joker game if the following condition holds: if Adam plays finitely many Jokers and his sequence of transitions on his token satisfies the parity condition, then Eve's run on her token is accepting.*

We say that Eve (resp. Adam) wins the Joker game on a parity automaton $\mathcal{A}$ if she (resp. he) has a winning strategy in the Joker game on $\mathcal{A}$.

Joker games, like 1-token games, can be explicitly represented as implication games too. For the Joker game on a parity automaton $\mathcal{A}$, assume that the priorities of $\mathcal{A}$ are in $[1, d]$ for some natural number $d > 1$: if they are in $[0, d']$ for some $d' > 0$, then we can increase every priority in $\mathcal{A}$ by 2, and then all priorities occur in $[1, d' + 2]$.

We can then give an explicit representation of the Joker game on $\mathcal{A}$ as a $(\pi_1 \Rightarrow \pi_2)$ implication game, where the labelling $\pi_1$ of edges in this game corresponds to priorities of Adam's token's transitions, and the Joker moves of Adam get priority 1 from $\pi_1$. The priorities given by $\pi_2$ correspond to the priorities of Eve's token's transitions. Thus, Proposition 3.19 below follows from Lemma 2.3 on implication games.

**Proposition 3.19.** *Joker games are finite-memory determined. Additionally, for every parity automaton $\mathcal{A}$, if Eve wins the Joker game on $\mathcal{A}$ then she has a positional winning strategy.*

We proceed with showing that Joker games are easier for Eve than 2-token games.

**Theorem 3.20** (Folklore). *For every parity automaton $\mathcal{A}$, if Eve wins the 2-token game on $\mathcal{A}$ then she wins the Joker game on $\mathcal{A}$.*

*Proof.* We prove that if Adam wins the Joker game on $\mathcal{A}$, then he also wins the 2-token game on $\mathcal{A}$. Note that this is the contrapositive of what we want to show due to finite-memory determinacy of both Joker games and 2-token games. Towards this, fix a finite-memory winning strategy $\sigma_J$ for Adam in the Joker game on $\mathcal{A}$.

We will that there is a natural number $K$ such that every play following $\sigma_J$ contains at most $K$ Jokers. To see this, let $K$ be 1 more than the product of the size of Adam's memory and the size of the arena of the Joker game on $\mathcal{A}$. Then the claim follows from pigeonhole principle; observe that any finite play in which Adam is playing according to $\sigma_J$ cannot have repetitions of Adam's memory and game's position with a Joker move occurring in the middle, since otherwise that segment can be repeated to yield a play following $\sigma_J$ that contains infinitely many Jokers and is hence losing for Adam. This contradicts the fact that $\sigma_J$ is a winning strategy for Adam.

For $K$ as above, we will show that Adam wins the $(K+1)$-token game on $\mathcal{A}$ by describing a winning strategy for him using his strategy $\sigma_J$. Adam keeps in his memory a play of the Joker game on $\mathcal{A}$ in which he is playing according to $\sigma_J$. He will select letters in the $(K+1)$-token game by selecting letters according to $\sigma_J$ against Eve's token. At the start of the play of the $(K+1)$-token game, his first token follows his token in the Joker game in his memory, while the rest of the tokens in his memory follow Eve's token.

In general, after he has played $i$ Jokers in the Joker game, Adam's $(i+1)^{th}$ token in the $(K+1)$-token game will copy the moves of Adam's token in the Joker game. His $(i+2)^{th}, (i+3)^{th}, \ldots$ tokens will copy Eve's token, while the rest of his tokens will take arbitrary letter. Adam will pick letters according to his strategy $\sigma_J$ in the Joker game against Eve's token that he has in his memory. When Adam plays a Joker in the Joker game in his memory, Adam starts copying the moves of his token in the Joker game in his $(i+2)^{th}$ token. Since Adam plays at most $K$ Jokers in the Joker game when playing according to $\sigma$, the run of one of Adam's tokens is eventually the same as the run of Adam's token in the Joker game in his memory. Thus, we have described a winning strategy for Adam in the $(K+1)$-token game on $\mathcal{A}$. By Lemma 3.15, we obtain that Adam wins the 2-token game on $\mathcal{A}$, as desired. □

Let us make the simple observation that if $\mathcal{A}$ is history-deterministic then Eve wins the 2-token game on $\mathcal{A}$ and the Joker game on $\mathcal{A}$.

**Lemma 3.21.** *Eve wins the Joker game and the 2-token game on every parity automaton that is history-deterministic.*

*Proof.* Let $\mathcal{A}$ be an HD parity automaton, and fix a winning strategy $\sigma$ for Eve in the HD game of $\mathcal{A}$. Consider the following strategy for Eve in the 2-token game and the Joker game on $\mathcal{A}$, in which she picks transition on her token according to $\sigma$ against the letters that Adam chooses, ignoring the moves of his tokens. Then Eve constructs an accepting run on her token whenever the word constructed by Adam is accepting. In particular, if Adam constructs an accepting run on one of his tokens in the 2-token game on $\mathcal{A}$, then his word must be in $L(\mathcal{A})$, implying that Eve's run on her token is accepting as well.

Similarly, in a play of the Joker game, suppose Adam plays finitely many Jokers and his token's sequence of transitions satisfies the parity condition. Let $i$ be the last round where Adam played a Joker. Then there is an accepting run on Adam's word, which can be obtained by concatenating Eve's run on her token until round $i-1$ with Adam's run on his token from round $i$. This implies that Adam's word is in $L(\mathcal{A})$, once again implying that Eve's run on her token is accepting. □

### 3.2.3 Safety and reachability automata

While Eve winning the 1-token game on a parity automaton does not imply that the automaton is history-deterministic (Lemma 3.16), it is known that this is the case for safety and reachability automata ([BL23a]). In fact, a safety or reachability automaton is history-deterministic if and only if it is determinisable-by-pruning, that is it has a language-equivalent deterministic subautomaton ([BKS17]).

**Theorem 3.22** ([BKS17, Theorem 17],[BL23a, Theorem 4.5 and 4.8]). *Let $\mathcal{A}$ be a safety or reachability automaton on which Eve wins the 1-token game. Then $\mathcal{A}$ is history-deterministic, and a positional winning strategy for Eve in the HD game on $\mathcal{A}$ can be computed in* PTIME. *Equivalently, $\mathcal{A}$ has a language-equivalent deterministic subautomaton $\mathcal{D}$, which can be constructed in* PTIME.

## 3.3 Token games between different automata

For our purposes, it will be useful for us to have $k$-token games where Eve's token and Adam's tokens are not necessarily in the same automaton.

**Definition 3.23** ($k$-token game). *For a natural number $k \geq 1$, parity automaton $\mathcal{A} = (Q, \Sigma, q_0, \Delta_A)$ and parity automata $\mathcal{B}_l = (P^l, \Sigma, p_0^l, \Delta_l)$ for each $l \in [k]$, the $k$-token game on $\mathcal{A}$ against $\mathcal{B}_1, \mathcal{B}_2, \ldots, \mathcal{B}_k$ starts with an Eve's token in the initial state of $\mathcal{A}$, and Adam's $k$ tokens that are in the initial states of $\mathcal{B}_1, \mathcal{B}_2, \ldots, \mathcal{B}_k$, respectively. For each $i \in \mathbb{N}$, the round $i$ starts with Eve's token at some state $q_i$ in $\mathcal{A}$ and Adam's $l^{th}$ token at some state $p_i^l$ in $\mathcal{B}_l$, for each $l \in [k]$. Round $i$ then proceeds as follows.*

1. *Adam selects a letter $a_i \in \Sigma$.*

2. *Eve moves her token along a transition $q_i \xrightarrow{a_i:c} q_{i+1} \in \Delta_A$.*

3. *Adam, for each $l \in [1,k]$, moves his token $l$ along the transition $p_i^l \xrightarrow{a_i:c_i^l} p_{i+1}^l$.*

*Thus, in a play of the k-token game on $\mathcal{A}$ against $\mathcal{B}_1, \mathcal{B}_2, \ldots, \mathcal{B}_k$, Eve constructs a run and Adam constructs k runs, all on the same word. Eve wins such a play if and only if the following condition holds: if at least one of Adam's k runs is accepting, then Eve's run is accepting.*

We use the shorthand $Gk(\mathcal{A}; \mathcal{B}_1, \mathcal{B}_2, \ldots, \mathcal{B}_k)$ to denote the $k$-token game on $\mathcal{A}$ against $\mathcal{B}_1, \mathcal{B}_2, \ldots, \mathcal{B}_k$. We say that Eve (resp. Adam) wins $Gk(\mathcal{A}; \mathcal{B}_1, \ldots, \mathcal{B}_k)$ if Eve (resp. Adam) has a winning strategy in the above game.

We will often care about the $k$-token game where both Eve and Adam's tokens are all in the same automaton $\mathcal{A}$ but start at different states. We will then write $Gk(q; p_1, p_2, \ldots, p_k)$ in $\mathcal{A}$ to denote the game $Gk((\mathcal{A}, q); (\mathcal{A}, p_1), \ldots, (\mathcal{A}, p_k))$. Finally, we use $Gk(\mathcal{A})$ to denote the game where Eve's token and Adam's $k$ tokens all start at the initial state of $\mathcal{A}$: this is the same as the $k$-token game on $\mathcal{A}$ we saw earlier in Section 3.2.

Similar to how we argued finite-memory determinacy for $k$-token games on the same automaton (Proposition 3.14), we can also argue finite-memory determinacy for $k$-token games between different automata. We can also argue positionality of Eve for 1-token games on different automata similar to how we showed the same for 1-token game on same automaton (Proposition 3.17).

**Proposition 3.24.** *For every $k \geq 1$ and parity automata $\mathcal{A}, \mathcal{B}_1, \mathcal{B}_2, \ldots, \mathcal{B}_k$, the game $Gk(\mathcal{A}; \mathcal{B}_1, \mathcal{B}_2, \ldots, \mathcal{B}_k)$ is finite-memory determined. Additionally, if Eve wins $G1(\mathcal{A}; \mathcal{B})$ for some parity automata $\mathcal{A}$ and $\mathcal{B}$, then Eve has a positional winning strategy in $G1(\mathcal{A}; \mathcal{B})$.*

We continue with observations on token games between different automata. Similar to how we showed transitivity for simulation, we can argue transitivity for $G1$ in a near identical way.

**Lemma 3.25.** *For every three parity automata $\mathcal{A}, \mathcal{B}$, and $\mathcal{C}$, if Eve wins $G1(\mathcal{A}; \mathcal{B})$ and $G1(\mathcal{B}; \mathcal{C})$, then Eve wins $G1(\mathcal{A}; \mathcal{C})$ as well.*

*Proof.* Let $\sigma_{AB}$ be a winning strategy for Eve in $G1(\mathcal{A}; \mathcal{B})$, and $\sigma_{BC}$ a winning strategy for Eve in $G1(\mathcal{B}; \mathcal{C})$. We describe a winning strategy $\sigma_{AC}$ for Eve in $G1(\mathcal{A}; \mathcal{C})$ as follows. Eve stores in her memory, a token that takes transitions in $\mathcal{B}$. In $G1(\mathcal{A}; \mathcal{C})$, she will choose transitions on her memory token in $\mathcal{B}$ according to her strategy $\sigma_{BC}$ against Adam's token in $\mathcal{C}$, and choose transitions on her token in $\mathcal{A}$ against her memory token in $\mathcal{B}$ via $\sigma_{AB}$. If Adam's run on his token in $\mathcal{C}$ is accepting then the run produced by Eve's memory token in $\mathcal{B}$ is accepting since $\sigma_{BC}$ is a winning strategy for Eve in $G1(\mathcal{B}; \mathcal{C})$. Since $\sigma_{AB}$ is a winning strategy for Eve in $G1(\mathcal{A}; \mathcal{B})$, it follows that the run on her token is accepting as well. □

We continue with the following results on token games, which either follows easily or is proved similarly to Lemma 3.25.

**Lemma 3.26.** *Let $\mathcal{A}, \mathcal{B}, \mathcal{B}', \mathcal{C}, \mathcal{C}'$ be nondeterministic parity automata. Then the following statements hold.*

1. *If Eve wins $G2(\mathcal{A}; \mathcal{B}, \mathcal{C})$, then Eve wins $G2(\mathcal{A}; \mathcal{C}, \mathcal{B})$.*

2. *If Eve wins $G2(\mathcal{A}; \mathcal{B}, \mathcal{C})$, then Eve wins $G1(\mathcal{A}; \mathcal{B})$ and Eve wins $G1(\mathcal{A}; \mathcal{C})$.*

3. *If Eve wins $G2(\mathcal{A}; \mathcal{B}, \mathcal{C})$ and $G1(\mathcal{B}; \mathcal{B}')$, then Eve wins $G2(\mathcal{A}; \mathcal{B}', \mathcal{C})$.*

4. *If Eve wins $G1(\mathcal{A}; \mathcal{A}')$ and $G2(\mathcal{A}'; \mathcal{B}, \mathcal{C})$, then Eve wins $G2(\mathcal{A}; \mathcal{B}, \mathcal{C})$.*

5. *If Eve wins $G1(\mathcal{A}; \mathcal{B})$, then $\mathcal{A}$ simulates $\mathcal{B}$.*

*Proof.* **Item 1.** This follows from symmetry.

**Item 2.** If Eve wins $G2(\mathcal{A}; \mathcal{B}, \mathcal{C})$ using strategy $\sigma$, then Eve can play $G1(\mathcal{A}; \mathcal{B})$ using $\sigma$ as if she is playing $G2(\mathcal{A}; \mathcal{B}, \mathcal{C})$ where Adam is picking transitions arbitrarily in $\mathcal{C}$. Then, if Adam's run on his token in $\mathcal{B}$ is accepting, so is the run of Eve's token in $\mathcal{A}$.

**Item 3.** Fix a winning strategy $\sigma_{ABC}$ for Eve in $G2(\mathcal{A}; \mathcal{B}, \mathcal{C})$ and another winning strategy $\sigma_{BB'}$ for Eve in $G1(\mathcal{B}; \mathcal{B}')$.

We describe a winning strategy $\sigma$ for Eve in $G2(\mathcal{A}; \mathcal{B}', \mathcal{C})$ as follows. Eve will keep in her memory a token in $\mathcal{B}$, in which she builds runs on Adam's word simultaneously. She chooses transitions on her memory token in $\mathcal{B}$ by using $\sigma_{BB'}$ against Adam's token in $\mathcal{B}'$, while she chooses transition on her token in $\mathcal{A}$ by playing $\sigma_{ABC}$ against her memory token in $\mathcal{B}$ and Adam's token in $\mathcal{C}$.

Then, if Adam's first token in $\mathcal{B}'$ constructs an accepting run, so does Eve's memory token in $\mathcal{B}$ (due to $\sigma_{BB'}$) and hence so does her token in $\mathcal{A}$ (due to $\sigma_{ABC}$). If Adam's token in $\mathcal{C}$ is accepting, then so does Eve's token in $\mathcal{A}$, due to $\sigma_{ABC}$.

**Item 4.** Fix a winning strategy $\sigma_{AA'}$ for Eve in $G1(\mathcal{A}; \mathcal{A}')$ and another winning strategy $\sigma_{A'BC}$ for Eve in $G2(\mathcal{A}'; \mathcal{B}, \mathcal{C})$.

We describe a winning strategy for Eve in $G2(\mathcal{A}; \mathcal{B}, \mathcal{C})$ as follows. Eve keeps in her memory a token in $\mathcal{A}'$ in which she builds runs on Adam's word simultaneously. She chooses transitions on her memory token in $\mathcal{A}'$ by playing $\sigma_{A'BC}$ against Adam's tokens in $\mathcal{B}$ and $\mathcal{C}$, and Eve chooses transitions on her token in $\mathcal{A}$ by playing $\sigma_{AA'}$ against her memory token in $\mathcal{A}'$. Then, if the runs of Adam in his token in $\mathcal{B}$ or $\mathcal{C}$ is accepting, so is the run of Eve's memory token in $\mathcal{A}'$, and hence so is the run of Eve's token in $\mathcal{A}$.

**Item 5.** Eve can use her winning strategy in $G1(\mathcal{A}; \mathcal{B})$ to play $\mathsf{Sim}(\mathcal{A}; \mathcal{B})$, where she ignores the additional information she gets by Adam picking a transition on his token earlier than Eve. □

In Lemma 3.26 above, we can have analogous versions of Items 1, 2, 3, and 4 for the $k$-token game instead of 2-token game, which can be proved very similarly. We will also use the following observation, which we state and not prove, but it can be proved similarly to Item 1 in Lemma 3.26.

**Proposition 3.27.** *If Eve wins $Gk(\mathcal{A}; \mathcal{B}_1, \ldots, \mathcal{B}_k)$ for parity automata $\mathcal{A}, \mathcal{B}_1, \ldots, \mathcal{B}_k$ and some $k > 1$, then Eve also wins $Gl(\mathcal{A}; \mathcal{B}_1, \ldots, \mathcal{B}_l)$ for each $l < k$.*

Equipped with these observations on token games, we move towards proving Theorem I. We first introduce some notions that Theorem I talks about.

**Weak coreachability**

Let $\mathcal{A}$ be a parity automaton. We say that the states $p$ and $q$ are coreachable in $\mathcal{A}$, denoted as $(p, q) \in \mathsf{CR}$, if there is a finite word $u$ such that there are runs from the initial state of $\mathcal{A}$ to $p$ and to $q$ on $u$. We call the transitive closure of the coreachability relation as *weak coreachability* and denote the weak coreachability relation by $\mathsf{WCR}$. That is, we say that the states $p$ and $q$ are weakly coreachable in $\mathcal{A}$, denoted $(p, q) \in \mathsf{WCR}$, if there is a sequence of states $q_1, q_2, \ldots, q_k$ such that $(p, q_1) \in \mathsf{CR}$, $(q_i, q_{i+1} \in \mathsf{CR})$ for each $i \in [1, k-1]$, and $(q_k, q) \in \mathsf{CR}$.

Note that $\mathsf{WCR}$ is an equivalence relation between the states of the automaton. We say that states $q_1, q_2, \ldots, q_k$ in $\mathcal{A}$ are weakly coreachable in $\mathcal{A}$ if any two states amongst $q_1, q_2, \ldots$ are weakly coreachable. For a state $p$, we use $\mathsf{WCR}(\mathcal{A}, p)$ to denote the set of states that are weakly coreachable to $p$ in $\mathcal{A}$. For a finite word $u$, we will use $\mathsf{WCR}(\mathcal{A}, u)$ to refer to the set $\mathsf{WCR}(\mathcal{A}, p)$, where $p$ is a state to which a run from the initial state of $\mathcal{A}$ on $u$. Note that the choice of $p$ does not matter here.

We make a couple of observations about the relations of coreachability and weak coreachability.

**Proposition 3.28.** *Let $p$ and $q$ be two coreachable states in a parity automaton $\mathcal{A}$. If $p'$ and $q'$ are states, such that there is a finite run from $p$ to $p'$ and from $q$ to $q'$ on a finite word $u$ then $p'$ and $q'$ are coreachable in $\mathcal{A}$.*

*Proof.* Let $v$ be a finite word, such that there are runs from the initial state of $\mathcal{A}$ to $p$ and $q$ on $v$. Then there are runs from the initial state of $\mathcal{A}$ to $p'$ and $q'$ on the word $vu$, and thus $p'$ and $q'$ are coreachable in $\mathcal{A}$. □

We can extend the above result to weak coreachability.

**Proposition 3.29.** *Let $p$ and $q$ be two weakly coreachable states in a parity automaton $\mathcal{A}$. If $p'$ and $q'$ are states, such that there is a finite run from $p$ to $p'$ and from $q$ to $q'$ on a finite word $u$ then $p'$ and $q'$ are weakly coreachable in $\mathcal{A}$.*

*Proof.* Let $q_1, q_2, \ldots, q_k$ be states, such that $p$ and $q_1$ are coreachable, $q_i$ and $q_{i+1}$ are coreachable for every $i \in [1, k-1]$, and $q_k$ and $q$ are coreachable. For each $i \in [1, k]$, let $q_i'$ be a state such that there is a run from $q_i$ to $q_i'$ on $u$; note that such a state exists since we assume our automata to be complete.

Then, from Proposition 3.28, we know that $p'$ is coreachable to $q_1'$, $q_i'$ is coreachable to $q_{i+1}'$ for each $i \in [1, k-1]$, and $q_k'$ and $q'$ are coreachable. Thus, $p'$ and $q'$ are weakly coreachable. □

**Winning token games from everywhere.** For a parity automaton $\mathcal{A}$ and $k \geq 1$, we say that *Eve wins the $k$-token game from everywhere on $\mathcal{A}$* if Eve wins the $k$-token game on $\mathcal{A}$ from every configuration of states that can be reached in the $k$-token game from the initial configuration where all tokens start at the initial state.

We next prove Theorem I. When trying to argue that a parity automaton $\mathcal{A}$ on which Eve wins the 2-token game is HD, Theorem I lets us assume without loss of generality that Eve wins the 2-token game from everywhere on $\mathcal{A}$ (Corollary 3.33).

**Theorem I.** *Let $\mathcal{A}$ be a parity automaton on which Eve wins the 2-token game. Then, there is a simulation-equivalent subautomaton $\mathcal{B}$ of $\mathcal{A}$ such that the following three conditions hold.*

1. *For each $k \geq 1$, Eve wins the $k$-token game from everywhere in $\mathcal{B}$.*

2. *For states $q, p_1, p_2, \ldots, p_k$ that are weakly coreachable in $\mathcal{B}$, Eve wins the $k$-token game from $(q; p_1, p_2, \ldots, p_k)$ in $\mathcal{B}$.*

3. *If $\mathcal{B}$ is history-deterministic, then so is $\mathcal{A}$.*

We start by showing the following result, which is the crux in our proof of Theorem I.

**Lemma 3.30.** *If Eve wins the 2-token game on $\mathcal{A}$ then there is a simulation-equivalent subautomaton $\mathcal{B}$ of $\mathcal{A}$, on which Eve wins the 2-token game from everywhere.*

*Proof.* Since Eve wins the 2-token game on $\mathcal{A}$, she wins the 3-token game on $\mathcal{A}$, say using a winning strategy $\sigma_3$. Consider the strategy $\sigma_2$ for Eve in the 2-token game obtained from $\sigma_3$, in which she plays using $\sigma_3$ supposing Adam's third token is copying her token. Observe that $\sigma_2$ is a winning strategy for Eve in the 2-token game. Let $\Delta'$ be the set of all transitions $\delta$ for which there is a play in which Eve is playing according to $\sigma_2$ and her token takes $\delta$. We let $\mathcal{B}$ be the subautomaton of $\mathcal{A}$ consisting of the transitions in $\Delta'$.

We will show that $\mathcal{B}$ is the subautomaton that satisfies the desired properties. Let $q_0$ be the initial state of $\mathcal{A}$. We start by proving the following claim.

**Claim 1.** *For every finite word $u$ and states $p, r_1, r_2$, such that there are runs on $u$ from $q_0$ to $p$ in $\mathcal{B}$ and from $q_0$ to $r_1$ and to $r_2$ in $\mathcal{A}$, Eve wins $G3((\mathcal{B}, p); (\mathcal{A}, r_1), (\mathcal{A}, r_2), (\mathcal{A}, p))$.*

We will show the claim by induction on the length of $u$. For the base case when $|u| = 0$, that is, $u = \varepsilon$, we need to show that Eve wins $G3((\mathcal{B}, q_0); (\mathcal{A}, q_0), (\mathcal{A}, q_0), (\mathcal{A}, q_0))$. We know that Eve wins $G2((\mathcal{B}, q_0); (\mathcal{A}, q_0), (\mathcal{A}, q_0))$ using the strategy $\sigma_2$; note that $\sigma_2$ only takes transitions in $\mathcal{B}$. Thus, in particular, Eve wins $G1((\mathcal{B}, q_0); (\mathcal{A}, q_0))$ (Lemma 3.26.2). Since Eve wins $G3(q_0; q_0, q_0, q_0)$ in $\mathcal{A}$, we get using an extension of Lemma 3.26.4 to 3-token games that Eve wins $G3((\mathcal{B}, q_0); (\mathcal{A}, q_0), (\mathcal{A}, q_0), (\mathcal{A}, q_0))$, as desired.

Thus, assume that the claimed statement holds for words $u_k$ of length $k$. We will show that the claimed statement also holds for all words $u_{k+1}$ of length $k + 1$. Let $u_{k+1} = u_k \cdot a$ such that length of $u_k$ is $k$ and $a$ is a letter in $\Sigma$. Let $p, r_1, r_2$ be states to which there is a run on $u_{k+1}$ in automata $\mathcal{B}, \mathcal{A}$, and $\mathcal{A}$ respectively from $q_0$. Then, there are states $\operatorname{pre}(p), \operatorname{pre}(r_1), \operatorname{pre}(r_2)$ such that $\operatorname{pre}(p)$ (resp. $\operatorname{pre}(r_1), \operatorname{pre}(r_2)$) is reachable in $\mathcal{B}$ (resp. $\mathcal{A}$) on the word $u_k$, and $\operatorname{pre}(p) \xrightarrow{a} p$ (resp. $\operatorname{pre}(r_i) \xrightarrow{a} r_i$ for $i = 1, 2$) is a transition in $\mathcal{B}$ (resp. $\mathcal{A}$).

We know using the induction hypothesis that

$$\text{Eve wins } G3((\mathcal{B}, \operatorname{pre}(p)); (\mathcal{A}, \operatorname{pre}(r_1)), (\mathcal{A}, \operatorname{pre}(r_2)), (\mathcal{A}, \operatorname{pre}(p))). \tag{3.1}$$

Thus, there is a state $p'$, such that there is a transition from $\operatorname{pre}(p)$ to $p'$ in $\mathcal{B}$ on $a$ and

$$\text{Eve wins } G3((\mathcal{B}, p'); (\mathcal{A}, r_1), (\mathcal{A}, r_2), (\mathcal{A}, p)). \tag{3.2}$$

Further, since $\operatorname{pre}(p) \xrightarrow{a} p$ is a transition in $\mathcal{B}$, we know there are states $q_1$ and $q_2$ that are weakly coreachable to $\operatorname{pre}(p)$ in $\mathcal{A}$, such that

$$\text{Eve wins } G3((\mathcal{B}, \operatorname{pre}(p)); (\mathcal{A}, q_1), (\mathcal{A}, q_2), (\mathcal{A}, \operatorname{pre}(p)))$$

and the transition $\operatorname{pre}(p) \xrightarrow{a} p$ preserves the winning region for Eve in $G3$ when Adam plays the letter $a$. Thus, in particular,

$$\text{Eve wins } G3((\mathcal{B}, p); (\mathcal{A}, \operatorname{suc}(q_1)), (\mathcal{A}, \operatorname{suc}(q_2)), (\mathcal{A}, \operatorname{suc}(\operatorname{pre}(p)))) \tag{3.3}$$

for states $\operatorname{suc}(q_1), \operatorname{suc}(q_2)$, and $\operatorname{suc}(\operatorname{pre}(p))$ that can be reached upon reading the letter $a$ from $q_1, q_2$ and $\operatorname{pre}(p)$ respectively in $\mathcal{A}$. In particular, Eve wins $G1((\mathcal{B}, p); (\mathcal{A}, p'))$. Since $\mathcal{B}$ is a subautomaton of $\mathcal{A}$, we note that Eve also wins $G1((\mathcal{B}, p); (\mathcal{B}, p'))$. Combining this with Eq. (3.2) using the extension of Lemma 3.26.4 to 3-token games, we get that

$$\text{Eve wins } G3((\mathcal{B}, p); (\mathcal{A}, r_1), (\mathcal{A}, r_2), (\mathcal{A}, p)), \tag{3.4}$$

thus completing our induction step. This completes the proof of our claim.

Note that since $\mathcal{B}$ is a subautomaton of $\mathcal{A}$, we know that $\mathcal{A}$ simulates $\mathcal{B}$. Furthermore, we know Eve wins $G1((\mathcal{B}, q_0); (\mathcal{A}, q_0))$ due to Lemma 3.26.2, which implies that $\mathcal{B}$ simulates $\mathcal{A}$ (Lemma 3.26.5). Therefore, $\mathcal{A}$ and $\mathcal{B}$ are simulation-equivalent.

The above claim also implies that Eve wins the 2-token game from everywhere in $\mathcal{B}$; If $p, r_1, r_2$ are states that can be reached upon reading the same word from $q_0$ in $\mathcal{B}$, then the same holds for $\mathcal{A}$. Claim 1 then tells us that Eve wins $G2((\mathcal{B}, p); (\mathcal{A}, r_1), (\mathcal{A}, r_2))$. Since $\mathcal{B}$ is a subautomaton of $\mathcal{A}$, we obtain that Eve wins $G2((\mathcal{B}, p); (\mathcal{B}, r_1), (\mathcal{B}, r_2))$, as desired. □

The following lemma strengthens Lemma 3.30 to triplets of weakly coreachable states.

**Lemma 3.31.** *Let $\mathcal{B}$ be an automaton such that Eve wins the 2-token game from everywhere in $\mathcal{B}$. Then, Eve wins the 2-token game from all triplets of weakly coreachable states in $\mathcal{B}$.*

*Proof.* We note from Lemma 3.26.2 that Eve wins the 1-token game from everywhere in $\mathcal{B}$,

that is, for all states $(p, q) \in \mathsf{CR}$, Eve wins $G1(p; q)$. Recall that the weak-coreachability relation $\mathsf{WCR}$ is a transitive closure of $\mathsf{CR}$, and since the relation of Eve winning $G1$ is transitive as well (Lemma 3.25), we obtain that Eve wins $G1(p; q)$ for all pairs $(p, q) \in \mathsf{WCR}$ of weakly coreachable states.

Thus, if $p, q, r$ are weakly coreachable states in $\mathcal{B}$, then Eve wins $G2(p; p, p)$, $G1(p; q)$, and $G1(p; r)$ in $\mathcal{B}$. Two applications of Lemma 3.26.3 gives us that Eve wins $G2(p; q, r)$ in $\mathcal{B}$, as desired. □

We next extend the above result to winning $k$-token games from all tuples of weakly coreachable states.

**Lemma 3.32.** *Let $\mathcal{B}$ be an automaton on which Eve wins the 2-token game from everywhere. Then, for all $k \geq 1$, Eve wins the $k$-token game from all tuples $(q; p_1, p_2, \ldots, p_k)$ in $\mathcal{B}$, where $q, p_1, \ldots, p_k$ are weakly coreachable states in $\mathcal{B}$.*

*Proof.* Since Eve wins $G2(q; q, q)$ in $\mathcal{B}$, Eve also wins $Gk(q; q, q, \ldots, q)$ in $\mathcal{B}$ (Lemma 3.15). Lemma 3.31 then tells us Eve wins $G1(q; p_i)$ for each $i \in [1, k]$. Combining this with an Lemma 3.26.3 to $k$-token games, we obtain that Eve wins $Gk(q; p_1, p_2, \ldots, p_k)$ in $\mathcal{B}$. □

Lemmas 3.30 to 3.32, together with the result that simulation-equivalence of two automata imply that either both automata are HD or neither are HD (Corollary 3.11), prove Theorem I.

**Theorem I.** *Let $\mathcal{A}$ be a parity automaton on which Eve wins the 2-token game. Then, there is a simulation-equivalent subautomaton $\mathcal{B}$ of $\mathcal{A}$ such that the following three conditions hold.*

1. *For each $k \geq 1$, Eve wins the $k$-token game from everywhere in $\mathcal{B}$.*

2. *For states $q, p_1, p_2, \ldots, p_k$ that are weakly coreachable in $\mathcal{B}$, Eve wins the $k$-token game from $(q; p_1, p_2, \ldots, p_k)$ in $\mathcal{B}$.*

3. *If $\mathcal{B}$ is history-deterministic, then so is $\mathcal{A}$.*

Let us make explicit a corollary of Theorem I above.

**Corollary 3.33.** *For a parity index $[i, j]$, the following two statements are equivalent.*

1. *Every $[i, j]$ automaton $\mathcal{A}$ on which Eve wins the 2-token game is HD.*

2. *Every $[i, j]$ automaton $\mathcal{A}$ on which Eve wins the 2-token game from everywhere is HD.*

*Proof.* It is clear that (1) implies (2). For the other direction, suppose (2) is true. Then, for every $[i, j]$ automaton $\mathcal{A}$ on which Eve wins the 2-token game, there is a simulation equivalent subautomaton $\mathcal{B}$ such that Eve wins the 2-token game from everywhere in $\mathcal{B}$ (Theorem I). Then (2) implies that $\mathcal{B}$ is HD, and due to $\mathcal{A}$ being simulation-equivalent to $\mathcal{B}$, we obtain due to Corollary 3.11 that $\mathcal{A}$ is HD as well. □

## 3.4 Joker games

Kuperberg and Skrzypczak introduced Joker games (see Definition 3.18) in 2015 to give a PTIME algorithm to decide history-determinism for coBüchi automata. However, they left open the question of whether (Eve winning) Joker games characterise history-determinism for coBüchi automata. We will answer this question positively for coBüchi as well as Büchi automata in Chapters 4 and 5.

To show this, we will use the following result which is similar in spirit to Theorem I.

**Theorem 3.34.** *Let $\mathcal{A}$ be a parity automaton on which Eve wins the Joker game. Then, there is a simulation-equivalent subautomaton $\mathcal{B}$ of $\mathcal{A}$, such that Eve wins the 1-token game from everywhere in $\mathcal{B}$.*

*Proof.* Suppose Eve wins the Joker game on $\mathcal{A}$, and let $\sigma$ be a positional winning strategy for Eve in the Joker game on $\mathcal{A}$; we know such a $\sigma$ exists due to Proposition 3.19. Let $\Delta'$ be the set of transitions $\delta$ that Eve picks on her token in some play of the Joker game in which she is playing according to $\sigma$. We let $\mathcal{B}$ be the subautomaton of $\mathcal{A}$ consisting of the transitions in $\Delta'$. We will show that $\mathcal{B}$ is the desired subautomaton. The crux of our argument is proving the following claim.

**Claim 2.** *For every finite word $u$, let $p$ and $q$ be states, such that there are runs from $q_0$ to $p$ in $\mathcal{B}$ and $q_0$ to $q$ in $\mathcal{A}$ on $u$. Then, Eve wins $G1((\mathcal{B}, p); (\mathcal{A}, q))$.*

We show the claim by induction on the length of $u$. When $|u| = 0$, it is clear that Eve wins $G1((\mathcal{B}, q_0); (\mathcal{A}, q_0))$, since Eve can play according to her strategy $\sigma$ in the Joker game on $\mathcal{A}$: this ensures Eve's token only takes transitions in $\mathcal{B}$ in $G1((\mathcal{B}, q_0); (\mathcal{A}, q_0))$.

Thus, assume that the claim holds for all words $u_k$ of length $k$, where $k \geq 0$. We will show that the claim also hold for every word $u_{k+1} = u_k \cdot a$ for every $a \in \Sigma$ and word $u_k$ of length $k$. Let $p$ and $q$ be states to which there is a run from $q_0$ on $u_{k+1}$ in $\mathcal{B}$ and $\mathcal{A}$ respectively. Then, there are states $\text{pre}(p), \text{pre}(q)$ to which there are runs from $q_0$ on the word $u_k$ in automata $\mathcal{B}$ and $\mathcal{A}$ respectively, and additionally, there are transitions $\text{pre}(p) \xrightarrow{a} p$ in $\mathcal{B}$ and $\text{pre}(q) \xrightarrow{a} q$ in $\mathcal{A}$. By the induction hypothesis, we know that

$$\text{Eve wins } G1((\mathcal{B}, \text{pre}(p)); (\mathcal{A}, \text{pre}(q))). \tag{3.5}$$

Thus, there must be a transition $\text{pre}(p) \xrightarrow{a} p'$ in $\mathcal{B}$ such that

$$\text{Eve wins } G1((\mathcal{B}, p'); (\mathcal{A}, q)). \tag{3.6}$$

Since $\text{pre}(p) \xrightarrow{a} p$ is a transition in $\mathcal{B}$, we know that there is a state $q'$ in $\mathcal{A}$, such that Eve wins the Joker game from $(\text{pre}(p); q')$ and the strategy $\sigma$ selects the transition $\text{pre}(p) \xrightarrow{a} p$ for Eve in Joker game from $(\text{pre}(p); q')$. If Adam then plays Joker and moves his token to $p'$, we note that the resulting new position must be still in the winning region for Eve in the Joker game: this implies

$$\text{Eve wins } G1((\mathcal{B}, p); (\mathcal{A}, p')). \tag{3.7}$$

Since $\mathcal{B}$ is a subautomaton of $\mathcal{A}$, Eq. (3.7) implies that Eve wins $G1((\mathcal{B},p);(\mathcal{B},p'))$. Combining this with Eq. (3.6) and the transitivity of $G1$ games (Lemma 3.25), we obtain that Eve wins $G1((\mathcal{B},p);(\mathcal{A},q))$. This proves our claim.

We note that since $\mathcal{B}$ is a subautomaton of $\mathcal{A}$, Claim 2 above implies that Eve wins the 1-token game from everywhere in $\mathcal{B}$. The automaton $\mathcal{B}$ being a subautomaton of $\mathcal{A}$ also implies that $\mathcal{A}$ simulates $\mathcal{B}$, while Claim 2 implies that Eve wins $G1(\mathcal{B};\mathcal{A})$, which gives us that $\mathcal{B}$ simulates $\mathcal{A}$ (Lemma 3.26.5). Thus, $\mathcal{A}$ and $\mathcal{B}$ are simulation-equivalent, and Eve wins the 1-token game from everywhere in $\mathcal{B}$. □

We note that if Eve wins the 1-token from everywhere in $\mathcal{A}$, then Eve wins the 1-token game from all pairs of weakly coreachable states in $\mathcal{A}$ as well. This follows due to transitivity of 1-token games (Lemma 3.25).

**Proposition 3.35.** *For every parity automaton $\mathcal{A}$, if Eve wins the 1-token game from everywhere, then Eve wins $G1(p;q)$ in $\mathcal{A}$ for all pairs $(p,q)$ of weakly coreachable states in $\mathcal{A}$.*

### 3.4.1 Semantic Determinism

For a parity automaton $\mathcal{A}$, we say that two states $p$ and $q$ in $\mathcal{A}$ are language-equivalent if $L(\mathcal{A},p) = L(\mathcal{A},q)$.

*Semantically-deterministic* (SD) automata are nondeterministic automata where from each state and each letter, all outgoing transitions from that state on that letter lead to language-equivalent states. This can easily be extended to finite words: if an automaton $\mathcal{A}$ is semantically deterministic, then all states that can be reached from $p$ upon reading $u$ are language-equivalent, for every state $p$ and finite word $u$. Automata on which Eve wins the 1-token game from everywhere is semantically-deterministic.

**Lemma 3.36.** *Every parity automaton $\mathcal{A}$ on Eve wins the 1-token game from everywhere is semantically-deterministic. Additionally, if $p$ and $q$ are weakly coreachable states in $\mathcal{A}$, then they are language equivalent.*

*Proof.* Observe that if Eve wins $G1((\mathcal{A},p);(\mathcal{A},q))$ for some states $p$ and $q$, then $(\mathcal{A},p)$ simulates $(\mathcal{A},q)$ and hence $L(\mathcal{A},q) \subseteq L(\mathcal{A},p)$. Thus, if states $p$ and $q$ are weakly coreachable, then we get that $L(\mathcal{A},p) = L(\mathcal{A},q)$ since Eve wins $G1(p;q)$ and $G1(q;p)$ in $\mathcal{A}$. It follows that $\mathcal{A}$ is semantically-deterministic. □

Kuperberg and Skrzypczak had originally called semantically-deterministic automata as residual automata [KS15]. We follow Abu Radi, Kupferman, and Leshkowitz [RKL21] and other recent works [RK23] in calling them SD automata instead.

## 3.5 Complexity of solving token games

In this section, we discuss the complexity of solving 1-token games and 2-token games.

### 3.5.1 1-token games

Recall that 1-token games can be explicitly represented as implication games, or equivalently, as 2-dimensional parity games. Thus, the upper bounds given by Chatterjee, Henzinger, and Piterman for 2-dimensional parity games apply to 1-token games as well.

**Lemma 3.37** ([CHP07, Theorem 3])**.** *There is an algorithm that computes Eve's (possibly empty) winning region and her uniform positional winning strategy from her winning region on an input 2-dimensional parity game $\mathcal{G} = (V, E)$ with n vertices and priority functions $\pi_1 : E \to [0, d_1]$ and $\pi_2 : E \to [0, d_2]$ in time*

$$\binom{d_1 + d_2}{d_1} \cdot \mathcal{O}(|E| \cdot |V|^{2(d_1+d_2)}).$$

In particular, if $d_1 = d_2 = d$ is fixed, then note that 2-dimensional parity games can be solved in PTIME. Thus, we obtain the following result.

**Lemma 3.38.** *Let $d > 0$ be a fixed natural number. Then there is a PTIME algorithm that, for given $[0, d]$ automata $\mathcal{A}$ and $\mathcal{B}$, either finds a positional winning strategy for Eve in $G1(\mathcal{A}; \mathcal{B})$ or determines that Adam wins $G1(\mathcal{A}; \mathcal{B})$.*

Recall that Joker games can be explicitly represented as 2-dimensional parity games too (Section 3.2.2). Thus, the upper bounds in Lemma 3.38 also applies to Joker games.

**Lemma 3.39.** *Let $d > 0$ be a fixed natural number. Then there is a PTIME algorithm that, for given $[0, d]$ automata $\mathcal{A}$, either finds a positional winning strategy for Eve in the Joker game on $\mathcal{A}$ or determines that Adam wins the Joker game on $\mathcal{A}$.*

**Remark 4.** *We note that 2-dimensional parity games, as well as the 1-token games can be solved faster than the running time presented in Lemma 3.37, by converting these games to Rabin games and using the state-of-the-art algorithms for solving Rabin games [MST24]. The focus of this thesis is not on obtaining the most efficient algorithms, however, so we avoid this detail for conciseness.*

### 3.5.2 2-Token games

We next show that 2-token games on parity automata can be solved in PSPACE, and if the priorities of the automata are from a fixed interval, then in PTIME.

**Theorem 3.40.** *There is an algorithm that solves the 2-token game on every input parity automaton $\mathcal{A}$ with priorities in $[0, d]$ in time*

$$d \cdot (2^{3d} \cdot \mathsf{poly}(\mathcal{A}))^{1+o(1)},$$

*where $\mathsf{poly}(\mathcal{A})$ is a fixed polynomial in the size of $\mathcal{A}$.*

**Theorem 3.41.** *There is a* PSPACE *algorithm to solve* 2-*token games on parity automata.*

We will show these results by constructing the Zielonka tree for the 2-token game. Recall that 2-token games can be represented as Muller games, as we had discussed for Proposition 3.14.

More concretely, suppose that the priorities of a parity automaton $\mathcal{A}$ are all in $[0, d]$ for some $d \geq 0$. Then the winning condition of the 2-token game on $\mathcal{A}$ can be represented as a $(C, \mathcal{F})$ Muller condition, where $C = [0, d] \times [0, d] \times [0, d]$. Here, the first component is for the priorities that occur in the transitions of Eve's token, while the second and third component are for the priorities that occur in the transitions of Adam's first token and second token, respectively. The set of accepting subsets $\mathcal{F} \subseteq \mathcal{P}(C)$ consists of nonempty subsets $S$ of $C$ that satisfies the following condition: if the least priority occurring amongst the second component of the elements of $S$ is even or if the least priority occurring amongst the third component of the elements of $S$ is even, then the least priority occurring amongst the first component of the element of $S$ is even.

We denote the Zielonka tree of the $(C, \mathcal{F})$-Muller condition as above as $\mathcal{Z}_{[0,d]}$. Then, $\mathcal{Z}_{[0,d]}$ has its root labelled by $[0, d] \times [0, d] \times [0, d]$. The rest of the tree can be generated via the following rules.

**R1.** If a node is labelled by $P = [e, d] \times [n_1, d] \times [n_2, d]$, for some even number $e$ and natural numbers $n_1, n_2$, then note that $P$ is in $\mathcal{F}$. The largest nonempty subset of $P$ that is not in $\mathcal{F}$ is $C = [e+1, d] \times [n_1, d] \times [n_2, d]$, if $e + 1 \leq d$. A node labelled $P$ therefore has at most one children, whose label is $C$.

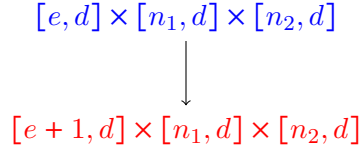

Figure 3.2: Rule R1

**R2.** If a node is labelled by $P = [o, d] \times [e_1, d] \times [e_2, d]$ for some odd number $o$ and even numbers $e_1, e_2$, then $P$ is not in $\mathcal{F}$. The largest nonempty subsets of $P$ that are in $\mathcal{F}$ are

$$C_1 = [o+1, d] \times [e_1, d] \times [e_2, d] \text{ if } o + 1 \leq d,$$

and $C_2 = [o, d] \times [e_1 + 1, d] \times [e_2 + 1, d]$ if $e_1 + 1, e_2 + 1 \leq d$.

A node labelled $P$ has therefore at most 2 children, with labels $C_1$ and $C_2$.

**R3.** If a node is labelled $P = [o, d] \times [o_1, d] \times [o_2, d]$ for some odd numbers $o, o_1, o_2$, then observe that $P$ is in in $\mathcal{F}$. The largest nonempty subsets of $P$ that are not in $\mathcal{F}$ then are

$$C_1 = [o, d] \times [o_1 + 1, d] \times [o_2, d] \text{ if } o_1 + 1 \leq d,$$

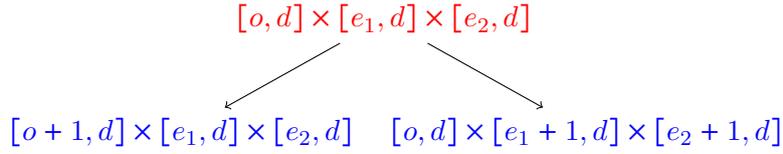

Figure 3.3: Rule R2

and $C_2 = [o, d] \times [o_1, d] \times [o_2 + 1, d]$, if $o_2 + 1 \leq d$.

Thus, a node labelled $P$ has at most two children that are labelled $C_1$ and $C_2$.

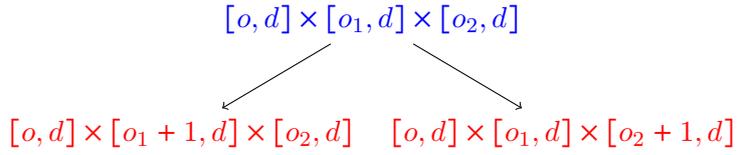

Figure 3.4: Rule R3

**R4.** If a node is labelled $P = [o, d] \times [e, d] \times [o', d]$, for some odd numbers $o, o'$ and even number $e$, then $P$ is not in $\mathcal{F}$. The largest nonempty subsets of $P$ that are in $\mathcal{F}$ are

$$C_1 = [o + 1, d] \times [e, d] \times [o', d] \text{ if } o + 1 \leq d,$$

and $C_2 = [o, d] \times [e + 1, d] \times [o', d]$ if $e + 1 \leq d$.

Thus, a node labelled $P$ has at most two children that are labelled $C_1$ and $C_2$.

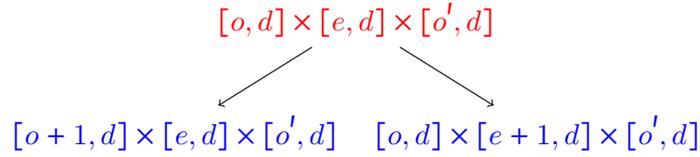

Figure 3.5: Rule R4

**R5.** Symmetrically to R4, if a node is labelled $P = [o, d] \times [o', d] \times [e, d]$, for some odd numbers $o, o'$ and even number $e$, then $P$ is not in $\mathcal{F}$. The largest nonempty subsets of $P$ that are in $\mathcal{F}$ are

$$C_1 = [o + 1, d] \times [o', d] \times [e, d] \text{ if } o + 1 \leq K,$$

and $C_2 = [o, d] \times [o', d] \times [e + 1, d]$ if $e + 1 \leq d$.

Thus, a node labelled $P$ has at most two children that are labelled $C_1$ and $C_2$.

The rules R1-5 above can be used to completely construct the Zielonka tree $\mathcal{Z}_{[0,d]}$. Let us make two observation that would help us with estimating the size of $\mathcal{Z}_{[0,d]}$. First, every node is labelled by a set of the form $[x, d] \times [y, d] \times [z, d]$, where $x, y, z \in [0, d]$.

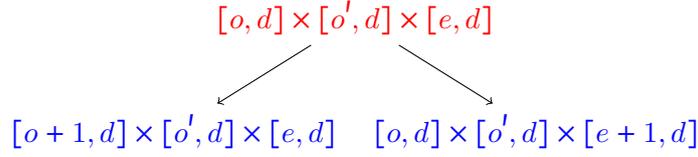

Figure 3.6: Rule R5

Thus, the total number of labels occurring in $\mathcal{Z}_{[0,d]}$ is at most $(d+1)^3$. Secondly, every node has at most two children, where the labels of the children are a strictly smaller subset of the label of the node. This tells us that the height of $\mathcal{Z}_{[0,d]}$ is at most $3(d+1)$, and hence the number of branches in $\mathcal{Z}_{[0,d]}$ is at most $2^{3(d+1)}$.

The 2-token game on a parity automaton $\mathcal{A} = (Q, \Sigma, q_0, \Delta)$ can be explicitly represented as a Muller game with $\mathsf{poly}(\mathcal{A})$ many vertices, where $\mathsf{poly}(\mathcal{A})$ is a fixed polynomial in the size of $\mathcal{A}$. If $\mathcal{A}$ has priorities in $[0,d]$, then this Muller game can be converted into a parity game with $\mathcal{O}(2^{3(d+1)} \cdot \mathsf{poly}(\mathcal{A}))$ many vertices and edges and $3(d+1)$ many priorities using $\mathcal{Z}_{[0,d]}$ (see Section 2.4).

Parity games with $n$ vertices, $m$ edges, and $k$ priorities can be solved in time $km^{1+o(1)}$, when $k = o(\log n)$ [JL17]. This is true for the parity game that we have converted the 2-token game into, and thus we obtain the following result.

**Theorem 3.40.** *There is an algorithm that solves the 2-token game on every input parity automaton $\mathcal{A}$ with priorities in $[0,d]$ in time*

$$d \cdot (2^{3d} \cdot \mathsf{poly}(\mathcal{A}))^{1+o(1)},$$

*where $\mathsf{poly}(\mathcal{A})$ is a fixed polynomial in the size of $\mathcal{A}$.*

Theorem 3.40 implies that 2-token games can be solved in PTIME when the number of priorities is fixed. We also remark that the upper bound in Theorem 3.40 is better than the upper bound of $d^2 \cdot (2^{d^2 \log d} \mathsf{poly}(\mathcal{A}))^{1+o(1)}$ given by Boker, Kuperberg, Lehtinen, and Skrzypczak in 2020 for solving 2-token games [BKLS20, Proposition 24].

We move towards showing that 2-token games can be solved in PSPACE. For this, we will use the structure called Zielonka DAGs, which are constructed as a Zielonka tree except that the nodes that have the same labels are merged, making it a directed acyclic graph (DAG). For the winning condition of the 2-token game on an automaton with priorities in $[0,d]$, note that the Zielonka DAG $\mathcal{Y}_{[0,d]}$ has at most $(d+1)^3$ many vertices since there are at most $(d+1)^3$ distinct labels in $\mathcal{Z}_{[0,d]}$. Furthermore, $\mathcal{Y}_{[0,d]}$ can be constructed in time polynomial in $d$ using rules R1-5. Since Muller games whose winning condition is given by a Zielonka DAG can be solved in PSPACE ([HD05, Corollary 4.3 in the full version]), we obtain that 2-token games can be solved in PSPACE as well.

**Theorem 3.41.** *There is a PSPACE algorithm to solve 2-token games on parity automata.*

# Part II

# Warm-up

# Chapter 4

# CoBüchi Automata

In 2020, Boker, Kuperberg, Lehtinen, and Skrzypczak showed that for every coBüchi or $[1,2]$ automaton $\mathcal{A}$, Eve wins the 2-token game on $\mathcal{A}$ if and only if $\mathcal{A}$ is history-deterministic [BKLS20]. Their argument is fairly involved, and uses insights from the polynomial time algorithm that Kuperberg and Skrzypcak gave in 2015 [KS15] and Bagnol and Kuperberg's result of 2018 [BK18] that showed the 2-token game characterisation of history-determinism on Büchi automata.

The argument in their work [BKLS20] roughly goes as follows. Suppose that Eve wins the 2-token game on a coBüchi automaton $\mathcal{A}$. They than argue that the automaton $\mathcal{A}$ satisfies a certain property, a (simplification of) which we will call safe-coverage. They assume, towards a contradiction, that $\mathcal{A}$ is not history-deterministic, and therefore, Adam has a finite-memory strategy in the HD game. They then show that Adam, using the safe-coverage of $\mathcal{A}$ and his finite-memory strategy in the HD game, can win the $N$-token game for some $N$ that is doubly exponential in the size of $\mathcal{A}$. Since, the 2-token game and the $N$-token game on $\mathcal{A}$ have the same winner, however, this is a contradiction.

In this chapter, we give a simpler proof for the 2-token game characterisation of history-deterministic coBüchi automata and show the following stronger statement.

**Theorem 4.1.** *For every coBüchi automaton $\mathcal{A}$, Eve wins the Joker game on $\mathcal{A}$ if and only if $\mathcal{A}$ is history-deterministic.*

The implication that Eve wins the Joker game on $\mathcal{A}$ if $\mathcal{A}$ is HD is clear (Lemma 3.21). We thus focus on the other implication, which is that every coBüchi automaton on which Eve wins the Joker game is HD.

Recall that if Eve wins the Joker game on a parity automaton $\mathcal{A}$ then there is a simulation equivalent subautomaton of $\mathcal{A}$ on which Eve wins the 1-token game from everywhere (Theorem 3.34). Thus, in order to prove Theorem 4.1, it suffices to show the following result.

**Theorem 4.2.** *Let $\mathcal{A}$ be a coBüchi automaton on which Eve wins the 1-token game from everywhere. Then, $\mathcal{A}$ is history-deterministic.*

We also observe that Theorem 4.1 implies the 2-token game characterisation of history-determinism on coBüchi automata, since if Eve wins the 2-token game on a parity automaton then Eve also wins the Joker game on that automaton (Theorem 3.20).

**Corollary 4.3.** *For every coBüchi automaton $\mathcal{A}$, Eve wins the 2-token game on $\mathcal{A}$ if and only if $\mathcal{A}$ is history-deterministic [BKLS20].*

We move towards proving Theorem 4.2. For the rest of this chapter, fix a coBüchi (or equivalently, a $[1, 2]$) automaton $\mathcal{A}$ on which Eve wins the 1-token game from everywhere. We first perform the following relabelling of priorities on $\mathcal{A}$, which preserves the acceptance of each infinite run: *Consider the graph $G_{>1}$ with its vertices as states of $\mathcal{A}$ and edges as (priority) 2-transitions of $\mathcal{A}$. For every 1-transition in $\mathcal{A}$ that is not in any SCC in $G_{>1}$, we change its priorities to 1.*

We claim that this modification preserves the acceptance of each run. This follows from the following sequence of equivalent statements, where the equivalence of consecutive statements are easy to verify.

1. A run $\rho$ in $\mathcal{A}$ is accepting.

2. $\rho$ contains finitely many 1-transitions in $\mathcal{A}$.

3. $\rho$ eventually stays in an SCC in $G_{>1}$.

4. $\rho$ sees finitely many 1-transitions in the modified automaton.

5. $\rho$ is an accepting run in the modified automaton.

Thus, this modification does not change the fact that Eve wins the 1-token game from everywhere. We therefore make the following assumption without loss of generality in this chapter.

**Assumption 4.4.** *The coBüchi automaton $\mathcal{A}$ is such that all 2-transitions occur in some SCC of the graph $G_{>1}$ that consists of only 2-transitions of $\mathcal{A}$.*

Let $\mathcal{A}_{\texttt{safe}}$ be the safety automaton obtained from $\mathcal{A}$ as follows. The automaton $\mathcal{A}_{\texttt{safe}}$ has all the states of $\mathcal{A}$ along with an additional rejecting sink state $q_\perp$ that has a priority 1 transition on every letter in the alphabet of $\mathcal{A}$. The transitions of $\mathcal{A}$ that have priority 2 are preserved in $\mathcal{A}_{\texttt{safe}}$, and transitions of priority 1 are redirected to the rejecting sink state.

We write that the automaton $\mathcal{A}$ has safe-coverage if for each state $q$, there is a state $p$ weakly coreachable to $q$ in $\mathcal{A}$ such that Eve wins $G1(p; q)$ in $\mathcal{A}_{\texttt{safe}}$.

> **Automata with safe-coverage.** We say that a coBüchi automaton $\mathcal{A}$ has the property of safe-coverage if for each state $q$, there is another state $p \in \mathsf{WCR}(\mathcal{A}, q)$ such that Eve wins $G1(p; q)$ in $\mathcal{A}_{\texttt{safe}}$.

We will use a similar property called the 1-safe double-coverage when proving the even-to-odd induction step (Theorem A) in Chapter 6.

We next show that $\mathcal{A}$ has safe-coverage.

**Lemma 4.5.** *For each state $q$ in $\mathcal{A}$, there is another state $p$ that is weakly coreachable to $p$ in $\mathcal{A}$ such that Eve wins $G1(p;q)$ in $\mathcal{A}_{\mathtt{safe}}$.*

*Proof.* Suppose, towards a contradiction, that there is a state $q$, such that Adam wins $G1(p;q)$ in $\mathcal{A}_{\mathtt{safe}}$ for all states $p \in \mathsf{WCR}(\mathcal{A}, q)$. We will describe a winning strategy of Adam for $G1(q;q)$ in $\mathcal{A}$, which is a contradiction to the fact that Eve wins the 1-token game from everywhere in $\mathcal{A}$.

Adam, in $G1(q;q)$ in $\mathcal{A}$, chooses letters and transitions on his token according to his strategy in $G1(q;q)$ in $\mathcal{A}_{\mathtt{safe}}$. This ensures that Eve is eventually forced to take a 1-transition in $\mathcal{A}$, while Adam's token has reached some state $q'$ via only priority 2-transitions. Due to our relabelling of priorities, there is a finite run back from $q'$ to $q$ via only priority 2-transitions, and so Adam plays letters and picks transitions on his token in the 1-token game according to this finite run. When Adam's token is back at state $q$, Eve's is at some state $p'$ that is weakly coreachable to $q$ in $\mathcal{A}$ (Proposition 3.29). By assumption, Adam wins $G1(p';q)$ in $\mathcal{A}_{\mathtt{safe}}$, and Adam thus repeats his strategy as above, ensuring that the run on his token takes only priority 2 transitions and hence, is accepting, while the run on Eve's token takes infinitely many priority 1 transitions and hence, is rejecting. □

The following result then follows by transitivity of $G1$ games (Lemma 3.25) and the fact that $\mathcal{A}$ has finitely many states.

**Lemma 4.6.** *For each state $q$ in $\mathcal{A}$, there is a state $p$ in $\mathsf{WCR}(\mathcal{A}, q)$ such that Eve wins $G1(p;q)$ and $G1(p;p)$ in $\mathcal{A}_{\mathtt{safe}}$.*

*Proof.* For each state $q$ in $\mathcal{A}$, consider the directed graph $G$ whose vertices consist of the states in $\mathsf{WCR}(\mathcal{A}, q)$. There is an edge in $G$ from $r$ to $s$ if Eve wins $G1(s;r)$ in $\mathcal{A}_{\mathtt{safe}}$.

Since $\mathcal{A}$ has safe-coverage (Lemma 4.5), every vertex in $G$ has outdegree at least 1. Thus, for every vertex $r$, there is a vertex $s$ such that there is a path from $r$ to $s$ and a cycle consisting of the vertex $s$ in $G$. It then follows from transitivity of 1-token games (Lemma 3.25) that for every state $r$, there is another state $s \in \mathsf{WCR}(\mathcal{A}, r)$ such that Eve wins $G1(s;r)$ and $G1(s;s)$ in $\mathcal{A}_{\mathtt{safe}}$. This concludes our proof. □

Recall that Eve wins the 1-token game on a safety automaton if and only it is determinisable-by-pruning, i.e., contains a language-equivalent deterministic subautomaton (Theorem 3.22). We thus call the states $p$ such that Eve wins $G1(p;p)$ in $\mathcal{A}_{\mathtt{safe}}$ as safe-deterministic. The above lemma can then be restated as the following result.

**Lemma 4.7.** *For every state $q$ in $\mathcal{A}$, there is another state $p$ in $\mathsf{WCR}(\mathcal{A}, q)$, such that Eve wins $G1(p;q)$ in $\mathcal{A}_{\mathtt{safe}}$ and $p$ is safe-deterministic.*

Using Lemma 4.7, we now prove that $\mathcal{A}$ is history-deterministic.

**Theorem 4.2.** *Let $\mathcal{A}$ be a coBüchi automaton on which Eve wins the 1-token game from everywhere. Then, $\mathcal{A}$ is history-deterministic.*

*Proof.* We assume, without loss of generality, that $\mathcal{A}$ is such that all 2-transitions occur in some SCC in the graph $G_{>1}$ consisting of only 2-transitions (Assumption 4.4). Fix $\sigma$ to be an uniform positional strategy in the HD game on $\mathcal{A}_{\texttt{safe}}$ from states that are safe-deterministic (Theorem 3.22). Let $\mathcal{D}_{\texttt{safe}}$ be the deterministic safety automaton obtained as the subautomaton of $\mathcal{A}_{\texttt{safe}}$ consisting of states that are safe-deterministic, and transitions that are in $\sigma$. Note that $L(\mathcal{D}_{\texttt{safe}}, p) = L(\mathcal{A}_{\texttt{safe}}, p)$ for every state $p$ that is safe-deterministic.

For a finite word $u$, we call a state $p$ *active* on $u$ if $p$ is in $\mathsf{WCR}(\mathcal{A}, u)$ and $p$ is safe-deterministic. Note that from Lemma 4.7, for every finite word $u$, there is some state that is active on $u$.

Let $w$ be a (finite or infinite) word. Consider then the DAG $D_w$ consisting of states of the form $(u, q)$, where $u$ is a finite prefix of $w$ and $q$ is a state in $\mathcal{A}$ that is active on $u$. We add an edge from $(u, q)$ to $(ua, q')$ in $D_w$ if $a \in \Sigma$, $u$ and $ua$ are prefixes of $w$, and $q \xrightarrow{a} q'$ is a transition in $\mathcal{D}_{\texttt{safe}}$ such that $q'$ is not the rejecting sink state. Observe that each vertex in the DAG $D_w$ has outdegree at most 1 since $\mathcal{D}_{\texttt{safe}}$ is deterministic.

We claim that if $w$ is an infinite word in $L(\mathcal{A})$ then the DAG $D_w$ contains an infinite path. Indeed, let $\rho$ be an accepting run of $\mathcal{A}$ on $w = uw'$ such that $\rho$ ends in some state $p$ upon reading the word $u$, after which $\rho$ does not contain any 1-transition. Then, note that $w' \in L(\mathcal{A}_{\texttt{safe}}, p)$. Let $q \in \mathsf{WCR}(\mathcal{A}, p)$ be a state, such that Eve wins $G1(q; p)$ in $\mathcal{A}_{\texttt{safe}}$ and $q$ is safe-deterministic. Then,

$$w' \in L(\mathcal{A}_{\texttt{safe}}, p) \subseteq L(\mathcal{A}_{\texttt{safe}, q}) = L(\mathcal{D}_{\texttt{safe}}, q).$$

It is then clear that in DAG $D_w$, the unique path from $(u, q)$ is of infinite length.

Fix a positional strategy $\sigma_{G1}$ for Eve in the 1-token game on $\mathcal{A}$ from all pairs of weakly coreachable states. Eve then has the following winning strategy in the HD game on $\mathcal{A}$.

At the start of the HD game on $\mathcal{A}$, Eve *tracks* a state $r_0$ in $\mathsf{WCR}(\mathcal{A}, q_0)$, such that $r_0$ is safe-deterministic. In general, after Adam has played $u$ in the HD game on $\mathcal{A}$, Eve will be at some state $q_u$, and she will be *tracking* some state $r_u$ that is weakly coreachable to $q_u$ in $\mathcal{A}$ and is safe-deterministic. After Adam plays the letter $a$, Eve will then choose the $a$-transition on her token given by the strategy $\sigma_{G1}$ from the position of the 1-token game on $\mathcal{A}$ where Eve's token is at $q_u$ and Adam's token is at $r_u$. If $r_u$ has a transition to a state $r_{ua}$ in $\mathcal{D}_{\texttt{safe}}$ that is not the rejecting sink state, or equivalently, there is an edge from $(u, r)$ to $(ua, r_{ua})$ in $D_{ua}$, then we let Eve track $r_{ua}$ for the next round. If there is no outgoing edge from $(u, r)$ in $D_{ua}$, then Eve *resets* to instead track a state $r'_{ua}$ such that the following holds: the path ending at $(ua, r'_{ua})$ has the longest length amongst the paths ending at vertices of the form $(ua, s)$ in $D_{ua}$ where $s$ is an active state of $ua$.

We claim that the strategy thus described is a winning strategy for Eve in the HD game on $\mathcal{A}$. Indeed, if Adam produces the word $w$ that is in $L(\mathcal{A})$, then the DAG $D_w$ contains an infinite path. Thus, after some point, Eve never resets the states she is tracking, and the states she is tracking constitutes an accepting run in $\mathcal{A}$. Since $\sigma_{G1}$ is a winning strategy for Eve from all pairs of weakly coreachable states, it follows that the run on Eve's token in the HD game is accepting run as well. This concludes our proof. □

From Theorem 4.2, it follows that Eve wins the Joker game on $\mathcal{A}$ if and only if it is HD (Theorem 4.1).

**Theorem 4.1.** *For every coBüchi automaton $\mathcal{A}$, Eve wins the Joker game on $\mathcal{A}$ if and only if $\mathcal{A}$ is history-deterministic.*

*Proof.* We focus on showing that if Eve wins the Joker game on $\mathcal{A}$ then $\mathcal{A}$ is HD, the converse being easy due to Lemma 3.21. From Theorem 3.34, we know that there is a simulation-equivalent subautomaton $\mathcal{B}$ of $\mathcal{A}$, such that Eve wins the 1-token game from everywhere. From Theorem 4.2 we just proved, $\mathcal{B}$ is HD and, therefore, so is $\mathcal{A}$ due to simulation-equivalence of $\mathcal{A}$ and $\mathcal{B}$ (Corollary 3.11). □

We also note that Theorem 4.1 also proves the 2-token game based characterisation of history-determinism on coBüchi automata, since if Eve wins the 2-token game on a parity automaton, then Eve also wins the Joker game on that automaton.

**Corollary 4.3.** *For every coBüchi automaton $\mathcal{A}$, Eve wins the 2-token game on $\mathcal{A}$ if and only if $\mathcal{A}$ is history-deterministic [BKLS20].*

# Chapter 5

# Büchi Automata

In this chapter, we extend our Joker-game based characterisation of history-determinism on coBüchi automata to Büchi automata.

**Theorem 5.1.** *For every Büchi automaton $\mathcal{A}$, Eve wins the Joker game on $\mathcal{A}$ if and only if $\mathcal{A}$ is history-deterministic.*

We will give two proofs of Theorem 5.1. For the first proof, starting with a winning strategy for Eve in the Joker game on $\mathcal{A}$, we will construct a strategy for Eve in the HD game on $\mathcal{A}$ that requires linear memory. This construction takes polynomial time. This strategy thus describes a language-equivalent deterministic Büchi automaton that has quadratic size as compared to the original Büchi automaton we started with.

**Theorem D.** *There is a polynomial-time procedure that converts every HD Büchi automaton with $n$ states into a language-equivalent deterministic Büchi automaton with $n^2$ states.*

Kuperberg and Skrzypczak in 2015 also gave a determinisation procedure for history-deterministic Büchi automata with a quadratic state-space blowup [KS15, Theorem 8]. Their procedure is in NP, however [KS15, Theorem 10], and they left the question of whether there is a polynomial time determinisation procedure open [KS15, Page 310]. Theorem D answers this question in the positive.

The template of our (first) proof is similar to that of our proof for coBüchi automata. We show that if Eve wins the 1-token game from everywhere on a Büchi automaton, then that automaton is HD.

**Theorem 5.2.** *For every Büchi automaton $\mathcal{A}$, if Eve wins the 1-token game on $\mathcal{A}$ then $\mathcal{A}$ is HD.*

To show Theorem 5.2, we introduce an intermediate property called reach-covering for Büchi automata, which is analogous to the property of safe-coverage we had used for coBüchi automata.

We first show in Section 5.1 that any automaton on which Eve wins the 1-token game from everywhere and has reach-covering is history-deterministic, and can be determinised efficiently.

We then show, in Section 5.2, that any Büchi automaton on which Eve wins the 1-token game from everywhere can be modified, in PTIME, to a simulation-equivalent automaton (with the same number of states) on which Eve still wins the 1-token game from everywhere and has reach-covering. This will conclude our proof for Theorem D and Theorem 5.2, and thus, also of Theorem 5.1. In Section 5.3, we will show that unlike for coBüchi automata (Theorem 4.1) and Büchi automata (Theorem 5.1), Eve winning the Joker game on a parity automaton need not imply that the automaton is HD (Theorem 5.18).

We will also give a second proof of Theorem 5.1, for which we introduce the lookahead games. For each $k \geq 0$, the $k$-lookahead game on an automaton is like the 1-token game, but Adam delays transitions on his token by $k$ rounds, that is, he picks his token's transition on the $i^{th}$ letter in the word in round $(k+i)$. We show that the 1-token game and the $k$-lookahead game on the same automaton have the same winner, for all $k \geq 0$.

**Theorem E.** *For every parity automaton $\mathcal{A}$, Eve wins the $1$-token game on $\mathcal{A}$ if and only if she wins the $k$-lookahead game on $\mathcal{A}$ for all $k \geq 0$.*

Using this, we prove the following result that implies Theorem 5.2 and Theorem 5.1.

**Theorem F.** *For every semantically-deterministic Büchi automaton $\mathcal{A}$, $\mathcal{A}$ is history-deterministic if and only if Eve wins the $1$-token game on $\mathcal{A}$.*

Before proving Theorem F in Section 5.5, we give, in Section 5.4, a technique to reduce game-based characterisations of history-determinism on automata to showing the same characterisations for universal automata, i.e., automata that accept all infinite words over its alphabet.

**Theorem 5.3.** *The following two statements are equivalent.*

1. *Every semantically-deterministic Büchi automaton on which Eve wins the $1$-token game is history-deterministic.*

2. *Every semantically-deterministic Büchi automaton that accepts all infinite words over its alphabet and on which Eve wins the $1$-token game is history-deterministic.*

Theorem 5.3 will simplify our proof of Theorem F in Section 5.5. A similar statement also holds for showing the 2-token game characterisation of history-determinism on parity automata (Theorem 5.20).

## 5.1 Automata with reach-covering

In this section, we introduce the property of reach-covering and show that every Büchi automaton on which Eve wins the 1-token game from everywhere and that has reach-covering is HD and can be efficiently determinised.

Reach-covering for Büchi automata is defined similarly to safe-coverage for coBüchi automata. For a Büchi automaton $\mathcal{A}$, we define the 1-approximation of $\mathcal{A}$, denoted by the reachability automaton $\mathcal{A}_{\texttt{reach}}$, as the automaton where transitions of priority 1 in $\mathcal{A}$ are kept as is, while transitions of priority 0 are redirected to an accepting sink state $q_\top$.

The following observation is easy to see.

**Proposition 5.4.** *For every Büchi automaton $\mathcal{B}$, $L(\mathcal{B}) \subseteq L(\mathcal{B}_{reach})$.*

*Proof.* If $\rho$ is an accepting run on some word $w$ in $\mathcal{B}$, then it must contain a 0 priority transition. Consider the run $\rho_{\texttt{reach}}$ of $\mathcal{B}_{\texttt{reach}}$ on $w$ that has the same transitions as the largest prefix of $\rho$ that contains only priority 1 transitions, and $\rho_{\texttt{reach}}$ has a transition to the accepting sink state at the position corresponding to the first priority 0 transition in $\rho$. Then $\rho_{\texttt{reach}}$ is an accepting run, as desired. □

We say that a Büchi automaton $\mathcal{A}$ has *reach-covering* if for every state $q$ in $\mathcal{A}$, there is another state $p$ weakly coreachable to $q$ in $\mathcal{A}$, such that Eve wins $G1(q;p)$ in $\mathcal{A}_{\texttt{reach}}$.

> **Automata with reach-covering.** We say that a Büchi automaton $\mathcal{A}$ has *reach-covering* if for each state $q$, there is another state $p \in \mathsf{WCR}(\mathcal{A}, q)$ such that Eve wins $G1(q;p)$ in $\mathcal{A}_{\texttt{reach}}$.

Let us compare the definitions of reach-covering and safe-coverage. Recall that for a coBüchi automaton $\mathcal{A}$, we say that $\mathcal{A}$ has safe-coverage if for each $q$ there is a $p$ in $\mathsf{WCR}(\mathcal{A}, q)$ such that Eve wins $G1(p;q)$ in $\mathcal{A}_{\texttt{safe}}$. For reach-covering, however, note that the roles of $p$ and $q$ in the 1-token game to be other way around.

Reach-covering is a special case of the property of 0-reach double-covering that we will use for the odd-to-even induction step (Theorem B in Chapter 7).

The next two results are analogous to Lemmas 4.6 and 4.7 for coBüchi automata.

**Lemma 5.5.** *If a Büchi automaton $\mathcal{A}$ has reach-covering, then for each state $q$ there is a state $p$ weakly coreachable to $q$ in $\mathcal{A}$, such that Eve wins $G1(q;p)$ and $G1(p;p)$ in $\mathcal{A}_{reach}$.*

*Proof.* This follows from the definition of reach-covering, the transitivity of 1-token games (Lemma 3.25), and the fact that there are finitely many states in $\mathcal{A}$. Concretely, for each state $q$ in $\mathcal{A}$, consider the directed graph $G$ whose vertices consist of states in $\mathsf{WCR}(\mathcal{A}, q)$. There is an edge in $G$ from vertices $r$ to $s$ if and only if Eve wins $G1(r;s)$ in $\mathcal{A}_{\texttt{reach}}$.

Since $\mathcal{A}$ has reach-covering, every vertex in $G$ has outdegree at least 1. Thus, for every vertex $r$, there is another vertex $s$, such that there is a path from $r$ to $s$ and a cycle consisting of the vertex $s$ in $G$. Observe that if there is a path from $r$ to $s$ in $G$, then due to transitivity of $G1$ (Lemma 3.25), Eve wins $G1(r;s)$ in $\mathcal{A}_{\texttt{reach}}$. It follows that for each $r$ in $G$, there is an $s$ such that Eve wins $G1(r;s)$ and $G1(s;s)$ in $\mathcal{A}_{\texttt{reach}}$. This concludes our proof. □

Recall from Theorem 3.22 that if Eve wins the 1-token game on a reachability automaton, then that automaton is history-deterministic and determinisable by pruning. Thus, we call a state $p$ in $\mathcal{A}$ *reach-deterministic* if Eve wins $G1(p;p)$ in $\mathcal{A}_{\texttt{reach}}$. We can then rephrase Lemma 5.5 as the following result.

**Lemma 5.6.** *If $\mathcal{A}$ is a Büchi automaton with reach-covering, then for every state $q$ in $\mathcal{A}$, there is another state $p$ weakly coreachable to $q$ in $\mathcal{A}$, such that Eve wins $G1(q;p)$ in $\mathcal{A}_{reach}$ and $p$ is reach-deterministic.*

We now show that any Büchi automaton with reach-covering on which Eve wins the 1-token game from everywhere is HD, and can be determinised with a quadratic state-space blowup.

**Lemma 5.7.** *Let $\mathcal{A}$ be a Büchi automaton with $n$ states, such that Eve wins the 1-token game from everywhere in $\mathcal{A}$ and $\mathcal{A}$ has reach-covering. Then $\mathcal{A}$ is history-deterministic. Additionally, a language-equivalent deterministic automaton $\mathcal{D}$ with at most $n^2$ states can be constructed in* PTIME.

*Proof.* Fix $\sigma_{\texttt{reach}}$ to be an uniform positional strategy for Eve in the HD game on $\mathcal{A}_{\texttt{reach}}$ from states that are reach-deterministic. Such a strategy $\sigma_{\texttt{reach}}$ can be computed in PTIME (Theorem 3.22). Let $\mathcal{D}_{\texttt{reach}}$ be the deterministic reachability automaton obtained as the subautomaton of $\mathcal{A}_{\texttt{reach}}$ consisting of states that are reach-deterministic and transitions that the strategy $\sigma_{\texttt{reach}}$ takes. Note that for each state $p$ in $\mathcal{A}$ that is reach-deterministic, $L(\mathcal{D}_{\texttt{reach}}, p) = L(\mathcal{A}_{\texttt{reach}}, p)$.

Fix a positional winning strategy $\tau$ for Eve in the 1-token game on $\mathcal{A}_{\texttt{reach}}$ from all pairs of states $q, p$ that are weakly coreachable in $\mathcal{A}$, such that Eve wins $G1(q;p)$ in $\mathcal{A}_{\texttt{reach}}$ and $p$ is reach-deterministic. This strategy $\tau$ can be computed in PTIME too, by solving the 1-token game in $\mathcal{A}_{\texttt{reach}}$ from all pairs of weakly coreachable states in $\mathcal{A}$ and finding a uniform positional winning strategy for Eve from her winning region (see Lemma 3.38).

We next describe a deterministic Büchi automaton $\mathcal{D}$ that we will show to be language-equivalent to $\mathcal{A}$.

- The states of $\mathcal{D}$ are pairs $(q, r)$ where $q$ is a state in $\mathcal{A}$ and $r$ is a state in $\mathcal{A}_{\texttt{reach}}$ such that Eve wins $G1(q;r)$ in $\mathcal{A}_{\texttt{reach}}$ and either of the following two conditions hold.
  1. $r$ is reach-deterministic and in $\mathsf{WCR}(\mathcal{A}, q)$.
  2. $r$ is the accepting sink state in $\mathcal{A}_{\texttt{reach}}$.

  We let the initial state $d_0$ of $\mathcal{D}$ to be $(q_0, r_0)$ such that $r_0$ is reach-deterministic and Eve wins $G1(q_0; r_0)$ in $\mathcal{A}_{\texttt{reach}}$.

- We next describe the deterministic transition of $\mathcal{D}$ on reading the letter $a$ from the state $(q, r)$. Let $\delta = q \xrightarrow{a:c} q'$ be the transition given by $\tau$ from $(q;r)$ in $G1(\mathcal{A}_{\texttt{reach}})$.
  
  - If $\delta = q \xrightarrow{a:0} q'$ is an accepting transition in $\mathcal{A}$, then we have the transition $(q, r) \xrightarrow{a:0} (q', r')$ in $\mathcal{D}$, where $r'$ is a state in $\mathsf{WCR}(\mathcal{A}, q')$, such that $r'$ is reach-deterministic and Eve wins $G1(q'; r')$ in $\mathcal{A}_{\texttt{reach}}$.
  - Otherwise, we have the transition $(q, r) \xrightarrow{a:1} (q', r')$ in $\mathcal{D}$, where $r \xrightarrow{a} r'$ is the unique transition from $r$ on $a$ in $\mathcal{D}_{\texttt{reach}}$.

We claim that $\mathcal{D}$ and $\mathcal{A}$ are language-equivalent.

$L(\mathcal{D}) \subseteq L(\mathcal{A})$. Note that if $(q, r) \xrightarrow{a:0} (q', r')$ is a transition in $\mathcal{D}$, then $q \xrightarrow{a:0} q'$ is a transition in $\mathcal{A}$. Thus, if $\rho_D$ is an accepting run of a word $w$ in $\mathcal{D}$, then the projection of $\rho_D$ on the component of $\mathcal{A}$ is an accepting run in $\mathcal{A}$ as well.

$L(\mathcal{A}) \subseteq L(\mathcal{D})$. Note that $\mathcal{A}$ is semantically-deterministic since Eve wins the 1-token game from everywhere (Lemma 3.36). Let $w$ be an infinite word, and $\rho_D$ be the unique run of $\mathcal{D}$ on $w$. We will show that if $\rho_D$ is a rejecting run, then $w$ is not in $L(\mathcal{A})$.

If $\rho_D$ is not an accepting run, then there is a finite prefix $u$ of $w$ after which no accepting transitions occur in $\rho_D$. Suppose that the run $\rho_D$ is at the state $(q, p)$ after reading $u$, and let $w'$ be such that $w = uw'$. Then $p$ is not the accepting sink state in $\mathcal{D}_{\texttt{reach}}$; otherwise, the run $\rho_D$ from $(q, p)$ must contain an accepting transition since $\tau$ is a winning strategy for Eve. Thus, $q$ and $p$ are weakly coreachable in $\mathcal{A}$ and Eve wins $G1(q; p)$ in $\mathcal{A}_{\texttt{reach}}$. Since $\mathcal{A}$ is semantically deterministic, we know that $L(\mathcal{A}, q) = L(\mathcal{A}, p) = u^{-1} L(\mathcal{A}, q_0)$. Furthermore, observe that for every state $r$ in $\mathcal{A}$, we have $L(\mathcal{A}, r) \subseteq L(\mathcal{A}_{\texttt{reach}}, r)$ (Proposition 5.4). Since $p$ is reach-deterministic, we know that

$$L(\mathcal{A}, p) \subseteq L(\mathcal{A}_{\texttt{reach}}, p) = L(\mathcal{D}_{\texttt{reach}}, p). \tag{5.1}$$

Because $\tau$ is a winning strategy of $G1(q; p)$ in $\mathcal{A}_{\texttt{reach}}$, and the run $\rho_D$ from $(q, p)$ on $w'$ does not contain an accepting transition, we note that the deterministic run on $w'$ in $(\mathcal{D}_{\texttt{reach}}, p)$ is rejecting. Thus, $w'$ is not in $L(\mathcal{A}, p)$ (due to Eq. (5.1)), and hence $w = uw'$ is not in $L(\mathcal{A}, q_0)$, as desired.

Thus, we have proved that $\mathcal{D}$ is language-equivalent to $\mathcal{A}$.

Since every run of $\mathcal{D}$ constitutes a run of $\mathcal{A}$ in its first component with the same priorities, the deterministic automaton $\mathcal{D}$ also describes a winning strategy for Eve in the HD game on $\mathcal{A}$: for every word, Eve construct a run on her token on that word according to the first component of the unique deterministic run in $\mathcal{D}$ on that word. □

## 5.2 Towards automata with reach-covering

In this section, we given an algorithm to convert a Büchi automaton on which Eve wins the 1-token game from everywhere into another Büchi automaton which, in addition, has reach-covering.

**Lemma 5.8.** *There is a* PTIME *algorithm that computes, for each input Büchi automaton $\mathcal{A}$ on which Eve wins the 1-token game from everywhere, a simulation-equivalent automaton $\mathcal{A}_N$ with at most as many states as $\mathcal{A}$, such that $\mathcal{A}_N$ has reach-covering and Eve wins the 1-token game from everywhere in $\mathcal{A}_N$.*

Using this result together with Lemma 5.7, which we proved in the previous section, we will deduce Theorems 5.1 and 5.2 and Theorem D.

Let $\mathcal{A}$ be a Büchi automaton on which Eve wins the 1-token game from everywhere. We first describe a parity game $\mathcal{G}_1(\mathcal{A})$ with priorities in $[0, 2]$ that is an explicit representation of the 1-token game on $\mathcal{A}$. We then prove Lemma 5.8 by describing a procedure that iteratively modifies the automaton $\mathcal{A}$ based on ranks in $\mathcal{G}_1(\mathcal{A})$.

**Explicitly representing the 1-token games on Büchi automata**

**Definition 5.9.** *For the Büchi automaton $\mathcal{A} = (Q, \Sigma, q_0, \Delta)$, define the $[0, 2]$ parity game $\mathcal{G}_1(\mathcal{A}) = (V, E)$ as follows.*

- *The set of vertices $V$ is the union of the following three sets.*

  1. $V_1 = \{(q, p) \mid q, p \text{ are weakly coreachable states in } \mathcal{A}\}$
  2. $V_2 = \{(q, a, p) \mid (q, p) \in V_1 \text{ and } a \in \Sigma\}$
  3. $V_3 = \{(q', p, a) \mid (q, a, p) \in V_2 \text{ and } q \xrightarrow{a:c} q' \in \Delta\}$

  *Eve's vertices are $V_\exists = V_2$, while Adam's vertices are $V_\forall = V_1 \cup V_3$*

- *The set of edges $E$ is the union of the following three sets.*

  1. $E_1 = \{(q, p) \to (q, a, p) \mid a \in \Sigma\}$ *(Adam chooses a letter)*
  2. $E_2 = \{(q, a, p) \to (q', p, a) \mid q \xrightarrow{a:c_e} q' \in \Delta\}$ *(Eve chooses a transition on her token)*
  3. $E_3 = \{(q', p, a) \to (q', p') \mid p \xrightarrow{a:c_a} p' \in \Delta\}$ *(Adam chooses a transition on his token)*

- *The priorities on these edges are given as follows. All elements in $E_1$ have priority 2, while an edge $(q, a, p) \to (q', p, a)$ in $E_2$ has priority 0 if the corresponding transition $\delta = q \xrightarrow{a:c_e} q'$ is accepting (or equivalently, $c_e$ is 0), and 2 otherwise. The edge $(q', p, a) \to (q', p')$ in $E_3$ has priority 1 if the corresponding transition $p \xrightarrow{a:c_a} p'$ is accepting, and 2 otherwise.*

The next result shows that the 1-token game on $\mathcal{A}$ and $\mathcal{G}_1(\mathcal{A})$ have the same winner.

**Lemma 5.10.** *For any Büchi automaton $\mathcal{A}$ and its states $q, p$ that are weakly coreachable, Eve wins $G1(q; p)$ in $\mathcal{A}$ if and only if Eve wins $\mathcal{G}_1(\mathcal{A})$ from the vertex $(q, p)$.*

*Proof.* We note that the game arena of the 1-token game on $\mathcal{A}$ and the game arena of $\mathcal{G}_1(\mathcal{A})$ (without its priorities) are syntactically equivalent. Thus, it suffices to prove that the winning plays in $G1(q; p)$ on $\mathcal{A}$ correspond to the winning plays in $\mathcal{G}_1(\mathcal{A})$ from $(q, p)$ and vice-versa.

To show this, we note that $\rho$ is a losing play for Eve in $\mathcal{G}_1(\mathcal{A})$ from $(q, p)$, if and only if, the priority 1 edges occur infinitely often and priority 0 edges occur finitely often in $\rho$, if and only if, the run on Adam's token contain infinitely many priority 0-transitions and is accepting while the run on Eve's token contain finitely many priority 0-transitions and is rejecting, if and only if, $\rho$ corresponds to a losing play in the 1-token game on $\mathcal{A}$. □

Lemma 5.10 implies that if $\mathcal{A}$ is a Büchi automaton on which Eve wins the 1-token game from everywhere, then Eve wins the game $\mathcal{G}_1(\mathcal{A})$ from all vertices.

We will use the ranks in $\mathcal{G}_1(\mathcal{A})$ (see Section 2.5) to iteratively modify $\mathcal{A}$ into a simulation-equivalent automaton that has reach-covering. The next result connects ranks in $\mathcal{G}_1(\mathcal{A})$ to the 1-token game between states of $\mathcal{A}_{\text{reach}}$.

**Lemma 5.11.** *Let $\mathcal{A}$ be a Büchi automaton on which Eve wins the 1-token game from everywhere. If $q$ and $p$ are weakly coreachable states in $\mathcal{A}$, such that $\mathsf{rank}(q,p) = 0$ in $\mathcal{G}_1(\mathcal{A})$, then Eve wins $G1(q;p)$ in $\mathcal{A}_{\text{reach}}$.*

*Proof.* Let $\sigma$ be an optimal strategy for Eve in $\mathcal{G}_1(\mathcal{A})$ (see Lemma 2.9). Then this optimal strategy ensures that in every play from $\mathcal{G}_1(\mathcal{A})$ on $(q,p)$, Eve sees a 0 priority edge before a 1 priority edge. In the game $G1(q;p)$ in $\mathcal{A}$, this strategy can thus be used to ensure Eve's token takes an accepting transition in no later round than Adam's token, or equivalently, in $G1(q;p)$ in $\mathcal{A}_{\text{reach}}$, Eve's token reaches the accepting sink state in no later round than Adam's token. Thus, Eve wins $G1(q;p)$ in $\mathcal{A}_{\text{reach}}$. □

Let us fix a Büchi automaton $\mathcal{A}$ on which Eve wins the 1-token game from everywhere. For each state $q$ in $\mathcal{A}$, we define the optimal rank of $q$, denoted $\mathsf{opt}(q)$, as the minimum rank amongst $\mathsf{rank}(q,p)$ for states $p$ that are weakly coreachable to $p$ in $\mathcal{A}$, i.e.,

$$\mathsf{opt}(q) = \min\{\mathsf{rank}(q,p) \mid (q,p) \in \mathsf{WCR}(\mathcal{A})\}.$$

Note that, due to Lemma 5.11, if all states $q$ in $\mathcal{A}$ have optimal rank 0, then $\mathcal{A}$ has reach-covering. The iterative modification we present below, which we call the *rank-reduction procedure*, thus has the objective of making the optimal rank 0 for all states.

**Rank-reduction procedure.** Starting with a Büchi automaton $\mathcal{A}$ on which Eve wins the 1-token game from everywhere, we iteratively modify $\mathcal{A}$ to $\mathcal{A}_N$, so that we obtain that all states $q$ have optimal rank 0 in $\mathcal{A}_N$.

Set $\mathcal{A}_0 = \mathcal{A}$. For each $i \geq 0$, we perform the following three steps on $\mathcal{A}_i$ until $\mathcal{A}_{i+1} = \mathcal{A}_i$.

Step 1 For each state state $q$ in $\mathcal{A}_i$, compute the optimal rank of $q$ in $\mathcal{G}_1(\mathcal{A}_i)$, which we denote $\mathsf{opt}_i(q)$.

Step 2 Obtain $\mathcal{A}'_i$ from $\mathcal{A}_i$ by removing all transitions $q \xrightarrow{a:1} q'$ with $\mathsf{opt}_i(q) < \mathsf{opt}_i(q')$.

Step 3 Obtain $\mathcal{A}_{i+1}$ from $\mathcal{A}'_i$, by changing priorities of transitions $q \xrightarrow{a:1} q'$ such that $\mathsf{opt}_i(q) > \mathsf{opt}_i(q')$ to 0.

We will argue in Section 5.2.1 that the following invariants are preserved during each of the Steps of the above procedure.

**I1** Simulation-equivalence to the automaton $\mathcal{A}$.

**I2** Eve winning the 1-token game from everywhere.

The ranks for vertices in $\mathcal{G}_1(\mathcal{A})$ can be computed in PTIME, since the ranks for a $[0, 2]$ parity game with $|V|$ vertices and $|E|$ edges can be computed in time $\mathcal{O}(|V| \cdot |E|)$ ([Jur00, Theorem 11]). Furthermore, observe that each iteration of the rank-reduction procedure either deletes transitions, or changes the priority of certain transitions from 1 to 0. Thus, the rank-reduction procedure terminates after at most $|\Delta|$ many iterations, and each iteration takes polynomial time.

**Lemma 5.12.** *For every Büchi automaton $\mathcal{A}$ on which Eve wins the 1-token game from everywhere, the rank-reduction procedure on $\mathcal{A}$ terminates in PTIME.*

Let $\mathcal{A}_N$ then be the automaton obtained by the rank-reduction procedure. We will show in Section 5.2.2 that all states in $\mathcal{A}_N$ have their optimal rank as 0, and hence $\mathcal{A}_N$ has reach-covering. This will prove Lemma 5.8, and allow us to deduce Theorems 5.1 and 5.2, and Theorem D.

### 5.2.1 Invariants for the rank-reduction procedure

We now show that the invariants I1 and I2 are preserved during each of the Steps 2 and 3 of the rank-reduction procedure.

Step 2 removes certain transitions from the automaton. To show that invariants are preserved during Step 2, we will use the following result.

**Lemma 5.13.** *Let $\mathcal{P}$ be a nondeterministic parity automaton, and $\mathcal{P}'$ a subautomaton of $\mathcal{P}$ such that for every pair of states $p, q$ that are weakly coreachable in $\mathcal{P}$, Eve wins $G1((\mathcal{P}', p); (\mathcal{P}, q))$. Then the following statements hold.*

1. *$\mathcal{P}$ and $\mathcal{P}'$ are simulation equivalent.*

2. *Eve wins the 1-token game from everywhere in $\mathcal{P}'$.*

*Proof.* Proof of (1). Note that since $\mathcal{P}'$ is a subautomaton of $\mathcal{P}$, $\mathcal{P}$ simulates $\mathcal{P}'$: Eve in the simulation game of $\mathcal{P}'$ by $\mathcal{P}$ can simply copy the transitions of Adam's token in $\mathcal{P}'$ in her token in $\mathcal{P}$. For the other direction, note that Eve wins $G1(\mathcal{P}'; \mathcal{P})$, and hence $\mathcal{P}'$ simulates $\mathcal{P}$ (Lemma 3.26.5). Thus $\mathcal{P}$ and $\mathcal{P}'$ are simulation-equivalent.

Proof of (2). If two states $p$ and $q$ are weakly coreachable in $\mathcal{P}'$, then they are also weakly coreachable in $\mathcal{P}$. Therefore Eve wins $G1((\mathcal{P}', p); (\mathcal{P}, q))$, and since $\mathcal{P}'$ is a subautomaton of $\mathcal{P}$, Eve wins $G1(p; q)$ in $\mathcal{P}'$. It follows that Eve wins the 1-token game from everywhere in $\mathcal{P}'$. □

We use Lemma 5.13 to show that invariants are preserved during Step 2.

**Lemma 5.14.** *The invariants I1 and I2 are preserved during Step 2.*

*Proof.* Fix $\sigma_{G1}$ to be an optimal positional strategy for Eve in the parity game $\mathcal{G}_1(\mathcal{A}_i)$, that is also a strategy for Eve in the 1-token game from everywhere on $\mathcal{A}_i$. Consider the strategy $\sigma'_{G1}$ that takes a 0 priority transition on Eve's token whenever Adam's letter $a$ is such that there is an outgoing transition on $a$ with priority 0 from Eve's token, and otherwise follows $\sigma_{G1}$. Note that $\sigma'_{G1}$ is a positional winning strategy since Eve wins $G1$ from everywhere, and furthermore, is an optimal strategy since 0 priority transitions on Eve's token corresponds to 0 priority edges in $\mathcal{G}_1(\mathcal{A}_i)$. Thus we can assume, without loss of generality, that the strategy $\sigma_{G1}$ takes a 0 priority transition on Eve's token whenever possible.

We will describe a winning strategy $\sigma'$ of Eve in $G1((\mathcal{A}'_i, q); (\mathcal{A}_i, p))$ for all states $q, p$ that are weakly coreachable in $\mathcal{A}_i$. It will then follow from Lemma 5.13 that the invariants are preserved during Step 2.

At a high level, the strategy $\sigma'$ will require Eve to store as memory an additional token. Eve's memory token will choose transitions according to $\sigma_{G1}$ by playing the 1-token game against Adam's token while Eve's token will select transitions according to $\sigma_{G1}$ by playing the 1-token game against her memory token, until her token takes a priority 0 transition or the transition given by $\sigma_{G1}$ has been deleted in $\mathcal{A}'_i$, in which case she *resets* her memory.

We describe the strategy $\sigma'$ inductively, as rounds of the 1-token game proceed along.

At the start of the 1-token game, let $q_0 = q, p_0 = p$, and we let Eve's memory token be at a state $r_0 \in \mathsf{WCR}(\mathcal{A}_i, q_0)$, such that $\mathsf{opt}(q_0) = \mathsf{rank}(q_0, r_0)$. Throughout the play, we will preserve the invariant that Eve's token, Eve's memory token, and Adam's token are all weakly coreachable states in $\mathcal{A}_i$.

After $i$ rounds in the 1-token game for some $i \geq 0$, suppose the 1-token game is at the position $(q_j, p_j)$, and Eve's memory token is at $r_j$. Eve then plays as follows. Adam chooses the letter $a_j$, and let $\delta_j = q_j \xrightarrow{a_j : c_j} q'_j$ be the transition given by $\sigma_{G1}$ from $(q_j, a_j, r_j)$. We distinguish between the following three cases.

**Case 1.** $\delta_j$ is a 0-transition.

Then, Eve takes this transition on her token, thus setting $q_{j+1}$ to be $q'_j$. She resets her memory token to be at $r_{j+1}$ in $\mathsf{WCR}(\mathcal{A}_i, q_{j+1})$, such that $\mathsf{rank}(q_{j+1}, r_{j+1}) = \mathsf{opt}(q_{j+1})$.

**Case 2.** If $\delta_j$ is a transition of priority 1 that has not been removed in $\mathcal{A}'_i$.

Then Eve moves her token to $q_{j+1}$. She takes the transition on her memory token $r_{j+1}$ that is given by the 1-token game strategy $\sigma_{G1}$ against Adam's token at $p_j$.

**Case 3.** If $\delta_j$ is no longer a transition in $\mathcal{A}'_i$.

Eve then finds a $s_j$ so that $\mathsf{opt}(q_j) = \mathsf{rank}(q_j, s_j)$. She picks the transition $\delta'$ given by $\sigma_{G1}$ from $(q_j, a, s_j)$ on her token. Note that $\delta'$ must be a transition in $\mathcal{A}'_i$ due to monotonicity of ranks (Proposition 2.10). Eve's memory token is then reset to $s_j$, that then follows transitions given by $\sigma_{G1}$ against Adam's tokens at $p_j$.

This concludes the description of Eve's strategy $\sigma'$. We will argue that $\sigma'$ is a winning strategy for Eve. Towards this, let $\rho$ be a play for Eve in $\mathsf{G1}((\mathcal{A}'_i, q); (\mathcal{A}_i, p))$ where Eve is playing according to $\sigma'$.

Firstly, suppose that eventually only moves from Case 2 occur in $\rho$. Let $w$ be the word that Adam plays in $\rho$, and suppose the run of his token on $w$ is accepting. Then, since Eve's memory token are eventually not reset, the suffix of the moves on Eve's memory token constitute a run on some suffix of $w$ in $\mathcal{A}_i$, which is accepting since $\sigma_{G1}$ is a winning strategy. The run of Eve's token then is accepting as well, as desired.

If moves from Case 1 occur infinitely often in $\rho$, then the run in Eve's token contains infinitely many 0-transitions and it is accepting, implying Eve wins any such play. We will therefore conclude from the next claim that $\sigma'$ is a winning strategy.

**Claim 3.** *In any play following $\sigma'$, either moves from Case 1 are taken infinitely often, or only moves from Case 2 are taken eventually.*

It suffices to show that at most $K$ many moves from Case 3 can be taken before a move from Case 1 is taken, where $K$ is the number of vertices in $\mathcal{G}_1(\mathcal{A}_i)$. Recall that $\mathsf{rank}(q)$ for any vertex in $\mathcal{G}_1(\mathcal{A}_i)$ is bounded by $K$ (Proposition 2.11). We next show that the rank of the configuration between the state of Eve's token and the state of Eve's memory token strictly decreases whenever a move from Case 3 is taken; it is clear from monotonicity of ranks (Proposition 2.10) that a move from Case 2 does not increase the rank.

Suppose at the start of round $j$ in a play following $\sigma'$, Eve's token is at $q_j$ and her memory token is at $r_j$, and Adam chooses a letter $a_j$ such that the strategy $\sigma'$ takes a move from Case 3. Suppose then that at the start of round $j+1$, Eve's token is at $q_{j+1}$ and memory token is at $r_{j+1}$. Let $\delta_j = q_j \xrightarrow{a_j:1} q_j'$ be the transition given by $\sigma_{G1}$ from $(q_j, a_j, r_j)$. Since $\delta_j$ is a deleted transition, we note that $\mathsf{opt}(q_j) < \mathsf{opt}(q_j')$. We therefore have the following series of inequalities

$$\begin{aligned}\mathsf{rank}(q_j, r_j) &\geq \mathsf{opt}(q_j') \\ &> \mathsf{opt}(q_j) \\ &\geq \mathsf{rank}(q_{j+1}, r_{j+1}).\end{aligned}$$

Here, the first inequality holds due to monotonicity of ranks (Proposition 2.10). For the third inequality, note that the transition $q_j \xrightarrow{a:c} q_{j+1}$ cannot have priority 0 since otherwise $\sigma_{G1}$ would have picked a priority 0 transition (and in particular, not $\delta_j$), as we assumed at the start of this proof. Thus, the third inequality also follows from monotonicity of ranks (Proposition 2.10).

Therefore, the quantity $\mathsf{rank}(q_l, r_l)$ strictly decreases from round $l$ to round $(l+1)$ whenever a move from Case 3 is taken in round $l$, and is non-increasing on moves from Case 2. Since the ranks in $\mathcal{G}_1(\mathcal{A}_i)$ are bounded by $K$, this proves the claim, as desired. □

Having proved that Step 2 preserves the invariants I1 and I2, we show the same for Step 3.

**Lemma 5.15.** *The invariants I1 and I2 are preserved during Step 3.*

*Proof.* In Step 3, the automaton $\mathcal{A}_{i+1}$ is obtained by relabelling priorities of certain transitions in $\mathcal{A}_i'$. It suffices to show that a run is accepted in $\mathcal{A}_i'$ if and only if that same run is accepting in $\mathcal{A}_{i+1}$.

One direction is clear: if a run in $\mathcal{A}'_i$ is accepting, then the same run must be accepting in $\mathcal{A}_{i+1}$, since we only changed the priority of certain transitions in $\mathcal{A}'_i$ to 0 to obtain $\mathcal{A}_{i+1}$.

For the other direction, suppose that $\rho$ is an accepting run in $\mathcal{A}_{i+1}$, and consider the same run $\rho$ in $\mathcal{A}'_i$. Note that due to Step 2, the $\mathsf{opt}_i$ values are non-increasing across transitions with priority 1. Additionally, for every transition $\delta = q \xrightarrow{a:0} q'$ that has priority 0 in $\mathcal{A}_{i+1}$ but not in $\mathcal{A}'_i$, the quantity $\mathsf{opt}_i$ strictly decreases, i.e., $\mathsf{opt}_i(q) > \mathsf{opt}_i(q')$. Thus, if $\rho$ has infinitely many 0-transitions in $\mathcal{B}_{i+1}$, then $\rho$ must also have infinitely many 0-transitions in $\mathcal{A}'_i$ since $\mathsf{opt}_i$ is bounded by the number of vertices in $\mathcal{G}_1(\mathcal{A}_i)$. □

Lemmas 5.14 and 5.15 together imply that the rank-reduction procedure preserves the invariants I1 and I2.

### 5.2.2 Stabilisation

In Steps 2 and 3 of the rank-reduction procedure, we are either removing transitions or change the priorities of certain transitions from 1 to 0. Thus, starting with an automaton $\mathcal{A}$ on which Eve wins the 1-token game from everywhere, the rank-reduction procedure terminates after at most $|\Delta|$-many Steps. Let $\mathcal{A}_N$ be the automaton obtained after the rank-reduction procedure terminates. Due to Lemmas 5.14 and 5.15, we know that Eve wins the 1-token game from everywhere in $\mathcal{A}_N$ and $\mathcal{A}_N$ is simulation-equivalent to $\mathcal{A}$. We next show that $\mathcal{A}_N$ has reach-covering.

**Lemma 5.16.** *The automaton $\mathcal{A}_N$ is such that Eve wins the 1-token game from everywhere in $\mathcal{A}_N$, $\mathcal{A}_N$ is simulation-equivalent to $\mathcal{A}$, and $\mathcal{A}_N$ has reach-covering.*

*Proof.* Since each step of the rank-reduction procedure preserves the invariants of Eve winning the 1-token game from everywhere and simulation-equivalence to $\mathcal{A}$, so does the rank-reduction procedure. Thus, Eve wins the 1-token game from everywhere in $\mathcal{A}_N$ and $\mathcal{A}_N$ is simulation-equivalent to $\mathcal{A}$. We next prove that $\mathcal{A}_N$ has reach-covering, for which, due to Lemma 5.11, it suffices to prove that all states in $\mathcal{A}_N$ have optimal rank 0. For every state $q$ and $p$ in $\mathcal{A}_N$, let $\mathsf{rank}_N(q,p)$ be the rank of $(q,p)$ in $\mathcal{G}_1(\mathcal{A}_N)$, and $\mathsf{opt}_N(q)$ be the optimal rank of $q$ in $\mathcal{A}_N$.

Suppose that there exists a state $q$ such that $\mathsf{opt}_N(q) = \mathsf{rank}_N(q,p) > 0$. We will show that the rank-reduction procedure can then run for at least one more iteration on $\mathcal{A}_N$, which will be a contradiction to the fact that $\mathcal{A}_N$ is the automaton obtained after termination of the rank-reduction procedure.

Fix an optimal winning strategy $\tau$ for Eve in $\mathcal{G}_1(\mathcal{A}_N)$. Since $\mathsf{rank}_N(q,p) > 0$, there is a finite play $\rho$ of $\mathcal{G}_1(\mathcal{A}_N)$ starting from $(q,p)$ where Eve is playing according to $\tau$ such that $\rho$ contains an edge of priority 1 but no edge of priority 0. Note that in this play, Eve's token must take only priority 1 transitions since a priority 0 transition on her token corresponds to a priority 0 edge in $\mathcal{G}_1(\mathcal{A}_N)$.

By monotonicity of ranks (Proposition 2.10), we know that $\mathsf{rank}_N$ strictly decreases across $\rho$ at some point. Then there must be a transition on Eve's token in $\rho$ across which the quantity $\mathsf{opt}_N$ decreases as well, since $\rho$ started at $(q,p)$, such that $\mathsf{opt}_N(q) =$ $\mathsf{rank}_N(q, p)$. But since such transitions are made accepting in Step 3 of the iteration, we get that the rank-reduction procedure can run for at least one more iteration on $\mathcal{A}_N$, as desired. □

Lemmas 5.12 and 5.16 together imply Lemma 5.8.

**Lemma 5.8.** *There is a* PTIME *algorithm that computes, for each input Büchi automaton $\mathcal{A}$ on which Eve wins the 1-token game from everywhere, a simulation-equivalent automaton $\mathcal{A}_N$ with at most as many states as $\mathcal{A}$, such that $\mathcal{A}_N$ has reach-covering and Eve wins the 1-token game from everywhere in $\mathcal{A}_N$.*

We now use Lemma 5.8 to prove Theorems 5.1 and 5.2, and Theorem D.

**Theorem 5.2.** *For every Büchi automaton $\mathcal{A}$, if Eve wins the 1-token game on $\mathcal{A}$ then $\mathcal{A}$ is HD.*

*Proof.* Let $\mathcal{A}$ be a Büchi automaton on which Eve wins the 1-token game from everywhere. By Lemma 5.8, we know that there is a simulation-equivalent Büchi automaton $\mathcal{A}_N$ that has reach-covering and on which Eve wins the 1-token game from everywhere. Lemma 5.7 then tells us that $\mathcal{A}_N$ is HD, and hence so is $\mathcal{A}$ due to simulation-equivalence with $\mathcal{A}_N$ (Corollary 3.11). □

Recall that any parity automaton on which Eve wins the Joker game has a simulation-equivalent subautomaton on which Eve wins the 1-token game from everywhere (Theorem 3.34). Theorem 5.2 thus proves Theorem 5.1.

**Theorem 5.1.** *For every Büchi automaton $\mathcal{A}$, Eve wins the Joker game on $\mathcal{A}$ if and only if $\mathcal{A}$ is history-deterministic.*

If Eve wins the 2-token game on a parity automaton, then Eve also wins the Joker game on that automaton (Theorem 3.20). Thus, Theorem 5.1 implies the following result.

**Corollary 5.17** ([BK18]). *For every Büchi automaton $\mathcal{A}$, Eve wins the 2-token game on $\mathcal{A}$ if and only if $\mathcal{A}$ is history-deterministic.*

We conclude this section by proving Theorem D.

**Theorem D.** *There is a polynomial-time procedure that converts every HD Büchi automaton with n states into a language-equivalent deterministic Büchi automaton with $n^2$ states.*

*Proof.* Let $\mathcal{A}$ be a history-deterministic Büchi automaton. Then, Eve wins the Joker game on $\mathcal{A}$ (Lemma 3.21). In Theorem 3.34, we showed that every automaton on which Eve wins the Joker game has a simulation-equivalent subautomaton on which Eve wins the 1-token game from everywhere. In the proof of Theorem 3.34, this subautomaton was constructed by taking a positional strategy $\sigma$ for Eve in the Joker game and removing all transitions that are not used by $\sigma$. Since such a positional strategy for Eve in Joker games can be found in PTIME (Lemma 3.39), we deduce that for $\mathcal{A}$, we can find a simulation equivalent subautomaton $\mathcal{B}$ on which Eve wins the 1-token game from everywhere in PTIME.

Then $\mathcal{B}$ can be converted to a simulation-equivalent automaton $\mathcal{B}_N$ in PTIME, such that $\mathcal{B}_N$ has at most as many states as $\mathcal{B}$, such Eve wins the 1-token game from everywhere on $\mathcal{B}_N$ and $\mathcal{B}_N$ has reach-covering (Lemma 5.8). The automaton $\mathcal{B}_N$ can then be determinised with a quadratic state-space blowup in PTIME (Lemma 5.7). □

## 5.3 When Joker games are not enough

Note that Theorems 4.1 and 5.1 together imply Theorem C.

**Theorem C.** *For every Büchi or coBüchi automaton $\mathcal{A}$, Eve wins the Joker game on $\mathcal{A}$ if and only if $\mathcal{A}$ is history-deterministic.*

The Joker-game based characterisation does not extend to parity automata, however. Consider the automaton $\mathcal{A}$ with priorities in $[1,3]$ as shown in Fig. 5.1.

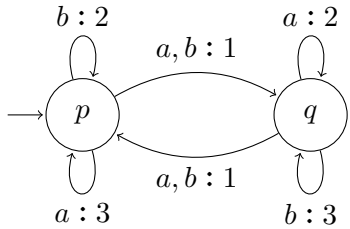

Figure 5.1: A $[1,3]$ automaton on which Eve wins the 1-Joker game and that is not history-deterministic.

Note that $\mathcal{A}$ accepts all words in $\{a,b\}^\omega$. The automaton $\mathcal{A}$ is not HD, since Adam can win the history-determinism game on $\mathcal{A}$ by choosing the letter $a$ when Eve's token is at $p$, and $b$ when her token is at $q$. This ensures that Eve's token never takes a transition of even priority, causing her run to be rejecting.

Eve wins the Joker game on $\mathcal{A}$, however. Consider the strategy of Eve where she switches states if her and Adam's tokens are at different states, and otherwise stays at the same state. For Adam to win, Adam must play only finitely many Jokers, and construct an accepting run, which requires him to eventually stay in the same state. But then, in such a play where Eve is playing according to this strategy, Eve's and Adam's runs on their tokens are identical, and hence Eve wins that play. Therefore, Eve wins the Joker game on $\mathcal{A}$.

**Theorem 5.18.** *There is a $[1,3]$ automaton that is not history-deterministic, and on which Eve wins the Joker game.*

We note that we do not know whether Eve winning Joker games characterises the history-determinism for $[0,2]$ automata.

## 5.4 Reducing to universal parity automata

In the next section, we will show that 1-token games characterise history-determinism on semantically-deterministic Büchi automata (Theorem F). This will give us a second proof of Theorems 5.1 and 5.2. In order to show Theorem F, we start by reducing its result to the restriction where our automata are universal (Theorem 5.3), i.e., recognise all words over its alphabet. A very similar reduction also shows that proving the 2-token theorem for universal parity automata is sufficient to conclude the 2-token theorem for parity automata (Theorem 5.20).

**Theorem 5.3.** *The following two statements are equivalent.*

1. *Every semantically-deterministic Büchi automaton on which Eve wins the $1$-token game is history-deterministic.*

2. *Every semantically-deterministic Büchi automaton that accepts all infinite words over its alphabet and on which Eve wins the $1$-token game is history-deterministic.*

To prove Theorem 5.3, we will use a result that is a strengthening of the equivalence of history-determinism and guidability for parity automata (Theorem 3.9). Recall that we say that a parity automaton $\mathcal{A}$ is guidable if it simulates all parity automata $\mathcal{B}$ with $L(\mathcal{B}) \subseteq L(\mathcal{A})$. We call a parity automaton $\mathcal{A}$ *safe-guidable* if $\mathcal{A}$ simulates all safety automata $\mathcal{S}$ with $L(\mathcal{S}) \subseteq L(\mathcal{A})$. We next show that an automaton is safe-guidable if and only if it history-deterministic.

**Lemma 5.19.** *For every nondeterministic parity automaton $\mathcal{A}$, $\mathcal{A}$ is history-deterministic if and only if it simulates all deterministic safety automaton $\mathcal{S}$ with $L(\mathcal{S}) \subseteq L(\mathcal{A})$.*

*Proof.* Let $\mathcal{A} = (Q, \Sigma, q_0, \Delta)$. If $\mathcal{A}$ is HD, then by Lemma 3.8, it simulates all safety automaton $\mathcal{S}$ with $L(\mathcal{S}) \subseteq L(\mathcal{A})$.

For the other direction, assume that $\mathcal{A}$ is not history-deterministic. Then, Adam has a finite-memory winning strategy in the HD game on $\mathcal{A}$ which can be represented by $\mathcal{M} = (M, m_0, \mu, \mathsf{nextmove})$, where $\mu : M \times \Delta \to M$ is the transition function for $\mathcal{M}$, and $\mathsf{nextmove} : Q \times M \to \Sigma$ is the function that dictates what letters Adam pick. When Eve's token in the HD game is at state $q$ and Adam's memory is $m$, Adam picks the letter given by $\mathsf{nextmove}(q, m)$, Eve responds by selecting a transition $\delta$ on her token, and Adam updates his memory to $m' = \mu(m, \delta)$.

Using $\mathcal{M}$, we will construct a nondeterministic safety automaton $\mathcal{M}_\Sigma$ that recognises all words that Adam can build in a play of the HD game on $\mathcal{A}$ when he is playing according to $\mathcal{M}$. More concretely, the states of $\mathcal{M}_\Sigma$ are elements of $Q \times M$ along with a rejecting sink state $q_\bot$. We have a transition $(q, m) \xrightarrow{a:0} (q', m')$ in $\mathcal{M}_\Sigma$ if $\mathsf{nextmove}(q, m) = a$ and there is a transition $\delta = q \xrightarrow{a:c} q'$ in $\mathcal{A}$ such that $\mu(m, \delta) = m'$. The initial state of $\mathcal{M}_\Sigma$ is $(q_0, m_0)$. We additionally have self-loops of priority 1 on all letters in $\Sigma$ on $q_\bot$, and we add an outgoing transition $(q, m) \xrightarrow{a:0} q_\bot$ for each letter $a$ in $\Sigma$, $q \in Q$ and $m \in M$.

Since $\mathcal{M}$ represents a finite-memory winning strategy for Adam, any play of the HD game in which he plays according to $\mathcal{M}$ produces a word in $L(\mathcal{A})$. Thus, $L(\mathcal{M}_\Sigma) \subseteq L(\mathcal{A})$.

Let $\mathcal{S}$ be the deterministic safety automaton obtained by determinising $\mathcal{M}_\Sigma$ using the subset construction.

We argue that $\mathcal{A}$ does not simulate $\mathcal{S}$, by describing a winning strategy for Adam in the simulation game of $\mathcal{S}$ by $\mathcal{A}$. Adam selects letters according to his strategy $\mathcal{M}$ in the HD game on $\mathcal{A}$ against Eve's token in $\mathcal{A}$, while Adam's token takes the deterministic transitions in $\mathcal{S}$. Then, in any play when Adam is playing according to the above strategy, Adam's word is in $L(\mathcal{M}_\Sigma) = L(\mathcal{S}) \subseteq L(\mathcal{A})$ and thus his token in $\mathcal{S}$ produces an accepting run, while the run on Eve's token is rejecting since $\mathcal{M}$ is a winning strategy for Adam in the HD game on $\mathcal{A}$. $\square$

Let us now prove Theorem 5.3.

*Proof of Theorem 5.3.* It is clear that 1 implies 2. For the other direction, suppose that for all SD Büchi automata $\mathcal{U}$ with $L(\mathcal{U}) = \Sigma^\omega$, $\mathcal{U}$ is HD if and only if Eve wins the 1-token game on $\mathcal{U}$.

Let $\mathcal{A} = (Q, \Sigma, \Delta, q_0)$ be a semantically-deterministic Büchi automaton. If $\mathcal{A}$ is HD, then is it clear that Eve wins the 1-token game on $\mathcal{A}$ (Lemma 3.21). For the converse, suppose that $\mathcal{A}$ is not HD. We will show that Adam wins the 1-token game on $\mathcal{A}$.

Due to Lemma 5.19, there is a deterministic safety automaton $\mathcal{S} = (S, \Sigma, \Delta_S, s_0)$ such that $L(\mathcal{S}) \subseteq L(\mathcal{A})$, and $\mathcal{A}$ does not simulate $\mathcal{S}$. We suppose that the priorities of $\mathcal{S}$ are in the interval $[0, 1]$. Consider the automaton $\mathcal{P} = (P, \Sigma, \Delta_P, p_0)$ that is given by taking the 'product' of $\mathcal{A}$ and $\mathcal{S}$, where

- The set of states $P$ is given by $Q \times S$, and the initial state $p_0$ is $(q_0, s_0)$
- $(q, s) \xrightarrow{a:c} (q', s')$ is a transition in $\Delta_P$ if and only if $q \xrightarrow{a:c} q'$ is a transition in $\mathcal{A}$, and $s \xrightarrow{a:0} s'$ is a transition in $\mathcal{S}$.

We note that $L(\mathcal{P}) = L(\mathcal{S}) \cap L(\mathcal{A}) = L(\mathcal{S})$. Consider the automaton $\mathcal{U}$ obtained by completing $\mathcal{P}$ with an accepting sink state $f$. That is, for all states $p$ in $\mathcal{P}$, $a \in \Sigma$ such that there is no outgoing transition of priority 0 from $p$ on $a$, we add the transition $p \xrightarrow{a:0} q_\top$, and the self-loops $q_\top \xrightarrow{a:0} q_\top$.

1. *$\mathcal{U}$ recognises all words in $\Sigma^\omega$.* Any word $w \in L(\mathcal{S})$ is accepted by $\mathcal{P}$ and hence by $\mathcal{U}$. Otherwise if $w \notin L(\mathcal{S})$, then any run of $\mathcal{U}$ on $w$ ends at $q_\top$, and hence $w \in L(\mathcal{U})$.

2. *$\mathcal{U}$ is semantically-deterministic.* Indeed, for each state $q$, we can argue similarly that $L(\mathcal{U}, q) = \Sigma^\omega$.

3. *$\mathcal{U}$ is not HD.* Adam can use a winning strategy for the simulation game of $\mathcal{S}$ by $\mathcal{A}$, to select letters in the HD game on $\mathcal{U}$. Since Adam will construct a word in $\mathcal{S}$ using this strategy, Eve's token will not reach the accepting sink state, neither can the run on her token in the HD game on $\mathcal{U}$ be accepting.

4. *If Adam wins the 1-token game on $\mathcal{U}$, then he wins the 1-token game on $\mathcal{A}$.* Let $\sigma$ be a winning strategy for Adam in the 1-token game on $\mathcal{U}$. Note that at any point in the 1-token game on $\mathcal{U}$, if Eve's token is at a state $(q, s)$, then Adam's token must be at a state of the form $(q', s)$ since $\mathcal{S}$ is deterministic, where $q'$ is another state in $\mathcal{A}$. Thus, $\sigma$ will never choose a letter $a$ in the 1-token game on $\mathcal{U}$ whenever Eve's token is at a state $(q, s)$ where there is no 0 priority transition on $a$ from $s$ in $\mathcal{S}$, since then Eve can move her token to $q_\top$ and win. Similarly, Adam will never be able to move his token to $q_\top$ if he plays according to $\sigma$ as well. Thus, any play of the 1-token game in $\mathcal{U}$ where Adam plays according to his $\sigma$ strategy corresponds to a play of 1-token game on $\mathcal{A}$, since $\mathcal{S}$ is deterministic. It follows that Adam can win the 1-token game on $\mathcal{A}$ by playing according to the strategy $\sigma$ and additionally keeping in his memory states of $\mathcal{S}$.

Since Adam wins the 1-token game on $\mathcal{U}$ if $\mathcal{U}$ is not history-deterministic by the hypothesis, 4 gives us that Adam wins the 1-token game on $\mathcal{A}$, as desired. □

An almost word-by-word identical proof to above also shows that the the proof of the 2-token game characterisation for nondeterministic parity automata can be reduced to the case where the automata recognise all words over its alphabet.

**Theorem 5.20.** *The following statements are equivalent:*

1. *For every non-deterministic parity automaton $\mathcal{A}$, Eve wins the 2-token game on $\mathcal{A}$ if and only if $\mathcal{A}$ is history-deterministic.*

2. *For every non-deterministic parity automaton $\mathcal{U}$ with $L(\mathcal{U}) = \Sigma^\omega$, Eve wins the 2-token game on $\mathcal{U}$ if and only if $\mathcal{U}$ is history-deterministic.*

Theorem 5.20 reduces proving the 2-token game based characterisation of history-determinism on parity automata to parity automata that recognise all words over its alphabet. But the history-determinism game on such an automaton is just a parity game, since Adam's word is always accepting. Proving the 2-token theorem thus reduces to showing that this parity game is equivalent to the 2-token game. Since parity games are well-understood games in which both players have positional strategies, it is natural to think that there should be a relatively easy proof of the 2-token theorem using Theorem 5.20. For now, we are unaware of such a proof.

## 5.5 Lookahead games

In this section, we introduce lookahead games and use them to show the following result.

**Theorem F.** *For every semantically-deterministic Büchi automaton $\mathcal{A}$, $\mathcal{A}$ is history-deterministic if and only if Eve wins the $1$-token game on $\mathcal{A}$.*

Before defining the lookahead game, let us briefly recall how a round of the 1-token game on a parity automaton $\mathcal{A}$ works. In each round, Adam selects a letter, then Eve selects a transition on that letter on her token, and then Adam selects a transition on that letter on his token. The winning condition for Eve is that either Eve's run on her token is accepting or Adam's run on his token is rejecting. This is very close to the simulation game of $\mathcal{A}$ by itself, except that the order of the moves in which Eve and Adam select transitions has been reversed. One can, however, see the 1-token game as a simulation game, where Adam picks the transition for round $i$ in round $(i + 1)$. Or equivalently, we can construct an automaton $\mathsf{Delay}(\mathcal{A})$ such that the nondeterminism on $\mathcal{A}$ is 'delayed' by one step, and then the 1-token game on $\mathcal{A}$ is equivalent to the simulation game of $\mathsf{Delay}(Ac)$ by $\mathcal{A}$. The automaton $\mathsf{Delay}(\mathcal{A})$ is constructed so that each state of $\mathsf{Delay}(\mathcal{A})$ is a pair $(q, a)$ consisting of state $q$ of $\mathcal{A}$ and a letter $a \in \Sigma$. When a letter $b$ is read, the transition on the $\mathcal{A}$ component in $\mathsf{Delay}(\mathcal{A})$ is based on an $a$-transitions from $q$ in $\mathcal{A}$, and the target of such a transition in $\mathsf{Delay}(\mathcal{A})$ now has the letter $b$ in its second component.

**Definition 5.21.** *For every nondeterministic parity automaton $\mathcal{A} = (Q, \Sigma, q_0, \Delta)$, we construct the automaton $\mathsf{Delay}(\mathcal{A}) = (Q', \Sigma, s, \Delta')$, where $Q' = Q \times \Sigma \cup \{s\}$, and $s$ is the initial state. The set of transitions $\Delta'$ is the union of the following sets of transitions.*

1. $\{(s \xrightarrow{a:0} (q_0, a)) \mid a \in \Sigma\}$.

2. $\{((p, a) \xrightarrow{b:c} (q, b)) \mid (p \xrightarrow{a:c} q) \in \Delta, b \in \Sigma\}$.

Observe that $\mathsf{Delay}(\mathcal{A})$ accepts the same language as $\mathcal{A}$. The following result follows since the 1-token game on $\mathcal{A}$ and the simulation game between $\mathcal{A}$ and $\mathsf{Delay}(\mathcal{A})$ are syntactically equivalent.

**Lemma 5.22.** *For every non-deterministic parity automaton $\mathcal{A}$, Eve wins the 1-token game on $\mathcal{A}$ if and only if $\mathcal{A}$ simulates $\mathsf{Delay}(\mathcal{A})$.*

Furthermore, Eve wins the 1-token game on $\mathsf{Delay}(\mathcal{A})$ if Eve wins the 1-token game on $\mathcal{A}$, by simply 'delaying' her winning strategy in the 1-token game on $\mathcal{A}$.

**Lemma 5.23.** *If Eve wins the 1-token game on an automaton $\mathcal{A}$, then Eve wins the 1-token game on $\mathsf{Delay}(\mathcal{A})$.*

For an automaton $\mathcal{A}$ and a natural number $k$, we define the automaton $\mathsf{Delay}^k(\mathcal{A})$ as $\mathcal{A}$ if $k = 0$, and $\mathsf{Delay}(\mathsf{Delay}^{k-1}(\mathcal{A}))$ if $k \geq 1$. We note that Lemma 5.23 implies that if Eve wins $G1(\mathcal{A})$, then Eve wins $G1(\mathsf{Delay}^k(\mathcal{A}))$ for all natural numbers $k$. Thus, an iterative application of Lemmas 5.22 and 5.23 gives us the following corollary.

**Corollary 5.24.** *If Eve wins the 1-token game on a parity automaton $\mathcal{A}$, then $\mathcal{A}$ simulates $\mathsf{Delay}^k(\mathcal{A})$ for all $k \geq 0$.*

*Proof.* If $k = 0$, it is clear that $\mathcal{A}$ simulates $\mathsf{Delay}^0(\mathcal{A}) = \mathcal{A}$. Otherwise, suppose that $k \geq 1$ and that Eve wins the 1-token game on $\mathcal{A}$. Then, due to Lemma 5.23, Eve wins the 1-token game on $\mathsf{Delay}^k(\mathcal{A})$ for all $k \in \mathbb{N}$. From Lemma 5.22, we note that $\mathsf{Delay}^k(\mathcal{A})$ simulates $\mathsf{Delay}^{k+1}(\mathcal{A})$ for all $k \in \mathbb{N}$. Combining this with the transitivity of simulation (Lemma 3.6), we get that $\mathcal{A}$ simulates $\mathsf{Delay}^k(\mathcal{A})$ for all $k \in \mathbb{N}$. □

For a natural number $k$ and a parity automaton $\mathcal{A}$, we define the *k-lookahead game* on $\mathcal{A}$ to be the simulation game of $\mathsf{Delay}^{k+1}(\mathcal{A})$ by $\mathcal{A}$, or equivalently, as the 1-token game $G1(\mathcal{A}; \mathsf{Delay}^k(\mathcal{A}))$. Note that the 1-token game of $\mathcal{A}$ is syntactically equivalent to the 0-lookahead game of $\mathcal{A}$. Corollary 5.24 can thus be restated as the following theorem.

**Theorem E.** *For every parity automaton $\mathcal{A}$, Eve wins the 1-token game on $\mathcal{A}$ if and only if she wins the k-lookahead game on $\mathcal{A}$ for all $k \geq 0$.*

Theorem E is analogous to Bagnol and Kuperberg's result that the 2-token games and $k$-token games have the same winner on the same automaton for all $k \geq 2$ (Lemma 3.15). We now proceed to show Theorem F. From Theorem 5.3, we know that it suffices to only consider SD Büchi automata that are universal. The following lemma shows that every universal SD Büchi automaton is history-deterministic with sufficient lookahead.

**Lemma 5.25.** *Let $\mathcal{U}$ be a semantically-deterministic Büchi automaton such that $L(\mathcal{U}) = \Sigma^\omega$. Then, there is a $K$ such that $\mathsf{Delay}^K(\mathcal{U})$ is history-deterministic.*

*Proof.* We let $K = 2^n + 1$, where $n$ is the number of states of $\mathcal{A}$. The key observation is the following claim.

**Claim 4.** *For every state $q$ in $\mathcal{U}$ that is reachable from the initial state, and for every finite word $u$ of length $K$, there is a run from $q$ on $u$ in $\mathcal{U}$ that visits an accepting transition.*

Note that since $\mathcal{U}$ is semantically-deterministic, we have that $L(\mathcal{U}, q) = \Sigma^\omega$. Consider the sequence of sets of states that can be visited on reading the prefixes of $u$:

$$\{q\} \xrightarrow{a_1} S_1 \xrightarrow{a_2} S_1 \ldots \xrightarrow{a_K} S_K,$$

where $u = a_1 a_2 \ldots a_K$ and $S_{l+1}$ is the set of all states to which there is a transition from a state is $S_l$ on the letter $a_l$. By the pigeonhole principle, observe that two subsets $S_i$ and $S_j$ for some $i < j$ must be the same.

Let $u'$ and $v$ be the finite words $u' = a_1 a_2 \ldots a_i$ and $v = a_{i+1} a_{i+2} \ldots a_j$. Consider the word $w = u'v^\omega$. Then, the sequence of sets of states visited on prefixes of $w$ cycles from $S_i$ upon reading a $v$ to $S_i$. Since $w$ is in $L(\mathcal{A}, q)$, there must be an accepting transition seen from a state in $S_i$ on a run on the finite word $v$. This concludes the proof of the claim.

Using Claim 4, we note that Eve has a winning strategy in the history-determinism game on $\mathsf{Delay}^K(\mathcal{U})$, where she exploits the lookahead of $K$ to take at least one accepting transition on her token in every $K$ consecutive rounds. The run on Eve's token then has infinitely many accepting transitions and hence is accepting, as desired. □

The proof of Theorem F now follows easily.

**Theorem F.** *For every semantically-deterministic Büchi automaton $\mathcal{A}$, $\mathcal{A}$ is history-deterministic if and only if Eve wins the 1-token game on $\mathcal{A}$.*

*Proof of Theorem F.* If $\mathcal{A}$ is history-deterministic, then it is clear that Eve wins the 1-token game on $\mathcal{A}$ (Lemma 3.21). For the converse, suppose that Eve wins the 1-token game on $\mathcal{A}$. Due to Theorem 5.3, we assume without loss of generality that $\mathcal{A}$ is universal. Then, from Lemma 5.25, there is a $K$ such that $\mathsf{Delay}^K(\mathcal{A})$ is history-deterministic. Since Eve wins the 1-token game on $\mathcal{A}$, we note that $\mathcal{A}$ simulates $\mathsf{Delay}^K(\mathcal{A})$ (Lemma 5.22). But since $\mathsf{Delay}^K(\mathcal{A})$ is language-equivalent to $\mathcal{A}$, we get from Corollary 3.10 that $\mathcal{A}$ is HD as well. $\square$

We note that Theorem F gives us a second proof of Theorem 5.2, since every Büchi automaton on which Eve wins the 1-token game from everywhere is SD as well (Lemma 3.36).

# Part III

# Climb

# Chapter 6

# The Even to Odd Induction Step

In this chapter and the next chapter, we will prove the 2-token theorem, that is, the result that Eve winning the 2-token game on a parity automaton implies its history-determinism. Our proof goes by induction on the parity index hierarchy, and involves two induction steps: one to add a most significant odd priority, which we will discuss in this chapter, and the other to add a most significant even priority, which we will discuss in Chapter 7.

More concretely, we will show in this chapter that if the 2-token game characterises history-determinism for $[0, K]$ automata, or equivalently $[2, K + 2]$ automata, then we can add a most important odd priority and extend the 2-token game characterisation of history-determinism to $[1, K + 2]$ automata.

**Theorem A.** *Let $K > 0$ be a natural number, such that for every $[0, K]$ (or equivalently, $[2, K+2]$) automaton $\mathcal{A}$, Eve wins the 2-token game on $\mathcal{A}$ if and only if $\mathcal{A}$ is HD. Then, for every $[1, K+2]$ automaton $\mathcal{A}$, Eve wins the 2-token game on $\mathcal{A}$ if and only if $\mathcal{A}$ is HD.*

Let us fix a $K \geq 1$ for the rest of this chapter. Note that if an automaton is HD, then Eve wins the 2-token game on that automaton, since Eve can pick transitions in the 2-token game using her strategy in the HD game, ignoring transitions of Adam's tokens (Lemma 3.21). We need to show that if Eve wins the 2-token game on a $[1, K + 2]$ automaton then that automaton is HD, provided that the 2-token game characterisation holds for $[0, K]$ automata. We proceed in the rest of this chapter with this hypothesis.

**Hypothesis 6.1.** *For every $[0, K]$ automaton $\mathcal{A}$, Eve wins the 2-token game on $\mathcal{A}$ if and only if $\mathcal{A}$ is HD.*

We will prove Theorem A by using a very similar template to that of coBüchi automata. We will start in Section 6.1 by introducing a convenient relabelling of priorities for $[1, K+2]$ automata, so that the automata we deal with are '2-priority reduced'. This relabelling of priorities will preserve the acceptance of each run.

In Section 6.2, we will introduce a property called 1-safe double-coverage, which is an extension of the property of safe-coverage we introduced for coBüchi automata. We will show that every $[1, K + 2]$ automaton on which Eve wins the 2-token game from everywhere and is 2-priority reduced has 1-safe double-coverage (Lemma 6.4). We will then show in Section 6.3 that every $[1, K + 2]$ automaton with 1-safe double-coverage and on which Eve wins the 2-token game is history-deterministic (assuming Hypothesis 6.1), thus concluding the proof of Theorem A.

## 6.1 Automata with 2-priority reduction

For every $[1, K + 2]$ automaton $\mathcal{A}$, define the 2-approximation of $\mathcal{A}$, denoted by $\mathcal{A}_{>1}$, as the automaton obtained by modifying $\mathcal{A}$, so that transitions of priority at least 2 are preserved, while transitions of priority 1 are redirected to a rejecting sink state. We say that $\mathcal{A}$ is *2-priority reduced* if all transitions in $\mathcal{A}_{>1}$ occur in some SCC in $\mathcal{A}_{>1}$, and every SCC in $\mathcal{A}_{>1}$ contains at least one transition of priority 2.

We will show that we can relabel the priorities of transitions of any $[1, K + 2]$ automaton while preserving the acceptance of each run so that the new automaton is 2-priority reduced. This relabelling of priorities is done via the iterative procedure presented below.

**2-priority reduction.** Let $\mathcal{A}$ be a $[1, K + 2]$ automaton, and let $\mathcal{A}_0 = \mathcal{A}$. For each $i \geq 0$, perform the following three steps till $\mathcal{A}_{i+1} = \mathcal{A}_i$.

1. Consider the graph $G^i_{>1}$ consisting of all transitions of $\mathcal{A}_i$ whose priorities are at least 2, and consider the strongly connected components of $G^i_{>1}$.

2. For each transition in $G^i_{>1}$ that does not appear in any SCC, change its priority to 1.

3. For every SCC in which the priorities of all transitions in it is strictly greater than 2, decrease the priority of every transition in that SCC by 2. Let $\mathcal{A}_{i+1}$ be the resulting $[1, K + 2]$ automaton.

Observe that in each iteration of 2-priority reduction, we are only decreasing the priorities of certain transitions. Hence, the above procedure terminates. We first show that the 2-priority reduction procedure preserves the acceptance of each run, by showing that each iteration of the 2-priority reduction procedure preserves the acceptance of each run.

**Lemma 6.2.** *For each $i \geq 0$, each infinite run $\rho$ in $\mathcal{A}_i$ is accepting if and only if $\rho$ is accepting in $\mathcal{A}_{i+1}$.*

*Proof.* Let $\rho$ be a run in $\mathcal{A}_i$. We distinguish between the cases of whether the number of priority 1 transitions occurring in $\rho$ is finite or infinite.

If $\rho$ contains finitely many transitions of priority 1 in $\mathcal{A}_i$, then $\rho$ eventually only visits transitions that are all a part of the same strongly connected component in $G^i_{>1}$. Since the priorities of transitions in this strongly connected component are either decreased for all transitions by 2 or unchanged in $\mathcal{A}_{i+1}$, the lowest priority appearing infinitely often in $\rho$ in $\mathcal{A}_i$ is even if and only if the lowest priority appearing infinitely often in $\rho$ in $\mathcal{A}_{i+1}$ is even. It follows that in this case, $\rho$ in an accepting run in $\mathcal{A}_i$ if and only if $\rho$ is an accepting run in $\mathcal{A}_{i+1}$.

If $\rho$ contains infinitely many transitions of priority 1 in $\mathcal{A}_i$, then it does the same in $\mathcal{A}_{i+1}$, since the priorities of transitions with 1 are unchanged. Since 1 is the lowest priority occurring in $\rho$ in both $\mathcal{A}_i$ and $\mathcal{A}_{i+1}$, $\rho$ is a rejecting run in both automata. $\square$

Let $\mathcal{A}\Downarrow$ be the automaton obtained via the 2-priority reduction procedure. It then follows from Lemma 6.2 that for every run $\rho$, $\rho$ is accepting in $\mathcal{A}$ if and only if $\rho$ is accepting in $\mathcal{A}\Downarrow$. We next show that $\mathcal{A}\Downarrow$ is 2-priority reduced.

**Lemma 6.3.** *For the automaton $\mathcal{A}\Downarrow$, consider the graph $G_{>1}$ consisting of transitions of $\mathcal{A}\Downarrow$ with priority at least 2. Then the following three conditions holds.*

1. *$\mathcal{A}$ is HD if and only if $\mathcal{A}\Downarrow$ is, and Eve wins the 2-token game from everywhere in $\mathcal{A}\Downarrow$ if and only if Eve wins the 2-token game from everywhere in $\mathcal{A}$.*

2. *Every SCC of $G_{>1}$ contains at least one transition of priority 2.*

3. *Any transition that does not belong to some SCC of $G_{>1}$ has priority 1.*

*Proof.* Statement 1 follows from the fact that a run is accepting in $\mathcal{A}$ if and only if that run is accepting in $\mathcal{A}\Downarrow$. Since an iteration of the 2-priority reduction procedure has no effect on $\mathcal{A}\Downarrow$, statements 2 and 3 follow. $\square$

## 6.2 Obtaining 1-safe double-coverage

Similar to how we defined safe-coverage for coBüchi automata, we define an analogous property of 1-safe double-coverage for $[1, K+2]$ automata, which is based on the 2-token game relations on the 2-approximations of $[1, K+2]$ automata.

For a $[1, K+2]$ automaton $\mathcal{A}$, we say that $\mathcal{A}$ has *1-safe double-coverage* if for every state $q$ in $\mathcal{A}$, there is another state $p$ weakly coreachable to $q$ in $\mathcal{A}$, such that Eve wins $G2(p; q, q)$ in $\mathcal{A}_{>1}$.

> **Automata with 1-safe double-coverage.** We say that a $[1, K+2]$ automaton $\mathcal{A}$ has 1-safe double-coverage if for every state $q$ in $\mathcal{A}$, there is another state $p$ in $\mathsf{WCR}(\mathcal{A}, q)$, such that Eve wins $G2(p; q, q)$ in $\mathcal{A}_{>1}$.

In this section, we will focus on proving the following result.

**Lemma 6.4.** *Every $[1, K+2]$ automaton that is 2-priority reduced and on which Eve wins the 2-token game from everywhere has 1-safe double-coverage.*

To show Lemma 6.4, we will also use the property of safe-coverage that we had defined for coBüchi automata, and which we now define for $[1, K+2]$ automata. For a $[1, K+2]$ automaton $\mathcal{A}$, define the automaton $\mathcal{A}_{\mathtt{safe}}$ as the safety automaton in which all transitions of $\mathcal{A}$ with priority at least 2 are made safe, i.e., relabelled to have priority 2, while transitions of priority 1 in $\mathcal{A}$ are redirected to a rejecting sink state in $\mathcal{A}_{\texttt{safe}}$.

The following observation is easy to see.

**Proposition 6.5.** *For every $[1, K+2]$ automaton $\mathcal{C}$, $L(\mathcal{C}_{>1}) \subseteq L(\mathcal{C}_{safe})$ and $L(\mathcal{C}_{>1}) \subseteq L(\mathcal{C})$.*

*Proof.* If $\rho$ is an accepting run over some word $w$ in $\mathcal{C}_{>1}$, then the same run is also an accepting run in $\mathcal{C}_{\texttt{safe}}$ and $\mathcal{C}$. □

We say that a $[1, K+2]$ automaton $\mathcal{A}$ has *safe-coverage* if for each state $q$ in $\mathcal{A}$, there is another state $p$ weakly coreachable to $q$ in $\mathcal{A}$ such that Eve wins $G1(p; q)$ in $\mathcal{A}_{\texttt{safe}}$.

**Remark 5.** *Let us compare the definitions of safe-coverage and 1-safe double-coverage. Firstly, the former deals with 1-token games between safety automata, while the latter deals with 2-token games on $[2, K+2]$ or $[0, K]$ automata. The reasoning behind going safety to $[0, K]$ automata and to involve 2 tokens instead of 1-token games for our even-to-odd induction step is natural: similar to how we used the 1-token game characterisation of history-determinism on safety automata to obtain Theorem 4.2 for coBüchi automata, we would like to use the 2-token game characterisation of history-determinism on $[0, K]$ automata (Hypothesis 6.1) to show the 2-token game characterisation of history-determinism on $[1, K+2]$ automata.*

We next prove that if Eve wins the 1-token game from everywhere in a $[1, K+2]$ automaton $\mathcal{A}$ that is 2-priority reduced, then $\mathcal{A}$ has safe-coverage. This is analogous to Lemma 4.5 for coBüchi automata, and is proved similarly: if $q$ is a state in $\mathcal{A}$ such that Adam wins $G1(p; q)$ in $\mathcal{A}_{\texttt{safe}}$ for all $p$ in $\mathsf{WCR}(\mathcal{A}, q)$, then we show that Adam wins $G1(q; q)$ in $\mathcal{A}$. The only difference to the proof of Lemma 4.5 is in the construction of Adam's strategy in $G1(q; q)$ in $\mathcal{A}$, where whenever Eve's token takes a priority 1-transition and his token has not seen a priority 1-transition, Adam's token returns to $q$ via a priority 2-transition without seeing a priority 1-transition. This is possible because $\mathcal{A}$ is 2-priority reduced.

**Lemma 6.6.** *Let $\mathcal{A}$ be a $[1, K+2]$ automaton that is 2-priority reduced and on which Eve wins the 1-token game from everywhere. Then for every state $q$ in $\mathcal{A}$, there is a state $p$ weakly coreachable to $q$ in $\mathcal{A}$ such that Eve wins $G1(p; q)$ in $\mathcal{A}_{safe}$.*

*Proof.* Let $q$ be a state such that Adam wins $G1(p; q)$ in $\mathcal{A}_{\texttt{safe}}$ for all states $p$ that are weakly coreachable to $q$ in $\mathcal{A}$. We will show that Adam wins $G1(q; q)$ in $\mathcal{A}$ by showing a winning strategy $\sigma$ for Adam, which contradicts the fact that Eve wins the 1-token game from everywhere in $\mathcal{A}$.

From the position $(q, q)$ in $G1$ on $\mathcal{A}$, Adam chooses letters and transitions on his token according to his winning strategy for $G1(q; q)$ in $\mathcal{A}_{\texttt{safe}}$, which will cause Eve's token to eventually take a transition of priority 1 in $\mathcal{A}$, while all priorities taken by Adam's token have priority at least 2. Due to Lemma 6.3, we know that the current position of Adam's token is in the same SCC in $\mathcal{A}_{>1}$ as $p$, and this SCC contains a priority 2 transition. Adam thus picks letters and transitions on his token so that his token returns to $q$ and the run on his token witnesses a priority 2 transition but no priority 1-transition. Eve's token then would be at some state $p$ that is weakly coreachable to $q$, and Adam's token is at $q$. Adam can repeat the same strategy as above, since Adam wins $G1(p;q)$ in $\mathcal{A}_{\text{safe}}$. Repeating this strategy infinitely many time causes the run of Eve's token to contain infinitely many 1-priority transitions, implying the run on her token is rejecting. The lowest priority occurring infinitely often in Adam's token is 2 and hence Adam wins $G1(q;q)$ in $\mathcal{A}$. □

The next result follows from the transitivity of 1-token games (Lemma 3.25), the fact that there are finitely many states in $\mathcal{A}$, and the 1-token game characterisation of history-determinism on safety automata (Theorem 3.22).

**Lemma 6.7.** *Let $\mathcal{A}$ be a $[1, K+2]$ automaton with safe-coverage. Then for every state $q$, there is another state $p$ that is weakly coreachable to $q$ in $\mathcal{A}$, such that $(\mathcal{A}_{\text{safe}}, p)$ is HD and Eve wins $G1(p;q)$ in $\mathcal{A}_{\text{safe}}$.*

*Proof.* We start by showing that for each $q$ in $\mathcal{A}$, there is a state $p$ in $\text{WCR}(\mathcal{A}, p)$, such that Eve wins $G1(p;p)$ and $G1(p;q)$ in $\mathcal{A}_{\text{safe}}$. The result then follows from the fact that safety automata on which Eve wins the 1-token game are history-deterministic (Theorem 3.22).

Let $q$ be a state in $\mathcal{A}$, and consider the directed graph $G$ whose vertices consist of the states in $\text{WCR}(\mathcal{A}, q)$. There is an edge in $G$ from vertices $r$ to $s$ if Eve wins $G1(s;r)$ in $\mathcal{A}_{\text{safe}}$.

Since $\mathcal{A}$ has safe-coverage, every vertex in $G$ has outdegree at least 1. Thus, for every vertex $r$, there is a vertex $s$ such that there is a path from $r$ to $s$ and a cycle consisting of the vertex $s$ in $G$. It then follows from transitivity of $G1$ (Lemma 3.25) that Eve wins $G1(s;r)$ and $G1(s;s)$ in $\mathcal{A}_{\text{safe}}$. This concludes the proof of Lemma 6.7. □

We now prove Lemma 6.4.

**Lemma 6.4.** *Every $[1, K+2]$ automaton that is 2-priority reduced and on which Eve wins the 2-token game from everywhere has 1-safe double-coverage.*

*Proof.* Let $\mathcal{A}$ be a $[1, K+2]$ automaton that is 2-priority reduced, on which Eve wins the 1-token game from everywhere, and that does not have 1-safe double-coverage. We will show that Eve does not win the 3-token game from everywhere in $\mathcal{A}$, which will imply that Eve does not win the 2-token game from everywhere (due to Lemma 3.32). Note that this will prove Lemma 6.4.

Since $\mathcal{A}$ does not have 1-safe double-coverage, there is a state $q$ in $\mathcal{A}$, such that for all states $p$ in $\text{WCR}(\mathcal{A}, q)$, Adam wins $G2(p; q, q)$ in $\mathcal{A}_{>1}$. Let $r$ be a state in $\text{WCR}(\mathcal{A}, q)$, such that $(\mathcal{A}_{\text{safe}}, r)$ is HD and Eve wins $G1(r;q)$ in $\mathcal{A}_{\text{safe}}$; we know that such a $r$ exists due to Lemma 6.7. Note that this implies $(\mathcal{A}_{\text{safe}}, r)$ simulates $(\mathcal{A}_{\text{safe}}, q)$ (Lemma 3.26.5) and hence $L(\mathcal{A}_{\text{safe}}, q) \subseteq L(\mathcal{A}_{\text{safe}}, r)$ (Lemma 3.5).

Fix a positional winning strategy $\sigma_{G1}$ for Eve in the 1-token game on $\mathcal{A}$ from all pairs of weakly coreachable states in $\mathcal{A}$, and a positional winning strategy $\sigma_{\text{safe}}$ in the HD game on $(\mathcal{A}_{\texttt{safe}}, r)$. We will show that Adam wins $G3(p; q, q, r)$ in $\mathcal{A}$, which implies that Eve does not win the 2-token game from everywhere in $\mathcal{A}$, as desired (Lemma 3.32).

**Claim 5.** *Adam wins $G3(p; q, q, r)$ in $\mathcal{A}$.*

We will prove the claim by describe a winning strategy $\sigma$ for Adam in $G3(q; q, q, r)$ in $\mathcal{A}$, where Adam stores two virtual tokens $m_1$ and $m_2$ in his memory, which take transitions in $\mathcal{A}_{>1}$. At the start of each round, Eve's token and Adam's tokens will all be at states that are weakly coreachable in $\mathcal{A}$. Adam's virtual tokens will be at either the rejecting sink state in $\mathcal{A}_{>1}$ or at some state in $\mathcal{A}_{>1}$ that is weakly coreachable to the state of Eve's token in $\mathcal{A}$.

Adam's tokens $t_1$ and $t_2$ will play in $G1$ against the virtual tokens $m_1$ and $m_2$, respectively, using $\sigma_{G1}$, whenever the corresponding virtual token is not in the rejecting sink state $q_\perp$ of $\mathcal{A}_{>1}$. If the token $m_i$ for $i = 1$ or $2$ is in the rejecting sink state of $\mathcal{A}_{>1}$, then Adam will pick transition on token $t_i$ arbitrarily. We now describe how Adam chooses letters, the behaviour of these two virtual tokens, and the transitions on his third token.

Initially and at each *reset*, the first two virtual tokens $m_1$ and $m_2$ will be at the state $q$ in $\mathcal{A}_{>1}$, Eve's token will be at a state $p$ in $\mathsf{WCR}(\mathcal{A}, p)$ while Adam's third token will be at the state $r$. Adam's first and second tokens will also be at some states $q_1$ and $q_2$ that are in $\mathsf{WCR}(\mathcal{A}, p)$.

Until Eve's token takes a priority 1 transition in $\mathcal{A}$, Adam chooses transitions on his virtual tokens and letters according to a winning strategy for Adam in $G2(p; q, q)$ in $\mathcal{A}_{>1}$ against Eve's token. Adam chooses transitions on his third token $t_3$ using the strategy $\sigma_{\texttt{safe}}$. This guarantees that the run on his third token does not take a 1-transition in $\mathcal{A}$: recall that $L(\mathcal{A}_{\texttt{safe}}, q) \subseteq L(\mathcal{A}_{\texttt{safe}}, r)$ since Eve wins $G1(r; q)$ in $\mathcal{A}_{\texttt{safe}}$, and the finite word played by Adam so far must be the prefix of a word in $L(\mathcal{A}_{>1}, q) \subseteq L(\mathcal{A}_{\texttt{safe}}, q)$ (Proposition 6.5), since Adam is choosing letters and transitions on his token according to his winning strategy in $G2(p; q, q)$ in $\mathcal{A}_{>1}$.

Whenever Eve's token takes a 1-transition, Adam changes his strategy and plays a word that allows $t_3$ to return to $r$ via a priority 2 transition and without seeing a priority 1 transition; this is possible due to the fact that $\mathcal{A}$ is 2-priority reduced. Then his strategy *resets* his virtual tokens $m_1$ and $m_2$ to be at $q$. Note that since $r$ and $p$ are weakly coreachable in $\mathcal{A}$, Eve's token must also at this point be in some state $p'$ such that is in $WCR(\mathcal{A}, q)$ (Proposition 3.29), which allows Adam to again play a winning strategy in $G2(p'; q, q)$ in $\mathcal{A}_{>1}$ at each reset. Adam's original tokens will also be in some states $q_1, q_2$ that are in $\mathsf{WCR}(\mathcal{A}, q)$, and hence Adam can again choose transitions on his token $t_1$ and $t_2$ by playing according to $\sigma_{G1}$ against his virtual tokens $m_1$ and $m_2$ respectively.

This concludes the description of Adam's strategy. In any play where Adam is playing according to the above strategy, if the run on Eve's token contains infinitely many priority 1 transitions and therefore is rejecting, then Adam's token $t_3$ contains infinitely many priority 2 transitions and no priority 1 transition, and hence is accepting. Otherwise, if Eve's run eventually does not contain a priority 1-transition, then finitely many resets occur. Since Adam is playing according to a winning strategy in $G2(\mathcal{A}_{>1})$ on his memory tokens against Eve's token, the run on Eve's token is rejecting, while one of the memory tokens $m_i$ of Adam eventually produces an accepting run in $\mathcal{A}_{>1}$. The corresponding token of Adam where Adam is choosing transitions using $\sigma_{G1}$ against this memory token then constructs an accepting run in $\mathcal{A}$. Thus, Adam wins any play of $G3(q;q,q,r)$ in $\mathcal{A}$ where Adam is playing according to the above strategy, as desired. □

## 6.3 Automata with 1-safe double-coverage are HD

We prove in this section that every $[0, K+2]$ automaton that has 1-safe double-coverage and on which Eve wins the 2-token game from everywhere is history-deterministic.

**Lemma 6.8.** *Every $[1, K+2]$ automaton that has 1-safe double-coverage and on which Eve wins the 2-token game from everywhere is history-deterministic.*

Lemma 6.4 together with Lemma 6.8 will then imply Theorem A. We start by proving the following result using Lemma 6.4, that is analogous to Lemma 6.7 for safe-coverage, and is proved similarly.

**Lemma 6.9.** *Let $\mathcal{A}$ be a $[1, K+2]$ automaton that is 2-priority reduced and on which Eve wins the 2-token game from everywhere. Then for every state $q$ in $\mathcal{A}$, there is another state $p$ weakly coreachable to $q$ in $\mathcal{A}$, such that Eve wins $G2(p;q,q)$ in $\mathcal{A}_{>1}$ and $(\mathcal{A}_{>1}, p)$ is HD.*

*Proof.* Note that if Eve wins $G2(\mathcal{A}_1; \mathcal{A}_2, \mathcal{A}_2)$ and $G2(\mathcal{A}_2; \mathcal{A}_3, \mathcal{A}_3)$ for some parity automata $\mathcal{A}_1, \mathcal{A}_2, \mathcal{A}_3$, then Eve also wins $G2(\mathcal{A}_1; \mathcal{A}_3, \mathcal{A}_3)$ (Lemma 3.26.4). Thus, it follows, from the facts that $\mathcal{A}$ has finitely many states and that $\mathcal{A}$ has 1-safe double-coverage, that for every state $q$ in $\mathcal{A}$, there is another state $p$ in $\mathsf{WCR}(\mathcal{A}, q)$, such that Eve wins $G2(p;q,q)$ and $G2(p;p,p)$ in $\mathcal{A}_{>1}$. Hypothesis 6.1 then implies that $(\mathcal{A}_{>1}, p)$ is HD, from which the lemma follows. □

For a $[1, K+2]$ automaton $\mathcal{A}$ that has 1-safe double-coverage, call a state $q$ in $\mathcal{A}$ as 1-safe HD if $(\mathcal{A}_{>1}, q)$ is HD. We will use the winning strategies for Eve in the HD game on $(\mathcal{A}_{>1}, q)$ from states $q$ that are 1-safe HD to construct a winning strategy for Eve in the HD game on $\mathcal{A}$.

For coBüchi automata, recall that we had constructed, in the proof of Theorem 4.2, a winning strategy for Eve in the HD game by playing the 1-token game against safe-deterministic states. This approach does not work immediately: we do not have the luxury to play the 1-token game against a memory token following the HD strategy in $\mathcal{A}_{>1}$ from a 1-safe HD state that we 'track' until a 1-transition, since, unlike for coBüchi automaton, a run in $\mathcal{A}_{>1}$ for a $[1, K+2]$ automaton $\mathcal{A}$ that does not end at a rejecting sink state can still be rejecting.

We will therefore use *explorability*, introduced by Hazard and Kuperberg [HK23], as an intermediate step. For a natural number $k \geq 1$, a nondeterministic parity automaton is $k$-explorable if Eve wins the following $k$-HD game on it, where she has $k$ tokens instead of one in the HD game, on which she constructs $k$ runs. Her objective is to produce an accepting run on at least one of her tokens if Adam's word is in the language of the automaton.

**Definition 6.10** ($k$-Explorability)**.** *For a nondeterministic parity automaton $\mathcal{B}$ and $k \geq 1$ a natural number, the $k$-HD game on $\mathcal{B}$ is played with $k$ tokens of Eve in $\mathcal{B}$. At the start of the game, Eve has $k$ tokens $t_1, t_2, \ldots, t_k$ in the initial state of $\mathcal{B}$. In round $i$ for each $i \in \mathbb{N}$,*

1. *Adam selects a letter $a_i$;*
2. *Eve moves each of her $l$ tokens along an $a_i$-transition from the tokens' current positions.*

*Eve wins a play of the HD-game if and only if the following condition holds: if Adam's word is in $L(\mathcal{B})$, then at least one of the runs on Eve's $k$ tokens are accepting. We say that $\mathcal{B}$ is $k$-explorable if Eve has a winning strategy in the $k$-HD game on it. We say that $\mathcal{B}$ is* explorable *if $\mathcal{B}$ is $m$-explorable for some positive natural number $m$.*

Note that an automaton is HD if and only if it is 1-explorable. Hazard and Kuperberg observed that 2-token game characterises history-determinism on explorable automata, which was also their original motivation to study explorable automata: in the hope that it helps towards the resolution of the 2-token conjecture.

**Lemma 6.11** ([HK23, Pages 8-9])**.** *Let $\mathcal{B}$ be an explorable parity automata. If Eve wins the 2-token game on $\mathcal{B}$, then $\mathcal{B}$ is HD.*

*Proof.* Let $\mathcal{B}$ be $k$-explorable for some $k \geq 1$. Since Eve wins the 2-token game on $\mathcal{B}$, Eve wins the $k$-token game on $\mathcal{B}$ as well (Lemma 3.15). Eve then has the following winning strategy in the HD game on $\mathcal{B}$. She stores in her memory $k$ tokens that are following a winning strategy in the $k$-HD game on $\mathcal{B}$, and constructs a run on her token in the HD game on $\mathcal{B}$ by choosing transitions according to a winning strategy in the $k$-token game on $\mathcal{B}$ against these $k$-memory tokens. □

Due to the above result, for proving Lemma 6.8, it suffices to show the following result.

**Lemma 6.12.** *Let $\mathcal{A}$ be a $[1, K+2]$ automaton that has 1-safe double-coverage and on which Eve wins the 2-token game from everywhere. Then $\mathcal{A}$ is explorable.*

*Proof.* For each state $q$ in $\mathcal{A}$, we know, due to Lemma 6.9, that there is another state $p$ in $\mathsf{WCR}(\mathcal{A}, q)$, such that $(\mathcal{A}_{>1}, p)$ is HD and Eve wins $G2(p; q, q)$ in $\mathcal{A}_{>1}$.

Let $\sigma$ be a uniform finite-memory strategy for Eve in the HD game of $\mathcal{A}_{>1}$, which is winning from all states $p$ that are HD in $\mathcal{A}_{>1}$, and let $\mathcal{M}$ be the memory of $\sigma$. Let $\mathcal{D}_{>1}$ be the deterministic automaton that is obtained by taking the product of $\mathcal{A}_{>1}$ with the strategy $\sigma$. That is, the states of $\mathcal{D}_{>1}$ are pairs $(p, m)$ such that $p$ is a state in $\mathcal{A}_{>1}$, such that $(\mathcal{A}_{>1}, p)$ is HD, and $m$ is a state in $\mathcal{M}$. There is a transition $(p, m) \xrightarrow{a:c} (p', m')$ in $\mathcal{D}_{>1}$ if in the HD game on $\mathcal{A}_{>1}$, when Eve's token is at $p$, her memory is $m$, and Adam plays the letter $a$, then $\sigma$ takes the transition $q \xrightarrow{a:c} q'$ in $\mathcal{A}_{>1}$ and updates its memory to $m'$.

We say that a state $(p, m)$ in $\mathcal{D}_{>1}$ is *active* on $u$ if $p$ is a state in $\mathsf{WCR}(\mathcal{A}, u)$. Note that if $(p, m)$ is an active state in $\mathcal{D}_{>1}$, then $p$ is 1-safe HD.

Fix a positional winning strategy $\sigma_{G1}$ for Eve in the 1-token game on $\mathcal{A}$ from all pairs of weakly coreachable states. We will show that $\mathcal{A}$ is $k$-explorable, where $k$ is the number of states in $\mathcal{D}_{>1}$, by describing a winning strategy for Eve in the $k$-HD game on $\mathcal{A}$. At a high-level, each of Eve's tokens in the $k$-HD game on $\mathcal{A}$ will play in the 1-token game on $\mathcal{A}$ according to $\sigma_{G1}$, against a corresponding memory token that is taking transitions in $\mathcal{A}$ according to the projection of a corresponding run in $\mathcal{D}_{>1}$ on the $\mathcal{A}$ component.

In more details, let us denote Eve's tokens in this game by $t_1, t_2, \ldots, t_k$, and we suppose that her $k$-tokens $t_1 < t_2 < \cdots < t_k$ are ordered. We describe her strategy in the $k$-HD game on $\mathcal{A}$ inductively as the rounds proceed.

After Adam has read a finite word $u$ in the $k$-HD game on $\mathcal{A}$, we will have exactly one token of Eve *tracking* every state $(p, m)$ of $\mathcal{D}_{>1}$ such that $(p, m)$ is active on $u$. The tokens that are not tracking any active states are said to be *idle*.
When Adam plays the letter $a \in \Sigma$, Eve's tokens take transitions as follows:

1. If Eve's token $t$ at state $q$ is tracking $(p, m)$, then Eve, on the token $t$, takes the transition given by $\sigma_{G1}$ in $G1(q; p)$ in $\mathcal{A}$ when Adam plays the letter $a$. Eve then updates the state that the token $t$ is tracking to be $(p', m')$, where $(p, m) \xrightarrow{a} (p', m')$ is the unique transition from $(p, m)$ on $\mathcal{A}$ in $\mathcal{D}_{>1}$. If $p'$ is the rejecting sink state in $\mathcal{A}_{>1}$, then token $t$ becomes idle and stops tracking $(p', m')$.

2. Otherwise if Eve's token $t$ is idle, then Eve takes an arbitrary outgoing transition on $a$ from the current state of $t$.

Eve then further updates the states of $\mathcal{D}_{>1}$ that her tokens are tracking as follows.

1. If $(p', m')$ is an active state on $ua$ in $\mathcal{D}_{>1}$, such that there are multiple tokens tracking $(p', m')$, we make all but the smallest token idle.

2. If there are active states of $\mathcal{D}_{>1}$ on $ua$ that are not tracked by any token, we pick one idle token for each of them and let them *track* those active states. Note that this is possible because Eve has as many tokens as there are states in $\mathcal{D}_{>1}$.

Thus, at the end of each round, exactly one Eve's token tracks an active state on $ua$. This concludes the description of Eve's strategy in the $k$-HD game on $\mathcal{A}$.

We claim that this strategy is winning for Eve in the $k$-HD game on $\mathcal{A}$. To prove this, suppose that Adam plays a word $w$ in $L(\mathcal{A})$ in a play of the $k$-HD game on $\mathcal{A}$ where Eve is playing according to the above strategy. Then, there is a run of $\mathcal{A}$ over $w$ that contains finitely many priority 1 transitions. Let $w = uw'$ be a decomposition of $w$, such that there is state $r$ of $\mathcal{A}$ reachable by $u$ from the initial state of $\mathcal{A}$ and $w' \in L(\mathcal{A}_{>1}, r)$. Then, by Lemma 6.9, there is a state $p$ in $\mathsf{WCR}(u)$, such that Eve wins $G1(p; r)$ in $\mathcal{A}_{>1}$ and $(\mathcal{A}_{>1}, p)$ is HD. Since $L(\mathcal{A}_{>1}, p) \supseteq L(\mathcal{A}_{>1}, r)$, $w' \in L(\mathcal{A}_{>1}, p)$. Thus, from some state $(p, m)$ in $\mathcal{D}_{>1}$, the run $\rho_D$ of $\mathcal{D}_{>1}$ on $w'$ is accepting, and $(p, m)$ is active after $u$. Furthermore, for all prefixes $v$ of $w'$, the state $(p_v, m_v)$ that $\rho_D$ reaches after the prefix $v$ is not of the form $(q_\perp, m)$ where $q_\perp$ is the sink state in $\mathcal{A}_{>1}$, and thus $(p_v, m_v)$ is active.

Since the tokens are ordered, and whenever more than one tokens are tracking the same state in $\mathcal{D}_{>1}$ we let the smallest token continue to track that state, we note that eventually the same token $t$ uninterruptedly tracks states occurring in $\rho_D$. Then, since $\rho_D$ is accepting, the run on $t$ is also accepting because $\sigma_{G1}$ is a winning strategy. It follows that the strategy we have described is a winning strategy for Eve in the $k$-HD game on $\mathcal{A}$ and, therefore, $\mathcal{A}$ is explorable. □

Since 2-token games characterise history-determinism on explorable automata, we have therefore proved Lemma 6.8.

**Lemma 6.8.** *Every $[1, K+2]$ automaton that has 1-safe double-coverage and on which Eve wins the 2-token game from everywhere is history-deterministic.*

We conclude with the proof of Theorem A.

**Theorem A.** *Let $K > 0$ be a natural number, such that for every $[0, K]$ (or equivalently, $[2, K+2]$) automaton $\mathcal{A}$, Eve wins the 2-token game on $\mathcal{A}$ if and only if $\mathcal{A}$ is HD. Then, for every $[1, K+2]$ automaton $\mathcal{A}$, Eve wins the 2-token game on $\mathcal{A}$ if and only if $\mathcal{A}$ is HD.*

*Proof.* We will show that if Eve wins the 2-token game on a $[1, K+2]$ automaton $\mathcal{A}$ then $\mathcal{A}$ is history-deterministic, under the assumption that the same holds for $[0, K]$ automata (Hypothesis 6.1). By Theorem I, we know that $\mathcal{A}$ has a simulation-equivalent subautomaton $\mathcal{B}$, such that Eve wins the 2-token game from everywhere in $\mathcal{A}$. Then Eve wins the 2-token game from everywhere in $\mathcal{B} \Downarrow$ (Lemma 6.3). By Lemma 6.4, $\mathcal{B} \Downarrow$ has 1-safe double-coverage, and it thus follows from Lemma 6.8 that $\mathcal{B} \Downarrow$ is HD. Thus, $\mathcal{B}$ is HD (Lemma 6.3) and due to the simulation-equivalence of $\mathcal{A}$ and $\mathcal{B}$, $\mathcal{A}$ is HD as well (Corollary 3.11). □

# Chapter 7

# The Odd to Even Induction Step

In this chapter, we will finish our proof of the 2-token theorem by proving our second induction step that adds a most significant even priority (Theorem B). This is done in Sections 7.1 to 7.5. We will then discuss some implications of the 2-token theorem in Section 7.6.

**Theorem B.** *Let $K > 1$ be a natural number, such that for every $[1, K]$ automaton $\mathcal{A}$, Eve wins the $2$-token game on $\mathcal{A}$ if and only if $\mathcal{A}$ is HD. Then, for every $[0, K]$ automaton $\mathcal{A}$, Eve wins the $2$-token game on $\mathcal{A}$ if and only if $\mathcal{A}$ is HD.*

For the rest of this chapter, we fix a natural number $K > 1$ and we suppose that the following hypothesis holds.

**Hypothesis 7.1.** *For every $[1, K]$ automaton $\mathcal{A}$, Eve wins the $2$-token game on $\mathcal{A}$ if and only if $\mathcal{A}$ is HD.*

We will show, assuming Hypothesis 6.1, that if Eve wins the 2-token game on a $[0, K]$ automata then that automaton is HD, which will prove Theorem B. As with our proof for Theorem A, as well as our proof for the Joker-game characterisation of history-determinism on Büchi and coBüchi automata (Chapters 4 and 5), we will use an intermediate property to prove Theorem B. For $[0, K]$ automata, we call this property as *0-reach double-covering*, which is analogous to the property of 1-safe double-coverage we introduced for $[1, K + 2]$ automata, and an extension of reach-covering for Büchi automata.

We introduce 0-reach double-covering in Section 7.1 and show that every $[0, K]$ automaton on which Eve wins the 2-token game from everywhere and that has 0-reach double-covering is history-deterministic.

Then, in Sections 7.2 to 7.5, we will give a normalisation procedure to convert a $[0, K]$ automaton on which Eve wins the 2-token game from everywhere into a simulation-equivalent subautomaton that has 0-reach double-covering and on which Eve still wins the 2-token game from everywhere. This will conclude our proof of Theorem B.

Our normalisation procedure to obtain $[0, K]$ automata with 0-reach double-covering is more technical than that of obtaining 1-safe double-coverage for $[1, K + 2]$ automata in Chapter 6, and involves iteratively modifying the automata using ranks in the 2-token game. This is similar to how we obtained reach-covering for Büchi automata, but unlike 1-token games on Büchi automata, the 2-token game on $[0, K]$ automata are not parity games but instead Muller games. We will thus first convert 2-token games into parity games using Zielonka trees. The normalisation procedure then involves iterative removal of transitions and relabelling of priorities based on the ranks in this parity game. We will give an overview of this iterative procedure in Section 7.2. In Sections 7.3 and 7.4, we will describe the subprocedures involves in the normalisation procedure, and in Section 7.5, we will conclude the correctness of our normalisation procedure.

## 7.1 Automata with 0-reach double-covering

For every $[0, K]$ automaton $\mathcal{A}$, we define the *1-approximation of $\mathcal{A}$*, denoted $\mathcal{A}_{>0}$ as the automaton obtained by preserving all transitions of $\mathcal{A}$ that have priority at least 1, and redirecting any priority 0 transitions to an additional accepting sink state that has self loops with priority 2 on all letters in the alphabet of $\mathcal{A}$. Note that $\mathcal{A}_{>0}$ is a $[1, K]$ automaton (since $K > 1$).

The following observation is easy to see.

**Proposition 7.2.** *For every $[0, K]$ automaton $\mathcal{B}$, $L(\mathcal{B}) \subseteq L(\mathcal{B}_{>0})$.*

*Proof.* Let $\rho$ be an accepting run on some word $w$ in $\mathcal{B}$. Consider the run $\rho_{>0}$ of $\mathcal{B}_{>0}$ on $w$ that has the same transitions as the largest prefix of $\rho$ that contains only transitions of priority at least 1, and $\rho_{>0}$ has a transition to the accepting sink state at the position corresponding to the first priority 0 transition in $\rho$ (if such a transition exists). Then $\rho_{>0}$ is an accepting run in $\mathcal{B}_{>0}$, as desired. □

We say that a $[0, K]$ automaton $\mathcal{A}$ has *0-reach double-covering* if for each state $q$ in $\mathcal{A}$, there is another state $p$ weakly coreachable to $q$ in $\mathcal{A}$, such that Eve wins $G2(q; p, p)$ in $\mathcal{A}_{>0}$.

> **Automata with 0-reach double-covering.** We say that a $[0, K]$ automaton $\mathcal{A}$ has 0-reach double-covering if for each state $q$, there is another state $p \in \mathsf{WCR}(\mathcal{A}, q)$, such that Eve wins $G2(q; p, p)$ in $\mathcal{A}_{>0}$.

**Remark 6.** *Let us compare the definition of 0-reach double-covering with 1-safe double-coverage we had defined for $[1, K+2]$ automata in Chapter 6. Recall that a $[1, K+2]$ automaton $\mathcal{B}$ has 1-safe double-coverage if for each state $q$ in $\mathcal{B}$, there is another state $p$ in $\mathsf{WCR}(\mathcal{B}, q)$, such that Eve wins $G2(p; q, q)$ in $\mathcal{B}_{>1}$. Thus, the roles of $q$ and $p$ in the 2-token game in the definitions of 1-safe double-coverage and 0-reach double-covering are reversed.*

*For Büchi automata, recall that we had defined the similar property of reach-covering; a Büchi automaton $\mathcal{A}$ has reach-covering if for each state $q$ in $\mathcal{A}$, there is another state $p$*

in $\mathcal{A}$, such that Eve wins $G1(q;p)$ in $\mathcal{A}_{reach}$. Note that $\mathcal{A}_{reach}$ and $\mathcal{A}_{>0}$ are the same for Büchi automata.

In this section, we will show the following result.

**Lemma 7.3.** *For every $[0, K]$ automaton $\mathcal{A}$, if $\mathcal{A}$ has 0-reach double-covering and Eve wins the 2-token game from everywhere in $\mathcal{A}$ then $\mathcal{A}$ is history-deterministic.*

The proof of Lemma 7.3 is very similar to that of Lemma 5.7. We start by showing results that are analogous to Lemmas 5.5 and 5.6 for Büchi automata, and also to Lemma 6.9 for $[1, K + 2]$ automata.

**Lemma 7.4.** *Suppose that $\mathcal{A}$ is a $[0, K]$ automaton that has 0-reach double-covering. Then for each state $q$ in $\mathcal{A}$, there is another state $p$ weakly coreachable to $q$ in $\mathcal{A}$, such that Eve wins $G2(q;p,p)$ and $G2(p;p,p)$ in $\mathcal{A}_{>0}$.*

*Proof.* Note that for any three nondeterministic parity automata $\mathcal{A}_1, \mathcal{A}_2$, and $\mathcal{A}_3$, if Eve wins $G2(\mathcal{A}_1; \mathcal{A}_2, \mathcal{A}_2)$ and $G2(\mathcal{A}_2; \mathcal{A}_3, \mathcal{A}_3)$, then Eve wins $G2(\mathcal{A}_1; \mathcal{A}_3, \mathcal{A}_3)$ (Lemma 3.26). For each state $q$ in $\mathcal{A}$, consider the directed graph $G$ whose vertices consist of the states in $\mathsf{WCR}(\mathcal{A}, q)$. There is an edge in $G$ from vertices $r$ to $s$ if Eve wins $G2(r;s,s)$ in $\mathcal{A}_{>0}$.

Since $\mathcal{A}$ has 0-reach double-covering, every vertex in $G$ has outdegree at least 1. Thus, for every vertex $r$, there is a vertex $s$ such that there is a path from $r$ to $s$ and a cycle consisting of the vertex $s$. Observe that if there is a path from $r$ to $s$ in $G$, then Eve wins $G2(r;s,s)$, in $\mathcal{A}_{>0}$. It follows that for each state $r$ in $G$, there is another state $s$ such that Eve wins $G2(r;s,s)$ and $G2(s;s,s)$ in $\mathcal{A}_{>0}$. This concludes our proof. $\square$

For a $[0, K]$ automaton $\mathcal{A}$, we call a state $p$ in $\mathcal{A}$ as *0-reach HD* if $(\mathcal{A}_{>0}, p)$ is HD. Note that since $\mathcal{A}_{>0}$ is a $[1, K]$ automaton, if Eve wins $G2((\mathcal{A}_{>0}, p))$ then, assuming Hypothesis 6.1, $(\mathcal{A}_{>0}, p)$ is HD. Lemma 7.4 thus can be rephrased as the following result.

**Lemma 7.5.** *Suppose that $\mathcal{A}$ is a $[0, K]$ automaton that has 0-reach double-covering. Then for each state $q$ in $\mathcal{A}$, there is another state $p$ weakly coreachable to $q$ in $\mathcal{A}$ such that Eve wins $G2(q;p,p)$ in $\mathcal{A}_{>0}$ and $(\mathcal{A}_{>0}, p)$ is HD.*

We now prove Lemma 7.3.

*Proof of Lemma 7.3.* Fix $\sigma_{>0}$ to be a winning HD strategy in $\mathcal{A}_{>0}$ from all 0-reach HD states $p$, i.e., states $p$ such that $(\mathcal{A}_{>0}, p)$ is HD. We also fix a positional winning strategy $\sigma_{G1}$ for Eve in the $G1$ game on $\mathcal{A}_{>0}$ from all pairs of weakly coreachable states $p, q$ in $\mathcal{A}$, such that Eve wins $G1((\mathcal{A}_{>0}, p); (\mathcal{A}_{>0}, q))$ and $q$ is 0-reach HD. We will construct a winning strategy $\sigma_{HD}$ for Eve in the HD-game on $\mathcal{A}$ using $\sigma_{>0}$ and $\sigma_{G1}$.

In strategy $\sigma_{HD}$, when Eve's token is at a state $q$, she stores in her memory a token that will be at some state $p$ in $\mathcal{A}_{>0}$, such that Eve wins $G1(q;p)$ in $\mathcal{A}_{>0}$ and either of the following two conditions hold.

1. $p$ is 0-reach HD and in $\mathsf{WCR}(\mathcal{A}, q)$.

2. $p$ is the accepting sink state in $\mathcal{A}_{>0}$.

Note that such a state $p$ exists for all states $q$ due to Lemma 7.5.

When Adam chooses a letter $a$, the strategy $\sigma_{HD}$ in the HD game on $\mathcal{A}$ then picks the transition given by $\sigma_{G1}$ on $G1(q;p)$ in $\mathcal{A}_{>0}$ if it is feasible to do so, i.e., the transition $\delta$ is not to the accepting sink state in $\mathcal{A}_{>0}$. The memory token $p$ takes the transition on $a$ given by $\sigma_{>0}$ in $\mathcal{A}_{>0}$.

Otherwise, if the transition $\delta$ leads to an accepting sink state in $\mathcal{A}_{>0}$, then we know there is a priority 0 transition from $q$ on the letter $a$ in $\mathcal{A}$. The strategy $\sigma_{HD}$ picks this transition $q \xrightarrow{a:0} q'$. We then reset the memory token to be at $p'$, such that $p'$ is weakly coreachable to $q'$ in $\mathcal{A}$, Eve wins $G1(q';p')$ in $\mathcal{A}_{>0}$ and $p'$ is 0-reach HD.

We will show that the above strategy $\sigma_{HD}$ is winning for Eve in the HD game on $\mathcal{A}$. Indeed, let $\rho$ in any play of the HD game where Eve is playing according to $\sigma_{HD}$. Then, if the memory token is reset infinitely often in $\rho$, then the run on Eve's token contains infinitely many priority 0 transitions and is accepting.

Otherwise, suppose that the memory token is reset finitely often. Furthermore, suppose that after Adam has played a finite word $u$, Eve's token in the HD game is at state $q$, Eve's memory token is at state $p$, and the memory token is not reset after this point in the play. If the state $p$ is the accepting sink state in $\mathcal{A}_{>0}$, then the run on Eve's token is accepting since $\sigma_{G1}$ is accepting. Otherwise if $p$ is in $\mathsf{WCR}(\mathcal{A}, q)$, then note that $L(\mathcal{A}, q) = L(\mathcal{A}, p) = u^{-1}L(\mathcal{A}, q_0)$ since $\mathcal{A}$ is semantically-deterministic (see Lemma 3.36). Furthermore, observe that $L(\mathcal{A}, p) \subseteq L(\mathcal{A}_{>0}, p)$, due to Proposition 7.2. Thus, if Adam's word $w$ from this point is such that $uw \in L(\mathcal{A})$, or equivalently, $w \in L(\mathcal{A}, p) \subseteq L(\mathcal{A}_{>0}, p)$, then the run on Eve's memory token in $\mathcal{A}_{>0}$ is accepting. Since Eve's token is following transitions given by the winning strategy $\sigma_{G1}$, Eve's token either eventually takes a priority 0-transition and resets her memory token or she builds an accepting run on her token in $\mathcal{A}_{>0}$ that is also an accepting run in $\mathcal{A}$. Since Eve's memory token is not reset after Adam has played $u$ in the HD game on $\mathcal{A}$, we are in the latter case, and therefore, the run on Eve's token is accepting. We conclude that this strategy of Eve in the HD game on $\mathcal{A}$ is winning, and thus, $\mathcal{A}$ is HD. □

## 7.2 Towards 0-reach double-covering: an overview

Having shown that $[0, K]$ automata on which Eve wins the 2-token game from everywhere and that have 0-reach double-covering are HD in the previous section (Lemma 7.3), we focus on showing the following result next.

**Lemma 7.6.** *Let $\mathcal{A}$ be a $[0, K]$ automaton on which Eve wins the 2-token game from everywhere. Then, there is a $[0, K]$ automaton $\mathcal{A}^I$, such that $\mathcal{A}^I$ is simulation-equivalent to $\mathcal{A}$, Eve which Eve wins the 2-token game from everywhere on $\mathcal{A}^I$, and $\mathcal{A}^I$ has 0-reach double-covering.*

Lemmas 7.3 and 7.6 together would then prove Theorem B. To prove Lemma 7.6, we will describe a normalisation procedure for $[0, K]$ automata on which Eve wins the 2-token game from everywhere. This normalisation procedure is based on the ranks of the games obtained by converting 2-token games on $[0,K]$ automata into winner-equivalent parity games via Zielonka trees.

Recall that the winning condition for the 2-token game on a $[0,K]$ automaton can be given by a $(C,\mathcal{F})$ Muller condition (Section 3.5.2). The colours $C$ are given by $C = [0,K] \times [0,K] \times [0,K]$, where the first components are for the priorities that occur along the transitions of Eve's token, while the second and the third component are for the priorities that occur along the transitions of Adam's first and second token, respectively. The set of accepting subsets $\mathcal{F}$ consists of nonempty subsets $S$ of $C$ in which the following condition holds: if the least priority occurring amongst the second components of the elements of $S$ is even or if the least priority occurring amongst the third components of the elements of $S$ is even, then the least priority occurring amongst the first components of the elements of $S$ is also even.

We will denote the Zielonka tree of the condition $(C,\mathcal{F})$ above as $\mathcal{Z}_{[0,K]}$, and the deterministic parity automaton recognising the $(C,\mathcal{F})$-Muller language as $\mathcal{D}_{[0,K]}$. That is, the automaton $\mathcal{D}_{[0,K]}$ has its alphabet as $C$, and a word $w$ in $C^\omega$ is accepted by $\mathcal{D}_{[0,K]}$ if and only if the set of colours occurring infinitely often in $w$ is a set in $\mathcal{F}$. The construction of Zielonka trees and corresponding deterministic automata for a given Muller condition can be found in Section 2.4. The construction of $\mathcal{Z}_{[0,K]}$ can be found in Section 3.5.2.

We will use $\mathcal{D}_{[0,K]}$ to construct parity games that have the same winner as the 2-token games on $[0,K]$ automata.

**Definition 7.7.** *For a $[0,K]$ automaton $\mathcal{A} = (Q, \Sigma, q_0, \Delta)$, we define the parity game $\mathcal{G}_2(\mathcal{A}) = (V,E)$ as follows.*

- *The set of vertices $V$ consists of the set $V = V_1 \cup V_2 \cup V_3$, where:*

  1. $V_1 = \{(q, p_1, p_2, \beta) \mid q, p_1, p_2$ *are weakly coreachable states and $\beta$ is a branch in $\mathcal{Z}_{[0,K]}$, or equivalently, a state in $\mathcal{D}_{[0,K]}\}$*
  2. $V_2 = \{(q, a, p_1, p_2, \beta) \mid (q, p_1, p_2, \beta) \in V_1, a \in \Sigma\};$
  3. $V_3 = \{(\delta, p_1, p_2, a, \beta) \mid (q, a, p_1, p_2, \beta) \in V_2$ *and* $\delta = q \xrightarrow{a:c_1} q' \in \Delta\}.$

  *Eve's vertices are $V_\exists = V_2$, while Adam's vertices are $V_\forall = V_1 \cup V_3$*

- *The set of edges $E$ is the union of following sets:*

  1. $E_1 = \{(q, p_1, p_2, \beta) \to (q, a, p_1, p_2, \beta) \mid a \in \Sigma\}$ *(Adam chooses a letter)*
  2. $E_2 = \{(q, a, p_1, p_2, \beta) \to (\delta, p_1, p_2, a, \beta) \mid q \xrightarrow{a:c_1} q' \in \Delta\}$ *(Eve chooses a transition on her token)*
  3. $E_3 = \{(\delta, p_1, p_2, a, \beta) \xrightarrow{d} (q', p'_1, p'_2, \beta') \mid \delta = q \xrightarrow{a:c_1} q', p_1 \xrightarrow{a:c_2} p'_1, p_2 \xrightarrow{a:c_3} p'_2$ *are transitions in $\Delta$ and $\beta \xrightarrow{(c_1,c_2,c_3):d} \beta'$ is a transition in $\mathcal{D}_{[0,K]}\}$ (Adam chooses transitions on his tokens)*

- *The priority of edges is as follows. All elements in $E_1$ and $E_2$ are assigned priority $h+1$, where $h$ is the height of the Zielonka tree $\mathcal{Z}_{[0,K]}$. The edges in $E_3$ as above are assigned priorities according to the transition from the current branch in $\mathcal{D}$ on the colour of the three transitions on Eve's token and Adam's two tokens.*

The correctness of the conversion from Muller to parity games via Zielonka trees (Lemma 2.7) implies that for every $[0,K]$ automaton $\mathcal{A}$ and states $q,p,r$ that are weakly coreachable, Eve wins the 2-token game on $\mathcal{A}$ from $(q;p,r)$, if and only if, Eve wins $\mathcal{G}_2(\mathcal{A})$ from $(q,p,r,\beta)$ for some branch $\beta$, if and only if, Eve wins $\mathcal{G}_2(\mathcal{A})$ from $(q,p,r,\beta)$ for all branches $\beta$. In particular, Eve wins the 2-token game on $\mathcal{A}$ from everywhere if and only if Eve wins $\mathcal{G}_2(\mathcal{A})$ from everywhere.

**Lemma 7.8.** *For every $[0,K]$ automaton $\mathcal{A}$, Eve wins the 2-token game on $\mathcal{A}$ from everywhere if and only if Eve wins $\mathcal{G}_2(\mathcal{A})$ from everywhere.*

We will thus use the games $G2(\mathcal{A})$ and $\mathcal{G}_2(\mathcal{A})$ interchangeably. For the rest of this section and for Sections 7.3 to 7.5, we fix a $[0,K]$ automaton $\mathcal{A}$ on which Eve wins the 2-token game from everywhere.

> We fix a $[0,K]$ automaton $\mathcal{A}$ on which Eve wins the 2-token game from everywhere.

Our normalisation procedure to convert $\mathcal{A}$ to a simulation-equivalent $[0,K]$ automaton that, in addition to Eve winning the 2-token game from everywhere, has 0-reach double-covering is based on the ranks of the parity game $\mathcal{G}_2(\mathcal{A})$. To explain the normalisation procedure, we first make some observations on the Zielonka tree $\mathcal{Z}_{[0,K]}$ and the automaton $\mathcal{D}_{[0,K]}$. For our purpose, we only require an understanding of the top three layers of $\mathcal{Z}_{[0,K]}$, which is as shown in Fig. 7.1.

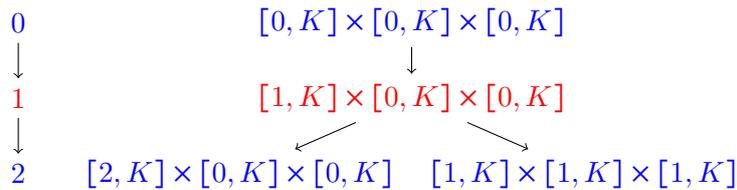

Figure 7.1: The first three layers of the Zielonka tree $\mathcal{Z}_{[0,K]}$

We say that a branch $\beta$ in $\mathcal{Z}_{[0,K]}$ is a *right-branch* if $\beta$ contains the node labelled $[1,K]\times[1,K]\times[1,K]$. Otherwise, we say that $\beta$ is a *left-branch* if $\beta$ contains the node labelled $[2,K]\times[0,K]\times[0,K]$. Note that every branch in $\mathcal{Z}_{[0,K]}$ is either a right-branch or a left-branch, and no branch is both.

Let us briefly recall some relevant details of how the automaton $\mathcal{D}_{[0,K]}$ is constructed from $\mathcal{Z}_{[0,K]}$ (Section 2.4). When there is a transition from branch $\beta$ to $\beta'$ on $(c_1,c_2,c_3)$ in $\mathcal{D}_{[0,K]}$, then the priority $d$ of that transition is same as the priority-depth of the deepest common node $\nu$ of $\beta$ and $\beta'$. For the tree $\mathcal{Z}_{[0,K]}$, the priority-depth of a node is same as its depth.

Additionally, the node $\nu$ is the support of $(c_1, c_2, c_3)$ in $\beta$, i.e., $\nu$ is the deepest node in $\beta$ that contains $(c_1, c_2, c_3)$. The following result on transitions between left and right branches then is clear, based on the first three layers of $\mathcal{Z}_{[0,K]}$ (Fig. 7.1).

**Proposition 7.9.** *Let $\beta \xrightarrow{(c_e, c_a^1, c_a^2):d} \beta'$ be a transition in $\mathcal{D}_{[0,K]}$. Then, the following statements are true.*

1. *If the priority $d$ is at least 2 then $\beta$ and $\beta'$ are either both right branches, or are both left branches.*

2. *Let $\beta$ be a left branch. Then $d$ is 0 if and only if $c_e$ is 0, and $d$ is 1 if and only if $c_e$ is 1.*

3. *Let $\beta$ be a right branch. Then $d$ is 0 if and only if $c_e$ is 0, and $d$ is 1 if and only if at least one of $c_a^1$ and $c_a^2$ is 0.*

Recall the concept of *ranks* introduced in Section 2.5. Specifically, recall that if $\mathsf{rank}(q, p, r, \beta) = 0$ for some vertex $(p, q, r, \beta)$ in $\mathcal{G}_2(\mathcal{A})$, then Eve's optimal winning strategy $\sigma$ ensures that a 0 priority edge occurs before the first occurrence of a 1 priority edge in any play (Lemma 2.9). Thus, the following result is true due to Proposition 7.9 and the fact that the first component in the elements of $C = [0, K] \times [0, K] \times [0, K]$ corresponds to the priorities of Eve's token, while the second and third components correspond to the priorities of Adam's first and second tokens, respectively.

**Proposition 7.10.** *Suppose $\mathsf{rank}(q, p, r, \beta) = 0$ in $\mathcal{G}_2(\mathcal{A})$ for some weakly coreachable states $q, p, r$ in $\mathcal{A}$ and a branch $\beta$ of the Zielonka tree. Fix $\sigma$ to be an optimal winning strategy for Eve in $\mathcal{G}_2(\mathcal{A})$. Then the following holds.*

1. *If $\beta$ is a left-branch, then the strategy $\sigma$ from $\mathcal{G}_2(q, p, r, \beta)$ in $\mathcal{A}$ ensures that a 0 priority transition occurs earlier than the first 1 priority transition in the run of Eve's token.*

2. *If $\beta$ is a right-branch, then the strategy $\sigma$ from $\mathcal{G}_2(q, p, r, \beta)$ in $\mathcal{A}$ ensures that a 0 priority transition in Eve's token occurs earlier or in the same round as the first occurrence of the 0 priority transition in any of Adam's tokens.*

A consequence of Proposition 7.10 above is that if $\mathsf{rank}(q, p, r, \beta) = 0$ for some right-branch $\beta$, then Eve wins $G2(q; p, r)$ in $\mathcal{A}_{>0}$. Indeed the optimal winning strategy $\sigma$ for Eve in $\mathcal{G}_2(\mathcal{A})$ from $(q, p, r, \beta)$ ensures that a 0 priority transition occurs in Eve's token no later than the first occurrence of a 0 priority transition in any of Adam's tokens. Thus, Eve can use $\sigma$ to play $G2(q; p, r)$ in $\mathcal{A}_{>0}$, keeping the branch of the corresponding vertices in $\mathcal{G}_2(\mathcal{A})$ in her memory, to make sure her token reaches the accepting sink state in $\mathcal{A}_{>0}$ in no later round than any of Adam's tokens. If the run on Eve's token does not reach the accepting sink state in $\mathcal{A}_{>0}$, then neither of Adam's token reach the accepting sink state, and since $\sigma$ is a winning strategy, the run on Eve's token in $\mathcal{A}_{>0}$ is accepting if any of the runs on Adam's token in $\mathcal{A}_{>0}$ is accepting.

**Lemma 7.11.** *If* $\mathsf{rank}(q,p,r,\beta) = 0$ *for weakly coreachable states $q,p,r$ in $\mathcal{A}$ and some right-branch $\beta$, then Eve wins $G2(q;p,r)$ in $\mathcal{A}_{>0}$.*

For a $[0,K]$ automaton $\mathcal{B}$ on which Eve wins the 2-token game from everywhere, we define the *optimal rank* of a state $q$ in $\mathcal{B}$ as

$$\mathsf{opt}(q) = \min\{\mathsf{rank}(q,p,r,\beta) \mid q,p,r \in \mathsf{WCR}(\mathcal{B},q) \text{ and } \beta \text{ is a branch in } \mathcal{Z}_{0,K}\}.$$

If $\mathsf{opt}(q) = 0$ and there is a right-branch $\beta$ such that $\mathsf{rank}(q,p,r,\beta) = 0$, then we say that $q$ is a *right state*. We define the *non-right states* as states $q$ for which $\mathsf{opt}(q) = 0$, but that are not right states.

Inspired by Lemma 7.11, the objective of the normalisation procedure we present next is to make all states right states, while preserving some reasonable invariants (I1 and I2).

**The normalisation procedure.** Set $\mathcal{A}^0 = \mathcal{A}$. For each $i \geq 0$, we do the following three steps on $\mathcal{A}^i$ until $\mathcal{A}^i = \mathcal{A}^{i+1}$ via the following three subprocedures.

1. *Rank-reduction.* We modify the automaton $\mathcal{A}^i$ into a simulation equivalent automaton $\mathcal{B}^i$, so that the optimal rank for all states $q$ in $\mathcal{G}_2(\mathcal{B}^i)$ in 0. This is done similar to the rank-reduction procedure for Büchi automata (Section 5.2), and we will elaborate upon the details of this subprocedure later in Section 7.3.

2. *Branch-separation.* For the automaton $\mathcal{B}^i$, we remove the following transitions to obtain $\mathcal{C}^i$.

    (a) Transitions $q \xrightarrow{a:c} q'$ from a right state $q$ to a non-right state $q'$ in which the priority $c$ is at least 1.
    
    (b) Transitions of priority 1 that are outgoing from non-right states.

3. *Priority-reduction.* The branch-separation step might change the fact that all states have their optimal rank as 0. We nevertheless use the right (resp. non-right) states of $\mathcal{C}^i$ to refer to the states that were originally right (resp. non-right) in $\mathcal{B}^i$. In automaton $\mathcal{C}^i$, for any transition $p \xrightarrow{a:c} p'$ where $p$ is an non-right state, if the priority $c$ is not 0 or 1, then we decrease the priority of that transition by 2. We let $\mathcal{A}^{i+1}$ be the automaton thus obtained.

Each of the above subprocedures, as we will describe them in detail, only remove certain transitions or decrease the priorities of certain transitions. Thus, the normalisation procedure eventually stabilises to an automaton $\mathcal{A}^I$. We will show that the automaton $\mathcal{A}^I$ has 0-reach double-covering in Section 7.5. Additionally, we will argue that $\mathcal{A}^I$ is simulation-equivalent to $\mathcal{A}$ and Eve wins the 2-token game from everywhere on $\mathcal{A}^I$ by showing that each of the subprocedures in the normalisation procedure preserve the following invariants.

I1 Simulation-equivalence to $\mathcal{A}$.

I2 Eve winning the 2-token game from everywhere.

We will prove that the rank-reduction procedure preserves the above invariants in Section 7.3, and in Section 7.4, we will prove the same for branch-separation and priority-reduction.

## 7.3 Rank-reduction procedure

Our rank-reduction procedure is iterative in itself, and is similar to the rank-reduction procedure we presented for Büchi automata (Section 5.2). Starting with a $[0, K]$ automaton $\mathcal{B}$ on which Eve wins the 2-token game from everywhere and that is simulation-equivalent to $\mathcal{A}$, we iteratively modify $\mathcal{B}$ to another $[0, K]$ automaton $\mathcal{B}_N$, so that the optimal rank is 0 for all states in $\mathcal{B}_N$.

**Rank-reduction procedure.** Set $\mathcal{B}_0 = \mathcal{B}$. For each $i \geq 0$, we perform the following three steps on $\mathcal{B}_i$ until $\mathcal{B}_{i+1} = \mathcal{B}_i$.

Step 1 For each state state $q$ in $\mathcal{B}_i$, compute the optimal rank of $q$ in $\mathcal{G}_2(\mathcal{B}_i)$, which we denote by $\mathsf{opt}_i(q)$.

Step 2 Obtain $\mathcal{B}'_i$ from $\mathcal{B}_i$ by removing all transitions $q \xrightarrow{a:c} q'$ with $\mathsf{opt}_i(q) < \mathsf{opt}_i(q')$ and $c > 0$.

Step 3 Obtain $\mathcal{B}_{i+1}$ from $\mathcal{B}'_i$, by changing priorities of transitions $q \xrightarrow{a:c} q'$ with $c > 0$ and $\mathsf{opt}_i(q) > \mathsf{opt}_i(q')$ to $c = 0$.

We will argue in Lemmas 7.14 and 7.15 that the invariants I1 and I2 are preserved through Steps 2 and 3 respectively. Assuming that these invariants hold, observe that in Steps 2 and 3, we are either removing transitions, or decreasing the priority of some transitions to 0. Thus, this procedure terminates after at most $|\Delta|$ many iterations. Let $\mathcal{B}_N$ be the automaton obtained after this procedure terminates. We argue that in $\mathcal{B}_N$, all opt values of states $q$, denoted $\mathsf{opt}_N(q)$, are all 0. The proof is nearly identical to that of Lemma 5.16.

**Lemma 7.12.** *For all states $q$ in $\mathcal{B}_N$, we have that $\mathsf{opt}_N(q) = 0$.*

*Proof.* Assume, towards a contradiction, that there is a state $q$, such that

$$\mathsf{opt}_N(q) = \mathsf{rank}_N(c_0) > 0$$

for some $c_0 = (q, p_1, p_2, \beta) \in \mathcal{G}_2(\mathcal{B}_N)$, where $\mathsf{rank}_N$ is the function that assigns vertices in $\mathcal{G}_2(\mathcal{B}_N)$ their ranks in $\mathcal{G}_2(\mathcal{B}_N)$. We will show that the rank-reduction procedure then can run for at least one more iteration on $\mathcal{B}_N$, which contradicts our assumption that the rank-reduction procedure has stabilised on $\mathcal{B}_N$.

Fix an optimal winning strategy $\sigma$ for Eve in $\mathcal{G}_2(\mathcal{B}_N)$. Since $\mathsf{rank}_N(c_0) > 0$, there is a finite play $\rho$ of $\mathcal{G}_2(\mathcal{B}_N)$ in which Eve is playing according to $\sigma$ and that contains a 1 priority edge but no 0 priority edge. Since a 0 priority edge in $\mathcal{G}_2(\mathcal{B}_N)$ corresponds to a 0 priority transition on Eve's token, it follows that Eve's token only takes transitions of priority at least 1 in $\rho$.

Since $\mathsf{rank}_N$ does not increase across edges of priority at least 1 and strictly decreases when an edge of priority 1 is seen (Proposition 2.10), we get that $\mathsf{rank}_N$ strictly decreases across $\rho$. Then, there must be a transition on Eve's token in $\rho$ across which the quantity $\mathsf{opt}_N$ decreases as well, since we started with $\mathsf{opt}_N(q) = \mathsf{rank}_N(c_0)$. This implies that there must be a transition in $\mathcal{B}_N$ that has priority greater than 0, and across which the quantity $\mathsf{opt}_N$ strictly decreases. But since Step 3 of the rank-reduction procedure changes the priority of such transitions to 0, we get a contradiction to the fact that the rank-reduction procedure has stabilised. $\square$

We now proceed to prove the invariants for Steps 2 and 3, which is proved similarly to the analogous results of Lemmas 5.14 and 5.15 for the rank-reduction procedure for Büchi automata.

Step 2 removes certain transitions from the automaton. To show that the invariants are preserved during Step 2, we will use the following result, which is an easy adaptation of Lemma 5.13.

**Lemma 7.13.** *Let $\mathcal{P}$ be a nondeterministic parity automaton, and $\mathcal{P}'$ a subautomaton of $\mathcal{P}$ such that for every three states $q, p, r$ that are weakly coreachable in $\mathcal{P}$, Eve wins $G2((\mathcal{P}', q); (\mathcal{P}, p), (\mathcal{P}, r))$. Then the following statements hold.*

1. *$\mathcal{P}$ and $\mathcal{P}'$ are simulation equivalent.*

2. *Eve wins the 2-token game from everywhere in $\mathcal{P}'$.*

*Proof.* Proof of (1). Note that since $\mathcal{P}'$ is a subautomaton of $\mathcal{P}$, $\mathcal{P}$ simulates $\mathcal{P}'$: Eve can win the simulation game of $\mathcal{P}'$ by $\mathcal{P}$ by simply copying the transitions of Adam's token in $\mathcal{P}'$ in her token in $\mathcal{P}$. For the other direction, note that Eve wins $G1(\mathcal{P}'; \mathcal{P})$, and hence $\mathcal{P}'$ simulates $\mathcal{P}$ (Lemma 3.26). Thus $\mathcal{P}$ and $\mathcal{P}'$ are simulation-equivalent.

Proof of (2). Note that if states $q, p, r$ are weakly coreachable in $\mathcal{P}'$, then they are also weakly coreachable in $\mathcal{P}$. Therefore Eve wins $G2((\mathcal{P}', q); (\mathcal{P}, p), (\mathcal{P}, r))$, and since $\mathcal{P}'$ is a subautomaton of $\mathcal{P}$, Eve wins $G2(q; p, r)$ in $\mathcal{P}'$. It follows that Eve wins the 2-token game from everywhere in $\mathcal{P}'$. $\square$

**Lemma 7.14** (Invariants for Step 2)**.** *If Eve wins the 2-token game from everywhere on $\mathcal{B}_i$ and $\mathcal{B}_i$ is simulation-equivalent to $\mathcal{A}$, then the same holds for $\mathcal{B}'_i$.*

*Proof.* Fix $\sigma_{G2}$ to be an optimal positional strategy for Eve in the parity game $\mathcal{G}_2(\mathcal{B}_i)$, and fix $\sigma_{G1}$ to be a winning positional strategy for Eve in $G1$ from all states $q, p$ that are weakly coreachable in $\mathcal{A}$. Consider the strategy $\sigma'_{G2}$ in $\mathcal{G}_2(\mathcal{B}_i)$ that takes a 0 priority transition on Eve's token whenever Adam's letter $a$ is such that there is an outgoing transition on $a$ with priority 0 from Eve's token, and otherwise follows $\sigma_{G2}$. Note that $\sigma'_{G2}$ is a winning strategy since Eve wins $G2$ from everywhere, and is an optimal positional strategy since 0 priority transitions on Eve's token corresponds to 0 priority edges in $\mathcal{G}_2(\mathcal{B}_i)$. Thus we assume, without loss of generality, that the strategy $\sigma_{G2}$ takes a 0 priority transition on Eve's token whenever possible.

We will describe a winning strategy $\sigma'$ for Eve in $G2((\mathcal{B}'_i, q); (\mathcal{B}_i, p^1), (\mathcal{B}_i, p^2))$, for all triplets of states $q, p^1, p^2$ are weakly coreachable in $\mathcal{A}_i$. It will then follow from Lemma 7.13 that the invariants I1 and I2 are preserved during Step 2.

At a high level, the strategy $\sigma'$ will require Eve to store as memory two additional tokens in $\mathcal{B}_i$, and a branch $\beta$ of $\mathcal{Z}_{[0,K]}$—or equivalently a state in $\mathcal{D}_{[0,K]}$. Eve's two memory tokens will each choose transitions according to $\sigma_{G1}$ by playing the 1-token game against the two Adam's tokens. Eve's token will select transitions according to $\sigma_{G2}$ by playing in $\mathcal{G}_2(\mathcal{B}_i)$ against her memory tokens and memory branch, till she takes a 0 priority transition or the transition given by $\sigma_{G2}$ had been deleted, in which case she *resets* her memory.

We will describe the strategy $\sigma'$ for Eve inductively, as rounds of the 2-token game $G2((\mathcal{B}'_i, q); (\mathcal{B}_i, p^1), (\mathcal{B}_i, p^2))$ proceed along.

At the start of the 2-token game, let $q_0 = q, p_0^1 = p^1, p_0^2 = p_2$. We let Eve's memory tokens be at states $r_0^1$ and $r_0^2$ and her memory branch be at $\beta_0$, such that

$$\mathsf{opt}(q_0) = \mathsf{rank}(q_0, r_0^1, r_0^2, \beta_0).$$

Throughout the play, we will preserve the invariant that the states of Eve's token, Eve's memory tokens, and Adam's tokens are all weakly coreachable in $\mathcal{B}_i$.

After $j$ rounds in the 2-token game for some $j \geq 0$, suppose that the 2-token game is at the position $(q_j, p_j^1, p_j^2)$; i.e., Eve's token is at $q_j$, and Adam's tokens are at $p_j^1$ and $p_j^2$. Suppose that Eve's memory is $(r_j^1, r_j^2, \beta_j)$. Eve then plays as follows. Adam chooses the letter $a_j$, and let $\delta_j = q_j \xrightarrow{a_j : c_j} q'_j$ be the transition given by $\sigma_{G2}$ from $(q_j, a_j, r_j^1, r_j^2, \beta_j)$. We distinguish between the following three cases.

**Case 1.** $\delta_j$ is a 0-transition.

Then, Eve takes this transition on her token, thus setting $q_{j+1}$ to be $q'_j$, and she updates her memory to be $(r_{j+1}^1, r_{j+1}^2, \beta_{j+1})$ so that

$$\mathsf{rank}(q_{j+1}, r_{j+1}^1, r_{j+1}^2, \beta_{j+1}) = \mathsf{opt}(q_{j+1}).$$

**Case 2.** If $\delta_j$ is a transition of priority at least 1 that hasn't been removed in $\mathcal{B}'_i$.

Then Eve moves her token to $q_{j+1}$. She updates her memory tokens at $r_{j+1}^1, r_{j+1}^2$ to take the transitions given by her strategy $\sigma_{G1}$ in $G1(\mathcal{B}_i)$ against Adam's tokens at $p_j^1, p_j^2$, respectively. The branch $\beta_{j+1}$ is updated according to the automata $\mathcal{D}_{[0,K]}$, based on the priorities that Eve's token and Eve's memory tokens take in the 2-token game of Eve's token against Eve's memory tokens.

**Case 3.** If $\delta_j$ is no longer a transition in $\mathcal{B}'_i$.

Eve then finds states $s_j^1, s_j^2 \in \mathsf{WCR}(\mathcal{B}_i, q_j)$ and a branch $\beta_j'$ so that

$$\mathsf{opt}(q_j) = \mathsf{rank}(q_j, s_j^1, s_j^2, \beta_j').$$

She picks the transition given by $\sigma_{G2}$ from $(q_j, a, s_j^1, s_j^2, \beta_j')$. Eve's memory tokens are then reset to $s_j^1$ and $s_j^2$, and she then takes the $a$-transitions on her memory tokens given by her strategy $\sigma_{G1}$ in the 1-token game against Adam's tokens at $p_j^1$ and $p_j^2$, respectively. The branch $\beta_{j+1}$ is obtained by the unique outgoing transition in $\mathcal{D}_{[0,K]}$ from $\beta_j'$, on the priorities of the transitions that Eve's token and Eve's memory tokens took (from $s_j^1$ and $s_j^2$).

This concludes the description of Eve's strategy $\sigma'$. Let $\rho$ be a play for Eve in the game $G2((\mathcal{B}_i', q); (\mathcal{B}_i, p^1), (\mathcal{B}_i, p^2))$, where Eve is playing according to $\sigma'$. We will show that $\rho$ is a winning play for Eve.

Firstly, suppose that eventually only moves from Case 2 occur in $\rho$. Let $w$ be the word that Adam plays in $\rho$, and suppose that the run of his token $t_j$ on $w$ is accepting, for some $j = 1$ or $2$. Then, since Eve's memory tokens are eventually not reset, and the suffix of the moves on the Eve's memory token $j$ constitute a run on some suffix of $w$ in $\mathcal{A}_i$, which is accepting since $\sigma_{G1}$ is a winning strategy. Because Eve's token is taking transitions according to a winning strategy in the 2-token game against her memory tokens, the run of Eve's token then is accepting as well. Thus, Eve wins $\rho$ in this case.

If moves from Case 1 are taken infinitely often in $\rho$, then the run on Eve's token contains infinitely many 0 priority transitions and it is accepting, implying Eve wins $\rho$.

We will therefore conclude from the next claim that $\rho$ is a winning play for Eve, which will finish our proof of Lemma 7.14.

**Claim 6.** *In the play $\rho$, either moves from Case 1 are taken infinitely often, or eventually only moves from Case 2 are taken.*

In order to prove the claim, it suffices to prove that at most $m$ many moves from Case 3 can be taken before a move from Case 1 is taken, where $m$ is the number of vertices in $\mathcal{G}_2(\mathcal{B}_i)$. Note that the ranks of vertices in $\mathcal{G}_2(\mathcal{B}_i)$ are bounded by $m$ (Proposition 2.11). We will show that the rank of the configuration between the state of Eve's token, the state of Eve's memory tokens, and Eve's memory branch strictly decreases whenever a move from Case 3 is taken; it is clear from Proposition 2.10 that a move from Case 2 does not increase the rank.

Suppose at the start of round $j$ in $\rho$, Eve's token is initially at $q_j$ and her memory is $(r_j^1, r_j^2, \beta_j)$, and Adam chooses a letter $a_j$ such that the strategy $\sigma'$ takes a move from Case 3. Suppose that the new Eve's state and memory then is $q_{j+1}$ and $(r_{j+1}^1, r_{j+1}^2, \beta_{j+1})$, respectively. Let $\delta_j = q_j \xrightarrow{a_j} q_j'$ be the transition given by $\sigma_{G2}$ from $(q_j, r_j^1, r_j^2, \beta_j)$. Since $\delta_j$ is a deleted transition, we note that $\mathsf{opt}(q_j) < \mathsf{opt}(q_j')$. We therefore have the following

series of inequalities

$$\begin{aligned}\mathsf{rank}(q_j,r_j^1,r_j^2,\beta_j) &\geq \mathsf{opt}(q_j') \\ &> \mathsf{opt}(q_j) \\ &\geq \mathsf{rank}(q_{j+1},r_{j+1}^1,r_{j+1}^2,\beta_{j+1}).\end{aligned}$$

Here, the first inequality holds due to monotonicity of ranks (Proposition 2.10). For the third inequality, note that the transition $\delta_j = q_j \xrightarrow{a_j} q_{j+1}$ cannot have priority 0 since otherwise $\sigma_{G2}$ would have picked a priority 0 transition, as we assumed in the start of this proof. Thus, the third inequality also follows from monotonicity of ranks (Proposition 2.10). Thus, the quantity $\mathsf{rank}(q_l, r_l^1, r_l^2, \beta_l)$ strictly decreases from round $l$ to round $(l+1)$ whenever a move from Case 3 is taken, and is non-increasing on moves from Case 2 due to Proposition 2.10. Since the ranks of vertices in $\mathcal{G}_2(\mathcal{B}_i)$ are bounded by $m$, we conclude the proof of the claim and hence of Lemma 7.14. □

Showing invariants are preserved during Step 3 is relatively easier.

**Lemma 7.15** (Invariants for Step 3). *If Eve wins the 2-token game from everywhere on $\mathcal{B}_i'$ and $\mathcal{B}_i'$ is simulation-equivalent to $\mathcal{A}$, then the same holds for $\mathcal{B}_{i+1}$.*

*Proof.* Recall that $\mathcal{B}_{i+1}$ is obtained by relabelling priorities of certain transitions in $\mathcal{B}_i'$. Hence, it suffices to show that a run is accepting in $\mathcal{B}_{i+1}$ if and only if that same run is accepting in $\mathcal{B}_i'$.

One direction is clear: if a run in $\mathcal{B}_i'$ is accepting, then the same run must be accepting in $\mathcal{B}_{i+1}$, since we only changed the priority of certain transitions to 0.

For the other direction, suppose that $\rho$ is an accepting run in $\mathcal{B}_{i+1}$, and consider the same run $\rho$ in $\mathcal{B}_i'$.

If $\rho$ in $\mathcal{B}_{i+1}$ is such that it contains finitely many transitions of priority 0 in $\mathcal{B}_{i+1}$, then because $\rho$ in $\mathcal{B}_{i+1}$ is accepting, it eventually only contains transitions of priority at least 2. Then note that the set of priorities visited infinitely often in $\rho$ in $\mathcal{B}_i'$ is the same as that in $\mathcal{B}_{i+1}$, and hence, $\rho$ is also accepting in $\mathcal{B}_{i+1}$.

Otherwise, suppose that $\rho$ contains infinitely many priority 0 transitions in $\mathcal{B}_{i+1}$. Note that due to Step 2, the $\mathsf{opt}_i$ values are non-increasing across transitions with priority greater than 0, and, due to Step 3, for every transition $\delta = q \xrightarrow{a:c} q'$ that has priority 0 in $\mathcal{B}_{i+1}$ but not in $\mathcal{B}_i'$, the $\mathsf{opt}_i$ value strictly decreases, i.e., $\mathsf{opt}_i(q) > \mathsf{opt}_i(q')$. Since $\mathsf{opt}_i$ is bounded by the size of the arena of $\mathcal{G}_2(\mathcal{B}_i)$, it follows that $\rho$ must also have infinitely many 0 priority transitions in $\mathcal{B}_i'$, and therefore is accepting. □

We have thus shown that the first subprocedure of an iteration of the normalisation step, the rank-reduction subprocedure, preserves the invariants I1 and I2. By Lemma 7.12, we conclude that in the automaton obtained after the rank-reduction procedure, all states have optimal rank 0.

## 7.4 Branch-separation and priority-reduction

We next prove that the subprocedures of branch-separation and priority-reduction involved in the normalisation procedure invariants also preserve the invariants I1 and I2.

**Branch-separation.** Let $\mathcal{B}$ be a $[0, K]$ automaton, such that the optimal ranks of all states in $\mathcal{B}$ are 0, Eve wins the 2-token game from everywhere in $\mathcal{B}$, and $\mathcal{B}$ is simulation equivalent to $\mathcal{A}$. The branch-separation procedure on automaton $\mathcal{B}$ removes the following transitions from $\mathcal{B}$, to obtain $\mathcal{C}$.

1. Transitions of priority at least 1 from right states to non-right states.

2. Transitions of priority 1 that are outgoing from non-right states.

Recall that we call a state $q$ as a right state if there is a right-branch $\beta$ and states $p, r \in \mathsf{WCR}(\mathcal{B}, q)$ such that $\mathsf{rank}(q, p, r, \beta) = 0$, and *non-right states* as states $q$ for which $\mathsf{opt}(q) = 0$, but that are not right states. Since $\mathsf{opt}(q) = 0$ for all states $q$ in $\mathcal{B}$, every state is either a right state or a non-right state.

We next show that the invariants I1 and I2 are preserved during the branch-separation procedure.

**Lemma 7.16.** *The branch separation procedure preserves the invariants I1 and I2.*

*Proof.* Let $\mathcal{B}$ be a $[0, K]$ automaton on which Eve wins the 2-token game from everywhere, and in which all states have optimal rank 0. Let $\mathcal{C}$ be the $[0, K]$ automaton obtained by the branch-separation procedure on $\mathcal{B}$. We will show that for all states $q, p, r$ that are weakly coreachable in $\mathcal{B}$, Eve wins $G2((\mathcal{C}, q); (\mathcal{B}, p), (\mathcal{B}, r))$. Due to Lemma 7.13, we conclude the proof of Lemma 7.16.

Fix an optimal strategy $\sigma_{G2}$ for Eve in $\mathcal{G}_2(\mathcal{B})$ from all triplets of weakly coreachable states (Lemma 2.9), as well as a winning positional strategy $\sigma_{G1}$ for Eve on the 1-token game from all pairs of weakly coreachable states. We will describe a winning strategy $\sigma$ for Eve in $G2((\mathcal{C}, q); (\mathcal{B}, p), (\mathcal{B}, r))$ using $\sigma_{G2}$ and $\sigma_{G1}$.

We describe $\sigma$ inductively, as rounds of G2 proceed along. Suppose that after $i$ rounds, the game is at the position $((\mathcal{C}, q_i); (\mathcal{B}, p_i), (\mathcal{B}, r_i))$. Eve will store, in her memory, a branch $\beta_i$ of $\mathcal{Z}_{[0,K]}$ and two additional tokens that are at states $s_i$ and $t_i$, such that $s_i, t_i \in \mathsf{WCR}(\mathcal{B}, q)$ and $\mathsf{rank}(q_i, s_i, t_i, \beta_i) = 0$ in $\mathcal{G}_2(\mathcal{B})$. Additionally, the branch $\beta_i$ is a right (resp. left) branch whenever $q_i$ is a right (resp. non-right) state. Since $\mathsf{opt}(q_i) = 0$ in $\mathcal{B}$, we know that such states $s_i, t_i$ and $\beta_i$ exist.

Informally, similar to the proof of Lemma 7.14, Eve's token will play the 2-token game against her memory tokens, and her memory tokens will play the 1-token game against Adam tokens.

When Adam picks a letter $a_i$, let $\delta = q_i \xrightarrow{a_i : c_i} q'$ be the transition that is given by $\sigma_{G2}$ at the position $(q_i, a_i, s_i, t_i, \beta_i)$ in $\mathcal{G}_2(\mathcal{B})$. We distinguish between the following four cases.

**Case** 1. The priority of $\delta = q_i \xrightarrow{a_i : c_i} q'$ is not 0, and $q$ is a right state.

We first observe that $q'$ must be a right state too due to the structure of $\mathcal{D}_{[0,K]}$ and $\sigma_{G2}$ being an optimal strategy. Indeed, from the position $v = (\delta, s_i, t_i, \beta_i)$ in $\mathcal{G}_2(\mathcal{A})$ that is an Adam vertex, any outgoing edge from $v$ to $(q', s', t', \beta')$ must be such that $\mathsf{rank}(q', s', t', \beta') = 0$ (Proposition 2.10), and note that if $\beta$ is a right branch, and $\beta \xrightarrow{(c_i, c', c''):d} \beta'$ is the corresponding transition in $\mathcal{D}$, then $\beta'$ is a left branch only if $d$ is 0 or 1 (Proposition 7.9). Note that $d$ cannot be 1 because $\sigma_{G2}$ is an optimal strategy, and $d$ is 0 if and only if $c_i$ is 0, which we assumed not to be the case. Thus, $q'$ is a right state, and $\beta'$ is a right branch.

It follows the transition $\delta$ has not been deleted in $\mathcal{C}$. Eve thus picks the transition $\delta$ on her token, and she updates her memory tokens at $s_i$ and $t_i$ to take the transitions on $a_i$ given by the strategy $\sigma_{G1}$ against Adam's tokens at $p_i$ and $r_i$, respectively. The memory branch $\beta_i$ is updated to $\beta_{i+1}$ in $\mathcal{D}_{[0,K]}$, according to the priorities of the transitions that Eve's tokens and Eve's memory tokens took in the 2-token game of Eve's token against Eve's memory tokens. Note that $\beta_{i+1}$ is a right branch, due to our argument in the previous paragraph.

**Case** 2. The priority of $\delta = q_i \xrightarrow{a_i : c_i} q'$ is not 0, and both $q_i$ and $q'$ are non-right states.

If $q_i$ is a non-right state, then $\delta$ cannot have priority 1 since $\sigma_{G2}$ is a rank-optimal strategy (Proposition 7.10), and thus, $\delta$ is a transition in $\mathcal{C}$. We let Eve move her token along the transition $\delta$. She updates her memory tokens at $s_i$ and $t_i$ to take transitions given by $\sigma_{G1}$ against Adam's tokens at $q_i$ and $r_i$ respectively. The memory $\beta_i$ is updated to $\beta_{i+1}$ in $\mathcal{D}_{[0,K]}$, according to the priorities of the transitions that Eve's tokens and Eve's memory tokens took in the $\mathcal{G}_2$ game between Eve's token against Eve's memory tokens.

Observe that $\beta_{i+1}$ is a left branch. The corresponding transition $\delta_D$ from $\beta_i$ to $\beta_{i+1}$ does not priority 0 since $\delta$ does not have priority 0, and since we argued that $\delta$ does not have priority 1, the transition $\delta_D$ does not have priority 1 too (Proposition 7.9). Since transitions of priority at least 2 from left states lead to left states (Proposition 7.9) in $\mathcal{D}_{[0,K]}$, we get that $\beta_{i+1}$ is a left branch.

**Case** 3. The priority of $\delta = q_i \xrightarrow{a_i : c_i} q'$ is not 0, $q_i$ is a non-right state, and $q'$ is a right state.

Since $\sigma_{G2}$ is a rank-optimal strategy, $\delta$ does not have priority 1 (Proposition 7.10). Eve thus takes the transition $\delta$ on her token to $q'$. Eve then *resets* her memory tokens to be at the state $s_{i+1}, t_{i+1}$ that are weakly coreachable to $q'$ in $\mathcal{B}$, and her memory branch to be a right branch $\beta_{i+1}$, such that $\mathsf{rank}(q', s_{i+1}, t_{i+1}, \beta_{i+1}) = 0$ in $\mathcal{G}_2(\mathcal{B})$.

**Case** 4. The priority of $\delta = q_i \xrightarrow{a_i : c_i} q'$ is 0.

Then Eve takes the transition $\delta$ on her token. Eve then *resets* her memory tokens to be at $s_{i+1}, t_{i+1} \in \mathsf{WCR}(\mathcal{B}, q')$ and her memory branch to be $\beta_{i+1}$, such that $\mathsf{rank}(q', s_{i+1}, t_{i+1}, \beta_{i+1}) = 0$ in $\mathcal{G}_2(\mathcal{B})$ and $\beta_{i+1}$ is a right (resp. left) branch if $q'$ is a right (resp. non-right) state.

Note that at the end of the round, in each of the above four cases, the configuration of the state of Eve's token, the states of Eve's memory token, and her memory branch has rank 0 in $\mathcal{G}_2(\mathcal{B})$. For moves from Case 1 and 2, this follows from the monotonicity of ranks (Proposition 2.10), and for moves from Cases 3 and 4, this is true because that is how we reset Eve's memory. Additionally, observe that if Eve's token is at a right (resp. non-right) state, then her memory branch is a right (resp. left) branch. Thus, Eve can continue playing similarly in the next round, and this concludes our inductive description of Eve's strategy.

We claim that the above strategy is winning for Eve. Indeed, if her token visits both right states and non-right states infinitely often, then since the only transitions from right states to non-right states have priority 0, Eve's run on her token contains infinitely many priority 0 transitions and hence is accepting.

Otherwise, Eve's token eventually stays only in right states, or only in non-right states. If this is the case, then eventually, Eve's memory tokens are never reset. Then eventually, Eve is choosing transitions on her token according to a $G2$ winning strategy against her memory tokens, which in turn are playing according to a $G1$ winning strategy against Adam's tokens. Thus, if any of the runs of Adam's tokens are accepting, then some suffix of the moves on the corresponding Eve's memory token constitute an accepting run, and therefore, the run on Eve's token is accepting as well. It follows that we have described a winning strategy for Eve, as desired. □

We now proceed to show that the invariants I1 and I2 are preserved during the priority-reduction subprocedure.

**Priority-reduction** We note that the removal of transitions from $\mathcal{B}$ to $\mathcal{C}$ in the branch separation procedure described above might change the fact that all states had their optimal rank as 0 in $\mathcal{B}$. We still use the terminology of right states and non-right states to denote the states in $\mathcal{C}$ that were originally right states and non-right states in $\mathcal{B}$, respectively.

Due to branch-separation, we know that there are no transitions of priority at least 1 from right states to non-right states. We do the following priority relabelling on $\mathcal{C}$ to obtain $\mathcal{A}'$: for every transition $\delta = p \xrightarrow{a:c} p'$ in $\mathcal{C}$ where $p$ is a non-right state and $c$ has priority greater than 0, change the priority of $\delta$ to be $c - 2$. Observe that $c$ cannot be 1, since we removed outgoing transitions of priority 1 from non-right states in the branch-separation procedure. We claim that relabelling priorities this way does not change the acceptance of each run.

Indeed, let $\rho$ be an infinite run. If $\rho$ eventually only stays in right states or only in non-right states, then it is clear that $\rho$ is accepting in $\mathcal{C}$ if and only if $\rho$ is accepting in $\mathcal{A}'$. Otherwise, suppose $\rho$ visits both right and non-right states infinitely often. Then, $\rho$ is accepting in $\mathcal{C}$, since any transition from a right state to non-right state must have priority 0. For the same reasoning, $\rho$ is also accepting in $\mathcal{A}'$. Thus, we conclude that a run in $\mathcal{A}'$ is accepting if and only if that same run is accepting in $\mathcal{C}$. We obtain that $\mathcal{A}'$ and $\mathcal{C}$ are simulation equivalent, and Eve wins the 2-token game from everywhere in $\mathcal{A}'$.

We have thus shown that the subprocedures of our normalisation procedure: rank-reduction, branch-separation, and priority-reduction each preserve the invariants I1 and I2.

## 7.5 Stabilisation

Observe that each of the three subprocedures of rank-reduction, branch separation, and priority-reduction either decrease the priorities of some transitions, or remove certain transitions. Thus, the normalisation procedure terminates after at most $(K+1) \times |\Delta|$ many iterations on $\mathcal{A}$, and suppose that the automaton then obtained is $\mathcal{A}^I$. Since we have shown that each of the subprocedures of the normalisation procedure preserves the invariants I1 and I2, so does the entire normalisation procedure, and thus $\mathcal{A}^I$ is simulation-equivalent to $\mathcal{A}$ and Eve wins the 2-token game from everywhere in $\mathcal{A}^I$. We will now show that $\mathcal{A}^I$ has 0-reach double-covering (Lemma 7.18), and therefore it is HD (Lemma 7.3). We start by showing that all states in $\mathcal{A}^I$ are right states.

**Lemma 7.17.** *Every state in the automaton $\mathcal{A}^I$ has optimal rank $0$ and is a right state.*

*Proof.* If the optimal rank of $q$ is not 0 for some state $q$ in $\mathcal{A}^I$, then we can run the normalisation procedure for one more step to obtain a different automaton, since the rank-reduction procedure would then modify the automaton so that all states have optimal ranks 0. But since the normalisation procedure terminates at $\mathcal{A}^I$, we deduce that the optimal ranks of all states in $\mathcal{A}^I$ is 0.

We next show that each state $q$ is a right state in $\mathcal{A}^I$. Assume, towards a contradiction, that $q$ is a non-right state for some state $q$ in $\mathcal{A}^I$. If $q$ had any outgoing transition of priority at least 1, then either the branch-separation or the priority-modification step would have changed the automaton $\mathcal{A}^I$. Thus, all transitions outgoing from $q$ have priority 0. But then it is clear that $q$ is a right-state, which is a contradiction. □

We now show that the automaton $\mathcal{A}^I$ has 0-reach double-covering.

**Lemma 7.18.** *For each state $q$ in $\mathcal{A}^I$, there is a state $p$ weakly coreachable to $q$ in $\mathcal{A}^I$ such that Eve wins $G2(q; p, p)$ in $(\mathcal{A}^I)_{>0}$.*

*Proof.* From Lemmas 7.11 and 7.17, we know that for each state $q$, there are weakly coreachable states $p_1, p_2$ such that Eve wins $G2(q; p_1, p_2)$ in $(\mathcal{A}^I)_{>0}$. We need to show that for each $q$, there is a $p$ weakly coreachable to $q$ such that Eve wins $G2(q; p, p)$ in $(\mathcal{A}^I)_{>0}$. We prove the existence of such a $p$ as follows. Construct a directed graph $G$ whose vertices are states of $\mathcal{A}^I$. For each state $q$, we add two edges $q \to p_1$ and $q \to p_2$ such that Eve wins $G2(q; p_1, p_2)$ in $(\mathcal{A}^I)_{>0}$, and $p_1$ and $p_2$ are weakly coreachable to $q$ in $\mathcal{A}^I$. We will then find a $p$ such that there are two distinct paths from $q$ to $p$ in $\mathcal{G}$, which would imply that Eve wins $G2(q; p, p)$ in $(\mathcal{A}^I)_{>0}$ due to the following claim.

**Claim 7.** *Suppose there are two distinct paths from $q$ to $p_1, p_2$. Eve then wins $G2(q; p_1, p_2)$ in $(\mathcal{A}^I)_{>0}$.*

We begin by recalling the following observations from Lemma 3.26, which we will use repeatedly to show the claim.

**Observation 0.** If Eve wins $G2(\mathcal{B}; \mathcal{C}_1, \mathcal{C}_2)$, then Eve wins $G2(\mathcal{B}; \mathcal{C}_2, \mathcal{C}_1)$.

**Observation 1.** If Eve wins $G2(\mathcal{B}; \mathcal{C}_1, \mathcal{C}_2)$, then Eve wins $G1(\mathcal{B}; \mathcal{C}_1)$ and $G1(\mathcal{B}; \mathcal{C}_2)$.

**Observation 2.** If Eve wins $G2(\mathcal{B}; \mathcal{C}_1, \mathcal{C}_2)$ and $G1(\mathcal{C}_1; \mathcal{C}_1')$ (resp. $G1(\mathcal{C}_2; \mathcal{C}_2')$), then Eve wins $G2(\mathcal{B}; \mathcal{C}_1', \mathcal{C}_2)$ (resp. $G2(\mathcal{B}; \mathcal{C}_1, \mathcal{C}_2')$).

**Observation 3.** If Eve wins $G1(\mathcal{B}; \mathcal{B}')$ and $G2(\mathcal{B}'; \mathcal{C}_1, \mathcal{C}_2)$, then Eve wins $G2(\mathcal{B}; \mathcal{C}_1, \mathcal{C}_2)$.

Let $P_1$ and $P_2$ be paths from $q$ to $p_1, p_2$ respectively, and let $P_0$ be the longest common prefix of $P_1$ and $P_2$. Let $r$ be the vertex at the end of $P_0$, and let $P_1', P_2'$ be paths, such that $P_1 = P_0 P_1'$ and $P_2 = P_0 P_2'$. From Observation 1, it is clear that if there is an edge from $q_u$ to $q_v$, then Eve wins $G1(q_u; q_v)$ in $(\mathcal{A}^I)_{>0}$. By transitivity of $G1$ (Lemma 3.25), this extends to paths, i.e., for $q$ and $r$ as above,

$$\text{Eve wins } G1(q; r) \text{ in } (\mathcal{A}^I)_{>0}. \tag{7.1}$$

The paths $P_1'$ and $P_2'$ are both distinct and start at $r$. Let $r_1, r_2$ be states, such that there are $P_1' = e_1 P_1''$ starts with the edge $e_1 = r \to r_1$ and $P_2' = e_2 P_2''$ starts with the edge $e_2 = r \to r_2$. Then,

$$\text{Eve wins } G2(r; r_1, r_2) \text{ in } (\mathcal{A}^I)_{>0}. \tag{7.2}$$

The transitivity of $G1$ across paths $P_1''$ and $P_2''$ implies Eve wins $G1(r_1; p_1)$ and $G1(r_2; p_2)$ in $(\mathcal{A}^I)_{>0}$. Combining this with Eq. (7.2) using Observation 2, we get that Eve wins $G2(r; p_1, p_2)$ in $(\mathcal{A}^I)_{>0}$. This together with the fact that Eve wins $G1(q; r)$ in $(\mathcal{A}^I)_{>0}$ (Eq. (7.1)), we get using Observation 3 that Eve wins $G2(q; p_1, p_2)$ in $(\mathcal{A}^I)_{>0}$, as desired. This concludes the proof of our claim.

Due to Claim 7, it suffices to show that for each $q$, there is a $p$, such that there are two distinct paths from $q$ to $p$ in $G$. Consider the SCC-decomposition of $G$, and consider the SCCs of $G$ that do not have an outgoing edge to another SCC—call these SCCs as end SCCs. For each state $p$ in an end SCC, there are edges from $p$ to $p_1$ and $p_2$, and there are paths from $p_1$ to $p$ and $p_2$ to $p$. Thus, each state in an end SCC has two distinct paths from itself to itself. Since any other state $q$ has a path to a vertex in an end SCC, we get the desired result. □

We have thus proved Lemma 7.6.

**Lemma 7.6.** *Let $\mathcal{A}$ be a $[0, K]$ automaton on which Eve wins the 2-token game from everywhere. Then, there is a $[0, K]$ automaton $\mathcal{A}^I$, such that $\mathcal{A}^I$ is simulation-equivalent to $\mathcal{A}$, Eve which Eve wins the 2-token game from everywhere on $\mathcal{A}^I$, and $\mathcal{A}^I$ has 0-reach double-covering.*

The proof of Theorem B now follows easily.

**Theorem B.** *Let $K > 1$ be a natural number, such that for every $[1, K]$ automaton $\mathcal{A}$, Eve wins the 2-token game on $\mathcal{A}$ if and only if $\mathcal{A}$ is HD. Then, for every $[0, K]$ automaton $\mathcal{A}$, Eve wins the 2-token game on $\mathcal{A}$ if and only if $\mathcal{A}$ is HD.*

*Proof.* We will show that $\mathcal{A}$ is a $[0, K]$ automaton on which Eve wins the 2-token game, then $\mathcal{A}$ is HD under the assumption of Hypothesis 7.1. By Theorem I, we know that $\mathcal{A}$ has a simulation-equivalent subautomaton $\mathcal{B}$ on which Eve wins the 2-token game from everywhere. From Lemma 7.6, $\mathcal{B}$ has a simulation-equivalent $[0, K]$ automaton $\mathcal{B}^I$, such that Eve wins the 2-token game from everywhere and $\mathcal{B}^I$ has 0-reach double-covering. By Lemma 7.3, $\mathcal{B}^I$ is HD. Since $\mathcal{A}$ is simulation-equivalent to $\mathcal{B}^I$, it follows from Corollary 3.11 that $\mathcal{A}$ is HD as well. □

## 7.6 The 2-Token theorem and its applications

Our two induction steps of Theorem A and Theorem B, together with the 2-token game characterisation of history-determinism on coBüchi automata (Corollary 4.3) imply that for any nondeterministic parity automaton $\mathcal{A}$, Eve wins the 2-token game on $\mathcal{A}$ if and only if $\mathcal{A}$ is history-deterministic. Recall that we had discussed the complexity of solving 2-token games on parity automata in Theorems 3.40 and 3.41. Combining these results, we get the 2-token theorem.

> **The 2-Token Theorem.** *For every nondeterministic parity automaton $\mathcal{A}$, Eve wins the 2-token game on $\mathcal{A}$ if and only if $\mathcal{A}$ is history-deterministic. Thus, the problem of deciding history-determinism is in* PTIME *for parity automata with a fixed number of priorities, and in* PSPACE *if the number of priorities is part of the input.*

More concretely, we have the following upper bound on the problem of deciding history-determinism on parity automata due to Theorem 3.40.

**Theorem 7.19.** *There is an algorithm to decide history-determinism for $[0, d]$ parity automata $\mathcal{A}$ in time*
$$d \cdot (2^{3d} \cdot \mathsf{poly}(\mathcal{A}))^{1+o(1)},$$
*where $\mathsf{poly}(\mathcal{A})$ is a fixed polynomial in the size of $\mathcal{A}$ that is the size of the 2-token game on $\mathcal{A}$.*

Boker, Kuperberg, Lehtinen, and Skrzypczak showed in 2020 that if the 2-token conjecture is true for nondeterministic parity automata, then this implies the correctness of a similar algorithm utilising 2-token games for deciding history-determinism of alternating parity automata as well [BKLS20, Page 11]. We do not expand on what history-deterministic alternating parity automata are in this thesis, but we remark that the upper bounds for the problem of checking history-determinism in nondeterministic parity automata also extends to alternating parity automata.

**Theorem 7.20.** *History-determinism for alternating parity automata with a fixed parity index can be checked in* PTIME, *and in* PSPACE *if the parity index is part of the input.*

We also note that the 2-token game characterisation of history-determinism for nondeterministic parity automata can be easily extended to nondeterministic $\omega$-regular automata.

**Theorem 7.21.** *Let $\mathcal{A}$ be an $\omega$-regular automaton. Then Eve wins the 2-token game on $\mathcal{A}$ if and only if $\mathcal{A}$ is history-deterministic.*

*Proof.* Let $\mathcal{A}$ be an $\omega$-regular automaton. If $\mathcal{A}$ is HD, then Eve wins the 2-token game on $\mathcal{A}$ by picking transitions on her token based on a winning strategy for Eve in the HD game on $\mathcal{A}$, ignoring the moves of Adam's tokens.

The harder direction is showing that if Eve wins the 2-token game on $\mathcal{A}$, then $\mathcal{A}$ is HD. Suppose that the winning condition of $\mathcal{A}$ is given by a nondeterministic Muller automaton $\mathcal{M}$. Then, we know that there is a deterministic parity automaton $\mathcal{D}$ that is language equivalent to $\mathcal{A}$.

Consider the nondeterministic parity automaton $\mathcal{P}$ that is language-equivalent to $\mathcal{A}$, obtained by 'composing' $\mathcal{D}$ with $\mathcal{A}$. More concretely, the states of $\mathcal{P}$ are pairs of states $(q, d)$, where $q$ is a state in $\mathcal{A}$ and $d$ is a state in $\mathcal{D}$. We have a transition $(q, d) \xrightarrow{a:n} (q', d')$, where $n$ is a natural number and $a \in \Sigma$, if there are transitions $q \xrightarrow{a:c} q'$ in $\mathcal{A}$ and a transition $d \xrightarrow{c:n} d'$ in $\mathcal{D}$. The initial state of $\mathcal{P}$ is the pair of initial states of $\mathcal{A}$ and $\mathcal{D}$.

We first argue that Eve wins $G2(\mathcal{P})$ using a winning strategy for Eve in the $G2(\mathcal{A})$. Indeed, note that Eve can ignore the $\mathcal{D}$ component (since it is deterministic) in $\mathcal{P}$ and choose transitions on her token in $G2(\mathcal{P})$ based on the transitions her winning strategy chooses on her token in $G2(\mathcal{A})$. By the 2-token theorem, we know that $\mathcal{P}$ is history-deterministic. A winning strategy for Eve in the HD game on $\mathcal{P}$ can then be used to construct a winning strategy for Eve in the HD game on $\mathcal{A}$, where she uses $\mathcal{D}$ as extra memory. Thus, we obtain that $\mathcal{A}$ is HD. □

The exact complexity of solving 2-token games for an $\omega$-regular automata depends on the representation of the acceptance condition (e.g., nondeterministic/ deterministic Muller automata, nondeterministic Büchi automata), but it is almost always more efficient to decide history-determinism by solving the 2-token game than via a direct approach that involves determinising the automaton.

We end this chapter with an application of 2-token theorem to the good-enough synthesis problem for when the specifications are given by nondeterministic parity automata.

### 7.6.1 Good-Enough synthesis and realisability

Recall that in the Church synthesis problem (Section 1.2.2), we have the objective of constructing a program that represents a winning strategy in an infinite game that models the interaction of a system with its environment.

In this game, in each round $l$, the environment selects an input letter $i_l$ from an input alphabet $I$, to which the program must respond with an output letter $o_l$ from an output alphabet $O$. Then, the goal of the program is to guarantee that in the limit, the word $(i_0 o_0 i_1 o_1 \dots)$ that describes the result of this interaction is in the specification $S \subseteq (I \times O)^\omega$.

One issue with Church synthesis is that a specification $S \subseteq (I \times O)^\omega$ can only be realisable if for each input sequence $w_I = i_1 i_2 i_3 \dots$ in $I^\omega$, there is an output sequence $w_O = o_1 o_2 o_3 \dots$ in $O^\omega$, such that $i_1 o_1 i_2 o_2 \dots$ is in $S$. For example, the specification of a coffee machine that requires it to produce a coffee whenever a user presses the button might turn out to be unrealisable because it is possible for users to refuse to refill the water canister. Then, no program behaviour could guarantee the satisfaction of the specification for *all* environment behaviours, and therefore we can not synthesise a coffee machine. Instead of giving up, we might like to synthesise a coffee machine that is guaranteed to work *as long as* the users are being reasonable, i.e. refill the water canister when required to do so. This is captured by a version of the Church synthesis problem, called *good-enough synthesis* [AK20], in which the program is only required to guarantee the specification for input words in the projection of $S$ onto the first component. As with Church synthesis, good-enough realisability is the problem of deciding whether there is such a program, while the good-enough synthesis problem also requires to producing a winning strategy, if one exists. A naive solution is to encode the condition on the environment into the specification, but this might make the specification exponentially larger.

The complexity of the good-enough realisability and synthesis problems from an LTL specification, as studied in [AK20], is dominated by the double-exponential determinisation procedure. However, if the specification is given directly as a deterministic parity automaton, then the complexity of the good-enough realisability problem is exactly the complexity of deciding whether a non-deterministic parity automaton is history-deterministic [BL23a, Page 23]. Our results therefore directly imply a novel procedure to decide the good-enough realisability problem for when the specification is given by a deterministic parity automaton.

**Corollary 7.22.** *Deciding whether the language of a deterministic parity automaton is good-enough realisable is in* PTIME *if the number of priorities is fixed, and in* PSPACE *otherwise.*

# Chapter 8

# When the Parity Index is not Fixed

The results of the previous chapters sum up to give a PSPACE upper bound for the problem of deciding history-determinism in parity automata (2-token theorem), when the parity index is part of the input. In this chapter, we supplement this result by proving an NP-hardness lower-bound for this problem. Along the way, we will also show NP-hardness for the problem of deciding simulation between two parity automata, and of solving token games on parity automata.

**Theorem G.** *The following problems are NP-hard.*

1. *Given a parity automaton $\mathcal{A}$, decide if $\mathcal{A}$ is HD.*

2. *Given a parity automaton $\mathcal{A}$, decide if Eve wins the 2-token game on $\mathcal{A}$.*

*The following problems are NP-complete.*

1. *Given a parity automaton $\mathcal{A}$, decide if Eve wins the 1-token game on $\mathcal{A}$.*

2. *Given parity automata $\mathcal{A}$ and $\mathcal{B}$, decide if $\mathcal{B}$ simulates $\mathcal{A}$.*

Kuperberg and Skrzypczak, in 2015, showed that deciding history-determinism for parity automata is at least as hard as the problem of finding the winner of a parity game [KS15]—a problem that can be solved in quasipolynomial time [CJK$^+$22] and is in NP ∩ coNP (and even in UP ∩ coUP [Jur98]). Theorem G is, therefore, an improvement upon this lower bound.

Our reduction for showing NP-hardness of the problem of deciding history-determinism is from solving implication games (Lemma 2.3). The reduction we use for deciding history-determinism also shows that both the problems of deciding if Eve wins 1-token games or of deciding if Eve wins 2-token games is NP-hard for parity automata.

Recall that for history-deterministic parity automata, the relation of language inclusion is equivalent to simulation (Lemma 3.8). This gives us an immediate NP upper bound for checking language inclusion of a nondeterministic parity automaton in an HD-parity automata, as was observed by Schewe [Sch20]. We show that we can do better, by showing the problem to be decidable in quasipolynomial time.

**Theorem H.** *There is an algorithm that on given input parity automaton $\mathcal{A}$ with $n_1$ states and $d_1$ priorities and an HD parity automaton $\mathcal{H}$ with $n_2$ states and $d_2$ priorities, both over the same alphabet $\Sigma$, decides whether $L(\mathcal{A}) \subseteq L(\mathcal{H})$ in time*

$$(n_1 \cdot d_1 \cdot n_2 \cdot d_2 \cdot |\Sigma|)^{\mathcal{O}(\log d_2)}.$$

The rest of this chapter is organised as follows. We start by giving a reduction from solving implication games (Lemma 2.3) to the problem of deciding simulation between two parity automata.

We then show in Section 8.2 that the problem of solving good implication games—a technical subclass of implication games—is also NP-hard. We then show that modifying our reduction from implication games to the problem of deciding simulation to take as input good implication games yields NP-hardness for the problem of checking history-determinism of parity automata as well as for the problem of deciding if Eve wins the 1-token games (resp. 2-token game) on parity automata.

Finally, in Section 8.3, we give a quasipolynomial algorithm to check whether the language of a nondeterministic parity automaton is contained in the language of a history-deterministic parity automaton. Our algorithm reduces this problem to the problem of finding the winner in a parity game.

## 8.1 NP-hardness of simulation

We call the problem of checking whether a given parity automaton simulates another parity automaton as SIMULATION.

SIMULATION: Given two parity automata $\mathcal{A}$ and $\mathcal{B}$, does $\mathcal{A}$ simulate $\mathcal{B}$?

In this section, we will show that SIMULATION is NP-complete.

**Theorem 8.1.** *Given two parity automata $\mathcal{A}$ and $\mathcal{B}$, deciding if $\mathcal{A}$ simulates $\mathcal{B}$ is NP-complete.*

Recall that SIMULATION is in NP, since simulation games can be explicitly represented as implication games (Proposition 3.4).

Recall that an implication game $\mathcal{G}$ consists of a game arena $G = (V, E)$ together with two priority functions $\pi_1 : E \to [0, d_1]$ and $\pi_2 : E \to [0, d_2]$. The winning condition $L \subseteq E^\omega$ for $\mathcal{G}$ consists of paths that satisfy the $(\pi_1 \Rightarrow \pi_2)$ condition. That is, Eve wins a play $\rho$ of $\mathcal{G}$ if and only if the following condition holds: if $\rho$ satisfies the $\pi_1$-parity condition then $\rho$ satisfies the $\pi_2$-parity condition.

We call the problem of deciding if Eve wins a given implication game as IMPLICATION GAME.

IMPLICATION GAME: Given an implication game $\mathcal{G}$, does Eve win $\mathcal{G}$?

Implication games are LOGSPACE-interreducible to 2-dimensional parity games, which are games where the winning condition is given by a disjunction of two parity conditions. This follows from the fact that the logical statements $(A \vee B)$ is equivalent to $(\neg A \Rightarrow B)$. Chatterjee, Henzinger, and Piterman, in 2007, argued that deciding if Eve wins a given 2-dimensional parity game is NP-complete, and thus, we get that the problem of IMPLICATION GAME is also NP-complete (Lemma 2.3).

In the rest of this section, we will give a LOGSPACE reduction from IMPLICATION GAME to SIMULATION.

**Overview of the reduction.**

Let $\mathcal{G}$ be a $(\pi_1 \Rightarrow \pi_2)$ implication game played on the arena $G = (V, E)$, with two priority functions $\pi_1$ and $\pi_2$. We shall construct two parity automata $\mathcal{H}$ and $\mathcal{D}$ such that $\mathcal{H}$ simulates $\mathcal{D}$ if and only if Eve wins $\mathcal{G}$. The automata $\mathcal{H}$ and $\mathcal{D}$ are over the alphabet $E \cup \{\$\}$, where $\$$ is a letter added for padding. The automaton $\mathcal{D}$ is deterministic, while the automaton $\mathcal{H}$ has nondeterminism on the letter $\$$ and contains a copy of $\mathcal{D}$.

Recall that $\mathcal{H}$ simulates $\mathcal{D}$ if and only if Eve wins the simulation game of $\mathcal{D}$ by $\mathcal{H}$, which we denote as $\mathsf{Sim}(\mathcal{H}, \mathcal{D})$ (Section 3.1). Adam, by his choice of letter in $\mathsf{Sim}(\mathcal{H}, \mathcal{D})$, will capture his moves from Adam vertices in $\mathcal{G}$. Similarly, Eve, by her choice of transitions on $\$$ in $\mathcal{H}$, which capture her moves from Eve vertices in $\mathcal{G}$. After each '$\$$-round' in $\mathsf{Sim}(\mathcal{H}, \mathcal{D})$, we require Adam to 'replay' Eve's choice as the next letter. Otherwise, Eve can take a transition to the same state as Adam (recall that $\mathcal{H}$ contains a copy of $\mathcal{D}$), from where she wins the play in $\mathsf{Sim}(\mathcal{H}, \mathcal{D})$ by copying Adam's transitions in each round from here onwards. The priorities of $\mathcal{D}$ are based on $\pi_1$, while the priorities of $\mathcal{H}$ are based on $\pi_2$. This way $\mathcal{D}$ and $\mathcal{H}$ roughly accept words that correspond to plays in $\mathcal{G}$ satisfying the $\pi_1$ and $\pi_2$-parity condition respectively.

**Remark 7.** *Unlike for automata in the rest of this thesis, the automata $\mathcal{H}$ and $\mathcal{D}$ we will describe next in this chapter are not complete, i.e., there might be a state and letter, such that there is no outgoing transition from that state on that letter. We note that we can complete them by adding a rejecting sink state, however. For convenience, we do not illustrate or describe this rejecting sink state in the following discussions. If Adam, in a play of $\mathsf{Sim}(\mathcal{H}, \mathcal{D})$, picks a letter that takes his token to this rejecting sink state, then Adam loses since then the run on his token is rejecting. Thus, we assume that Adam, in each round of $\mathsf{Sim}(\mathcal{H}, \mathcal{D})$, only picks letters on which there are outgoing transitions from the state his token is at.*

If, at the start of a round in $\mathsf{Sim}(\mathcal{H}, \mathcal{D})$, Adam's token is at a state $q_a$ in $\mathcal{D}$ and Eve's token is a state $q_e$ in $\mathcal{G}$, we say that the game is at the position $(q_a, q_e)$.

We first present our reduction on an example implication game whose sub-game consists of vertices $u, v, v', w, w'$ with edges between them as shown in Fig. 8.1. Let $d$ be the largest priority occurring in the priorities given by $\pi_1$ and $\pi_2$. For Adam's vertex $u$, we have corresponding states $u_D$ in $\mathcal{D}$ and $u_H$ in $\mathcal{H}$. An Adam move from $u$ in $\mathcal{G}$ corresponds to one round of $\mathsf{Sim}(\mathcal{H}, \mathcal{D})$ from the position $(u_D, u_H)$.

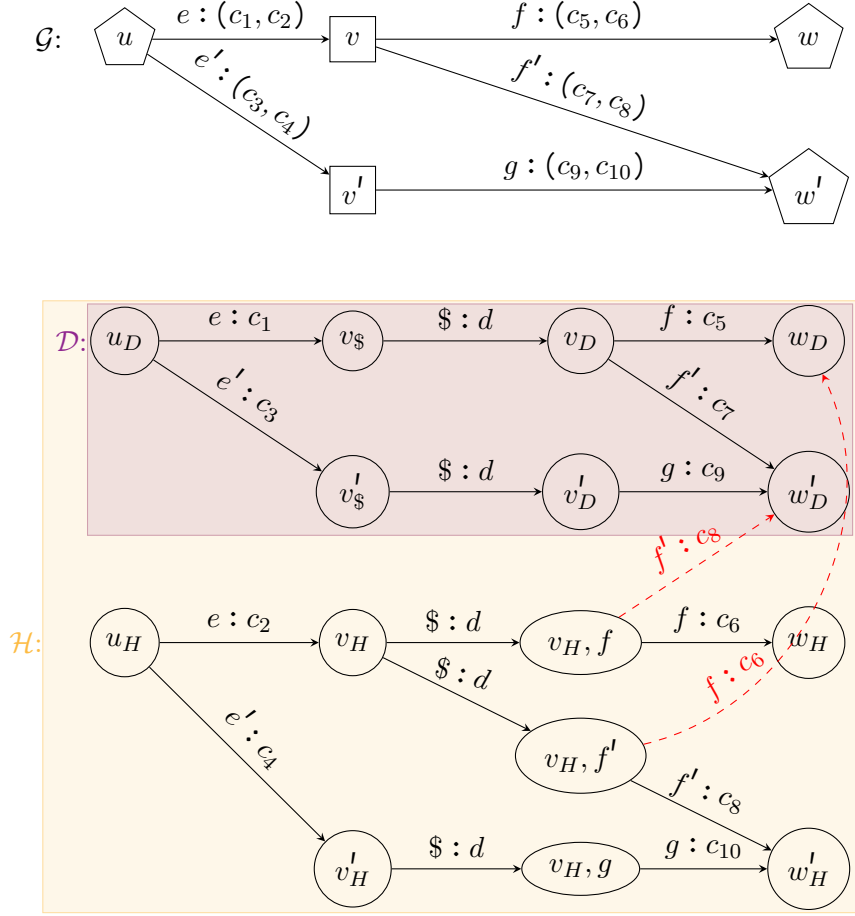

Figure 8.1: A snippet of a game $\mathcal{G}$, and the corresponding automata $\mathcal{D}$ and $\mathcal{H}$ constructed in the reduction. The Adam vertices are represented by pentagons and Eve vertices by squares. The automaton $\mathcal{D}$ is deterministic, and $\mathcal{H}$ contains a copy of $\mathcal{D}$.

In $\mathcal{G}$, Adam chooses an outgoing edge, say $e = u \xrightarrow{c_1, c_2} v$ from $u$. This corresponds to Adam choosing the letter $e$ in $\mathsf{Sim}(\mathcal{H}, \mathcal{D})$. We then have the corresponding unique transitions $u_D \xrightarrow{e:c_1} v_\$$ in $\mathcal{D}$ and $u_H \xrightarrow{e:c_2} v_H$ in $\mathcal{H}$, and hence, the simulation game is now at the position $(v_\$, v_H)$, i.e., Adam's token and Eve's token are at the states $v_\$$ and $v_H$, respectively.

An Eve move from $v$ in $\mathcal{G}$ corresponds to two rounds of $\mathsf{Sim}(\mathcal{H}, \mathcal{D})$ from $(v_\$, v_H)$. In $\mathsf{Sim}(\mathcal{H}, \mathcal{D})$, Adam selects the letter $\$$ and the unique transition $v_\$ \xrightarrow{\$:d} v_D$ in $\mathcal{D}$ on his token, where $d$ is the largest priority occurring in the priorities of $\pi_1$ and $\pi_2$. Eve must now select a transition on $\$$ from $v_H$ in her token. Suppose she picks the transition $v_H \xrightarrow{\$:d} (v_H, f)$ where $f = v \xrightarrow{c_5, c_6} w$ is an outgoing edge from $v$ in $\mathcal{G}$. This corresponds to Eve selecting the edge $f$ from her vertex $v$ in $\mathcal{G}$. The simulation game goes to the position $(v_D, (v_H, f))$.

From here, Adam may select any outgoing edge from $v$ as the letter. If he picks the edge $f' = v \xrightarrow{c_7, c_8} w'$ and the transition $v_D \xrightarrow{f':c_7} w'_D$ on his token, then Eve can pick the transition $(v_H, f) \xrightarrow{f':c_8} w'_D$ on her token and move to the same state as Adam's token: such transitions are indicated by dashed edges in Fig. 8.1. From here, Eve can win $\mathsf{Sim}(\mathcal{H}, \mathcal{D})$ by simply copying the transitions of Adam's tokens in her token. Otherwise, Adam picks the edge $f$ as the letter, same as Eve's 'choice' in the previous round. This result in the transition $v_D \xrightarrow{f:c_5} w_D$ on Adam's token in $\mathcal{D}$ and the transition $(v_H, f) \xrightarrow{f:c_6} w_H$ on Eve's token in $\mathcal{H}$. The simulation game then goes to the position $(w_D, w_H)$, from where the game continues similarly.

**The reduction.**

Let $\mathcal{G} = (V, E)$ be a $(\pi_1 \to \pi_2)$ implication game with priority functions $\pi_1$ and $\pi_2$. We assume that the arena of $\mathcal{G}$ is *bipartite*, that is, every edge in $\mathcal{G}$ is either from an Eve-vertex to an Adam-vertex, or from an Adam-vertex to an Eve-vertex. Any implication game $\mathcal{G} = (V, E)$ can be converted to a winner-equivalent bipartite implication game $\mathcal{G}'$ by adding at most $|E|$ edges: for each edge $e = u \xrightarrow{c_1, c_2} v$ such that $u$ and $v$ belong to the Eve (resp. Adam), we add an Adam-vertex (resp. Eve-vertex) $v_e$, remove the edge $e$, and add the edges $u \xrightarrow{c_1, c_2} v_e$ and $v_e \xrightarrow{c_1 : c_2} v$. It is then clear that the resulting implication game is bipartite, has the same winner as $\mathcal{G}$, and can be constructed in LOGSPACE. Thus, we assume, without loss of generality, that the implication game $\mathcal{G}$ itself is bipartite.

**Assumption 8.2.** *Every edge in $\mathcal{G}$ is from an Eve-vertex to an Adam-vertex or from an Adam-vertex to an Eve-vertex.*

We will construct two parity automata $\mathcal{D}$ and $\mathcal{H}$, such that $\mathcal{H}$ simulates $\mathcal{D}$ if and only if Eve wins $\mathcal{G}$. We encourage the reader to refer to Fig. 8.1 while reading the construction of the automata described below.

Both automata $\mathcal{D}$ and $\mathcal{H}$ are over the alphabet $\Sigma = E \cup \{\$\}$. Let $d$ be the largest priority occurring in $\mathcal{G}$. The automaton $\mathcal{D}$ is given by $\mathcal{D} = (P, \Sigma, p_0, \Delta_D)$, where the set $P$ consists of the following states.

- States $u_D$ for each Adam-vertex $u \in V_\forall$.
- States $v_\$$ and $v_D$ for each Eve-vertex $v \in V_\exists$.

The state $p_0 = \iota_D$ is the initial vertex, where $\iota$ is the initial vertex of the game $\mathcal{G}$. The set $\Delta_D$ consists of the following transitions.

- Transitions $u_D \xrightarrow{e:\pi_1(e)} v_\$$ for every edge $e = (u, v)$ in $\mathcal{G}$, such that $u \in V_\forall$ is an Adam-vertex in $\mathcal{G}$.
- Transitions $v_\$ \xrightarrow{\$:d} v_D$ for every $v \in V_\exists$ that is an Eve-vertex in $\mathcal{G}$.
- Transitions $v_D \xrightarrow{f:\pi_1(f)} w_D$ for every edge $f = (v, w)$ in $\mathcal{G}$, such that $v \in V_\exists$ is an Eve-vertex in $\mathcal{G}$.

The automaton $\mathcal{H}$ is given by $\mathcal{H} = (Q, \Sigma, q_0, \Delta_H)$, where the set $Q$ consists of the following states.

- States $u_H$ for each Adam-vertex $u \in V_\forall$.

- States $v_H$ for each Eve-vertex $v \in V_\exists$.

- States $(v_H, f)$ for each edge $f = (v, w)$ in $\mathcal{G}$ that is outgoing from an Eve-vertex $v \in V_\exists$.

- All states of $\mathcal{D}$.

The state $q_0 = \iota_H$ is the initial vertex. The set $\Delta_H$ consists of the following transitions.

- Transitions $u_H \xrightarrow{e:\pi_2(e)} v_H$ for every edge $e = (u,v)$ in $\mathcal{G}$ such that $u \in V_\forall$ is an Adam-vertex in $\mathcal{G}$.

- Transitions $v_H \xrightarrow{\$:d} (v_H, f)$ for every edge $f = (v, w)$ in $\mathcal{G}$ that is outgoing from an Eve-vertex $v \in V_\exists$.

- Transitions $(v_H, f) \xrightarrow{f:\pi_2(f)} w_H$ for every edge $f = (v, w)$ in $\mathcal{G}$ outgoing from an Eve-vertex $v \in V_\exists$.

- Transitions $(v_H, f) \xrightarrow{f':\pi_2(f')} w'_D$ for every edge $f' = (v, w') \neq f$ in $\mathcal{G}$ outgoing from an Eve-vertex $v \in V_\exists$.

- All transitions of $\mathcal{D}$.

Note that $\mathcal{H}$ contains a copy of $\mathcal{D}$.

**Correctness of the reduction.**

We now show that Eve wins the simulation game $\mathsf{Sim}(\mathcal{H}, \mathcal{D})$ if and only if Eve wins the game $\mathcal{G}$. We call a play of the simulation game *uncorrupted* if the following holds: whenever Eve's token in $\mathcal{H}$ is at $(v_H, f)$ at the start of a round of $\mathsf{Sim}(\mathcal{H}, \mathcal{D})$, Adam plays the letter $f$. If Adam plays a letter $f' \neq f$, then we call such a move *corrupted*. Any play consisting of a corrupted move is called a *corrupted play*.

It is clear that Eve wins any play of $\mathsf{Sim}(\mathcal{H}, \mathcal{D})$ that is corrupted, since a corrupted move causes Eve's token in $\mathcal{H}$ and Adam's token in $\mathcal{D}$ to be at the same state in $\mathsf{Sim}(\mathcal{H}, \mathcal{D})$. Then, the runs of Eve's and Adam's tokens from that point identical and determined by the choices of Adam's letters. In particular, the run of Eve's token is accepting if the run of Adam's token is accepting.

We observe that the following invariants are preserved throughout any uncorrupted play of $\mathsf{Sim}(\mathcal{H}, \mathcal{D})$.

> *Invariant:* At the start of any round of the simulation game $\mathsf{Sim}(\mathcal{H}, \mathcal{D})$ following an uncorrupted play:
>
> - Adam's state is at $u_D$ for some $u \in V_\forall$ if and only if Eve's state is at $u_H$
> - Adam's state is at $v_\$$ for some $v \in V_\exists$ if and only if Eve's state is at $v_H$



- Adam's state is at $v_D$ for some $v \in V_\exists$ if and only if Eve's state is at $(v_H, f)$ for some edge $f$ that is outgoing from $v$.

This invariant is easy to observe from the construction, and can be shown by a routine inductive argument.

Let us assume that the initial vertex $\iota$ of $\mathcal{G}$ is an Adam vertex; if not, we can add an Adam vertex $u$ and have an outgoing edge from $u$ to $\iota$, and make $u$ the initial vertex.

Note that then Adam, in any play of $\mathsf{Sim}(\mathcal{H}, \mathcal{D})$, constructs a word of the form

$$w = e_0 \$ f_0 e_1 \$ f_1 \ldots,$$

which we denote by $(e_i \$ f_i)_{i \geq 0}$ for succinctness, where $e_i$'s are outgoing edges from Adam's vertices while $f_i$'s are outgoing edges from Eve's vertices in $\mathcal{G}$. If the above play is an uncorrupted play of $\mathsf{Sim}(\mathcal{H}, \mathcal{D})$, then note Eve's run on $\mathcal{H}$ is uniquely determined, since the letter $f_i$ indicates how nondeterminism on $\mathcal{H}$ was resolved by Eve on the $i^{th}$ occurrence of $\$$ in $\mathsf{Sim}(\mathcal{H}, \mathcal{D})$. Thus, any uncorrupted play in the simulation can be thought of as Adam selecting the $e_i$'s and Eve selecting the $f_i$'s, resulting in the word $w = (e_i \$ f_i)_{i \geq 0}$ being constructed in the simulation game. Note that then, by construction, $(e_i f_i)_{i \geq 0}$ is a play in $\mathcal{G}$. Conversely, if $(e_i f_i)_{i \geq 0}$ is a play in $\mathcal{G}$, then there is an uncorrupted play of $\mathsf{Sim}(\mathcal{H}, \mathcal{D})$ whose word is $w = (e_i \$ f_i)_{i \geq 0}$.

Observe that the transition on each letter $e$ in $E$ in any uncorrupted play has the priorities $\pi_1(e)$ and $\pi_2(e)$ in $\mathcal{D}$ and $\mathcal{H}$ respectively, while transitions on $\$$ have priority $d$. Thus, in a uncorrupted play of $\mathsf{Sim}(\mathcal{H}, \mathcal{D})$ whose word is $(e_i \$ f_i)_{i \geq 0}$, the least priorities occurring infinitely often in the run of Adam's token in $\mathcal{D}$ and Eve's token in $\mathcal{H}$ are the same as the least $\pi_1$-priority and the least $\pi_2$-priority occurring infinitely often in the play $(e_i f_i)_{i \geq 0}$ of $\mathcal{G}$, respectively. Thus, an uncorrupted play in $\mathsf{Sim}(\mathcal{H}, \mathcal{D})$ whose word is $w = (e_i \$ f_i)_{i \geq 0}$ is winning for Eve if and only if the play $(e_i f_i)_{i \geq 0}$ in $\mathcal{G}$ is winning for Eve.

Since Eve wins any corrupted play, it follows easily that the games $\mathcal{G}$ and $\mathsf{Sim}(\mathcal{H}, \mathcal{D})$ have the same winner. If Eve has a winning strategy in $\mathcal{G}$, she can use her strategy to select transitions so that the word $w = (e_i \$ f_i)_{i \geq 0}$ that is constructed in any uncorrupted play $\rho$ of $\mathsf{Sim}(\mathcal{H}, \mathcal{D})$ corresponds to a winning play for her in $\mathcal{G}$, and hence $\rho$ is winning in $\mathsf{Sim}(\mathcal{H}, \mathcal{D})$. If Adam ever makes a corrupted move, she wins trivially. Conversely, if she has a winning strategy in $\mathsf{Sim}(\mathcal{H}, \mathcal{D})$, then she can use her strategy to choose moves in $\mathcal{G}$ so that the play $(e_i f_i)_{i \geq 0}$ corresponds to a winning uncorrupted play of $\mathsf{Sim}(\mathcal{H}, \mathcal{D})$ in which the word $(e_i \$ f_i)_{i \geq 0}$ is constructed, thus resulting in the play $(e_i f_i)_{i \geq 0}$ to also be winning for Eve.

This concludes our proof of correctness for the reduction, and hence our proof of Theorem 8.1.



## 8.2 Deciding history-determinism is NP-hard

Let us denote the problem of deciding if a given parity automaton is history-deterministic as HISTORY-DETERMINISTIC.

> HISTORY-DETERMINISTIC: Given a nondeterministic parity automaton $\mathcal{A}$, is $\mathcal{A}$ history-deterministic?

In this section, we will prove that HISTORY-DETERMINISTIC is NP-hard. We will reduce from the problem of deciding if Eve wins a restricted version of implication games, which we call good implication games. We first show in Section 8.2.1 that the problem of deciding if Eve wins a given good implication game is NP-hard. In Section 8.2.2, we use this result to show NP-hardness for the three problems of deciding if Eve wins HD games, 1-token games, and 2-token games on parity automata.

### 8.2.1 Good implication games

**Definition 8.3** (Good implication game). *A $(\pi_1 \Rightarrow \pi_2)$ implication game $\mathcal{G}$ with the priority functions $\pi_1$ and $\pi_2$ is called* good *if any play in $\mathcal{G}$ that satisfies the $\pi_2$-parity condition also satisfies $\pi_1$-parity condition.*

We call the problem of deciding whether Eve wins a good implication game as GOOD IMPLICATION GAME. Chatterjee, Henzinger, and Piterman's reduction from SAT to deciding if Eve wins 2-D parity games [CHP07] can also be seen as a reduction to GOOD IMPLICATION GAME, as we show below.

**Lemma 8.4.** *The following problem is NP-hard: given a good implication game $\mathcal{G}$, decide if Eve wins $\mathcal{G}$.*

*Proof.* We reduce from the problem of SAT. Let $\phi$ be a Boolean formula over the variables $X = \{x_1, x_2, \ldots, x_M\}$ that is a conjunction of terms $t_i$ for each $i \in [1, N]$, where each term $t_i$ is a finite disjunction of *literals*—elements of the set

$$L = \{x_1, x_2, \ldots, x_M, \neg x_1, \neg x_2, \ldots, \neg x_M\}.$$

We will construct a good implication game $\mathcal{G}_\phi$ such that Eve wins $\mathcal{G}_\phi$ if and only if $\phi$ has a satisfying assignment.

Let $T = \{t_1, t_2, \ldots, t_N\}$ be the set of all terms in $\phi$. The game $\mathcal{G}_\phi$ has the set $T \cup L$ as its set of vertices. The elements of $L$ are Adam vertices, while the elements of $T$ are Eve vertices. We set the element $x_1$ in $L$ to be the initial vertex. Each Adam-vertex $l$ in $L$ has an outgoing edge $l \to t$ to every term $t$ in $T$, and each Eve-vertex $t \in T$ has an outgoing edge $t \to l$ to a literal $l$ if $l$ is a disjunct in $t$. Thus, each play in the game $\mathcal{G}_\phi$ can be seen as Adam and Eve choosing a term and a literal in that term in alternation, respectively.

The game $\mathcal{G}_\phi$ has two priority functions $\pi_1$ and $\pi_2$, with the winning condition given by $(\pi_1 \Rightarrow \pi_2)$. To every edge $e = l \to t$ that is outgoing from an Adam vertex, both



priority functions $\pi_1$ and $\pi_2$ assign $e$ the priority $2M$, i.e., $\pi_1(e) = \pi_2(e) = 2M$. Every edge $e = t \to l$ that is outgoing from an Eve-vertex is assigned priorities as follows:

$$\pi_1(e) = \begin{cases} 2j - 2 & \text{if } l = x_j \\ 2j - 1 & \text{if } l = \neg x_j \end{cases} \qquad \pi_2(e) = \begin{cases} 2j & \text{if } l = x_j \\ 2j - 1 & \text{if } l = \neg x_j \end{cases}$$

This concludes our description of the game $\mathcal{G}_\phi$. We now show that $\mathcal{G}_\phi$ is a good implication game that Eve wins if and only if $\phi$ is satisfiable.

**$\mathcal{G}_\phi$ is a good implication game.** Let $\rho$ be a play in $\mathcal{G}_\phi$ that satisfies the $\pi_2$-parity condition. If $2c$ is the least $\pi_2$-priority occurring infinitely often in $\rho$, then by construction, $2c - 2$ is the least $\pi_1$-priority occurring infinitely often in the $\rho$. Thus, $\rho$ satisfies the $\pi_1$-parity condition.

**If $\phi$ is satisfiable, then Eve wins $\mathcal{G}_\phi$.** Let $f : \{x_1, x_2, \ldots, x_M\} \to \{\top, \bot\}$ be a satisfying assignment of $\phi$. Let $\sigma$ be a function which assigns, to each term $t_i$, a literal $l \in t_i$ that is assigned $\top$ in $f$. Consider the Eve strategy $\sigma_\exists$ in $\mathcal{G}_\phi$, where from the term $t$ she picks the edge $t \to \sigma(t)$.

We claim that $\sigma_\exists$ is a winning strategy. Indeed, let $\rho$ be a play in $\mathcal{G}_\phi$ following $\sigma_\exists$, and consider the least $i$ such that $x_i$ or $\neg x_i$ appear infinitely often in $\rho$. Since $\sigma_\exists$ is obtained from a satisfying assignment, we know that either only $x_i$ appears infinitely often, or only $\neg x_i$ appears infinitely often. In the former case, the least $\pi_2$ priority appearing infinitely often is $2i$, which is even, and hence $\rho$ is winning for Eve. In the latter case, the least $\pi_1$ priority appearing infinitely often is $2i - 1$, which is odd, and hence the $\pi_1$-parity condition is not satisfied, again implying that $\rho$ is winning for Eve.

**If Eve wins $\mathcal{G}_\phi$, then $\phi$ is satisfiable.** If Eve wins $\mathcal{G}_\phi$, she has a positional winning strategy (Lemma 2.3). Let $\sigma_\exists : T \to L$ be such a strategy, where Eve chooses the edge $t \to \sigma_\exists(t)$ at a vertex $t$. If there are no two terms $t$ and $t'$, such that $\sigma_\exists(t) = x_i$ and $\sigma_\exists(t') = \neg x_i$ for some $x_i$, then consider the assignment $\sigma$ for $\phi$ defined as follows. The assignment $\sigma$ maps all variables $x$ that are in the image of $\sigma_\exists$ to $\top$, while any term $x_j$ such that neither $x_j$ or $\neg x_j$ appear in the image of $\sigma_\exists$ is assigned $\top$. It is clear then that $\sigma$ is a satisfying assignment, since each term $t$ in $\phi$ evaluates to $\top$.

Otherwise, if there are terms $t, t'$ with $\sigma_\exists(t) = x_i$ and $\sigma_\exists(t') = \neg x_i$, then we claim that Adam wins the game $\mathcal{G}_\phi$. Adam can alternate between picking $t$ and $t'$, and then the least $\pi_1$ priority appearing infinitely often is $2i - 2$ (even) while the least $\pi_2$ priority appearing infinitely often is $2i - 1$ (odd). This implies that the play is winning for Adam, which is a contradiction to the fact that $\sigma_\exists$ is a winning strategy for Eve. □

### 8.2.2 NP-hardness for HD games and token games

We now show that deciding the history-determinism, whether Eve wins the 1-token game, and whether Eve wins the 2-token game of a given parity automaton is NP-hard. Much of



the work towards this has already been done in the reduction from IMPLICATION GAME to SIMULATION given in Section 8.1. We show that the automaton $\mathcal{H}$ that is constructed when using this reduction from a good implication game $\mathcal{G}$ is such that Eve wins $\mathcal{G}$ if and only if $\mathcal{H}$ is history-deterministic. Since GOOD IMPLICATION GAME is NP-hard (Lemma 8.4), we get that HISTORY-DETERMINISTIC is NP-hard as well.

**Theorem 8.5.** *The following problems are* NP*-hard:*

1. *Given a parity automaton $\mathcal{A}$, is $\mathcal{A}$ history-deterministic?*

2. *Given a parity automaton $\mathcal{A}$, does Eve win the 2-token game of $\mathcal{A}$?*

*Additionally, the following problem is* NP*-complete: Given a parity automaton $\mathcal{A}$, does Eve win the 1-token game of $\mathcal{A}$?*

*Proof.* Recall that deciding if Eve wins the 1-token game on a given parity automaton is in NP (Proposition 3.17). We thus focus on showing NP-hardness for the problems mentioned above.

Let us consider a good implication game $\mathcal{G}$. Recall the construction of the automata $\mathcal{H}$ and $\mathcal{D}$ in Section 8.1, which is such that Eve wins $\mathcal{G}$ if and only if $\mathcal{H}$ simulates $\mathcal{D}$. We will show that if $\mathcal{G}$ is a good implication game, then the following statements are equivalent.

1. Eve wins $\mathcal{G}$.

2. $\mathcal{H}$ simulates $\mathcal{D}$.

3. $\mathcal{H}$ is history-deterministic.

4. Eve wins the 1-token game on $\mathcal{H}$.

5. Eve wins the 2-token game on $\mathcal{H}$.

The equivalence of 1 with 3, 4, and 5 would then conclude the proof. The equivalence of 1 and 2 has already been shown in the proof of Theorem 8.1.

(2) $\iff$ (3). Let $\Sigma = E \cup \{\$\}$, and for $j = 1, 2$, consider the languages $L_j$ over $\Sigma$ consisting of the words $(e_i \$ f_i)_{i \geq 0}$ such that $(e_i f_i)_{i \geq 0}$ is a play in $\mathcal{G}$ that satisfies $\pi_j$-parity condition. By construction, we know $L(\mathcal{D}) = L_1$, and $L(\mathcal{H}) = L_1 \cup L_2$. Furthermore, since $\mathcal{G}$ is good, we know that $L_1 \supseteq L_2$ and hence $L(\mathcal{D}) = L(\mathcal{H})$. Observe that by construction, $\mathcal{D}$ is deterministic. Then, it follows from Corollary 3.10 that if $\mathcal{H}$ simulates $\mathcal{D}$, then $\mathcal{H}$ is HD. Conversely, from Lemma 3.8, we obtain that if $\mathcal{H}$ is HD, then $\mathcal{H}$ simulates $\mathcal{D}$.

(3) $\iff$ (5). This follows from the 2-token theorem.

(2) $\iff$ (4). If $\mathcal{H}$ simulates $\mathcal{D}$, then $\mathcal{H}$ is HD (due to equivalence of (2) and (3)) and hence Eve wins the 1-token game on $\mathcal{H}$ (Lemma 3.21). We will show that if Adam wins the simulation game of $\mathcal{D}$ by $\mathcal{H}$, then Adam wins the 1-token game on $\mathcal{H}$.

Fix a winning strategy $\sigma_\forall$ of Adam in the simulation game of $\mathcal{D}$ by $\mathcal{H}$. We now describe how Adam can win the 1-token game of $\mathcal{H}$ by using $\sigma_\forall$. At a high level, Adam will exploit the nondeterminism on \$ to ensure his token eventually moves to $\mathcal{D}$ and traces



out an accepting run, while picking letters according to $\sigma_\forall$ and ensuring Eve's run on her token is rejecting.

In more detail, Adam, in $G1(\mathcal{H})$, will keep in his memory a token in $\mathcal{D}$, and he will pick the letters and transitions in the 1-token game of $\mathcal{H}$ using his winning strategy in $\mathsf{Sim}(\mathcal{H}, \mathcal{D})$ as if Eve's token in is taking transitions in $\mathcal{H}$ and his memory token is taking transitions in $\mathcal{D}$. Note that since $\mathcal{D}$ is deterministic, Adam's transitions on his memory token in $\mathsf{Sim}(\mathcal{H}, \mathcal{D})$ depends solely on Adam's choice of letters. For Adam's token in $G1(\mathcal{H})$, he copies Eve's transitions on her token until Eve hasn't had to resolve nondeterminism, i.e., there has been a unique transition from the state of Eve's token on Adam's letter in every round so far. Eventually, however, there must be a round where Eve has a nondeterministic choice on her token; otherwise, since $L(\mathcal{D}) = L(\mathcal{H})$, Eve would construct an accepting run on her token if the word is accepting.

Thus, let Eve's token in $G1(\mathcal{H})$ be at the state $q$ when she needed to resolve the nondeterminism for the first time in $\mathsf{Sim}(\mathcal{H}, \mathcal{D})$. Then, Adam's token is also at $q$, and by construction, we know that $q = v_H$ for some $v \in V_\exists$. In the corresponding play in $\mathsf{Sim}(\mathcal{H}, \mathcal{D})$, Adam's memory token is at $v_\$$. Now, Adam selects the letter $, and his memory token in the simulation game is then moved to $v_D$. Eve, on her token, then picks a transition $v_H \xrightarrow{\$:0} (v_H, f)$ where $f = v \to w$ is an outgoing edge from $v$. Adam will then pick a transition $v_H \xrightarrow{\$:0} (v_H, f')$ for some edge $f' \neq f$ on his token.

In the next round, Adam must pick the letter $f$ according to $\sigma_\forall$, or Adam's memory token and Eve's token in $\mathsf{Sim}(\mathcal{H}, \mathcal{D})$ would both be the same and in $\mathcal{D}$, causing Eve to win $\mathsf{Sim}(\mathcal{H}, \mathcal{D})$, which is a contradiction since $\sigma_\forall$ is a winning strategy in $\mathsf{Sim}(\mathcal{H}, \mathcal{D})$. Thus, Adam picks the letter $f = v \to w$, and Adam's memory token in $\mathsf{Sim}(\mathcal{H}, \mathcal{D})$ moves to the state $w_D$. In the 1-token game, Eve's token moves to $w_H$ while Adam's token moves to $w_D$—same as the state of memory token in $\mathsf{Sim}(\mathcal{H}, \mathcal{D})$. From here, Adam can continue choosing letters according to $\sigma_\forall$, while the transitions on his token on $G1(\mathcal{H})$ are uniquely determined. Since $\sigma_\forall$ is a winning strategy, Eve's run on her token in $G1(\mathcal{H})$ is rejecting, while Adam's run in $\mathcal{D}$ on her memory token in $\mathsf{Sim}(\mathcal{H}, \mathcal{D})$ is accepting. Since the run of Adam's token eventually coincides with the run of his memory token, the run on his token in the $G1(\mathcal{H})$ is accepting as well, and hence Adam wins the 1-token game on $\mathcal{H}$.

We have thus shown that the statements 1, 2, 3, 4, and 5 are all equivalent, as desired. □

Theorems 8.1 and 8.5 together prove Theorem G.

**Theorem G.** *The following problems are* NP*-hard.*

1. *Given a parity automaton $\mathcal{A}$, decide if $\mathcal{A}$ is HD.*

2. *Given a parity automaton $\mathcal{A}$, decide if Eve wins the 2-token game on $\mathcal{A}$.*

*The following problems are* NP*-complete.*

1. *Given a parity automaton $\mathcal{A}$, decide if Eve wins the 1-token game on $\mathcal{A}$.*

2. *Given parity automata $\mathcal{A}$ and $\mathcal{B}$, decide if $\mathcal{B}$ simulates $\mathcal{A}$.*

## 8.3 Language containment

In this section, we will consider the following problem.

> HD-AUTOMATON CONTAINMENT: Given two parity automata $\mathcal{A}$ and $\mathcal{B}$ such that $\mathcal{B}$ is history-deterministic, is $L(\mathcal{A}) \subseteq L(\mathcal{B})$?

While the problem of checking language inclusion between two nondeterministic parity automata is PSPACE-complete (regardless of whether the parity index is fixed or not) [KV98], the same for deterministic parity automata is NL-complete [AF20, Theorem 5]. For history-deterministic parity automata, recall that language inclusion reduces to deciding simulation (Lemma 3.8). This shows that HD-AUTOMATON CONTAINMENT is in NP. We improve upon this upper bound by reducing the problem to solving parity games, which yields a quasipolynomial time algorithm.

**Theorem H.** *There is an algorithm that on given input parity automaton $\mathcal{A}$ with $n_1$ states and $d_1$ priorities and an HD parity automaton $\mathcal{H}$ with $n_2$ states and $d_2$ priorities, both over the same alphabet $\Sigma$, decides whether $L(\mathcal{A}) \subseteq L(\mathcal{H})$ in time*

$$(n_1 \cdot d_1 \cdot n_2 \cdot d_2 \cdot |\Sigma|)^{\mathcal{O}(\log d_2)}.$$

Towards proving Theorem H, let us fix a nondeterministic parity automaton $\mathcal{A}$ and a history-deterministic parity automaton $\mathcal{H}$ over the alphabet $\Sigma$ throughout the rest of this section, for which we want to decide if $L(\mathcal{A}) \subseteq L(\mathcal{H})$. Suppose that $\mathcal{A}$ has $n_1$ states and $d_1$ priorities, and that $\mathcal{H}$ has $n_2$ states and priorities in $[i, i + d_2]$ for some $i = 0$ or $1$.

It is well known that every such parity automaton $\mathcal{A}$ can be converted efficiently to a language-equivalent nondeterministic Büchi automaton $\mathcal{B}$ that has at most $(n_1 \cdot d_1)$ states [Cho74]. Then, to decide if $L(\mathcal{B}) \subseteq L(\mathcal{H})$, it suffices to check if Eve wins $\mathsf{Sim}(\mathcal{H}, \mathcal{B})$ (Lemma 3.8).

The game $\mathsf{Sim}(\mathcal{H}, \mathcal{B})$ can be explicitly represented as a $(\pi_1 \Rightarrow \pi_2)$ implication game $\mathcal{G}$ with $\mathcal{O}(n_1 \cdot d_1 \cdot n_2 \cdot |\Sigma|)$ many vertices that has the priority functions $\pi_1 : V \to [0, 1]$ and $\pi_2 : V \to [i, i + d_2]$, where $V$ is the set of vertices of $\mathcal{G}$.

Equivalently, the game $\mathcal{G}$ can also be represented as a Muller game with the same number of vertices and the Muller condition $(C, \mathcal{F})$, where $C = [0, 1] \times [i, i + d_2]$ and the set of accepting subsets $\mathcal{F}$ consists of sets $S \subseteq C$, such that if the least priority in the first components of the elements of $S$ is even then the least priority in the second components of the elements of $S$ is even.

We denote the Zielonka tree (Section 2.4) of this Muller condition as $\mathcal{Y}_{[i, i+d_2]}$. We next bound the size of $\mathcal{Y}_{[i, i+d_2]}$.

**Lemma 8.6.** *The Zielonka tree $\mathcal{Y}_{[i, i+d]}$ has height $d$. It has $1 + \lceil \frac{d}{2} \rceil$ many leaves if $i = 1$, and $1 + \lfloor \frac{d}{2} \rfloor$ many leaves if $i = 0$.*

*Proof.* We prove the lemma by induction on $d$. When $d$ is 1, the tree $\mathcal{Y}_{[i, i+d]}$ for $i = 0, 1$ is as shown below in Fig. 8.2, and the induction hypothesis is clearly satisfied.

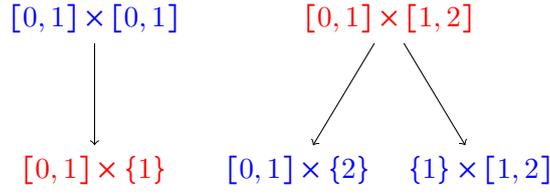

Figure 8.2: On the left, the tree $Y_{[0,1]}$; and on the right, the tree $Y_{[1,2]}$.

Suppose $d \geq 2$, and suppose that the lemma holds for all $d' \leq d$. We distinguish between the cases of when $i$ is 0 or 1.

If $i = 0$, then the root of $\mathcal{Y}_{[0,d]}$ is labelled by the set $[0,1] \times [0,d]$, which is in $\mathcal{F}$. The only maximal set which is a proper subset of $[0,1] \times [0,d]$ and not in $\mathcal{F}$ is $[0,1] \times [1,d]$. Thus, the child $c$ of the root is labelled $[0,1] \times [1,d]$ and $c$ is then the root of the tree $\mathcal{Y}_{[1,d]}$ itself (see Fig. 8.3). By induction hypothesis, the tree $\mathcal{Y}_{[1,d]}$ has height $(d-1)$ and $\lceil \frac{d-1}{2} \rceil$ leaves, and hence, the tree $\mathcal{Y}_{[0,d]}$ has height $(d-1) + 1 = d$ and $\lceil \frac{d-1}{2} \rceil = \lfloor \frac{d}{2} \rfloor$ many leaves, as desired.

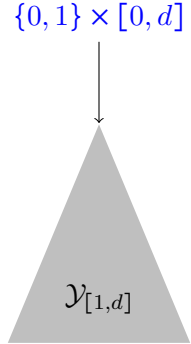

Figure 8.3: Zielonka tree $\mathcal{Y}_{[0,d]}$

If $i = 1$, then the root of $\mathcal{Y}_{[1,1+d]}$ is labelled by the set $[0,1] \times [1, 1+d]$, which is not in $\mathcal{F}$. There are two maximal proper subsets of $[0,1] \times [1, 1+d]$ that are in $\mathcal{F}$: the set $[0,1] \times [2, 1+d]$ and the set $\{1\} \times [1, 1+d]$, which then has no proper subsets that are not in $\mathcal{F}$.

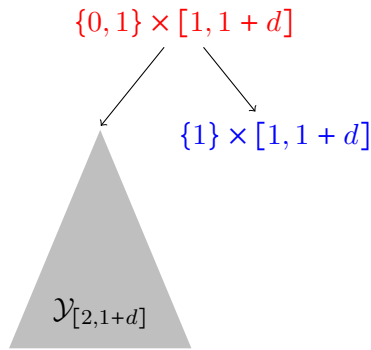

Figure 8.4: Zielonka tree $\mathcal{Y}_{[1,1+d]}$

Accordingly, the root of $\mathcal{Y}_{[1,1+d]}$ has two children, one labelled by $\{0,1\} \times [2, 1+d]$

that is then the the root of the tree $\mathcal{Y}_{[2,1+d]}$, and another labelled $\{1\} \times [1, 1 + d]$, as shown in Fig. 8.4.

The tree $\mathcal{Y}_{[2,1+d]}$ has the same size as $\mathcal{Y}_{[0,d-1]}$ and has height $(d-1)$ and $\lfloor \frac{d-1}{2} \rfloor$ leaves by the induction hypothesis. Therefore, the tree $\mathcal{Y}_{[1,1+d]}$ has height $d$ and $1 + \lfloor \frac{d-1}{2} \rfloor = \lceil \frac{d}{2} \rceil$ leaves, as desired. □

The game $\mathsf{Sim}(\mathcal{H}, \mathcal{B})$ can be converted to a winner-equivalent parity game $\mathcal{G}'$ by taking the product of the game with the deterministic parity automaton $\mathcal{D}_{(C,\mathcal{F})}$ corresponding to the Zielonka tree $\mathcal{Y}_{[i,i+d_2]}$ (see Section 2.4 for details). The automaton $\mathcal{D}_{(C,\mathcal{F})}$ has as many states as number of leaves in $\mathcal{Y}_{[i,i+d_2]}$, and the number of priorities in $\mathcal{D}_{(C,\mathcal{F})}$ is the same as the height of $\mathcal{Y}_{[i,i+d_2]}$. Thus, $\mathcal{G}'$ has at most

$$(n_1 \cdot d_1 \cdot n_2 \cdot |\Sigma| \cdot \lceil \tfrac{d_2}{2} \rceil)$$

many vertices and $d_2 + 1$ priorities. Since parity games with $n$ vertices and $d$ priorities can be solved in time $n^{\mathcal{O}(\log d)}$ [CJK$^+$22], Theorem H follows.

**Theorem H.** *There is an algorithm that on given input parity automaton $\mathcal{A}$ with $n_1$ states and $d_1$ priorities and an HD parity automaton $\mathcal{H}$ with $n_2$ states and $d_2$ priorities, both over the same alphabet $\Sigma$, decides whether $L(\mathcal{A}) \subseteq L(\mathcal{H})$ in time*

$$(n_1 \cdot d_1 \cdot n_2 \cdot d_2 \cdot |\Sigma|)^{\mathcal{O}(\log d_2)}.$$

# Chapter 9

# Open Problems

We end with a discussion on the problems that remain open, focusing on problems that might now be possible to attack due to this thesis.

**Recognising history-deterministic parity automata.** The most major problem that our thesis leaves open is the complexity of deciding history-determinism of parity automata when the parity index is part of the input. The 2-token theorem only gives us a PSPACE upper bound, and we showed a NP-hardness lower bound in Chapter 8.

**Open Problem 1.** *What is the complexity of deciding history-determinism for nondeterministic parity automata whose parity index is not fixed?*

We conjecture that this problem is NP-complete. More specifically, we conjecture that for every history-deterministic automaton $\mathcal{H}$, there is another equivalent history-deterministic automaton $\mathcal{H}'$ with at most as many states as $\mathcal{H}$, which satisfies certain properties that makes verification of history-determinism of $\mathcal{H}'$ possible in NP. An NP-upper bound to check the history-determinism of $\mathcal{H}$ would then involve guessing such an $\mathcal{H}'$ and the polynomial time certificate required to verify history-determinism of $\mathcal{H}'$, as well as positional strategies in simulation games of $\mathcal{H}$ by $\mathcal{H}'$ and of $\mathcal{H}'$ by $\mathcal{H}$.

**Conjecture 1.** *The problem of checking history-determinism for parity automata is NP-complete.*

Our work is a step towards such an approach, and in general, towards understanding history-deterministic parity automata. Indeed, our proofs shows that every $[0, K]$ (resp. $[1, K]$) history-deterministic automaton $\mathcal{A}$ has a normal which in which certain states are history-deterministic in $\mathcal{A}_{>0}$ (resp. $\mathcal{A}_{>1}$). We think that understanding and strengthening these normal forms could be the key to establishing NP-completeness for the problem of recognising history-deterministic parity automata.

**Determinisation of history-deterministic Büchi automata.** In Chapter 5, we gave a PTIME algorithm to determinise HD Büchi automata that involves a quadratic blowup in the number of states. However, we do not know whether this is tight. In fact, it is still open if HD Büchi automata can be strictly more succinct than determinstic Büchi automata.

**Open Problem 2.** *Is there a history-deterministic Büchi automaton that is strictly more succinct than any language-equivalent deterministic Büchi automaton?*

We conjecture that history-deterministic Büchi automata are no more succinct than deterministic Büchi automata. More precisely, we conjecture the following.

**Conjecture 2.** *Let $\mathcal{A}$ be a history-deterministic Büchi automaton that has reach-covering. Then, there is a language-equivalent deterministic Büchi automaton $\mathcal{D}$ consisting of the reach-deterministic states of $\mathcal{A}$, (a subset of) priority $1$ transitions that are outgoing from reach-deterministic states, and where priority $0$ transitions in $\mathcal{A}$ from reach-deterministic states are deterministically redirected to some other reach-deterministic state.*

The crux in proving this conjecture is in coming up with the right method for the redirection of priority 0 transitions in Conjecture 2. On the other hand, we think proving that such a redirection is not possible for some counterexample will likely lead to the construction of a HD Büchi automaton that is strictly more succinct than every language-equivalent deterministic Büchi automaton.

**Minimisation of history-deterministic parity automata.** While HD coBüchi automata (with transition-based acceptance) can be minimised in PTIME [AK22], we do not know if the same is true for history-deterministic parity automata, or even history-deterministic Büchi automata.

**Open Problem 3.** *Can history-deterministic parity automata be minimised in PTIME? What about history-deterministic Büchi automata?*

The proof of Abu Radi and Kupferman for the minimisation of history-deterministic coBüchi automata introduces a canonical form for HD coBüchi automata [AK22], that comes from a normal form of HD coBüchi automata that was proposed by Kuperberg and Skrzypczak in 2015 [KS15].

Accordingly, we think that the crux of Open Problem 3 lies in strengthening the normal forms that we have introduced (0-reach double-covering and 1-safe double-coverage). We are also hopeful that if we obtain a positive answer to Open Problem 3 for HD Büchi automata, then we can extend it to HD parity automata, by using techniques similar to our two induction steps in Chapters 6 and 7 for the 2-token theorem.

**Synthesising strategies in HD games.** While the 2-token theorem gives us an algorithm to solve the good-enough realisability problem for when the specification is given by a deterministic parity automata, we note that the 2-token theorem does not solve the good-enough synthesis problem, since this requires us to construct a winning strategy for Eve in the HD games on parity automata. Winning strategies for Eve in the HD game on parity automata require exponential memory in general, with an exponential lower bound already holding for HD coBüchi automata [KS15].

We can, nevertheless, ask whether it is possible to efficiently synthesise an on-the-fly strategy for Eve in HD games in that picks the next transition in PTIME, while maintaining a memory that is polynomial in the size of the automaton. We can answer this question positively for coBüchi, Büchi, and $[0,2]$ automata using our proof of the 2-token theorem, but we run into a hurdle for $[1,3]$ automata. Recall that in our even-to-odd induction step in Chapter 6, we used explorability as an intermediate step (Lemma 6.12) to construct a winning strategy for Eve in the HD game. Thus, the procedure from which we construct the strategy for Eve already takes exponential time for $[1,3]$ automata, and for $[1,2d]$ automata, takes time that is a tower of $d$ exponents. Therefore, obtaining a way of constructing strategies for Eve in the HD game which avoids this intermediate step of explorability is the hurdle behind having reasonably efficient algorithms to synthesize strategies for Eve in the HD games.

**Constructing history-deterministic parity automata.** While HD parity automata satisfy several good algorithmic properties, there are currently no (non brute-force) algorithms to construct history-deterministic parity automata that are not determinisable-by-pruning. This is a crucial missing link to make HD parity automata a viable tool in practice. We note that we have such an algorithm to construct HD coBüchi automata due to Abu Radi and Kupferman [AK22] for language that are recognised by coBüchi automata, however. Extending this to HD parity automata is a widely open problem. The structural insights we have gained on HD parity automata in our proof of the 2-token theorem might be the first steps towards such a construction.

**Open Problem 4.** *Are there algorithms which can construct history-deterministic parity automata that are not determinisable-by-pruning for a given ω-regular language?*

**Recognising history-deterministic labelled transition systems.** We showed in this thesis that Eve winning the 2-token game on a parity automaton is equivalent to that automaton being history-deterministic. We can ask whether the same holds for infinite labelled transition systems with parity acceptance conditions.

The previous proofs of the 2-token theorem for Büchi automata [BK18] and coBüchi automata [BKLS20] used finite-memory determinacy for Adam in HD games, so it was not possible to use their proofs to extend the result to Büchi or coBüchi labelled transition systems. Our proofs overcome this difficulty since they are constructive, i.e., starting with a strategy for Eve in the 2-token game on a parity automaton, we construct a strategy for Eve in the HD game on that automaton. However, we also use finiteness of states in Lemmas 5.6, 6.9 and 7.5. Proving these results without using finiteness of states is therefore the current hurdle in extending the 2-token theorem to infinite state systems with parity acceptance conditions. Or, in the case the 2-token theorem does not extend to infinite state systems, we think that this hurdle provides insights on how to build a LTS on which Eve wins the 2-token game but that is not HD.

**Open Problem 5.** *Does the 2-token theorem extend to labelled transition systems with parity acceptance conditions?*

# Alphabetical Index